\def\be{\begin{equation}}
\def\ee{\end{equation}}
\def\bea{\begin{eqnarray}}
\def\eea{\end{eqnarray}}
\def\bi{\begin{itemize}}
\def\ei{\end{itemize}}

\documentclass[]{tADP2e}

\begin{document}


\markboth{Dynamics of a Quantum Phase Transition and Relaxation to a Steady State}{Advances in Physics}

\articletype{Review}

\title{ Dynamics of a Quantum Phase Transition and Relaxation to a Steady State}

\author{ Jacek Dziarmaga$^{\rm a}$$^{\ast}$
         \thanks{$^\ast$ Email: dziarmaga@th.if.uj.edu.pl \vspace{6pt}} 
         \\\vspace{6pt} 
         $^{\rm a}${\em{
         Instytut Fizyki im. Mariana Smoluchowskiego and
         Mark Kac Complex Systems Research Centre, 
         Uniwersytet Jagiello\'nski, ul. Reymonta 4, 
         PL-30-059 Krak\'ow, Poland
         }}
         \\\vspace{6pt}\received{May 26, 2010}
}

\maketitle

\begin{abstract}

We review recent theoretical work on two closely related issues: excitation of an isolated 
quantum condensed matter system driven adiabatically across a continuous quantum phase 
transition or a gapless phase, and apparent relaxation of an excited system after a sudden 
quench of a parameter in its Hamiltonian. Accordingly the review is divided into two parts. The first part 
revolves around a quantum version of the Kibble-Zurek mechanism including also phenomena that go 
beyond this simple paradigm. What they have in common is that excitation of a gapless many-body 
system scales with a power of the driving rate. The second part attempts a systematic presentation 
of recent results and conjectures on apparent relaxation of a pure state of an isolated quantum many-body
system after its excitation by a sudden quench. This research is motivated in part by recent 
experimental developments in the physics of ultracold atoms with potential applications in the 
adiabatic quantum state preparation and quantum computation.

\bigskip


\bigskip

\newpage

\centerline{\bfseries Contents}\medskip
\hbox to \textwidth{\hsize\textwidth\vbox{\hsize48pc

\hspace*{-12pt} {1.}  Introduction\\
{2.}  Dynamics of a quantum phase transition\\
\hspace*{10pt} {2.1.}  Introduction\\
\hspace*{10pt} {2.2.}  Kibble-Zurek mechanism (KZM) in a classical phase transition\\
\hspace*{10pt} {2.3.}  KZM in a quantum phase transition \\
\hspace*{10pt} {2.4.}  KZM: adiabatic transition in a finite system \\
\hspace*{10pt} {2.5.}  KZM in a non-linear quench \\
\hspace*{10pt} {2.6.}  KZM: from a critical point into a gapped phase \\
\hspace*{10pt} {2.7.}  KZM in space \\
\hspace*{10pt} {2.8.}  KZM and dynamics of an inhomogeneous phase transition \\
\hspace*{10pt} {2.9.}  The Landau-Zener (LZ) model \\
\hspace*{10pt} {2.10.}  The LZ model in the adiabatic-impulse approximation \\
\hspace*{10pt} {2.11.}  KZM as a set of independent Landau-Zener transitions \\
\hspace*{10pt} {2.12.}  KZM from adiabatic perturbation theory \\
\hspace*{10pt} {2.13.}  Quantum Ising chain: transition across an isolated critical point\\
\hspace*{24pt} {2.13.1} Landau-Zener argument \\
\hspace*{24pt} {2.13.2} Transition in a finite chain \\
\hspace*{24pt} {2.13.3} Exact solution of the time-dependent Bogoliubov-de Gennes equations \\
\hspace*{24pt} {2.13.4} Entropy of a block of spins in the final state \\
\hspace*{24pt} {2.13.5} Impurity of the final state \\
\hspace*{24pt} {2.13.6} Correlation functions in the final state \\
\hspace*{24pt} {2.13.7} Fidelity between the final state and the final ground state\\
\hspace*{24pt} {2.13.8} Geometric phase of the final state \\
\hspace*{24pt} {2.13.9} Generalised entanglement in the final state \\
\hspace*{24pt} {2.13.10} Linear quench to $t\to\infty$: quantum dephasing after KZM\\
\hspace*{24pt} {2.13.11} Transition in space \\
\hspace*{24pt} {2.13.12} Inhomogeneous transition \\
\hspace*{10pt} {2.14.}  Quench across a multicritical point of the XY chain\\
\hspace*{10pt} {2.15.}  Kitaev model in 2D: quench across a gapless phase\\
\hspace*{10pt} {2.16.}  The random Ising chain: logarithmic dependence of excitation density 
                        on transition rate\\
\hspace*{10pt} {2.17.}  Topological insulators: anomalous excitation of edge modes\\
\hspace*{10pt} {2.18.}  The Lipkin-Meshkov-Glick model: KZM and infinite coordination number\\
\hspace*{10pt} {2.19.}  The Bose-Hubbard model: transition between gapped Mott insulator
                        and gapless superfluid \\
\hspace*{24pt} {2.19.1} Slow transition from Mott insulator to superfluid\\
\hspace*{24pt} {2.19.2} Fast transition from Mott insulator to superfluid at small density\\
\hspace*{24pt} {2.19.3} Fast transition from Mott insulator to superfluid at large density \\
\hspace*{24pt} {2.19.4} Fast transition from superfluid to Mott insulator at large density \\
\hspace*{10pt} {2.20.}  Loading a 1D Bose gas into and optical lattice: transition into the
                        gapped phase of the sine-Gordon model \\
\hspace*{10pt} {2.21.}  Spin-1 Bose-Einstein condensate: transition from the paramagnetic to the
                        ferromagnetic phase \\
\hspace*{10pt} {2.22.}  Adiabatic sweep across a gapless regime \\
\hspace*{10pt} {2.23.}  Many-particle Landau-Zener problem: adiabatic passage across a Feshbach resonance\\
{3.}  Apparent relaxation of an isolated quantum system after a sudden quench \\
\hspace*{10pt} {3.1.}   Introduction \\
\hspace*{10pt} {3.2.}   Quasiparticle light cone effect in dephasing after a sudden quench\\
\hspace*{10pt} {3.3.}   Generalised Gibbs ensemble (GGE) for a quadratic Hamiltonian\\
\hspace*{10pt} {3.4.}   Local relaxation to GGE with a quadratic Hamiltonian \\
\hspace*{10pt} {3.5.}   GGE and the transverse Ising chain \\
\hspace*{10pt} {3.6.}   GGE and the Luttinger model \\
\hspace*{10pt} {3.7.}   GGE and hard core bosons \\
\hspace*{10pt} {3.8.}   GGE and the Bose-Hubbard model \\
\hspace*{10pt} {3.9.}   GGE and a system solvable by Bethe ansatz \\
\hspace*{10pt} {3.10.}   Eigenstate thermalisation hypothesis (ETH) \\
\hspace*{10pt} {3.11.}   Dynamics of relaxation to a steady state \\
\hspace*{10pt} {3.12.}   Summary \\
{3.}  Conclusion \\
}}
\end{abstract}

\newpage

\section{ Introduction }

Experiments with ultracold atoms provide unprecedented opportunities to emulate precisely 
tailored model Hamiltonians at temperatures close to zero \cite{BodzioReview}. These model 
quantum systems are not only well isolated from their environment, but also their parameters 
can be changed in time in a prescribed way 
\cite{Greiner,Kinoshita,NewtonCradle,Paredes,superexchange,boshier,SGExp,IsingIon}. 
Consequently, the emerging field of non-equilibrium quantum systems driven out of their initial 
ground state by a time-dependent Hamiltonian is no longer of purely academic interest. Quite to 
the contrary, it becomes essential for such practical applications as adiabatic quantum state 
preparation for quantum simulation \cite{BodzioReview,iontraps,IsingIon,IsingNMR,QA,Kasevich} and, 
in particular, adiabatic quantum computation \cite{AdiabaticQC}. In this article we review recent 
theoretical work on 
two closely related issues: excitation of a condensed matter system driven across a continuous 
quantum phase transition or a gapless phase, and apparent relaxation of an excited system towards 
thermal equilibrium or a non-thermal steady state. 

The former issue had been intensively studied both theoretically and experimentally in different
systems at finite temperature including cosmological phase transitions, liquid crystals, superfluid 
helium, convection cells, superconductors, and Bose-Einstein condensation, where the excitation is 
described by the Kibble-Zurek mechanism (KZM). In this review we collected recent evidence that, 
although in many situations many details are different than in the finite temperature KZM, its key 
ingredients -- which are crossovers between adiabatic and non-adiabatic stages of time evolution -- 
remain the same in the zero temperature quantum limit. Consequently, the excitation of a system in 
general scales with a power of a transition rate. We also include examples where the KZM paradigm is 
not so predictive, like when driving the system across a gapless bosonic phase, even though the 
excitation is still a power of the driving rate, and an example of a disordered system where it is 
the KZM that predicts a logarithmic dependence on the rate instead of the usual power law. Finally, 
we mention an example where the evolution is essentially non-adiabatic, i.e., the excitation density 
diverges with increasing system size.   

Once a system got excited the latter issue becomes important: does an excited state of an isolated 
quantum system relax to any steady state? A quick ``global'' answer is ``no'', because unitary 
evolution cannot evolve an initial pure state into a mixed state. However, this does not exclude some 
sort of apparent {\it local} relaxation, where expectation values of local observables relax to become 
the same as in a global mixed state. With this distinction in mind, we can further ask what is the nature
of the global steady state? In this article we briefly review recent work on integrable and 
non-integrable quantum systems. General arguments together with exactly solvable examples suggest
that non-interacting quadratic bosonic or fermionic Hamiltonians appear to relax {\it locally} 
to a non-thermal generalised Gibbs ensemble (GGE). The ensemble is strongly constrained by integrals 
of motion which is this case are the numbers of Bogoliubov quasiparticles. In case of non-integrable
systems the evidence is less conclusive. When an excited state of a many-body system after a sudden 
quench has a narrow energy distribution then, according to the conjectured ``eigenstate thermalisation 
hypothesis'' (ETH), for simple few-body observables the state appears to relax to a microcanonical ensemble.    
This thermalisation mechanism is qualitatively different than the ergodicity required of classical 
non-integrable systems.  

Even though in practice the two issues cannot be always sharply ``disentangled'', because the 
relaxation begins already during the excitation of a system, this review is divided into two 
corresponding major parts in Sections \ref{DynamicsQPT} and \ref{Relaxation}.

\section{ Dynamics of a quantum phase transition }
\label{DynamicsQPT}

\subsection{ Introduction }

A quantum phase transition is a fundamental change in the ground state of a quantum 
system when one of parameters in its Hamiltonian is driven across a critical 
point \cite{Sachdev}. A continuous (or second order) quantum phase transition can often 
be characterised by an energy gap between the ground state and the first excited state
or, more generally, a relevant energy scale which vanishes at the critical point in the 
thermodynamic limit of infinite system size. 
Consequently, no matter how slowly the parameter is driven the quantum state of the system 
cannot follow adiabatically the instantaneous ground state near the critical point. It is 
the aim of this part to quantify the level of this inevitable excitation in different 
systems.

This research is motivated in part by adiabatic quantum computation \cite{AdiabaticQC},
where one would like to prepare a system in a simple ground state of an initial Hamiltonian
$H_i$, and then drive the system adiabatically to a final Hamiltonian $H_f$, whose ground state 
is a solution to a non-trivial computational problem. Unfortunately, since $H_i$ and $H_f$
are qualitatively different, they are generally separated by a critical point, where the
driving cannot be adiabatic in the limit of infinite system size. Consequently, there
is an upper limit to the size of the system (or number of qubits) which can cross the critical 
point adiabatically at a given transition rate and, consequently, the minimal necessary time of 
``adiabatic computation'' increases with the number of qubits. The same problem can arise 
in the context of quantum simulators \cite{Feynman,QA}, where the key idea 
is to emulate a model condensed matter Hamiltonian with, say, ultracold atoms in an optical 
lattice potential \cite{BodzioReview,Kasevich,SGExp}, ions in an ion trap \cite{iontraps,IsingIon}
or NMR molecules \cite{IsingNMR}. In this context, one usually prepares the system in the 
uncorrelated ground state of a simple initial Hamiltonian $H_i$, like e.g. a Bose-Einstein 
condensate, and then drives it into a correlated state by a slow change of a parameter
in the Hamiltonian \cite{Greiner,superexchange,IsingIon,IsingNMR,SGExp}.  

A quantum phase transition across an isolated quantum critical point between two gapped phases
turns out to be well described by a quantum version of the Kibble-Zurek mechanism (KZM). The essence
of the mechanism is an adiabatic-impulse-adiabatic approximation, where evolution of
a system driven slowly across a phase transition is assumed adiabatic in the gapped phases 
sufficiently far from criticality, and impulse in a close neighbourhood of the gapless 
critical point, where the state of the system does not change in this approximation. Consequently, 
the state after the transition can be argued to have a finite correlation length
$
\hat\xi~\sim~\tau_Q^{\frac{\nu}{1+\nu z}}~,
$
where $\tau_Q$ is a characteristic time of the adiabatic transition between the two phases
and $\nu,z$ are the critical exponents. This universal scale of length determines density of 
excitations and other physical quantities. For example, the density of excitations or excitation 
energy above the ground state scales as an inverse power of the transition time $\tau_Q$.

However, the world of quantum critical phenomena is rich: phase diagrams have gapless lines
and gapless phases, and a system driven across a gapless line or phase also gets excited. 
It turns out that in many of such non-standard situations the adiabatic-impulse-adiabatic 
approximation still applies and is able to predict the correct scaling of the density of 
excitations (or density of excitation energy) with the transition time, but there are also 
systems where this approximation is not practical or not predictive enough. These include some 
bosonic systems, which are non-adiabatic in the thermodynamic limit, i.e., their density of 
excitations diverges with increasing system size. Nevertheless, the adiabatic-impulse-adiabatic 
approximation is the leading motif of Section \ref{DynamicsQPT}.

Section \ref{DynamicsQPT} is organised as follows. In Section \ref{KZMclassical} we briefly 
review KZM in the historical context of finite temperature classical phase transitions, and 
then in Section \ref{KZargument} we generalise KZM to a zero temperature quantum phase transition
between two gapped phases. Once KZM is reformulated in this most ``canonical'' set-up its key 
ingredient, which is the adiabatic-impulse-adiabatic approximation, can be bend and twisted in many 
different ways as required by actual application. Some general examples are described in 
Sections \ref{KZAdiab}, \ref{halfKZ}, \ref{KZinspace}, \ref{KZinhom}, and \ref{non-linear}. 
In Section \ref{KZAdiab} we drive a finite system across the critical value of the parameter
in its Hamiltonian and estimate how slow the driving rate has to be for the transition to
be adiabatic. In Section \ref{halfKZ} we drive an infinite system into a gapped phase starting exactly at 
a critical point. In Section \ref{KZinspace} a quantum phase transition takes place not in time 
but in space, i.e., the parameter in the Hamiltonian driving the transition is time-independent 
but inhomogeneous in space, so that different parts of the system are in different phases. 
In Section \ref{KZinhom} the phase transition is driven in an inhomogeneous way such that some 
parts of the system cross the critical point earlier than the other. The inhomogeneous transition 
turns out to be a way to suppress excitations at the critical point. Another way to minimise 
excitations is a transition with a non-linear time-dependence of the driving parameter considered 
in Section \ref{non-linear}. 

After this incomplete review of non-standard generalisations, we return to the basics and in Sections
\ref{LZvKZ},\ref{LZvKZ2},\ref{LZargument} investigate relations between KZM and the Landau-Zener (LZ) model of 
level anti-crossing \cite{LZ}. We begin in Section \ref{LZvKZ} by a review of the standard LZ model 
together with more exotic transitions that begin or end at the anti-crossing centre. In Section \ref{LZvKZ2} 
we apply the adiabatic-impulse-adiabatic approximation, which is central to KZM, to obtain approximate 
solutions of the LZ model in some limits of parameters. Finally, in Section \ref{KZargument} we consider
a general quadratic fermionic Hamiltonian which can be mapped to sets of independent LZ transitions. 
This class of integrable models provides exact solutions supporting KZM or its generalisations. In the
following Section \ref{argumentPolkovnikov} we rederive KZM within the time-dependent perturbation 
theory. This perturbative treatment reproduces approximately the exact results in the integrable models, 
but it does not seem limited to the quadratic Hamiltonians. Section \ref{argumentPolkovnikov} completes 
the general part of Section \ref{DynamicsQPT}, and the following Sections describe applications of the 
general ideas to a number of specific integrable and non-integrable models.

The list of examples is opened in Section \ref{KZIsing} by one of the cornerstones of the theory
of quantum phase transitions -- the integrable quantum Ising chain in transverse magnetic field. 
It has a quantum phase transition between two gapped phases: paramagnetic and ferromagnetic.
Since the model can be mapped via the Jordan-Wigner transformation \cite{JW} to a quadratic 
fermionic Hamiltonian, virtually all the general ideas mentioned above are illustrated
in the subsections of Section \ref{KZIsing}. This is why the quantum Ising chain 
takes relatively large portion of this article.

Once the review of the Ising chain is completed, in Sections \ref{subsectionmulticritical}
and \ref{quenchKitaev} we review another two quadratic fermionic systems: the XY spin chain  
and the 2D Kitaev model. They provide exactly solvable examples where KZM requires careful 
generalisation to obtain correct scaling of the density of excitations with the 
transition rate. The next example in Section \ref{RandomIsing} is the random quantum 
Ising chain, where the density of excitations turns out to be a logarithmic function of the 
transition rate. There is no usual KZ power-law scaling in this model, but the logarithmic 
dependence can still be obtained from the adiabatic-impulse-adiabatic approximation like in KZM. 
This example is followed by topological insulators in Section \ref{Topological}, where the edge 
zero modes that must exist in a topological insulator have a different scaling of the 
excitation probability than in the bulk of the same insulator. In Section \ref{quenchLMG} the quantum 
KZM is extended to the Lipkin-Meshkov-Glick model with infinite coordination number, and in 
Section \ref{KZinBH} to the non-integrable Bose-Hubbard model. 

The last model is important because it is another cornerstone of the theory of quantum 
phase transitions and it was realised experimentally in Refs. \cite{Greiner,GreinerPRL,Kasevich}.
The 1D version of the model has the Berezinski-Kosterlitz-Thouless transition between the
gapped Mott insulator and gapless superfluid. Since the model is non-integrable, we review 
different approximate results for transitions between the two phases. Another closely
related model is considered in Section \ref{sinusGordon}, where a quasi-1D gas of bosonic
atoms is loaded into an optical lattice potential. The gas is a gapless (critical) Luttinger 
liquid for which a weak optical lattice potential is a relevant perturbation leading to
an effective sine-Gordon model with a finite gap in its excitation spectrum. This is an example of 
a transition starting at a critical point and going into a gapped phase introduced
in Section \ref{halfKZ}. In Section \ref{FerroKZ} we review mean-field treatment of the transition 
from a paramagnetic to a ferromagnetic phase in a spin-1 Bose-Einstein condensate. 

Finally, in Section \ref{sweep} we consider adiabatic sweep across a gapless 
phase in a quadratic model of harmonic oscillators, and in Section \ref{manyLZ} a semiclassical 
treatment of a many-body Landau-Zener model. The 1D gapless harmonic model turns out to be  
non-adiabatic in a sense that its density of excitations increases with the system size. The 
many-body LZ model is in turn an example, where the density of LZ anti-crossings is so large that 
they cannot be treated as independent transitions, so the whole problem has to be and can be 
treated by semiclassical methods. A summary of Section \ref{DynamicsQPT} is made in Section 
\ref{Summary1}.

\subsection{ KZM in a classical phase transition  }
\label{KZMclassical}

Phase transition is a fundamental change in the state of a system when one 
of its parameters passes through the critical point. In a second order phase 
transition, the fundamental change is continuous and the critical point is 
characterised by divergent correlation length and relaxation 
time. This critical slowing down implies that no matter how slowly a system 
is driven across the transition, its evolution cannot be adiabatic close to 
the critical point. As a result, ordering of the state after a transition
from a disordered symmetric phase to an ordered broken symmetry phase   
is not perfect: the state is a mosaic of ordered domains whose finite size $\hat\xi$ 
depends on the rate of the transition. This scenario was first described in 
the cosmological context by Kibble \cite{K} who appealed to relativistic 
causality to set the size of the domains. The dynamical mechanism relevant for 
second order phase transitions was proposed by Zurek \cite{Z} and it is briefly
as follows.

A dimensionless distance from the critical point at a finite temperature $T_c$ can
be defined as
\be
\epsilon~=~\frac{T-T_c}{T_c}~.
\label{epsilonT}
\ee
In the initial symmetric phase above $T_c$ there are finite thermal fluctuations of
the order parameter with diverging correlation length  
\be
\xi~\sim~|\epsilon|^{-\nu}~
\label{xifiniteT}
\ee
and relaxation time
\be
\tau~\sim~\xi^z~
\label{tauT}
\ee
describing reaction time of the system to external perturbations. This divergent 
relaxation time becomes infinitely slow at the critical point. 

\begin{figure}[t]
\begin{center}
\includegraphics[width=0.8\columnwidth,clip=true]{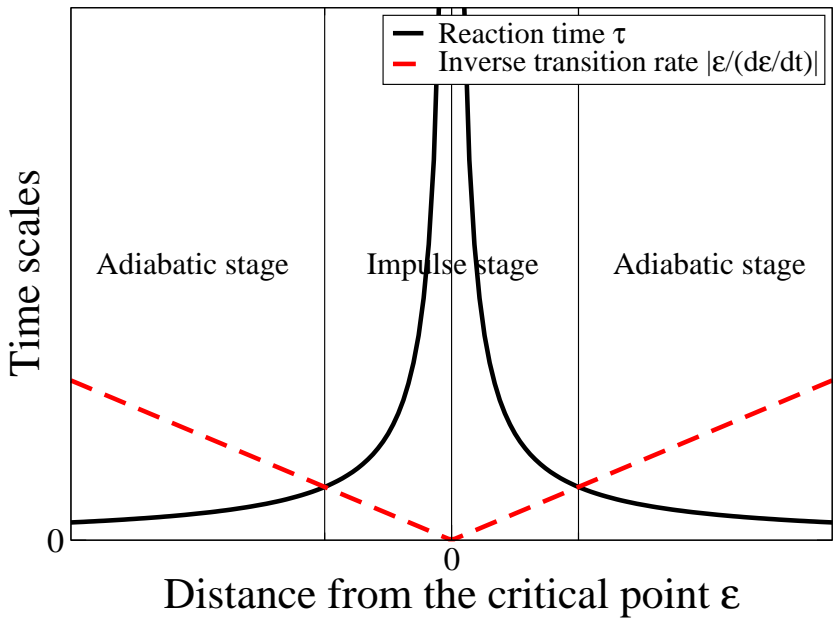}
\caption{ 
The reaction time $\tau$ and the inverse transition rate 
$\left|\epsilon/\dot\epsilon\right|$ as a function of dimensionless distance
$\epsilon$ from the critical point. The reaction time equals the inverse transition
rate at $+\hat\epsilon$ and $-\hat\epsilon$. These two points mark crossovers between
the impulse and adiabatic stages of time evolution.
}
\label{FigKZM}
\end{center}
\end{figure}

The system can be driven below $T_c$ either directly by cooling, or by increasing $T_c$
above a constant $T$ by varying another parameter such as pressure \cite{He4a,He4b}. 
In any case, there is a time-dependent $\epsilon(t)$ running continuously from 
an initial $\epsilon_i>0$ to a final $\epsilon_f<0$. Near the critical point it can be 
linearised as
\be
\epsilon(t)~\approx~-\frac{t}{\tau_Q}~,
\ee
where the coefficient $\tau_Q$ can be identified as a ``quench time''. This linearised form 
implies a relative transition rate 
$
\left|
\epsilon^{-1}
\frac{d\epsilon}{dt}
\right|=
t^{-1}
$
which diverges near the critical point. The evolution of the system with a time-dependent
$\epsilon(t)$ can be divided into three stages, see Fig. \ref{FigKZM}. Initially the transition 
rate is slower than the relaxation time $\tau$ and the system adiabatically follows the instantaneous 
state of thermal equilibrium for current $\epsilon(t)$. This adiabatic stage lasts until
\begin{equation}
\hat{\epsilon}~
\sim~
\tau_{Q}^{-\frac{1}{1+\nu z}}
\label{hatepsilonT}
\end{equation}
when the transition rate $t^{-1}$ equals the instantaneous relaxation rate $\tau^{-1}$ of the system
and long wavelength fluctuations begin to go out of equilibrium.
After $\hat\epsilon$ reactions of the system are too sluggish to follow the varying $\epsilon(t)$ 
and, in a first impulse approximation, the state does not change until $-\hat\epsilon$ when 
the reactions of the system become faster than the transition rate again. In this way
the system arrives at $-\hat\epsilon$, which is already in the symmetry broken phase, 
still remaining in the state of thermal equilibrium at the $+\hat\epsilon$ in the symmetric phase, where 
there are small thermal fluctuations of the order parameter with a finite correlation length
\be
\hat\xi~\sim~\hat\epsilon^{-\nu}~\sim~\tau_Q^{\frac{\nu}{1+\nu z}}~.
\label{hatxiT} 
\ee
These small fluctuations are the initial state for the last adiabatic stage of the evolution
after $-\hat\epsilon$.

Around $-\hat\epsilon$ the system ``realises'' at last that zero order parameter is no longer its state
of equilibrium. The proper equilibrium state has a finite order parameter whose eventual
variations in space can be characterised by a finite healing length $|-\hat\epsilon|^{\nu z}$, 
which is incidentally the same as the correlation length $\hat\xi$ (but see Ref. \cite{Rivers}). 
Near $-\hat\epsilon$ the small fluctuations of the order parameter with wave lengths longer than the 
healing length $\hat\xi$ blow up quasi-exponentially in time until magnitude of the order parameter 
becomes comparable to its finite equilibrium value. At this point the order parameter becomes a mosaic 
of ordered domains whose average size is set by $\hat\xi$. The orientation of the order parameter is
approximately constant in each domain, but uncorrelated between different domains. 
 
In the finite temperature context the focus was on topological defects. When a vacuum manifold
of the order parameter has a non-trivial homotopy group, then there are stable topological defects.
Their density after a transition is set by the size $\hat\xi$ of the correlated domains \cite{K}.
For example, the density of point-like monopoles in 3D is $\simeq\hat\xi^{-3}$, and vortices in 2D 
is $\simeq\hat\xi^{-2}$. Topologically stable defects are a robust and relatively 
easy to detect imprint of the non-equilibrium phase transition. However, 
most of them are not permanent because, for instance, pairs of vortices with opposite winding number 
attract each other and annihilate. This equilibration process, known as phase ordering 
kinetics \cite{Bray}, leads to gradual coarse-graining of the order parameter on increasing scale of 
length. What eventually remains is a net winding number of the initial mosaic. This is why in the finite 
temperature experiments one has to prevent defects from phase ordering by, e.g., stabilizing them 
in a rotating cryostat \cite{He3}, or trapping them in a superconducting ring \cite{lowTc}.   
 
KZM for classical phase transitions was confirmed by numerical simulations of 
the time-dependent Ginzburg-Landau model \cite{KZnum} and successfully tested by experiments 
in a wide range of condensed matter systems including liquid crystals \cite{LC}, superfluid 
$^3$He \cite{He3}, both high-$T_c$ \cite{highTc} and low-$T_c$ \cite{lowTc} superconductors, 
non-equilibrium convection systems \cite{ne}, and Bose-Einstein condensation driven by evaporative
cooling \cite{Anderson}. With the exception of superfluid $^4$He - where the early detection of defect 
formation \cite{He4a} was subsequently attributed to vorticity introduced by stirring \cite{He4b}, 
and the situation remains unclear - experimental results are consistent with KZM, but more 
quantitative experimental tests are needed to verify e.g. the KZ scaling of defect density with 
transition rate.

KZM is a universal theory of the dynamics of continuous symmetry-breaking phase transitions whose 
applications range from the low temperature Bose-Einstein condensation to the ultra high temperature 
transitions in the grand unified theories of high energy physics. However, the zero temperature quantum 
limit remained unexplored until recently and quantum phase transitions are in many respects qualitatively 
different from transitions at finite temperature. Most importantly time evolution is unitary, so 
there is no damping, and there are no thermal fluctuations to initialise the symmetry breaking. 

In the next Section, following Ref. \cite{KZIsing}, we rederive KZM at zero temperature where the 
scalings turn out to be formally the same but the underlying physics is different.

\subsection{ KZM in a quantum phase transition } 
\label{KZargument}

Near an isolated quantum critical point between two gapped phases both the reaction
time $\tau$ and the correlation (or healing) length $\xi$ diverge as
\bea
\tau &\sim& |\epsilon|^{-\nu z}~, \label{tau} \\
\xi  &\sim& |\epsilon|^{-\nu}~,   \label{xi}
\eea
where $\epsilon$ is a dimensionless parameter which measures distance from
the critical point. For instance, when the transition is driven by varying 
a parameter $g$ in the Hamiltonian across a critical point at a finite $g_c$, 
then 
\be
\epsilon ~=~ \frac{g-g_c}{g_c} ~.
\label{epsilong}
\ee  
The reaction time determines how fast the system can react to external 
perturbations and the healing length sets the scale on which the order 
parameter heals in space, i.e., returns to its ground state value. In a 
quantum phase transition the reaction time is set by an inverse of a
gap $\Delta$ between the ground state and the first relevant excited 
state
\be
\tau~\simeq~\Delta^{-1}~,
\label{tauDelta}
\ee 
because this is the shortest time scale on which the ground state can adjust adiabatically 
to a varying $\epsilon$. Near the critical point the gap vanishes as
\be
\Delta~\sim~|\epsilon|^{\nu z}~ 
\label{Delta}
\ee
implying that the evolution across the critical point cannot be adiabatic.

We consider a second order quantum phase transition that is crossed at a finite 
rate set by the quench timescale $\tau_Q$:
\be
\epsilon(t) ~=~ -\frac{t}{\tau_Q} ~. 
\label{tauQ}
\ee
In general, $\epsilon(t)$ does not need to be a linear function of $t$, but 
here we assume the generic case when $\epsilon(t)$ can be {\it linearised} near 
the critical point as
\be
\epsilon(t) ~=~ \frac{d\epsilon}{dt}(0) ~ t ~ + ~ {\cal O}(t^2) ~.
\label{linearise}
\ee
The quench time in Eq. (\ref{tauQ}) can be identified as $\tau_Q=|\frac{d\epsilon}{dt}(0)|^{-1}$. 
A more general, but not quite generic, non-linear quench which cannot be linearised near $\epsilon=0$ 
is considered in Section \ref{non-linear}.

Initially, at $t\to-\infty$, the system is prepared in the ground state. As long as 
the reaction time in Eq. (\ref{tau}) is fast enough or, equivalently, the gap in 
Eq. (\ref{Delta}) is large enough, the state of the system follows its adiabatic ground 
state. The adiabaticity fails near an instant $t=-\hat t$ when the transition rate 
$|\dot\epsilon/\epsilon|=1/|t|$ equals the gap $\Delta\sim|\epsilon|^{\nu z}=|t/\tau_Q|^{\nu z}$ 
or
\be
\hat t ~\sim~ \tau_Q^{\frac{\nu z}{1+\nu z}} ~,
\label{hatt}
\ee
From this time on long wavelength modes cease to be adiabatic.
In a first approximation, after $-\hat t$ the evolution becomes impulse, i.e., the 
state effectively freezes out between $-\hat t$ and $+\hat t$ when the reaction 
time $\tau$ is too slow for the system to follow the evolving parameter $\epsilon(t)$. 
The adiabatic evolution of the state restarts again at $+\hat t$ when the gap becomes 
less than the transition rate, see Fig. \ref{FigKZM}.

Near the freeze-out time $-\hat t$, corresponding to 
\be
\hat\epsilon ~\sim~ \tau_Q^{-\frac{1}{1+\nu z}} ~,
\label{hatepsilon}
\ee
the state is still in the instantaneous ground state with correlation length
\be
\hat\xi ~\sim~ \hat\epsilon^{-\nu} ~\sim~ \tau_Q^\frac{\nu}{1+\nu z}~.
\label{hatxi}
\ee 
In the adiabatic-impulse-adiabatic approximation, this state does not change between 
$-\hat t$ and $\hat t$. When the adiabatic evolution restarts near $\hat t$, 
corresponding to $-\hat\epsilon$, the ground state frozen at $\hat\epsilon$ becomes 
an initial excited state for the last adiabatic stage of the evolution. 

In this approximation, the impulse stage of the linear quench in Eq. (\ref{tauQ}) is equivalent 
to a sudden quench from $\hat\epsilon$ to $-\hat\epsilon$, where the initial ground state at 
$\hat\epsilon$ becomes an initial excited state for the following adiabatic evolution after 
$-\hat\epsilon$.  

Note that when the quench time $\tau_Q$ is large, then $\hat\epsilon$ in Eq. (\ref{hatepsilon}) 
is small and the linearisation in Eq. (\ref{linearise}) is self-consistent because all the 
non-trivial KZM physics happens in the narrow interval between $\hat\epsilon$ and $-\hat\epsilon$
which is very close to the critical point.

In the adiabatic-impulse-adiabatic approximation the state after the transition at $-\hat\epsilon$
is equal to the ground state of the system at $+\hat\epsilon$. In the adiabatic limit of large $\tau_Q$
this ground state is very close to the critical point so, by the usual scaling hypothesis of the
renormalisation group theory, expectation value of an operator $O$ in this state is proportional to a power
of the diverging correlation length $\hat\xi$ in the ground state at $\hat\epsilon$. Moreover,
since $\hat\xi$ itself scales with the transition time, see Eq. (\ref{hatxi}), the expectation 
value $\langle O\rangle$ also scales with a power of the transition time $\tau_Q$. For example, 
if the phase after the transition admits quasiparticle excitations, then their density scales like
\be
n_{\rm ex} ~\simeq~ \hat\xi^{-d} ~\sim~ \tau_Q^{-\frac{d\nu}{1+\nu z}} ~,
\label{nex}
\ee
where $d$ is the number of space dimensions. The same scale $\hat\xi$ determines a range of 
correlations and other physical quantities. 

In the same way as $\hat\xi$ is the universal scale of length, the freeze-out time $\hat t$ in 
Eq. (\ref{hatt}) is the universal scale of time. For instance, the density of excitations during 
the transition depends on time as
\be
n_{\rm ex}(t) ~\simeq~ 
\tau_Q^{-\frac{d\nu}{1+\nu z}} ~ 
F\left(t/\hat t\right) ~,
\ee
where $F$ is a non-universal scaling function. Examples can be found in 
Refs. \cite{Bishop,Meisner,Viola2} and in Section \ref{KZinBH}.

KZM predicts a correlation length $\hat\xi$ and a characteristic timescale $\hat t$. 
These are the basic quantities from which one can derive many other physical observables 
simply by dimensional analysis. This conjecture is confirmed by a series of examples, 
notably the Ising model, where quantities such as density of excitations, excitation 
energy, correlation functions (often equivalent by a Fourier transform to the momentum 
distribution that is usually measured in the time of flight experiments), residual 
magnetization, entropy of entanglement, generalized entanglement, Berry phase, penetration 
depth in a transition in space, or threshold velocity in an inhomogeneous transition 
are all constructed out of $\hat\xi$ and/or $\hat t$. These are the basic building blocks 
from which one can construct other physical quantities. 

In this Section we presented KZM in its most ``canonical'' form. However, the following 
Sections provide examples how the adiabatic-impulse-adiabatic approximation, which is the 
essence of KZM, can be bend and twisted in many different ways as required by actual 
application.

\subsection{ KZM: adiabatic transition in a finite system } 
\label{KZAdiab}

The discussion in the previous Section assumes an infinite system
where the correlation length (\ref{hatxi}) can diverge to infinity
as the transition time becomes infinitely slow. However, in a large but finite system of 
linear size $L$ the divergence must be terminated when 
$\hat\xi\simeq L$, or equivalently
\be
\tau_Q~\sim~L^{\frac{1+\nu z}{\nu}}~.
\label{tauQadiab}
\ee
For slower transitions the scaling (\ref{hatxi}) breaks down and the transition becomes adiabatic. 
This adiabatic regime is a consequence of a non-zero gap $\Delta_c$ at the critical 
point of a finite system. When the transition is slow enough, then the finite gap suppresses any 
excitation exponentially. This effect was discussed in Ref. \cite{KZIsing} and it is illustrated by 
an exact solution in the quantum Ising chain, see Section \ref{FiniteIsing} and Refs. \cite{Dziarmaga2005,Cincio}. 
A similar crossover to an adiabatic regime, which can be characterized by a steeper scaling of 
$n_{\rm ex}$ with $\tau_Q$ instead of an exponential decay, was discussed in 
Ref. \cite{infinitezimalPolkovnikov} in case of a ``half-quench'' considered in Section \ref{halfKZ} below.

\subsection{ KZM in a non-linear quench } \label{non-linear}

The adiabatic-impulse-adiabatic approximation of Section \ref{KZargument} can be 
also applied to the non-linear quench considered in Refs. 
\cite{nonlinSengupta,nonlinPolkovnikov}, where
\be
\epsilon(t) ~\approx~ -{\rm sign}(t) ~ \left| \frac{t}{\tau_Q} \right|^r ~
\label{nonlinearepsilon}
\ee
near the critical point $\epsilon=0$. This function cannot be linearised as in Eq.
(\ref{linearise}) but, nevertheless, essentially the same argument can be applied 
here as in the linear case. 

Indeed, the transition rate $|\dot\epsilon/\epsilon|=r/|t|$ equals the gap 
$\Delta\sim|\epsilon|^{\nu z}$
at $\hat\epsilon\sim(r/\tau_Q)^{r/(1+r\nu z)}$ corresponding to
the KZ correlation length and density of excitations
\bea
\hat\xi     ~\sim~ \tau_Q^{\frac{r\nu}{1+r\nu z}}~,~~
n_{\rm ex} ~\simeq~ \hat\xi^{-d} ~\sim~ \tau_Q^{-\frac{r\nu d}{1+r\nu z}} ~
\label{nexnonlin}
\eea
respectively. These equations reduce to the corresponding Eqs. (\ref{hatxi},\ref{nex}) for 
the linear quench (\ref{tauQ}) where $r=1$.

Equation (\ref{nexnonlin}) shows that the linear quench is not the best choice if 
we want to minimize $n_{\rm ex}$ for a given transition time $\tau_Q$, but it is better  
to take a non-linear exponent $r\gg 1/\nu z$ such that $n_{\rm ex}\sim\tau_Q^{-d/z}$. This density is 
less than the density after the linear quench by a factor 
\be
\frac{n_{\rm ex}(r\gg 1/\nu z)}{n_{\rm ex}(r=1)} ~\sim~
\tau_Q^{-\frac{d}{1+\nu z}}  ~
\ee
which tends to zero for large $\tau_Q$. 

The non-linear quench with a sufficiently large $r\gg1/\nu z$ is a good choice if we want 
to improve adiabaticity of the transition. However, the non-linear quench requires better 
experimental control over a system than the linear quench. Not only $\epsilon(t)$ has to be 
made non-linear, but it has to be non-linear precisely at the critical point. If not, i.e., 
when the critical point is misplaced by $\delta\epsilon$,    
\be
\epsilon(t)~=~\delta\epsilon-{\rm sign}(t)~\left|\frac{t}{\tau_Q}\right|^r~,
\label{epsilonr}
\ee
then the quench (\ref{epsilonr}) can be linearised as in Eq. (\ref{linearise}) and the linearisation
is self-consistent for large enough $\tau_Q$ when $\hat\epsilon$ in Eq. (\ref{hatepsilon})
is much less than $\delta\epsilon$. However, as we will see in the next Section and 
especially in Section \ref{sinusGordon} this technical problem does not arise 
in some ``half-quenches'' that are robust enough to begin precisely at a critical point
and go into a gapped phase.

\subsection{ KZM: from a critical point into a gapped phase  } 
\label{halfKZ}

In the previous Sections we considered adiabatic passage across an isolated quantum critical
point separating two gapped phases, where KZM was essentially the adiabatic-impulse-adiabatic
approximation. Here we consider a transition from a critical point into a gapped phase
and argue that it can be described by an impulse-adiabatic approximation. 

As above, the distance from the critical point is measured by the dimensionless parameter
\be
\epsilon(t)~=~\left(\frac{t}{\tau_Q}\right)^r~\geq~0~,
\label{halfepsilon}
\ee
which is in general a non-linear function of time running from $0$ to $\infty$.

The initial ground state at the critical $\epsilon=0$ has infinite correlation length while
the ground state at a finite $\epsilon>0$ has a finite correlation length $\xi\sim\epsilon^{-\nu}$. 
On one hand, in first approximation the two ground states look different on length scales longer 
than $\xi$, but they appear the same on distances less than the correlation length. On the other
hand, the critical ground state is a non-stationary excited state with a finite density of excitations
with respect to the Hamiltonian at the finite $\epsilon>0$. This excited state and the ground 
state at $\epsilon>0$ appear different when we look on a scale much longer than the 
average distance between the excitations, but they appear the same when we focus on a shorter scale. 
Consequently, we can identify the density of excitations as $n_{\rm ex}\simeq\xi^{-1}$.  

The evolution with the non-linear ramp (\ref{halfepsilon}) is initially non-adiabatic. It becomes 
adiabatic near $\hat t$ when the quench rate $|\dot\epsilon/\epsilon|=r/|\hat t|$ equals the gap 
$\Delta\sim\epsilon^{z\nu}$, or equivalently at 
\be
\hat \epsilon~\equiv~
\epsilon(\hat t)~\sim~
\tau_Q^{-\frac{r}{1+r\nu z}}~.
\label{halfhatepsilon}
\ee
In the crudest impulse-adiabatic approximation, the initial critical ground state at $\epsilon=0$ remains 
the state of the system until $\hat\epsilon$ when the evolution becomes adiabatic. Since there is 
a finite correlation length $\hat\xi\sim\hat\epsilon^{-\nu}$ in the ground state of the Hamiltonian 
at $\hat\epsilon$, the density of excitations is
\be
n_{\rm ex} ~\simeq~ \hat\xi^{-d} ~\sim~ \tau_Q^{-\frac{dr\nu}{1+r\nu z}} ~.
\label{halfnex}
\ee
It does not change in the following adiabatic evolution after $\hat\epsilon$. The density 
(\ref{halfnex}) was obtained for the first time in Refs. \cite{DeGrandiPolkovnikov,infinitezimalPolkovnikov}
within the adiabatic perturbation theory reviewed in Section \ref{argumentPolkovnikov}. 

The scaling exponent in Eq. (\ref{halfnex}) is the same as in the corresponding Eq. (\ref{nex}) 
for a full passage across a critical point, but there is in general a difference in numerical 
pre-factors omitted at the excuse of using ``$\sim$''. We can expect the $n_{\rm ex}$ in 
Eq. (\ref{halfnex}) to be less than that in Eq. (\ref{nex}) because in the ``half-quench'' starting 
from the critical point the impulse stage is shorter than in the full passage across this point. 

However, a quench from a critical point into a gapped phase is not always a mere half-quench, as 
illustrated in Section \ref{sinusGordon} by the process of loading a one-dimensional 
Bose gas (critical Luttinger liquid) into an optical lattice potential. An infinitesimally weak
potential induces an energy gap proportional to the potential strength. This is a realistic
experiment, see Ref. \cite{SGExp}, where a transition into a gapped phase begins exactly at a 
critical point.

We will make one more comment on the half-quench after Eq. (\ref{P2n}) below where we discuss 
``one-half'' of the Landau-Zener transition. In the next Section we describe a more unusual 
application of KZM to a phase transition that takes place in space rather than happens in time.

\subsection{ KZM in space } 
\label{KZinspace}

\begin{figure}[t]
\begin{center}
\includegraphics[width=0.8\columnwidth,clip=true]{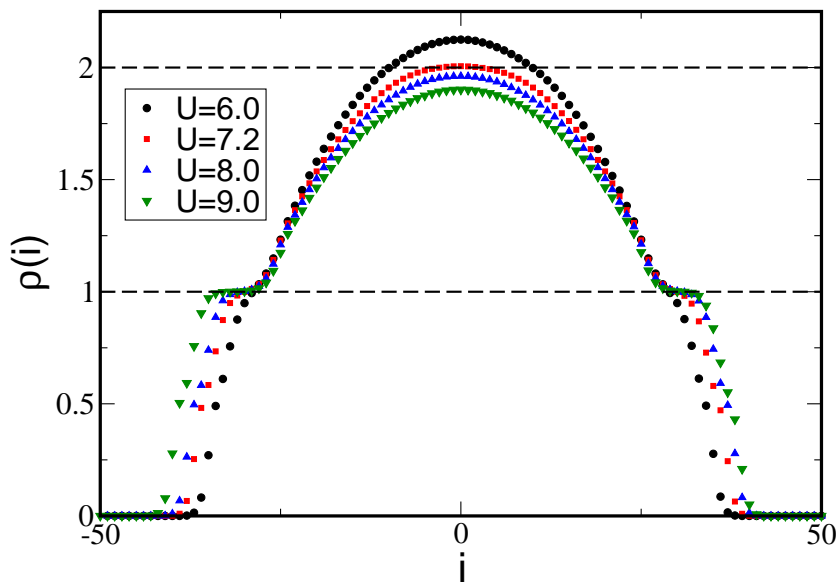}
\caption{ 
Density profiles in the ground state of the one-dimensional Bose-Hubbard model (\ref{HBH}) 
at different repulsion strengths $U$ \cite{Batrouni}. The model describes bosonic atoms in 
an optical lattice potential subject to additional harmonic trap confinement, compare 
Fig. \ref{FigGreiner} for a typical experimental set up. Here $\rho(i)$ is average number of 
atoms at site $i$. The dashed lines are to draw attention to density plateaux at the integer 
value of $\rho=1$ manifesting areas occupied by incompressible Mott insulator phases. 
These Mott phases coexist with superfluid phases at non-integer $\rho$. There are no sharp
boundaries between the ``integer'' Mott and ``non-integer'' superfluid phases. The phase 
boundaries are rounded off on a finite length scale $\hat\xi$. This scale is long enough to 
round off plateaux expected at $\rho=2$.  
(Figure from Ref.\cite{Batrouni}b)
}
\label{FigPancake}
\end{center}
\end{figure}

References \cite{Dornerinhom,Bodzioinspace,Ramsinhom,Karevski,z2} considered a symmetry breaking 
``phase transition in space'' where, instead of being time-dependent, a local parameter 
$\epsilon(\vec r)$ is a time-independent function of spatial coordinates $\vec r$. 
This is a generic scenario in ultracold atom gases confined in magnetic/optical 
traps, where the trapping potential results in an inhomogeneous density of atoms $\rho(\vec r)$ 
and, in general, the critical point $g_c$ depends on the local atomic density. Thus even a perfectly 
uniform parameter $g$ translates into an inhomogeneous 
\be
\epsilon(\vec r)~=~\frac{g-g_c[\rho(\vec r)]}{g_c[\rho(\vec r)]}~,      
\ee
with the critical point on the surface where $\epsilon(\vec r)=0$. The part of the atomic cloud 
where $\epsilon(\vec r)>0$ is in a different phase than the part where $\epsilon(\vec r)<0$. The 
phase transition between the two phases coexisting in a trap takes place near the critical surface 
$\epsilon(\vec r)=0$. A classic example of such a phase coexistence is shown in Fig. \ref{FigPancake}. 
Another plausible experimental scenario can be found in Section \ref{FerroKZ} and 
Fig. \ref{FigBodzioinspace}. 

\begin{figure}[t]
\begin{center}
\includegraphics[width=0.8\columnwidth,clip=true]{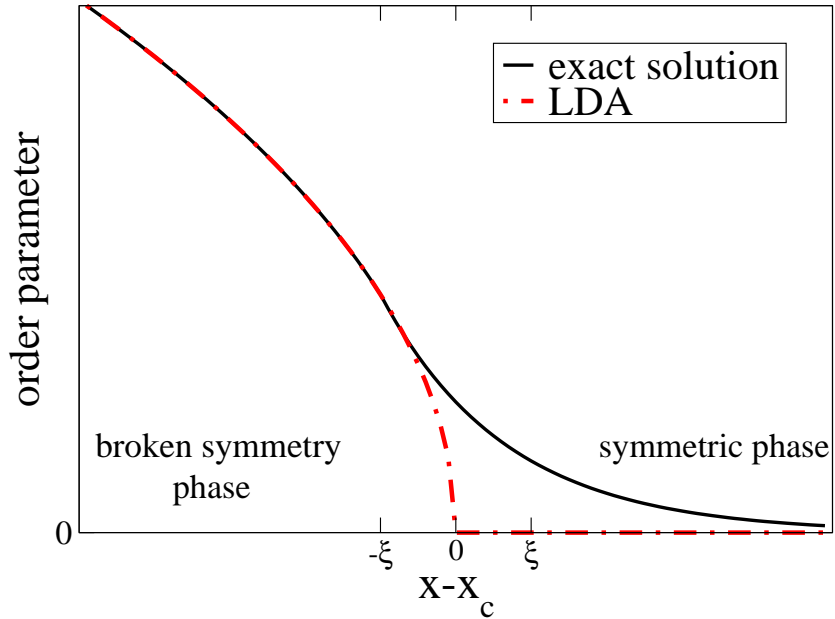}
\caption{ Phase transition in space from a broken symmetry phase to a symmetric phase.
Order parameter is shown as a function of position $x-x_c$ with respect to a critical point  
$x_c$ where $\epsilon(x_c)=0$. In the local density approximation (LDA) the order parameter is zero in the
symmetric phase and follows $(x_c-x)^\beta$ with the critical exponent $\beta$ in the symmetry broken
phase. It is non-analytic at the critical point $x_c$. In contrast, the exact order parameter
in the ground state of the system is rounded off on the length scale $\hat\xi$ in 
Eq. (\ref{tildexigeneral}). A non-zero order parameter penetrates into the symmetric phase to 
a depth $\simeq\hat\xi$.
}
\label{FigKZSpace}
\end{center}
\end{figure}

In order to apply KZM in this situation, we proceed in a similar way as in Eq. (\ref{linearise}) and 
linearise  
\be
\epsilon(x)~\approx~\alpha~(x-x_c)~,
\label{nc}
\ee
near the critical point at $x_c$ where $\epsilon(x_c)=0$. Here the gradient $\alpha$ of $\epsilon(\vec r)$ 
is along the $x$-axis. The system is in the broken symmetry phase where $x<x_c$ and in the symmetric phase 
where $x>x_c$. In the first ``local density approximation'' (LDA), we would expect that the order parameter behaves 
as if the system were locally uniform, i.e., it is non-zero only for $x<x_c$ and tends to zero as 
$(x_c-x)^\beta$ when $x\to x_c^-$ with the critical exponent $\beta$, see Fig. \ref{FigKZSpace}  .

However, the LDA is not consistent with the divergence of the healing length $\xi\sim|\epsilon|^{-\nu}$ 
near the critical point. The diverging $\xi$ is the shortest scale of length on which the order parameter 
can adjust to (or heal with) the varying $\epsilon(x)$. Consequently, when approaching $x_c$ from
the broken symmetry side, the local approximation $(x_c-x)^\beta$ must break down when 
the local correlation length $\xi\sim[\alpha(x_c-x)]^{-\nu}$ equals the distance 
$(x_c-x)$ remaining to the critical point. Solving this equality with respect to $\xi$ 
we obtain 
\be
\hat\xi ~\sim~ \alpha^{-\frac{\nu}{1+\nu}}~.
\label{tildexigeneral}
\ee 
From the point $(x-x_c)\simeq-\hat\xi$ the ``evolution'' of the order parameter in $x$ becomes ``impulse'', 
i.e, the order parameter does not change until $(x-x_c)\simeq+\hat\xi$ in the symmetric phase, where it begins to 
catch up with the local $\epsilon(x)>0$ and decays to zero on the length scale of $\hat\xi$. Thus the 
``adiabatic-impulse-adiabatic'' approximation in space, or the ``KZM in space'', predicts that the non-zero order 
parameter penetrates into the symmetric phase to the depth of $\hat\xi$ in Eq. (\ref{tildexigeneral}).
As compared to the non-analytic LDA prediction $(x_c-x)^\beta$, in KZM the order parameter is effectively 
``rounded off'' on the length scale of $\hat\xi$, see Fig. \ref{FigKZSpace}. 

We expect that other physical quantities, which are normally singular or discontinuous at the critical point, 
are also ``rounded off'' on the scale of $\hat\xi$. In particular, the energy gap that in a homogeneous system vanishes 
like $\Delta\sim\xi^{-z}$ when $\xi$ diverges near the critical point, here should remain finite and scale as 
\footnote{
A non-linear generalization of Eq. (\ref{nc}),
$
\epsilon(x)~=~{\rm sign}(x-x_c)\left|\alpha(x-x_c)\right|^r~,
$  
results in a critical point rounded off on a length scale 
$\hat\xi\sim\alpha^{-\nu r/(1+\nu r)}$ and a finite energy gap
scaling as $\hat\Delta\sim\alpha^{z\nu r/(1+\nu r)}$.
}
\be
\hat\Delta~\sim~\hat\xi^{-z}~\sim~\alpha^{\frac{z\nu}{1+\nu}}~.
\label{tildeDeltageneral}
\ee
This finite gap sharply contrasts with the LDA, where one might expect gapless quasiparticle excitations localized 
near the critical point at $x_c$.

Two examples supporting KZM in space are described in Sections \ref{IsingKZinspace} and \ref{FerroKZ} below. 
Encouraged by these examples, in the next Section we consider an inhomogeneous phase transition that, in a sense, 
takes place in both space and time.

\subsection{ KZM and dynamics of an inhomogeneous phase transition } 
\label{KZinhom}

As pointed out already in the finite temperature context \cite{Volovik,VolovikRecent}, 
in a realistic experiment it is difficult to make $\epsilon$ exactly homogeneous throughout a 
system. For instance, in the classic superfluid $^3$He experiments \cite{He3} a
phase transition was forced by neutron irradiation of helium 3. Heat released in each
fusion event, $n~+~^3{\rm He}~\to~^4{\rm He}$, created a bubble of normal fluid above the 
superfluid critical temperature $T_c$. Thanks to quasiparticle diffusion, 
the bubble was expanding and cooling with a local temperature 
$T(t,r)=\exp(-r^2/2Dt)/(2\pi Dt)^{3/2}$,
where $r$ is a distance from the centre of the bubble and $D$ is a diffusion 
constant. Since this $T(t,r)$ is hottest in the centre, the transition 
back to the superfluid phase, driven by an inhomogeneous parameter 
\be
\epsilon(t,r)~=~\frac{T(t,r)-T_c}{T_c}~,
\ee 
proceeded from the outer to the central part of the bubble with a critical front 
$r_c(t)$, where $\epsilon(t,r_c)=0$, shrinking with a finite velocity $v=dr_c/dt<0$.

A similar scenario is likely in the ultracold atom gases in magnetic/optical traps. 
The trapping potential results in an inhomogeneous density of atoms $\rho(\vec r)$
and, in general, a critical point $g_c$ depends on atomic density $\rho$. Thus even 
a transition driven by a perfectly uniform $g(t)$ is effectively inhomogeneous,
\be
\epsilon(t,\vec r)~=~\frac{g(t)-g_c[\rho(\vec r)]}{g_c[\rho(\vec r)]}~,      
\ee
with the surface of critical front, where $\epsilon(t,\vec r)=0$, moving with a finite
velocity. 
 
According to KZM, in a homogeneous symmetry breaking transition, a state after 
the transition is a mosaic of finite ordered domains of average size $\hat\xi$. Within 
each finite domain the orientation of the order parameter is constant, but uncorrelated 
to orientations in other domains. In contrast, in an inhomogeneous symmetry breaking 
transition \cite{Volovik} the parts of the system that cross the critical point 
earlier may be able to communicate their choice of orientation of the order parameter
to the parts that cross the transition later and bias them to make the same choice. 
Consequently, the final state may be correlated at a range longer than $\hat\xi$, or 
even end up being a ground state with long range order. In other words, the final
density of excited quasiparticles may be lower than the KZ estimate in 
Eq. (\ref{nex}) or even zero. 

From the point of view of testing KZM, this inhomogeneous scenario, when relevant, may 
sound like a negative result because an imperfect inhomogeneous transition suppresses 
system excitation by KZM. However, from the point of view of adiabatic quantum computation 
or adiabatic quantum state preparation, it is the KZM itself that is a negative result: no 
matter how slow the homogeneous transition is, there is a finite density of excitations 
(\ref{nex}) in the final state which decays only as a power of the 
transition time $\tau_Q$. From this perspective, the inhomogeneous transition may be 
the way to suppress KZ excitations and prepare the desired final ground state 
adiabatically.

In order to estimate if and when the inhomogeneity is actually relevant, we linearise 
the parameter $\epsilon(t,x)$ in both $t$ and $x$ near the critical front where 
$\epsilon(t,x)=0$,
as
\be
\epsilon(t,x)~\approx~\alpha~(x-vt)~,
\label{alphavt}
\ee
in a similar way as in Eq. (\ref{tauQ}). Here $\alpha$ is an inhomogeneity of the 
quench and $v$ is velocity of the critical front. When observed locally at 
a fixed $x$, the inhomogeneous transition in Eq. (\ref{alphavt}) looks like the 
homogeneous quench in Eq. (\ref{tauQ}) with
\be
\tau_Q~=~\frac{1}{\alpha v}~.
\label{tauQalphav}
\ee    
The part of the system where $x<vt$, or equivalently $\epsilon(t,x)<0$, is already
in the broken symmetry phase. The orientation of the order parameter chosen in this 
part can be communicated across the critical point not faster than a threshold 
velocity
\be
\hat v~\simeq~\frac{\hat\xi}{\hat t}~.
\label{hatv}
\ee 
When $v\gg\hat v$ the communication is too slow for the inhomogeneity to be relevant, 
but when $v\ll\hat v$ we can expect the final state to be less excited than predicted 
by KZM. 

Given the relation (\ref{tauQalphav}), the condition (\ref{hatv}) can be rewritten either 
as
\bea
\hat v &\sim& \tau_Q^{-\frac{(z-1)\nu}{z\nu+1}} ~, \label{hatvtauQ}\\
\hat v &\sim& \alpha^{\frac{\nu(z-1)}{1+\nu}}   ~, \label{hatvalpha}
\eea
or as a relation between the slope and the threshold transition time,
\be
\hat\tau_Q~\sim~\alpha^{-\frac{z\nu+1}{1+\nu}}~.
\label{hattauQ}
\ee
The last relation means that, for a given inhomogeneity $\alpha$, the transition is
effectively homogeneous when $\tau_Q\ll\hat\tau_Q$, but the inhomogeneity becomes
relevant when the transition is slow enough and $\tau_Q\gg\hat\tau_Q$. In the 
homogeneous limit of $\alpha\to0$ the threshold transition time $\hat\tau_Q\to\infty$.

An example of inhomogeneous transition is solved in detail in Section \ref{IsingKZinhom} 
and in Refs. \cite{Dornerinhom,Ramsinhom,Schaller,z2,inhomoKarevski}. 
In the next Section we return to the basics and review the Landau-Zener model.

\subsection{ The Landau-Zener (LZ) model }
\label{LZvKZ}

In this Section we briefly review the Landau-Zener (LZ) model \cite{LZ} because in 
a number of integrable models KZM can be derived exactly by mapping a model to a set 
of independent LZ anti-crossings, see Section \ref{LZargument} and the Sections that follow. 
After this brief review in Section \ref{LZvKZ2} we will attempt to invert the relation between 
the LZ model and KZM and estimate LZ excitation probabilities in the adiabatic-impulse-adiabatic
approximation essential for KZM \cite{Bodzio1,bodzioimpulse}.

\begin{figure}[t]
\begin{center}
\includegraphics[width=0.7\columnwidth,clip=true]{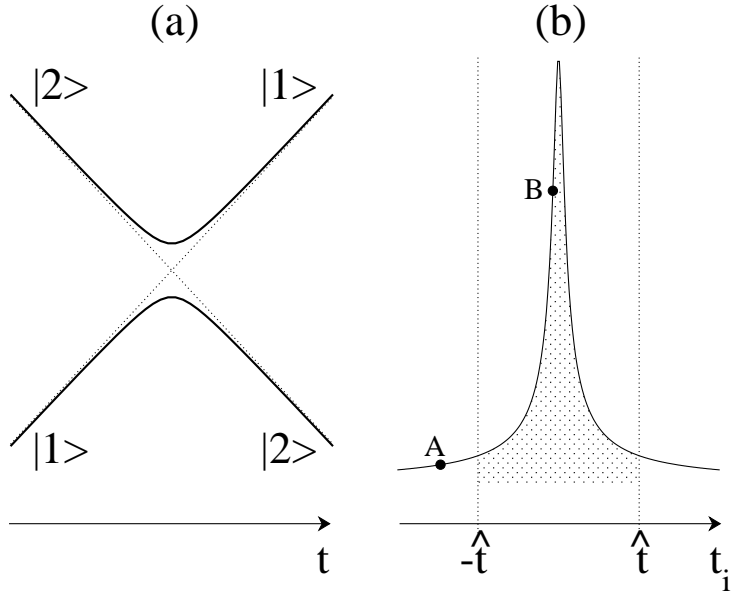}
\caption{ In panel A, the instantaneous energy spectrum of the two-level
system in Eq. (\ref{HLZ}) parametrised by the time $t$. In panel B, the adiabatic 
($t<-\hat t$ or $t>\hat t$) and impulse regimes ($-\hat t<t<\hat t$) in the two-level system dynamics. 
Note the similarity of panel B to Figure \ref{FigKZM}.
(Figure from Ref. \cite{bodzioimpulse})
}
\label{FigAdIm}
\end{center}
\end{figure}

\begin{figure}
\begin{center}
\includegraphics[width=0.8\columnwidth,clip=true]{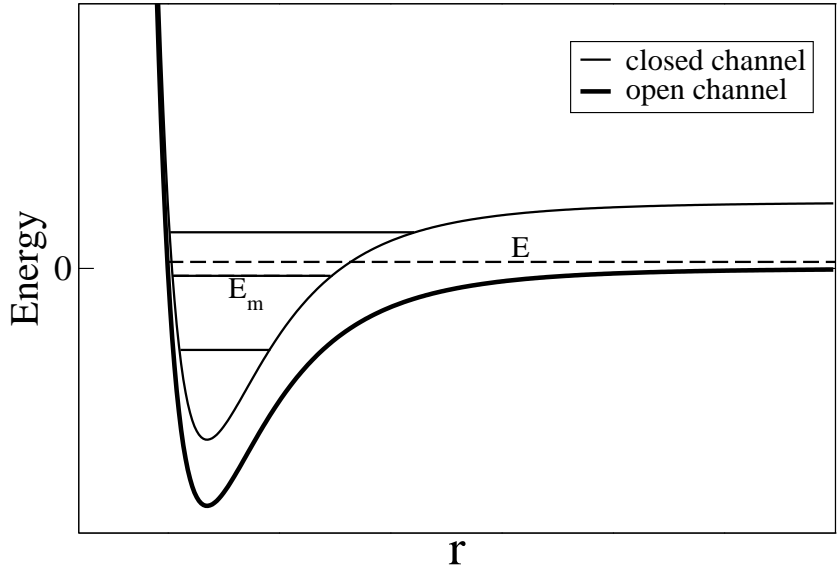}
\caption{ 
The plot shows a scattering potential for two atoms as a function of their separation $r$.
The scattering potential depends on internal states of the colliding atoms. For a small
scattering energy $E$ the atoms can be either in an open (thick solid line) or
closed (thin solid line) channel. When one of the bound states in the closed channel
with energy $E_m$ is close to the scattering energy $E\approx0$, then we have 
the Feshbach resonance \cite{eddi}. An external magnetic field coupled to magnetic
moments of the colliding atoms can shift the energy in the closed channel moving
the energy $E_m$ either up or down with respect to the open channel. The field can be used to tune 
the scattering 
atoms close to the Feshbach resonance \cite{eddi}. In Section \ref{manyLZ} we consider scattering
of two fermions, corresponding to fermionic annihilation operators $c_\uparrow$ and 
$c_\downarrow$, near resonance with a bound molecular state $E_m$, correspoding
to a bosonic annihilation operator $b$. A coupling between the two fermions
and the molecular state is decribed by a Landau-Zener Hamiltonian
$
H=
-\frac12\epsilon b^\dag b 
+ \frac14\epsilon \left(c^\dag_\uparrow c_\uparrow+c^\dag_\downarrow c_\downarrow\right)
+\frac12\left(b^\dag c_\downarrow c_\uparrow + c^\dag_\uparrow c^\dag_\downarrow b \right).
$
Here $\epsilon$ is a distance from the resonance proportional to a difference between 
the actual and resonant values of the magnetic field. The state with two fermions
corresponds to the state $|1\rangle$ and the state with a molecule to the state $|2\rangle$ 
in Eq. (\ref{HLZ}). In a scattering of two fermions far from the resonance a weakly occupied molecular state 
can be elliminated and in a second order perturbative expansion one obtains an effective 
Hamiltonian for fermions 
$
H_{\rm eff}=
\frac{1}{\epsilon}\left(c^\dag_\uparrow c_\uparrow c^\dag_\downarrow c_\downarrow\right).
$
Depending on the sign of $\epsilon$ this is additional repulsion or attraction, so the magnetic 
field can be used to change the strength and even sign of the interaction between the fermions. 
This effective interaction strength diverges at $\epsilon=0$, but this is where the occupation of 
the molecular state is large so it cannot be elliminated and has to be included explicitly in 
the calculations.  
} 
\label{FigFeshbach}
\end{center}
\end{figure}

In dimensionless variables, the LZ model is defined by a two-level time-dependent Hamiltonian
\be
H~=~\frac12
\left(
\begin{array}{cc}
\epsilon(t) & 1              \\
1           & -\epsilon(t) 
\end{array}
\right)
\label{HLZ}
\ee
where 
\be
\epsilon(t)~=~\frac{t}{\tau_Q}
\label{LZtauQ}
\ee
and $\tau_Q$ is a transition time. The Hamiltonian passes through an anti-crossing 
at $\epsilon=0$. At any fixed $\epsilon$, it has two instantaneous eigenstates: the 
ground state $|\downarrow(\epsilon)\rangle$ and the excited state 
$|\uparrow(\epsilon)\rangle$. In the time-independent basis $|1\rangle,|2\rangle$ 
of the Hamiltonian (\ref{HLZ}), the instantaneous eigenstates are
\bea
|\uparrow(\epsilon)\rangle~=~|1\rangle \cos\frac{\theta}{2}+|2\rangle \sin\frac{\theta}{2}~,~~
|\downarrow(\epsilon)\rangle~=~-|1\rangle \sin\frac{\theta}{2}+|2\rangle \cos\frac{\theta}{2}~,
\label{updown}
\eea
where $\cos\theta=\epsilon/\sqrt{1+\epsilon^2}$ and $\sin\theta=1/\sqrt{1+\epsilon^2}$.
Their eigenenergies are $\Delta/2$ and $-\Delta/2$ respectively, where 
\be
\Delta~=~\sqrt{1+\epsilon^2}
\label{DeltaLZ}
\ee
is an instantaneous energy gap. It is minimal at the anti-crossing centre at $\epsilon=0$. 
A good example of the Landau-Zener two level system is the Feshbach resonance in
Figure \ref{FigFeshbach}, see also Section \ref{manyLZ}.

In the LZ problem the system is prepared in the ground state at an initial time $t_i$
and we want to know the probability $P$ that it is excited at a final time $t_f$. 
A general solution to this problem was obtained in Ref. \cite{LZ} and then
analysed in Refs. \cite{LZmore,bodzioimpulse}:
\bea
|\psi(t)\rangle ~=~
\left[
|1\rangle
\left(2i\partial_t+\frac{t}{\tau_Q}\right)+
|2\rangle
\right]~
\left[
a~D_{-1-i\tau_Q/4}\left( iz \right) +
b~D_{-1-i\tau_Q/4}\left(-iz \right)
\right]
\eea
where $z=\frac{t}{\sqrt{\tau_Q}}e^{-i\pi/4}$, $D_m(s)$ is a Weber function \cite{Weber},
and $a,b$ are constants to be determined by initial conditions.

The excitation probability $P$ can be obtained in a closed form in a number of special cases:

\begin{enumerate}  

\item[(i)] In the textbook case of $t_i\to-\infty$ and $t_f\to\infty$ we obtain the 
exponential LZ formula
\be
P~=~e^{-\pi\tau_Q/2}~.
\label{PLZ}
\ee

\item[(ii)] When the evolution begins at $t_i=0$ in the ground state at the anti-crossing 
centre and runs to $t_f\to\infty$, then according to Refs. \cite{LZmore,bodzioimpulse}
\bea
P &=&
1-
\frac{2\sinh\left(\frac{\pi\tau_Q}{4}\right)}{\pi\tau_Q}
e^{-\pi\tau_Q/8}
\left|
\Gamma\left(1+\frac{i\tau_Q}{8}\right)+
e^{i\pi/4}\sqrt{\frac{\tau_Q}{8}}
\Gamma\left(\frac12+\frac{i\tau_Q}{8}\right)
\right|^2
\nonumber\\
&\approx&
\frac14~
\tau_Q^{-2}~,
\label{PhalfLZ}
\eea
where $\Gamma(x)$ is the gamma function \cite{Weber} and the last approximate form is 
the asymptote when $\tau_Q\gg1$. Note that, unlike in the full LZ transition (i), here 
the probability does not decay exponentially with $\tau_Q$, but only as $\tau_Q^{-2}$. 

\item[(iii)] When the evolution runs from $t_i\to-\infty$ to $t_f=0$, then the excitation 
probability $P$ is the same as in case (ii), see Ref. \cite{LZmore,bodzioimpulse}.

\item[(iv)] When the LZ problem is not symmetric with respect to the anti-crossing and 
\be
\epsilon(t)~=~
\left\{
\begin{array}{c}
\frac{t}{\tau_Q}        ~,~{\rm when}~ t  < 0~ \\              
\frac{t}{\tau_Q\delta}  ~,~{\rm when}~ t\geq0~
\end{array}
\right.
\ee
with $\delta\neq 1$, then the excitation probability after a passage from $t_i\to-\infty$ 
to $t_f\to+\infty$ is \cite{LZmore,bodzioimpulse} 
\bea
P &=&
1-
\frac12
\sinh\left(\frac{\pi\tau_Q\delta}{4}\right)
e^{-\frac{\pi(1+\delta)\tau_Q}{8}}
\left|
\frac{\Gamma\left(\frac12+\frac{i\tau_Q\delta}{8}\right)}
     {\Gamma\left(\frac12+\frac{i\tau_Q}{8}\right)}+
\sqrt{\frac{1}{\delta}}
\frac{\Gamma\left(1+\frac{i\tau_Q\delta}{8}\right)}
     {\Gamma\left(1+\frac{i\tau_Q}{8}\right)}
\right|^2
\nonumber\\
&\approx&
\frac14 \left(\frac{1-\delta}{\delta}\right)^2 \tau_Q^{-2}
~,
\label{PnonsymLZ}
\eea
where the last asymptote is accurate when $\tau_Q\gg1$. Note that when $\delta\to\infty$
the second half of the transition becomes adiabatic, the excitation probability does not 
change in this adiabatic part of the evolution, and the final $P$ becomes the same as 
in case (iii) where the transition ends at $t_f=0$.

\end{enumerate}

In the adiabatic limit of $\tau_Q\gg1$, the excitation probability $P$ is exponentially small 
in the standard LZ model (i), but in the non-symmetric cases (ii,iii,iv) there is much 
slower power-law decay $P\sim\tau_Q^{-2}$. In these cases an instantaneous rate of the transition 
$d\epsilon/dt$ is a discontinuous function of time. For instance, in case (ii) 
\be
\frac{d\epsilon}{dt}~=~
\left\{
\begin{array}{cl}
0~, {\rm when}~t<0~, \\
\frac{1}{\tau_Q}, {\rm when}~t\geq0~. 
\end{array}
\right.
\label{disc}
\ee
The discontinuity at $t=0$ translates into a fat high frequency tail in the Fourier transform of 
$\epsilon(t)$ which helps to excite the system despite its finite energy gap. More generally, 
as discussed in Refs. \cite{LZ,infinitezimalPolkovnikov}, any non-analytic behaviour of 
$\epsilon(t)$ results in a power law decay of the excitation probability. A discontinuity 
in the $r$-th derivative of $\epsilon(t)$ leads to
\footnote{
Indeed, in the basis of instantaneous eigenstates (\ref{updown}) the state is 
$|\psi(t)\rangle=
  \alpha|\uparrow(\epsilon)  \rangle e^{-\frac12i\int^t_{-\infty}dt'~\Delta[\epsilon(t')]}+
  \beta|\downarrow(\epsilon)\rangle e^{+\frac12i\int^t_{-\infty}dt'~\Delta[\epsilon(t')]/2}$.
In the adiabatic limit we have $|\alpha|\ll1,\beta\approx1$ and, to leading order in the
small $\alpha$, we obtain a perturbative excitation probability
$$ 
P=|\alpha|^2\approx
\left|
\int_{-\infty}^\infty dt~
\frac{d\epsilon}{dt}~
\langle\uparrow|\frac{d}{d\epsilon}|\downarrow\rangle~
e^{i\int_{-\infty}^t dt' \Delta(t')}
\right|^2.
$$
The discontinuity in Eq. (\ref{disc}) yields $P\simeq\tau_Q^{-2}$ for $\tau_Q\gg1$. A more general 
non-linear 
$$
\epsilon(t)=
\left\{
\begin{array}{cl}
0            & {\rm , when}~ t<0\\
(t/\tau_Q)^r & {\rm , when}~ t\geq0
\end{array}
\right.
$$
has a discontinuous $r$-th derivative at $t=0$ and we obtain $P\simeq\tau_Q^{-2r}$ for large $\tau_Q$.
}
\be
P~\sim~\tau_Q^{-2r}~
\label{P2n}
\ee 
for large enough $\tau_Q$.

Case (ii) above is similar to the half-quench in Section \ref{halfKZ}.
This connection will become more transparent in Section \ref{LZargument}, where KZM will be represented as 
a series of independent LZ transitions. The non-linear parameter $\epsilon(t)$ 
in Eq. (\ref{nonlinearepsilon}) has a discontinuous $r$-th time derivative at $t=0$.
Since on a microscopic level KZM excitations originate from LZ transitions, then
one might expect their density to decay as $n_{\rm ex}\sim\tau_Q^{-2r}$ for large
enough $\tau_Q$. However, in the thermodynamic limit the density of excitations $n_{\rm ex}$ is 
dominated by contributions from low frequency modes, with wavelength longer than $\hat\xi$, whose 
excitation probabilities are close to $1$ and for whom the asymptotic tail in Eq. (\ref{P2n}) does not apply. 

These comments complete our brief review of the LZ model. In Section \ref{LZargument} below 
the LZ model is applied to derive KZM in a class of exactly solvable models, but in the next 
Section we will attempt to rederive and reinterpret cases (i-iv) using the 
adiabatic-impulse-adiabatic approximation essential for KZM, see Refs. \cite{Bodzio1,bodzioimpulse}.

\subsection{ The Landau-Zener model in the adiabatic-impulse approximation }
\label{LZvKZ2}

A unitary evolution with the time-dependent Hamiltonian (\ref{HLZ}) is adiabatic when 
$t\to\mp\infty$ and the gap in Eq. (\ref{DeltaLZ}) is large enough, but it may be not 
adiabatic near the anti-crossing centre at $t=0$ where the gap is minimal. In the 
adiabatic-impulse-adiabatic approximation, the evolution is adiabatic before $-\hat t$ and 
after $+\hat t$, but it is impulse between $-\hat t$ and $+\hat t$, see panel (b) in 
Figure \ref{FigAdIm}. The crossover time $\hat t$ 
between the adiabatic and impulse stages is the time when the transition rate measured by 
$|\dot\epsilon/\epsilon|=1/|t|$ equals the instantaneous energy gap in Eq. (\ref{DeltaLZ}):
\be
\sqrt{1+\left(\frac{\hat t}{\tau_Q}\right)^2}~=~
\frac{\alpha}{|t|}~,
\ee
where $\alpha\simeq1$ is an adjustable parameter, compare Fig. \ref{FigAdIm}. The 
solution is
\be
\hat\epsilon~=~
\frac{\hat t}{\tau_Q}~=~
\frac{1}{\sqrt2}
\sqrt{\sqrt{1+\frac{4}{(\alpha\tau_Q)^2}}-1}~.
\label{hatepsilonLZ}
\ee
With this estimate at hand, we can reiterate the cases (i-iv) listed in Section \ref{LZvKZ}:

\begin{enumerate}

\item[(i)] The evolution (\ref{HLZ}) starts from the instantaneous ground state 
at $t_i\ll-\hat t$ and ends at $t_f\gg\hat t$. In the adiabatic stage before $-\hat t$ the 
state follows the instantaneous ground state. Then in the impulse stage between $-\hat t$ 
and $+\hat t$ the state does not change and remains
equal to the instantaneous ground state $|\downarrow(-\hat\epsilon)\rangle$ at $-\hat t$.
Thus in the adiabatic-impulse-adiabatic approximation, the excitation probability at $\hat t$ is
\be
P_{\rm AI}~=~|\langle\uparrow(\hat \epsilon)|\downarrow(-\hat \epsilon)\rangle|^2~=~
\frac{\hat\epsilon^2}{1+\hat\epsilon^2}
\ee
and it does not change in the following adiabatic evolution after $\hat t$. Its expansion
up to second power of $\tau_Q$,
\be
P_{\rm AI}~=~1-\alpha\tau_Q+\frac{(\alpha\tau_Q)^2}{2}+{\cal O}(\tau_Q^3)~,
\ee
matches the corresponding expansion of the exact exponent in Eq. (\ref{PLZ}) when 
we set $\alpha=\pi/2\simeq1$.

\item[(ii)] The evolution (\ref{HLZ}) begins at $t_i=0$ in the instantaneous ground 
state $|\downarrow(0)\rangle$ and ends at $t_f\gg\hat t$. Since this state does
not change in the initial impulse stage of the evolution until $\hat t$, the excitation
probability at $\hat t$ is
\be
P_{\rm AI}~=~|\langle\uparrow(\hat\epsilon)|\downarrow(0)\rangle|^2~=~
\frac12
\left(
1-\frac{1}{\sqrt{1+\hat\epsilon^2}}
\right)~,
\ee
and it does not change in the following adiabatic evolution after $\hat t$. Its expansion 
in powers of $\tau_Q$ is
\be
P_{\rm AI}~=~
\frac12-
\frac12\sqrt{\alpha\tau_Q}+
\frac18(\alpha\tau_Q)^{3/2}+{\cal O}(\tau_Q^{5/2})~,
\ee
while the corresponding expansion of the exact solution (\ref{PhalfLZ}) is
\be
P~=~
\frac12-
\frac{\sqrt{\pi\tau_Q}}{4}+
\frac{\pi^{3/2}}{64}\left(2-\frac{4\ln2}{\pi}\right)\tau_Q^{3/2}+
{\cal O}(\tau_Q^{5/2})~.
\label{PsmalltauQ}
\ee
Given that $2-4\ln2/\pi=1.1\approx1$, the two expansions are in good agreement when we 
choose $\alpha=\pi/4\simeq1$.

\item[(iii)] The evolution (\ref{HLZ}) begins at $t_i\ll-\hat t$ and ends at
$t_f=0$. It is adiabatic before $-\hat t$ and impulse from $-\hat t$ to $0$.
In the impulse stage the state remains the instantaneous ground state 
$|\downarrow(-\hat\epsilon)\rangle$ at $-\hat t$. Similarly as for the exact 
solution, the final excitation probability 
$P_{\rm AI}=|\langle\uparrow(0)|\downarrow(-\hat\epsilon)\rangle|^2$ is the same
as in case (ii). 

\item[(iv)] The evolution (\ref{HLZ}) begins in the ground state at $t_i\ll-\hat t$
and ends at $t_f\gg\hat t_\delta$. The transition is not symmetric: when $t>0$ the
transition time is $\tau_Q\delta$ instead of $\tau_Q$ for $t<0$. Consequently,
the crossover from the adiabatic to impulse stage takes place near $-\hat t$ as before,
but the second crossover from the impulse to adiabatic stage is at $\hat t_\delta$
given by Eq. (\ref{hatepsilonLZ}), but with $\tau_Q$ replaced by $\tau_Q\delta$.
Similarly as in case (i), the excitation probability is
\bea
P_{\rm AI} &=&
|\langle\uparrow(\hat\epsilon_\delta)|\downarrow(-\hat\epsilon)\rangle|^2
\nonumber\\
&=&
1-\frac14\left(1+\sqrt\delta\right)^2\alpha\tau_Q+
\frac{1}{16}(1+\delta)(1+\sqrt\delta)^2(\alpha\tau_Q)^2+
{\cal O}(\tau_Q^3)~,
\eea
while the corresponding expansion of the exact solution (\ref{PnonsymLZ}) is
\be
P~=~
1-\frac{\pi}{8}\left(1+\sqrt\delta\right)^2\tau_Q+
\frac{\pi^2}{64}(1+\delta)(1+\sqrt{\delta})^2\tau_Q^2+
{\cal O}(\tau_Q^3)~.
\ee
The two expansions match when we set $\alpha=\pi/2\simeq1$. 

\end{enumerate}

In all cases, after fitting only one adjustable parameter $\alpha\simeq1$, we obtain accurate 
leading and next-to-leading order terms of the expansion in powers of $\tau_Q$, so the 
adiabatic-impulse-adiabatic approximation is accurate when $\tau_Q\ll1$ and the excitation 
probability $P$ is substantial. Since the KZM excitation density $n_{\rm ex}$ in Eq. (\ref{nex}) 
is dominated by contributions from long wavelength modes, whose excitation probability is 
close to $1$, the adiabatic-impulse-adiabatic approximation can accurately predict $n_{\rm ex}$. 
The approximation is not accurate for long wavelength modes, but their contribution to the total
$n_{\rm ex}$ is negligible.

In view of the above conclusion, it is not quite surprising that in the next Section the LZ model 
provides exact solutions which are consistent with KZM.

\subsection{ KZM as a set of independent Landau-Zener transitions }
\label{LZargument}

The KZM argument in Section \ref{KZargument} was confirmed by exact solutions in a number of integrable 
models, see e.g. Ref. \cite{Dziarmaga2005,CherngLevitov,Cincio,XYHindusi,Viola2} and the following 
Sections. There is, however, an assumption in Section \ref{KZargument} that the transition passes 
through an isolated quantum critical point between two gapped phases. As demonstrated by exact solutions 
in a number of integrable models, KZM does require careful generalisation for a transition 
across a multicritical point \cite{multicritical,Viola3}, along a gapless line \cite{gaplessline}, 
or across a gapless phase \cite{Kitaev}. The generalisation was obtained in integrable 
spin models, such as the XY spin chain or the two-dimensional Kitaev model \cite{KitaevModel},
all of which can be mapped to non-interacting fermions by a Jordan-Wigner transformation \cite{JW}. 
A transition in a translationally invariant system of non-interacting fermions can be mapped
to a set of independent LZ anti-crossings. The general LZ argument summarized in this Section 
was gradually developed in Refs. \cite{Bodzio1,Dziarmaga2005,CherngLevitov,XYHindusi,multicritical,
gaplessline,Kitaev,Viola2,Viola3,HindusiReview,HindusiwariacjeLZ,semiDirac,Balazs}. 

We assume a general local Hamiltonian quadratic in fermionic creation/annihilation operators 
and translationally invariant on a $d$-dimensional lattice. Its quasimomentum representation is
\be
H~=~
\frac12
\sum_{\vec k}~
\left(c^\dag_{\vec k},c_{-\vec k}\right)
H_{\vec k}
\left(
\begin{array}{c}
c_{\vec k} \\
c^\dag_{-\vec k}
\end{array}
\right)~,
\label{HHk}
\ee 
where $\vec k$ is a quasimomentum running over the first Brillouin zone, $c_{\vec k}$ is fermionic 
annihilation operator, and $H_{\vec k}$ is a $2\times2$ Hermitian matrix. The Hamiltonian can be diagonalized 
by a Bogoliubov transformation
\be
c_{\vec k}~=~u_{\vec k}\gamma_{\vec k}+v^*_{-\vec k}\gamma^\dag_{-\vec k}~,
\label{cuv}
\ee
where $\gamma_{\vec k}$ annihilates a fermionic Bogoliubov quasiparticle, and $(u_{\vec k},v_{\vec k})$ 
is an eigenmode of stationary Bogoliubov-de Gennes equations
\be
\omega_{\vec k}
\left(
\begin{array}{c}
u_{\vec k} \\
v_{\vec k}
\end{array}
\right)~=~
H_{\vec k}~
\left(
\begin{array}{c}
u_{\vec k} \\
v_{\vec k}
\end{array}
\right)~
\label{statgeneralBdG}
\ee
with positive eigenfrequency $\omega_{\vec k}$. The diagonalized Hamiltonian is 
$H=\sum_{\vec k}\omega_{\vec k}\left(\gamma^\dag_{\vec k}\gamma_{\vec k}-\frac12\right)$.
Its ground state is a Bogoliubov vacuum $|0\rangle$ annihilated by all $\gamma_{\vec k}$. 
This ground state is an initial state for a dynamical quantum phase transition.

A time-dependent problem can be solved in the Heisenberg picture, where the state remains the initial 
Bogoliubov vacuum $|0\rangle$, but the operators $c_{\vec k}$ evolve in time. Their evolution can be 
conveniently described by time-dependent modes $(u_{\vec k},v_{\vec k})$ in Eq. (\ref{cuv}). Indeed, 
given that $\gamma_{\vec k}$ remains time-independent in the Heisenberg picture, a Heisenberg
equation $i\frac{d}{dt}c_{\vec k}=[c_{\vec k},H(t)]$ is equivalent to time-dependent Bogoliubov-de Gennes
equations
\be
i\frac{d}{dt}
\left(
\begin{array}{c}
u_{\vec k} \\
v_{\vec k}
\end{array}
\right)~=~
H_{\vec k}(t)~
\left(
\begin{array}{c}
u_{\vec k} \\
v_{\vec k}
\end{array}
\right)~.
\label{generalBdG}
\ee
Thus the problem was separated into a set of independent two-level systems enumerated by $\vec k$. 

For each $\vec k$, an initial state $[u_{\vec k}(-\infty),v_{\vec k}(-\infty)]$ is the positive-$\omega_{\vec k}$ 
eigenmode of Eq. (\ref{statgeneralBdG}) with the initial $H_{\vec k}(-\infty)$. This positive eigenmode represents
the initial vacuum ground state for the quasiparticle $\gamma_{\vec k}$. What we want to know is a probability 
$p_{\vec k}$ that after the dynamical transition a quasiparticle $\gamma_{\vec k}$ of the final Hamiltonian $H(\infty)$ 
is excited. Since this excited state is represented by the negative eigenmode of Eq. (\ref{statgeneralBdG})
with eigenfrequency $-\omega_{\vec k}$, then $p_{\vec k}$ is equal to the probability that the solution of 
Eq. (\ref{generalBdG}) ends in the negative eigenmode of the final $H_{\vec k}(\infty)$ or, simply, 
the initially excited two-level system becomes deexcited. This is essentially the same question as in the Landau-Zener 
(LZ) model in Section \ref{LZvKZ} so in order to answer this question we map Eq. (\ref{generalBdG}) to the LZ model 
in Eq. (\ref{HLZ}).   

Near a quantum phase transition driven by a parameter $\epsilon$ the two-level Hamiltonian can be linearised, or is
linear, in the small $\epsilon$: 
\be
H_{\vec k}(t) ~=~ \epsilon(t) ~ \sigma(\vec k) ~+~ \sigma'(\vec k) ~.
\label{Hveck}
\ee
Here $\sigma(\vec k)=\sum_i a_i(\vec k) \sigma^i$, $\sigma^i$ with $i=x,y,z$ are Pauli matrices,
$\sigma'(\vec k)=\sum_i a'_i(\vec k) \sigma^i$, $a_i$ and $a'_i$ are model-dependent
functions, and $\epsilon(t)=t/\tau_Q$ is the dimensionless parameter in Eq. (\ref{tauQ}) quenched from 
$t\to-\infty$ to $t\to+\infty$. Notice that in this general Hamiltonian we ignored a $c$-number
term proportional to the $2\times2$ identity matrix because it would contribute
to the global phase only, but see Ref. \cite{c-term}.
The instantaneous eigenvalues of the Hamiltonian $H_{\vec k}$ are $\pm\omega_{\vec k}$,
where
\be
\omega_{\vec k} ~=~ \sqrt{\left(\epsilon~\vec a+\vec a'\right)^2} ~.
\label{spectrumgenBdG}
\ee
The question is what is the probability for the initially excited state $+\omega_{\vec k}$
to become deexcited to the ground state $-\omega_{\vec k}$? 

To answer this question, we map the general Eqs. (\ref{generalBdG},\ref{Hveck}) to the LZ 
model (\ref{HLZ}). The main obstacle is that, unlike in the LZ model, the ``spin matrices'' $\sigma$ 
and $\sigma'$ in the Hamiltonian (\ref{Hveck}) are not necessarily pointing in orthogonal directions.
This minor problem can be easily fixed by orthonormalisation 
\footnote{Indeed, we rewrite the matrix $\sigma(\vec k)=a(\vec k)\hat\sigma(\vec k)$, where 
$a=\sqrt{\vec a^2}$ is the length of vector $\vec a$, and 
$\hat\sigma(\vec k)=\sum_i \hat a(\vec k)\sigma^i$ is a normalized Pauli spin matrix pointing along 
the unit vector $\hat a(\vec k)=\vec{a}(\vec k)/a(\vec k)$. Next, we decompose 
$\sigma'(\vec k)=b(\vec k)\hat\sigma(\vec k)+\Delta(\vec k)\hat\sigma_{\perp}(\vec k)$,
where $b(\vec k)=\hat a(\vec k)~\vec a'(\vec k)$ is the component of vector $\vec a'(\vec k)$ along 
the unit vector $\hat a(\vec k)$, $\vec a_{\perp}(\vec k)=\vec a'(\vec k)-b(\vec k)\hat a(\vec k)$ is 
the component of $\vec a'(\vec k)$ orthogonal to $\hat a(\vec k)$, 
$\Delta(\vec k)=\sqrt{\vec a_{\perp}(\vec k)^2}$ is the length of the orthogonal
component, and $\hat\sigma_{\perp}=\sum_i \frac{a^i_{\perp}(\vec k)}{\Delta(\vec k)}\sigma^i$
is a normalized Pauli spin matrix pointing perpendicular to $\hat\sigma(\vec k)$.}.
The orthonormalised Hamiltonian is 
\be
H_{\vec k}(t) ~=~ 
\left[ a(\vec k)~\epsilon(t)~+~b(\vec k) \right] ~ \hat\sigma(\vec k) ~+~ 
\Delta(\vec k) ~ \hat\sigma_{\perp}(\vec k) ~
\label{HabDelta}
\ee
with instantaneous quasiparticle spectrum
\be
\omega_{\vec k}~=\sqrt{\left[\epsilon~a(\vec k)+b(\vec k)\right]^2+\Delta^2(\vec k)}~.
\label{spectrumDelta}
\ee
We can eliminate $b(\vec k)$ by shifting the time variable in $\epsilon(t)=t/\tau_Q$ to 
$t'=t+\tau_Q~b(\vec k)/a(\vec k)$:
\be
H_{\vec k}(t') ~=~ 
\frac{t'}{\tau_Q}~a(\vec k)~\hat\sigma(\vec k) ~+~ 
\Delta(\vec k) ~ \hat\sigma_{\perp}(\vec k) ~.
\label{HLZt'}
\ee
Up to an unimportant rotation of ``spin'' quantization axes,  this is the Landau-Zener 
Hamiltonian (\ref{HLZ}) and we can use the LZ formula for the (de)excitation probability
\be
p_{\vec k} ~=~ \exp\left( -\frac{\pi~\tau_Q~\Delta^2(\vec k)}{a(\vec k)} \right)~.
\label{pveck}
\ee
Consequently, the density of quasiparticles is an average of $p_{\vec k}$ over the first
Brillouin zone 
\be
n_{\rm ex} ~=~
\frac{1}{N}
\sum_{\vec k}
p_{\vec k} ~\approx~
\int
\frac{d^d k}{(2\pi)^d}~
p_{\vec k}~,
\label{nexBZ}
\ee
where $N$ is a number of lattice sites and the integral becomes accurate in the thermodynamic
limit $N\to\infty$.

\begin{figure}[t]
\begin{center}
\includegraphics[width=0.7\columnwidth,clip=true]{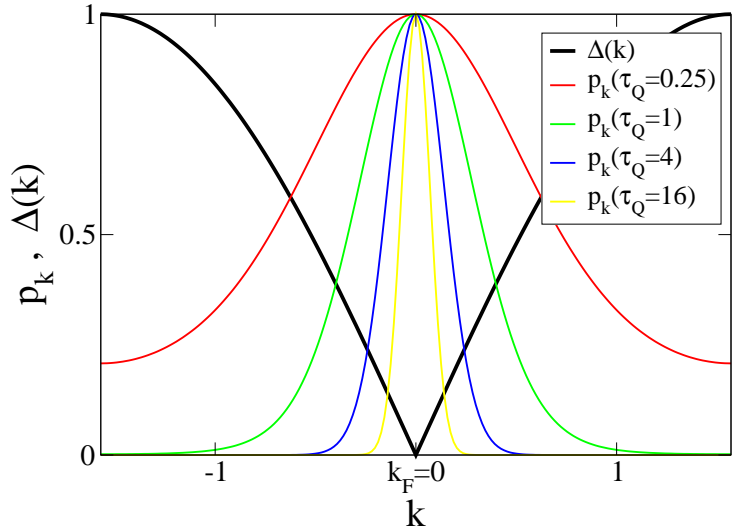}
\caption{ 
The minimal gap function $\Delta(k)=|2\sin(k)|$ and excitation probability 
$p_k=\exp\left(-\frac{\pi\tau_Q\Delta^2(k)}{2}\right)$ 
for different quench times $\tau_Q$ in the one-dimensional transverse field quantum Ising chain in Section \ref{KZIsing}. 
The gap function is zero at the Fermi point $k_F=0$. With increasing $\tau_Q$ the excitation probability becomes
a Gaussian $p_k\approx\exp(-2\pi\tau_Q k^2)$ localised around the Fermi point. Its integral
$n_{\rm ex}=\int \frac{dk}{2\pi} p_k=\frac{1}{2\sqrt{2}\pi}~\tau_Q^{-1/2}$ is the density of excited quasiparticles. 
}
\label{pkIsing}
\end{center}
\end{figure}

In order to evaluate the integral in Eq. (\ref{nexBZ}), we note that in the LZ Hamiltonian 
(\ref{HLZt'}) the minimal instantaneous gap at $t'=0$ is twice $\Delta(\vec k)$. It
vanishes on a ``Fermi manifold'' (point/line/surface) in quasimomentum space where 
\be
\Delta(\vec k)~=~0~. 
\label{Fermi}
\ee
Consequently, Eq. (\ref{pveck}) implies that $p_{\vec k}=1$ only on the Fermi manifold and, in the 
adiabatic limit $\tau_Q\to\infty$, the integrand $p_{\vec k}$ in Eq. (\ref{nexBZ}) remains 
non-negligible only very close to the Fermi manifold, see the examples in Figures \ref{pkIsing} and
\ref{pk_Kitaev}. In this limit the integral (\ref{nexBZ}) can be easily done in a few special cases:

\begin{enumerate}

\item[(i)]

When the quench runs through an isolated critical point between two gapped phases, 
as in Section \ref{KZargument}, then there is an isolated Fermi point $\vec k_F$ 
where $\Delta(\vec k_F)=0$. When the Fermi point is isotropic in $\vec k$-space, then the gap 
vanishes like
\be
\Delta^2(\vec k) ~\sim~ |\vec k-\vec k_F|^{z_\Delta}
\label{zDelta}
\ee
near $\vec k_F$, and the integral in Eq. (\ref{nexBZ}) yields
\be
n_{\rm ex} ~\sim~ \tau_Q^{-d/z_\Delta}~,
\label{LZisolated}
\ee
when $\tau_Q$ is large enough for $p_{\vec k}$ to be localised close enough to $\vec k_F$. 
Examples of case (i) are the transverse field quantum Ising chain in Section \ref{KZIsing}
and the quench across a multicritical point of the transverse field $XY$ chain in 
Section \ref{subsectionmulticritical}.

\item[(ii)]

More generally, when an isolated Fermi point in $d$ dimensions is not isotropic and
has different critical exponents in different directions, then the density of 
excitations scales as
\be
n_{\rm ex} ~\sim~ 
\tau_Q^{-\left(\frac{m\nu_{\-}}{1+\nu_{\-}z_{\-}}+
               \frac{(d-m)\nu_\perp}{1+\nu_\perp z_\perp}\right)}~
\ee
for large enough $\tau_Q$.
Here the exponents $z_{\-},\nu_{\-}$ and $z_\perp,\nu_\perp$ apply in respectively $m$ 
and $(d-m)$ directions. This is the case of the semi-Dirac points in optical lattices
considered in Ref. \cite{semiDirac}.

\item[(iii)]

When the Fermi manifold is an $m$-dimensional surface and the minimal gap 
depends on the distance from the Fermi surface as $\Delta^2(\vec k)\sim|\vec k-\vec k_F|^{z_\Delta}$, 
where $\vec k_F$ is the point on the Fermi surface closest to $\vec k$, then the integration in 
Eq. (\ref{nexBZ}) yields
\be
n_{\rm ex} ~\sim~ \tau_Q^{-(d-m)/z_\Delta}~
\ee
when $\tau_Q$ is large enough. An interesting example is the 2D Kitaev model in 
Section \ref{quenchKitaev} and Ref. \cite{Kitaev}, see Figure \ref{pk_Kitaev}.

\begin{figure}
\begin{center}
\begin{minipage}{100mm}
\subfigure[]{
\resizebox*{5cm}{!}{\includegraphics{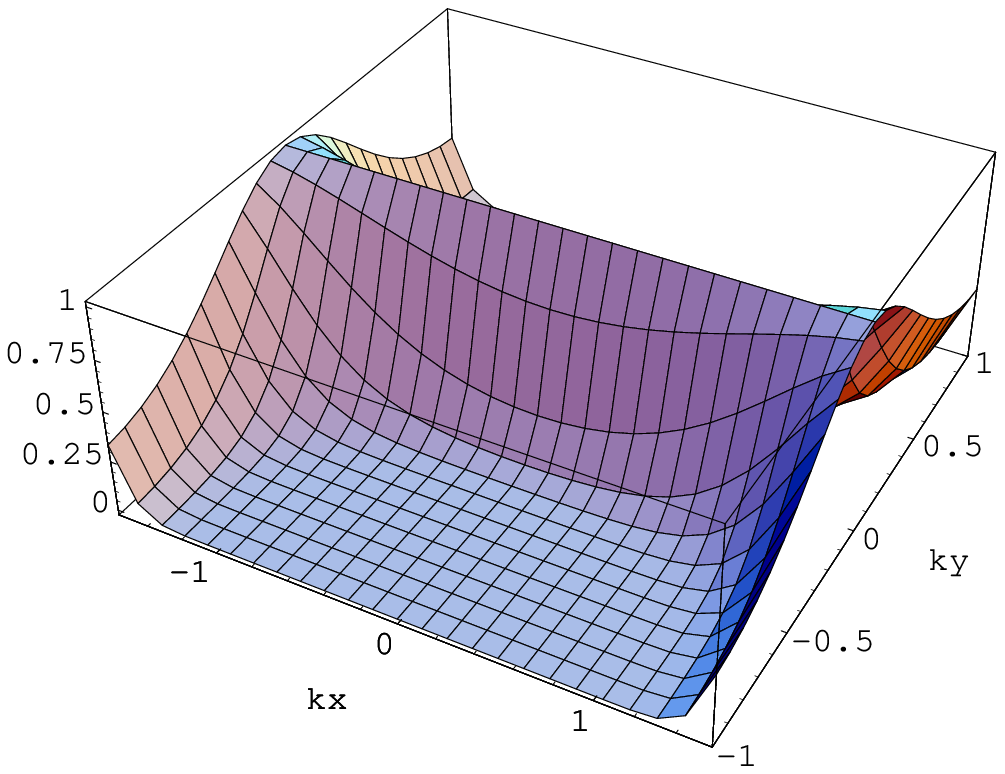}}}%
\subfigure[]{
\resizebox*{5cm}{!}{\includegraphics{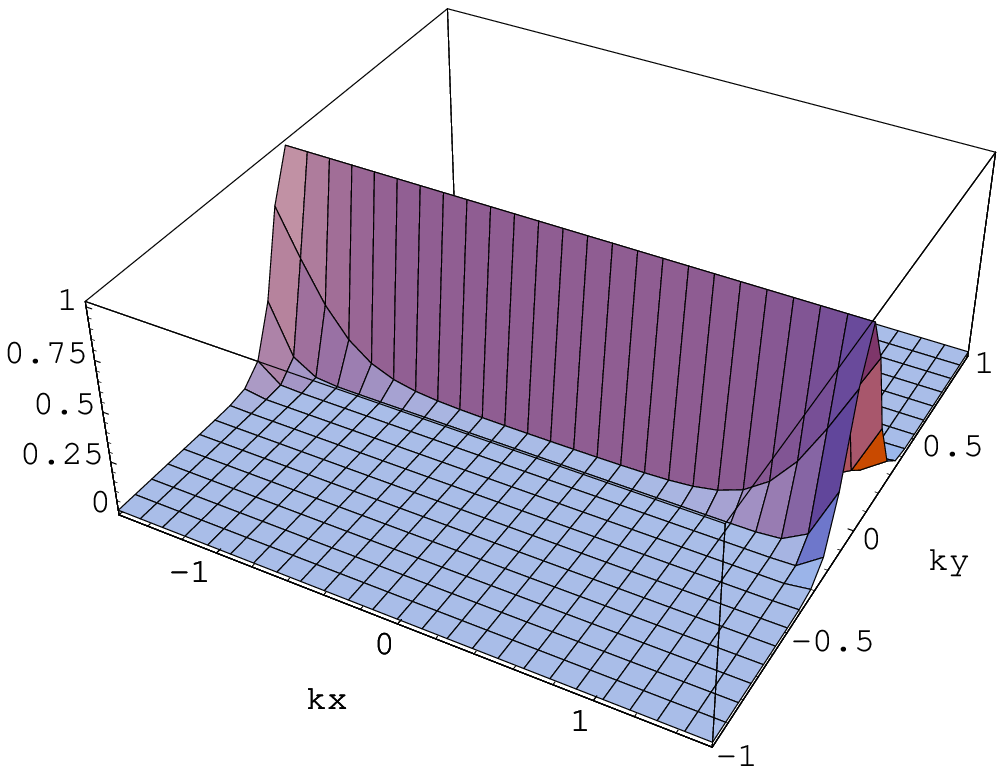}}}%
\caption{ 
In the two-dimensional Kitaev model on a honeycomb lattice in Section \ref{quenchKitaev} the minimal gap function is
$\Delta(\vec k)=2[J_1\sin(\vec k\vec M_1)-J_2\sin(\vec k\vec M_2)]$, where
$\vec M_1=\frac{\sqrt{3}}{2}\hat i+\frac32\hat j$ and $\vec M_2=\frac{\sqrt{3}}{2}\hat i-\frac32\hat j$ are spanning 
vectors of triangular reciprocal lattice. Here we assume $J_x=J_y=1$ so that the Fermi line is $k_y=0$.
The 3D plots show excitation probabilities $p_{\vec k}$ for $\tau_Q=1$ (left panel) and $\tau_Q=16$ (right panel).
With increasing $\tau_Q$ the excitation probability localizes on the Fermi line. 
}
\label{pk_Kitaev}
\end{minipage}
\end{center}
\end{figure}

\item[(iv)]

When, as in case (iii), the Fermi manifold is an $m$-dimensional surface with
$\Delta^2(\vec k)\sim|\vec k-\vec k_F|^{z_\Delta}$ near a generic point $\vec k_F$ 
on the surface, but there is an isolated ``dominant'' Fermi point $\vec k_F^D$ where
$\Delta^2(\vec k)\sim|\vec k-\vec k_F^D|^{z^D_\Delta}$ with $z^D_\Delta<z_\Delta$, 
then the integral in Eq. (\ref{nexBZ}) is dominated by a neighbourhood of the dominant 
Fermi point and
\be
n_{\rm ex} ~\sim~ \tau_Q^{-d/z^D_\Delta}~
\ee 
when $\tau_Q$ is large enough. An example can be found in Ref. \cite{Viola2}.

\end{enumerate}

Case (i) is the problem treated in Section \ref{KZargument} by the standard KZM, while cases (ii-iv) 
are generalisations that go beyond that simple argument. However, even in case (i) we obtain the 
density of excitations (\ref{LZisolated}) with an exponent $d/z_\Delta$ which in principle can be 
different than the exponent $d\nu/(1+\nu z)$ in the KZM equation (\ref{KZargument}), unless
\be
z_\Delta~\stackrel{?}{=}~\frac{1+\nu z}{\nu}~.
\label{zDelta?}
\ee   
Here $z_\Delta$ on the left hand side characterizes the minimal gap function $\Delta(\vec k)$ while 
the critical exponents $z,\nu$ on the right hand side characterize the critical point. 

To see if the relation (\ref{zDelta?}) holds, we return to the Hamiltonian (\ref{HabDelta}) with the 
quasiparticle spectrum (\ref{spectrumDelta}). Without loss of generality, we can assume that $b(\vec k_F)=0$ 
at the Fermi point 
\footnote{
This can be always achieved by a $\vec k$-independent shift $\epsilon'=\epsilon+b(\vec k_F)/a(\vec k_F)$.
}
and its asymptote near $\vec k_F$ is
\be
b^2(\vec k) ~\sim~ |\vec k-\vec k_F|^{z_b}~.
\label{zb}
\ee
On one hand, the critical exponents $z$ and $\nu$ are defined by the asymptotes near the Fermi 
point: $\omega_{\vec k}\sim|\vec k-\vec k_F|^z$ at $\epsilon=0$ and 
$\omega_{\vec k_F}\sim|\epsilon|^{\nu z}$ for small $\epsilon$, see Ref. \cite{Sachdev}. 
On the other hand, the actual asymptotes of the spectrum (\ref{spectrumDelta}) are
$\omega_{\vec k}\sim|\vec k-\vec k_F|^{\frac12{\rm min}(z_b,z_\Delta)}$ 
at $\epsilon=0$ and $\omega_{\vec k_F}\sim|\epsilon|^1$ for small $\epsilon$. 
Comparing the definitions with the actual asymptotes we obtain $z=\frac12{\rm min}(z_b,z_\Delta)$ and
$\nu z=1$. The equality (\ref{zDelta?}) holds if 
\be
z_\Delta ~\leq~  z_b~
\label{zDeltazb}
\ee
or, equivalently, near the Fermi point the function $b^2(\vec k)$ is negligible as compared to $\Delta^2(\vec k)$.
The inequality is the condition for the KZM argument in Section \ref{KZargument} to give 
the same scaling as the exact solution in case (i). 

This condition may be not quite surprising because when we neglect $b^2(\vec k)$, then for each $\vec k$ the 
instantaneous gap in Eq. (\ref{spectrumDelta}) is minimised at the critical point $\epsilon=0$. Since the critical 
point is the anti-crossing center for each $\vec k$, the dynamical excitation of each two-level system takes
place near the critical point, so the excitation probabilities $p_{\vec k}$ must be determined by the critical 
exponents that characterize the quasiparticle spectrum near this critical point. Given the accumulated evidence, the 
condition (\ref{zDeltazb}) seems to be satisfied by most quenches across an isolated quantum critical point, but 
it is not fulfilled by quenches across a multicritical point considered in Refs. \cite{multicritical,Viola3} and 
Section \ref{subsectionmulticritical}. 

For instance, in the quench considered in Section \ref{subsectionmulticritical} one finds $z_\Delta=6,z_b=4$
and $z=1/\nu=2$, the inequality (\ref{zDeltazb}) is not satisfied, and the exact scaling 
$n_{\rm ex}\sim\tau_Q^{-1/6}$ in Eq. (\ref{LZisolated}) is different than the $n_{\rm ex}\sim\tau_Q^{-1/4}$ 
predicted by Eq. (\ref{nex}). In this example $b^2(\vec k)$ is not negligible, so it shifts the anti-crossings 
away from the critical point, and the excitation probabilities $p_{\vec k}$ are dominated by non-critical energy 
modes away from $\epsilon=0$. The shifts, and the anomalous exponent $1/6$, originate in the fact that the dynamical 
excitation process takes place asymmetrically with respect to the multicritical point. As a consequence, dynamical 
scaling may require introducing new non-static exponents
\footnote{ Indeed, the effective Hamiltonian (\ref{HLZt'}) has
instantaneous spectrum $\omega'_{\vec k}=\sqrt{(\epsilon')^2+\Delta^2}$, where $\epsilon'=t'/\tau_Q$.
From this spectrum we formally obtain effective exponents $z'=z_{\Delta}/2$ and $\nu'z'=1$. 
With these effective exponents $z'$ and $\nu'$ in place of the ``canonical'' $z$ and $\nu$, the KZM 
Eq. (\ref{nex}) gives the exact scaling in Eq. (\ref{LZisolated}).}
  
In this Section we obtained exact scalings by mapping a class of integrable fermionic systems
to the LZ model. In the next Section we rederive the KZM scaling in Eq.(\ref{nex}) from 
the adiabatic perturbation theory. This approximate derivation does not seem limited to integrable 
models.

\subsection{ KZM from adiabatic perturbation theory }
\label{argumentPolkovnikov}

An alternative derivation of the KZ scaling in Eq. (\ref{nex}) was presented in 
Refs. \cite{Polkovnikov2005,infinitezimalPolkovnikov}. In the integrable models equivalent 
to non-interacting fermions considered in the previous Section, this derivation is 
a perturbative approximation to the exact LZ argument, but its validity does not seem 
limited to integrable models. The argument is as follows.

Let the set of functions $\phi_p(\epsilon)$ represent an eigenbasis of a Hamiltonian 
$H(\epsilon)$ depending on a parameter $\epsilon$. A wave function of the system can be 
expanded in the eigenbasis
\be
\psi ~=~ \sum_p a_p(r) \phi_p(\epsilon)~.
\ee
We assume the linear $\epsilon(t)=t/\tau_Q$ in Eq. (\ref{tauQ}) with a large  
$\tau_Q\to\infty$. The Schr\"odinger equation is equivalent to
\be
i\frac{da_p}{dt}+
\frac{i}{\tau_Q}\sum_{q}a_q
\langle p|\frac{d}{d\epsilon}|q\rangle=
\omega_p~a_p~,
\ee
where $\omega_p(\epsilon)$ is the eigenenergy of the Hamiltonian $H(\epsilon)$
corresponding to the eigenstate $\phi_p(\epsilon)$ (or $|p\rangle$).
After a unitary transformation eliminating the dynamical phase
\be
a_p(t)~=~
\tilde{a}_p(t)~e^{-i\int^t dt'~\omega_p[\epsilon(t')]}~=~
\tilde{a}_p(\epsilon)~e^{-i\tau_Q\int^\epsilon d\epsilon'~\omega_p(\epsilon')}~,  
\ee
the Schr\"odinger equation becomes
\be
\frac{d\tilde{a}_p}{d\epsilon}~=~
-\sum_q \tilde{a}_q(\epsilon)
\langle p|\frac{d}{d\epsilon}|q \rangle~
e^{i\tau_Q\int^\epsilon d\epsilon'~[\omega_p(\epsilon')-\omega_q(\epsilon')]}~. 
\ee
Since the system were initially prepared in the ground state $|0\rangle$, then in the 
adiabatic limit the single term $q=0$ dominates the sum and the excitation probability 
is given by
\be
P_{\rm ex} ~\approx~
\sum_{p\neq 0}
\left|
\int_{-\infty}^{\infty} d\epsilon~
\langle p|\frac{d}{d\epsilon}|0 \rangle~
e^{i\tau_Q\int^\epsilon d\epsilon' [\omega_p(\epsilon')-\omega_0(\epsilon')]}
\right|^2~.
\label{Pex1}
\ee
The derivative of the ground state $\frac{d}{d\epsilon}|0\rangle$ is finite at 
a continuous phase transition.

We assume a uniform $d$-dimensional system with a single (relevant) branch of
excitations which can be labelled by (quasi-)momentum $\vec k$. The 
branch has a dispersion relation with a gap $\Delta(\epsilon)$. The gap is 
non-zero everywhere except the critical point at $\epsilon=0$, where it vanishes like 
$\Delta\sim|\epsilon|^{\nu z}$ and the excitation $\vec k=0$ is gapless. 
Conservation of momentum implies that only pairs of excitations
with opposite momenta $(\vec k,-\vec k)$ can be excited by the time-dependent 
$\epsilon$. In this framework, Eq. (\ref{Pex1}) gives the following density of 
excitations
\bea
n_{\rm ex}    &\approx&
\int \frac{d^dk}{(2\pi)^d}~
p_{\vec k}~,                      
\label{nex0}\\
p_{\vec k} &\approx&
\left|
\int_{-\infty}^{\infty} d\epsilon~
\langle \vec k,-\vec k|\frac{d}{d\epsilon}|0 \rangle~
e^{i\tau_Q\int^\epsilon_{-\infty} d\epsilon'~ 
   [\omega_{\vec k}(\epsilon')-\omega_0(\epsilon')]}
\right|^2~,
\label{nex1}
\eea
Here $|\vec k,-\vec k\rangle$ is an excited state with one pair of excitations of
opposite momenta $\pm \vec k$. 

A general scaling argument implies that
\be
\omega_{\vec k}(\epsilon)-\omega_0(\epsilon) ~=~
\Delta ~ F(\Delta/k^z)     ~=~
|\epsilon|^{z\nu} ~ \tilde{F}(|\epsilon|^{z\nu}/k^z)~, 
\label{scalingF}
\ee
where $k=|\vec k|$ and $F$ (or $\tilde{F}$) is a non-universal function with a universal tail
$F(x)\sim 1/x$ for large $x$. Substituting Eq. (\ref{scalingF}) to Eq. (\ref{nex1}) we 
obtain
\be
p_{\vec k} \approx
\left|
\int_{-\infty}^{\infty} dx~
\langle \vec k,-\vec k|\frac{d}{dx}|0 \rangle~
e^{i\left( \tau_Q k^{\frac{1+z\nu}{\nu}} \right)
\int_{-\infty}^x dx'~|x'|^{z\nu}~\tilde{F}(|x'|^{z\nu})}
\right|^2~,
\label{pex1}
\ee
where $x=k^{-1/\nu}\epsilon$ and $x'=k^{-1/\nu}\epsilon'$. Another scaling argument
\be
\langle \vec k,-\vec k|\frac{\partial}{\partial\Delta}|0\rangle=
\frac{1}{k^z}G(\Delta/k^z)
~~\to~~
\langle \vec k,-\vec k|\frac{\partial}{\partial\epsilon}|0\rangle=
{\rm sign}(\epsilon)
\frac{|\epsilon|^{z\nu-1}}{k^z}\tilde{G}(|\epsilon|^{z\nu}/k^z)
\ee
transforms Eq. (\ref{pex1}) into
\be
p_{\vec k} \approx
\left|
\int_{-\infty}^{\infty} dx~
{\rm sign}(x)~|x|^{z\nu-1}~\tilde{G}\left(|x|^{z\nu}\right)
e^{i\left( \tau_Q k^{\frac{1+z\nu}{\nu}} \right)
\int_{-\infty}^x dx~|x'|^{z\nu}~\tilde{F}(|x'|^{z\nu})}
\right|^2~,
\label{pex2}
\ee
which depends on $k$ only through the dimensionless combination $\tau_Qk^{\frac{1+z\nu}{\nu}}$. This
means that there is a characteristic wave vector
\be
\hat k ~\simeq~ \tau_Q^{-\frac{\nu}{1+z\nu}}~,
\ee
corresponding to the KZ wavelength $\hat\xi\simeq \hat k^{-1}$ in Eq. (\ref{hatxi}).
The $\hat k$ marks a crossover between a regime of large $k\gg\hat k$, where
the adiabatic approximation is self-consistent and Eq. (\ref{pex2}) accurately 
predicts small $p_k$, and a regime of small $k\ll\hat k$ where we do not expect 
Eq. (\ref{pex2}) to be accurate. It is expected to be an overestimate for 
fermionic excitations and an underestimate for bosons. 

Since $\hat k$ is the only momentum scale in Eq. (\ref{nex0}), then for dimensional reasons
the integration yields
\be
n_{\rm ex}~\simeq~\hat k^d~\simeq~\hat\xi^{-d}~\sim~\tau_Q^{-\frac{d\nu}{1+z\nu}}
\label{nexPolkovnikov}
\ee
up to possible logarithmic corrections \cite{Polkovnikov2005}. This is again the 
KZ equation (\ref{nex}). 

This Section completes our general discussion of KZM. The following Sections provide
specific examples in a number of model systems.

\subsection{ Quantum Ising chain: 
             transition across an isolated critical point }
\label{KZIsing}

\begin{figure}[t]
\begin{center}
\includegraphics[width=0.5\columnwidth,clip=true]{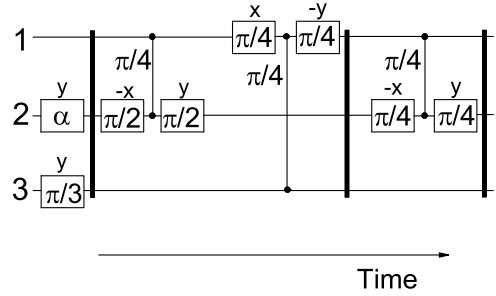}
\caption{
NMR gate sequence to prepare an effective pure state $|\rightarrow\rightarrow\rightarrow\rangle$
in the quantum Ising chain (\ref{Hsigma}) by spatial averaging from thermal equilibrium state.
(Figure from Ref. \cite{IsingNMR})
}
\label{FigIsingNMR}
\end{center}
\end{figure}

\begin{figure}[t]
\begin{center}
\includegraphics[width=0.5\columnwidth,clip=true]{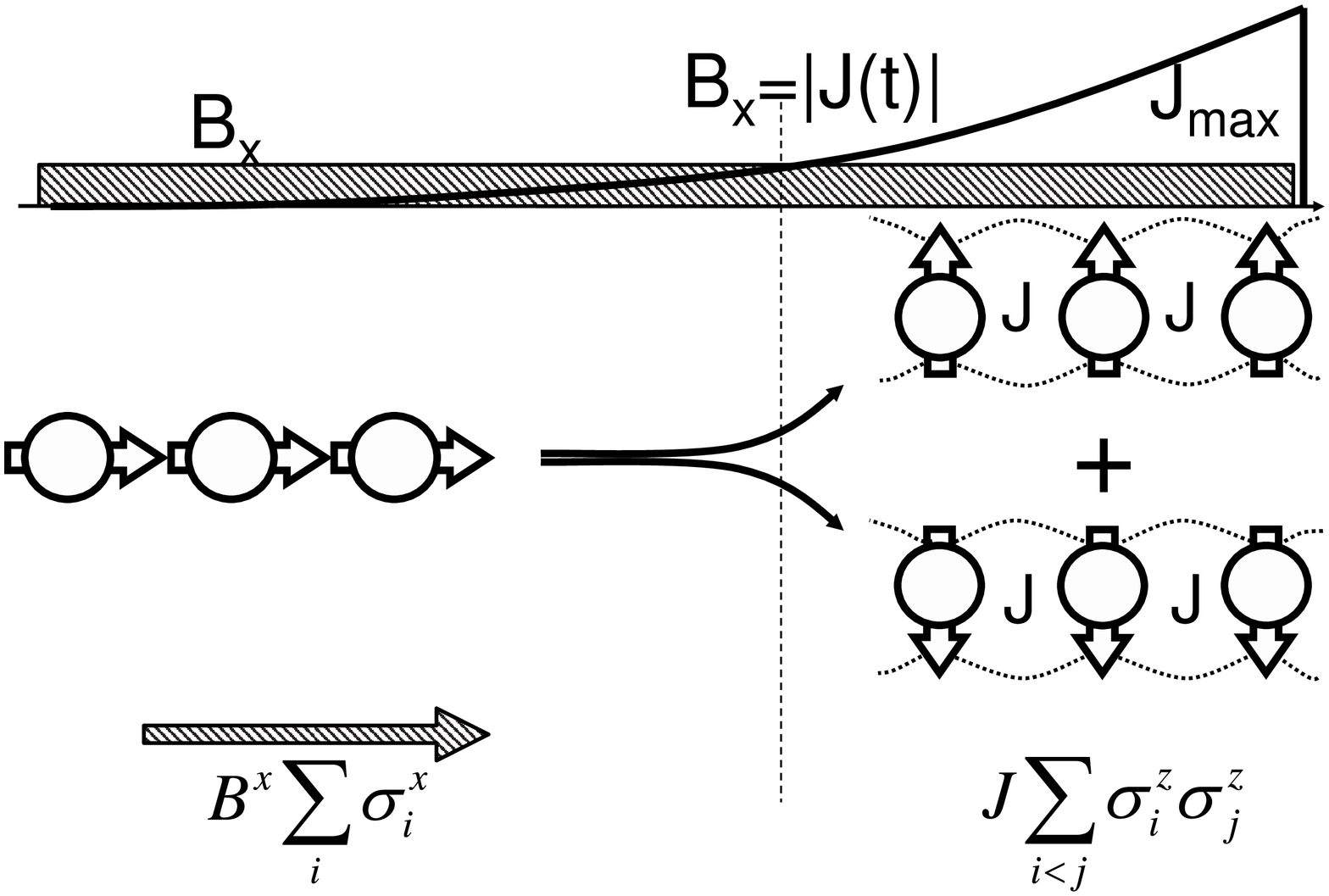}
\caption{
Schematic picture of the ion trap experiment in Ref. \cite{IsingIon}. By adiabatically
increasing effective spin-spin interaction an initial paramagnetic state is driven into
a superposition of two ferromagnetic states with opposite magnetisation. 
(Figure from Ref. \cite{IsingIon})
}
\label{FigIsingIon}
\end{center}
\end{figure}

Much of our understanding of quantum phase transitions is based on the 
prototypical integrable quantum Ising chain \cite{Sachdev} 
\be
H~=~-\sum_{n=1}^N \left( g~\sigma^z_n + \sigma^x_n\sigma^x_{n+1} \right)~
\label{Hsigma}
\ee
with periodic boundary conditions
$
\sigma_{N+1}~=~\sigma_1~.
$
In the thermodynamic limit $N\to\infty$ the model has two critical points at
$g_c=\pm 1$ between a ferromagnetic phase when $|g|<1$ and two paramagnetic
phases when $|g|>1$. Their critical exponents are $z=\nu=1$. Today the quantum Ising 
chain is not only an exactly solvable toy model, but it is also becoming a subject of
quantum simulation in experiments: the NMR simulation in Ref. \cite{IsingNMR} and the ion trap experiment in 
Ref. \cite{IsingIon}, see Figs. \ref{FigIsingNMR} and \ref{FigIsingIon} respectively.

Here, like in Refs. \cite{KZIsing,Dziarmaga2005,CherngLevitov,Cincio}, we consider a linear ramp 
\be
g(t)~=~-\frac{t}{\tau_Q}~
\label{glinear}
\ee
with $t$ running from $-\infty$ to $0$ across the critical point at $g_c=1$. The quench begins 
at $g\to\infty$ in the ground state $|\uparrow\uparrow\uparrow\dots\uparrow\rangle$ with all spins 
polarized along the $z$-axis. At the final $g=0$ there are two degenerate 
ferromagnetic ground states with all spins pointing either left or right along the $x$-axis:
$|\rightarrow\rightarrow\rightarrow\dots\rightarrow\rangle$ or
$|\leftarrow\leftarrow\leftarrow\dots\leftarrow\rangle$. 
In an adiabatic classical transition from the paramagnetic 
to ferromagnetic phase, the system would choose one of the two ferromagnetic states. In the 
analogous quantum case, any superposition of these two states is also a `legal'  
ground state providing it is consistent with other quantum numbers 
conserved by the transition from the initial paramagnetic state.

However, when $N\to\infty$, then energy gap at $g=1$ tends to zero (quantum version of the
critical slowing down) and it is impossible to pass the critical point at a finite 
speed without exciting the system. As a result, the system ends in a quantum 
superposition of states like
\be
|\dots
\rightarrow
\leftarrow\leftarrow\leftarrow\leftarrow\leftarrow
\rightarrow\rightarrow\rightarrow\rightarrow\rightarrow\rightarrow\rightarrow
\leftarrow\leftarrow\leftarrow\leftarrow
\rightarrow\rightarrow\rightarrow\rightarrow\rightarrow\rightarrow
\leftarrow
\dots\rangle
\label{domains}
\ee
with finite domains of spins pointing left or right and separated by kinks where the
polarisation of spins changes its orientation. Average size of the domains or,
equivalently, average density of kinks depends on the transition rate. When
the transition is slow, then the domain size is large, but when it is very fast, 
then orientation of individual spins can become random, uncorrelated with their 
nearest neighbours. 

When we define a dimensionless parameter
\be 
\epsilon ~=~ \frac{g-g_c}{g_c} ~=~ g-1~,
\ee
then according to KZM in Section \ref{KZargument} the evolution is
adiabatic before $\hat\epsilon\simeq\sqrt{\tau_Q}$ and after $-\hat\epsilon$, and impulse
near the critical point. The correlation length in the adiabatic ground 
state at $\hat\epsilon$ is proportional to
\be
\hat\xi~=~\tau_Q^{1/2}~.
\label{hatxiIsing}
\ee
This ground state is (approximately) the initial state for the last adiabatic 
stage of the evolution after $-\hat\epsilon$. This argument shows that when 
passing across the critical point, the state of the system gets imprinted with 
a finite KZ correlation length proportional to $\hat\xi$. In particular, this 
coherence length determines average density of kinks after the transition as 
\be
n_{\rm ex}~\simeq~\hat\xi^{-1}~=~\tau_Q^{-1/2}~.
\label{KZscalingIsing}
\ee
This is an order of magnitude estimate with a pre-factor $\simeq1$. 

In the following we derive an exact solution \cite{Dziarmaga2005,CherngLevitov,Cincio} 
not only to confirm the scaling exponent in Eq. (\ref{KZscalingIsing}) and provide an unknown 
numerical pre-factor, but also to see how much the state after the transition resembles the adiabatic 
ground state at $\hat\epsilon$. In the adiabatic limit $\tau_Q\to\infty$ this near-critical ground state 
has a divergent correlation length $\hat\xi$ which, by the standard scaling hypothesis of the
renormalisation group, determines all physical observables according to their scaling dimensions.

\subsubsection{ Landau-Zener argument }

Here we assume that $N$ is even for convenience. After the non-local
Jordan-Wigner transformation \cite{JW},
\bea
&&
\sigma^x_n~=~
 \left( c_n^\dagger + c_n \right)
 \prod_{m<n}(1-2 c^\dagger_m c_m)~, \\
&&
\sigma^y_n~=~
 i
 \left( c_n^\dagger - c_n \right)
 \prod_{m<n}(1-2 c^\dagger_m c_m)~, \\
&&
\sigma^z_n~=~1~-~2 c^\dagger_n  c_n~, 
\label{JordanWigner}
\eea
introducing fermionic operators $c_n$ which satisfy
$\left\{c_m,c_n^\dagger\right\}=\delta_{mn}$ and 
$\left\{ c_m, c_n \right\}=\left\{c_m^\dagger,c_n^\dagger \right\}=0$
the Hamiltonian (\ref{Hsigma}) becomes \cite{LSM}
\be
 H~=~P^+~H^+~P^+~+~P^-~H^-~P^-~.
\label{Hc}
\ee
Above
\be
P^{\pm}=
\frac12\left[1\pm P\right]
\label{Ppm}
\ee
are projectors on subspaces with even ($+$) and odd ($-$) parity
\be 
P~=~
\prod_{n=1}^N\sigma^z_n ~=~
\prod_{n=1}^N\left(1-2c_n^\dagger c_n\right) ~
\ee
and  
\bea
H^{\pm}~=~
\sum_{n=1}^N
\left( 
2g c_n^\dagger  c_n - c_n^\dagger  c_{n+1} - c_{n+1}^\dagger  c_n - c_{n+1}  c_n - c_n^\dag c_{n+1}^\dag - g
\right)~
\label{Hpm}
\eea
are corresponding reduced Hamiltonians. The $c_n$'s in $H^-$ satisfy
periodic boundary condition $c_{N+1}=c_1$, but the $c_n$'s in $H^+$
obey $c_{N+1}=-c_1$ - what we call ``antiperiodic'' boundary conditions. 

The parity $P$ of the number of $c$-quasiparticles is a good quantum number and the ground
state has even parity for any $g\neq0$. Assuming that a quench begins in 
the ground state, we can confine to the subspace of even parity. $H^+$ is diagonalised 
by a Fourier transform followed by a Bogoliubov transformation \cite{LSM}. The Fourier 
transform consistent with the antiperiodic boundary condition is  
\be
c_n~=~ 
\frac{e^{-i\pi/4}}{\sqrt{N}}
\sum_k c_k e^{ikn}~,
\label{Fourier}
\ee
where the pseudomomentum $k$ takes ``half-integer'' values
\be
k~=~
\pm \frac12 \frac{2\pi}{N},
\dots,
\pm \frac{N-1}{2} \frac{2\pi}{N}~.
\label{halfinteger}
\ee
The Hamiltonian (\ref{Hpm}) becomes
\bea
H^+~ ~=~
\sum_k
\left\{
2[g-\cos k] c_k^\dagger c_k +
\sin k
\left[ 
 c^\dagger_k c^\dagger_{-k}+
 c_{-k} c_k
\right]-
g
\right\}~.
\label{Hck}
\eea
Diagonalisation of $H^+$ is completed by a Bogoliubov transformation
\be
c_k ~=~ u_k  \gamma_k + v_{-k}^*  \gamma^\dagger_{-k}~,
\label{Bog}
\ee
provided that Bogoliubov modes $(u_k,v_k)$ are eigenstates of the stationary
Bogoliubov-de Gennes equations
\bea
\omega 
\left(
\begin{array}{c}
u_k \\
v_k
\end{array}
\right) &=&
\left[
\begin{array}{cc}
2(g-\cos k) &  2\sin k     \\
2\sin k     & -2(g-\cos k)
\end{array}
\right]
\left(
\begin{array}{c}
u_k \\
v_k
\end{array}
\right)~. 
\label{stBdG}
\eea
There are two eigenstates for each $k$ with energies $\omega=\pm\omega_k$, where  
\be
\omega_k~=~2\sqrt{(g-\cos k)^2+\sin^2 k}~.
\label{epsilonkIsing}
\ee
The positive energy eigenstate 
\be
(u_k,v_k)~=~\left(\cos\frac{\theta_k}{2},\sin\frac{\theta_k}{2}\right)~,
\label{uvplus}
\ee
where $\tan\theta_k=\sin k/(g-\cos k)$, defines the quasiparticle operator 
\be 
\gamma_k~=~u_k^* c_k + v_{-k}^* c_{-k}^\dagger~, 
\label{gammak}
\ee 
and the negative energy eigenstate 
$
(u^-_k,v^-_k)=(-v_k,u_k)~ 
\label{uv-}
$
defines 
$
\gamma_k^-~=~(u_k^-)^*c_k+(v_{-k}^-)^*c_{-k}^\dagger=-\gamma_{-k}^\dagger
$. 
After the Bogoliubov transformation, the Hamiltonian 
$
H^+=
\frac12
\sum_k \omega_k
\left( 
\gamma_k^\dagger \gamma_k - \gamma_k^{-\dagger} \gamma_k^-    
\right)
$
is equivalent to
\be
H^+~=~
\sum_k \omega_k~
\left(
\gamma_k^\dagger \gamma_k-\frac12
\right)~.
\label{Hgamma}
\ee
This is a simple-looking sum of quasiparticles with half-integer pseudomomenta. 
However, thanks to the projection $P^+~H^+~P^+$ in Eq.~(\ref{Hc}) only states 
with even numbers of quasiparticles belong to the spectrum of $H$.   

In the linear quench (\ref{glinear}), the system is initially in its ground state at 
large initial value of $g\gg 1$, but as $g$ is ramped 
down to zero, the system gets excited from its instantaneous ground state
and, in general, its final state at $t=0$ has finite number of kinks. Comparing 
the Ising Hamiltonian (\ref{Hsigma}) at $g=0$ with the Bogoliubov Hamiltonian 
(\ref{Hgamma}) at $g=0$ we obtain a simple expression for the operator of the 
number of kinks 
\bea
{\cal N} ~\equiv~ \frac12 \sum_{n=1}^{N} 
                  \left(1-\sigma^z_n\sigma^z_{n+1}\right)~=~
                  \sum_k \gamma_k^\dagger \gamma_k~.   
\label{calN}
\eea 
The number of kinks is equal to the number of quasiparticles excited at $g=0$. 
The excitation probability 
\be
p_k~=~\langle\gamma_k^\dagger\gamma_k\rangle~
\ee
in the final state at $g=0$ can be found with the time-dependent 
Bogoliubov method. 

The initial ground state at $g\to\infty$ is a Bogoliubov vacuum $|0\rangle$ annihilated by all 
quasiparticle operators $\gamma_k$ determined in Eq. (\ref{gammak}) by the positive-$\omega$ 
eigenmodes $(u_k,v_k)\approx(1,0)$ of Eq. (\ref{stBdG}). As $g(t)$ is ramped down, the 
quantum state gets excited from the instantaneous ground state. 
We work in the Heisenberg picture where the state remains the initial Bogoliubov vacuum
and the operators $\gamma_k$ do not depend on time, but the operators $c_k$ 
are time-dependent
\be
c_k ~=~ u_k(t) \gamma_k + v_{-k}^*(t) \gamma^\dagger_{-k}~,
\label{tildeBog}
\ee
with the initial condition $[u_k(-\infty),v_k(-\infty)]=(1,0)$. They satisfy Heisenberg equations
$
i\frac{d}{dt} c_k~=~\left[ c_k, H^+\right]
$
equivalent to a time-dependent version of the Bogoliubov-de Gennes equations (\ref{stBdG}):
\bea
i\frac{d}{dt} 
\left(
\begin{array}{c}
u_k \\
v_k
\end{array}
\right) &=&
\left[
\begin{array}{cc}
2[g(t)-\cos k] &  2\sin k        \\
2\sin k        & -2[g(t)-\cos k]
\end{array}
\right]
\left(
\begin{array}{c}
u_k \\
v_k
\end{array}
\right)~. 
\label{dynBdG}
\eea
These equations are an example of the general equations (\ref{generalBdG}). 

Comparing equations (\ref{generalBdG},\ref{HabDelta}) and (\ref{dynBdG}) with $g(t)=1+\epsilon(t)$,
we can identify $a(k)=2$, $b(k)=2(\cos k-1)$, 
\be
\Delta(k)~=~2\sin k~,
\ee 
$\hat\sigma(k)=-\sigma^x$, and $\hat\sigma_{\perp}(k)=\sigma^z$. There are two Fermi points 
$k_F=0,\pi$ where $\Delta(k_F)=0$ which correspond to the critical points $g_c=1,-1$ respectively. 
Modes near $k_F=\pi$ do not get excited in our quench from $g\to\infty$ through $g_c=1$ to $g=0$, so 
we can focus on $k_F=0$ only. Expansion in $k-k_F$ like in Eqs. (\ref{zDelta},\ref{zb}) 
yields $z_\Delta=2$ and $z_b=4$ and the LZ argument in Section \ref{LZargument} case (i) yields
\be
n_{\rm ex} ~\simeq~ \tau_Q^{-1/z_{\Delta}} ~=~ \tau_Q^{-1/2}~
\ee
for $\tau_Q\gg 1$. This is the same as the KZ prediction (\ref{KZscalingIsing}) because the condition 
$z_\Delta\leq z_b$ in Eq. (\ref{zDeltazb}) is satisfied. 

What is more, with the minimal gap function $\Delta\approx2|k|$ near $k_F=0$ we have
\be
p_k ~\approx~ e^{-2\pi\tau_Qk^2}~,
\label{LZpk}
\ee 
localised near $k=0$ when $\tau_Q\gg1$, see Figure \ref{pkIsing}, and the integral in Eq. (\ref{nexBZ}) yields
the density of excitations
\be
n_{\rm ex}~=~\frac{1}{2\sqrt{2}\pi}~\tau_Q^{-1/2}~.
\label{nexIsing}
\ee    
The numerical pre-factor is $\frac{1}{2\sqrt{2}\pi}=0.113$. For comparison, in Ref. \cite{Polkovnikov2005} the 
perturbative Eqs. (\ref{nex0},\ref{nex1}) were used to estimate the same density of defects as
$
n_{\rm ex}~\approx~0.124~\tau_Q^{-1/2}~.
$
The scaling is the same as in the exact solution (\ref{nexIsing}) but, as a result
of the perturbative approximation in Eq. (\ref{nex1}), the numerical pre-factor
is overestimated as expected for fermionic quasiparticles, see Section \ref{argumentPolkovnikov}.

\subsubsection{ Adiabatic transition in a finite chain }
\label{FiniteIsing}

No matter how long $\tau_Q$ is, a quench across a critical point 
in an infinite system is not adiabatic because the energy gap at the 
critical point is zero. However, in a finite system the gap is finite
and we can expect the quench to become adiabatic for a sufficiently long
$\tau_Q$, see Section \ref{KZAdiab}. For instance, in the periodic Ising chain the (relevant) energy 
gap at $g_c=1$ for excitation of the lowest two energy quasiparticles with $k=\pm\pi/N$ 
is $\omega_{+\pi/N}+\omega_{-\pi/N}=4\pi/N$ and decays like $1/N$.
Since $p_k$ is a probability to excite a pair of quasiparticles $(+k,-k)$
and different pairs evolve independently, a probability for the state
to follow the adiabatic ground state without any quasiparticles is a product
\be
P_{\rm GS}~=~
\prod_{k>0}\left(1-p_k\right)~
\label{calP}
\ee
over the positive ``half-integer'' $k=(m+1/2)2\pi/N$ in Eq. (\ref{halfinteger}). 
Well on the adiabatic side 
$p_{\pi/N}~\approx~\exp\left(-2\pi^3\frac{\tau_Q}{N^2}\right)\ll1$ 
is exponentially small and
$p_{(1+2m)\pi/N}~\approx~(p_{\pi/N})^{(1+2m)^2}~\ll~p_{\pi/N}$ for $m>0$,
so we can safely approximate
\be
P_{\rm GS}~=~
1-p_{\pi/N}~+~{\cal O}\left(p_{\pi/N}^9\right)\approx~
1-\exp\left(-2\pi^3\frac{\tau_Q}{N^2}\right)~.
\label{calPapprox}
\ee
This $P_{\rm GS}\approx 1$ and the quench in a finite chain is adiabatic when
\be
\tau_Q~\gg~\frac{N^2}{2\pi^3}~,
\label{nonadiabatic}
\ee
i.e., when even the lowest energy pair of quasiparticles with $k=\pm\pi/N$
is not likely to get excited. Reading this inequality from right to left, the 
size $N$ of a defect-free chain grows like $\tau_Q^{1/2}$. This is consistent 
with Eqs. (\ref{hatxiIsing}) and (\ref{nexIsing}).

\subsubsection{ Exact solution of the time-dependent Bogoliubov-de Gennes equations }

If we want to go beyond the simple result for $n_{\rm ex}$ in Eq. (\ref{nexIsing}), then we need 
an exact solution of Eq. (\ref{dynBdG}). When $k>0$ a new time variable
\be
\tau~=~4\tau_Q\sin k\left(\frac{t}{\tau_Q}+\cos k\right)
\ee
brings Eqs. (\ref{dynBdG}) to the canonical LZ form (\ref{HLZ}) 
\bea
i\frac{d}{d\tau}
\left(
\begin{array}{c}
u_k \\
v_k
\end{array}
\right) &=&
\frac12
\left[
\begin{array}{cc}
-R_k\tau & 1        \\
1        & R_k\tau
\end{array}
\right]
\left(
\begin{array}{c}
u_k \\
v_k
\end{array}
\right)~,
\label{BdGLZ}
\eea
with a transition rate $R_k=1/4\tau_Q\sin^2 k$. 
A general solution of the Landau-Zener equations (\ref{BdGLZ}), see Section \ref{LZvKZ}, is
\bea
v_k(\tau)&=&
a D_{-s-1}(-iz) + b D_{-s-1}(iz) ~, \nonumber\\
u_k(\tau)&=&
\left(\tau R_k-2i\frac{\partial}{\partial\tau}\right)
v_k(\tau) ~,
\label{general} 
\eea
with arbitrary complex parameters $a,b$. Here $D_m(x)$ is a Weber function, 
$s=\frac{1}{4iR_k}$, and $z=\sqrt{R_k}\tau e^{i\pi/4}$.
The parameters $a,b$ are fixed by the initial conditions 
$u_k(-\infty)=1$ and $v_k(-\infty)=0$. Using the asymptotes of the Weber
functions when $\tau\to-\infty$, we obtain 
\be
a=0~,~~|b|^2=\frac{e^{-\pi/8R_k}}{4R_k}~.
\label{solution}
\ee 
At the end of the quench at $t=0$ when $\tau=2\tau_Q \sin(2k)$, the argument of the Weber function is 
$iz=\sqrt{R_k}\tau e^{i\pi/4}=2\sqrt{\tau_Q}e^{i\pi/4}\cos(k){\rm sign}(k)$. When $\tau_Q\gg1$ the modulus 
of this argument is large for most $k$, except near $k=\pm\frac{\pi}{2}$, and we can again use asymptotes 
of Weber functions to obtain products
\bea
|u_k|^2&=&
\frac{1-\cos k}{2}+
e^{-2\pi\tau_Q k^2}~,\nonumber\\
|v_k|^2&=&
1-|u_k|^2~,\nonumber\\
u_k v_k^* &=&
\frac12\sin k+
{\rm sign}(k)~
e^{-\pi\tau_Q k^2}
\sqrt{1-e^{-2\pi\tau_Q k^2}}~
e^{i\varphi_k}~,\nonumber\\
\varphi_k &=&
\frac{\pi}{4}+
2\tau_Q-
(2-\ln 4)\tau_Q k^2+ 
k^2 \tau_Q\ln\tau_Q-
\arg\left[\Gamma\left(1+i\tau_Q k^2\right)\right]~.
\label{att0smallk}
\eea
These products depend on $k$ and $\tau_Q$ through two dimensionless combinations:
$\tau_Q k^2$, which implies the usual KZ correlation length
$\hat\xi=\sqrt{\tau_Q}$, and $k^2\tau_Q\ln\tau_Q$ which implies
a second scale of length 
$
\sqrt{\tau_Q\ln\tau_Q}.
$ 
The final quantum state at $g=0$ cannot be characterised by a unique scale of length. Physically, this
reflects a combination of two processes: KZM that sets up the initial post-transition state of the system 
at $-\hat\epsilon$, and the subsequent adiabatic evolution with quantum dephasing.

\subsubsection{Entropy of a block of spins after a quench}
\label{entropyIsing}

The Von Neumann entropy of a block of $L$ spins,
\be
S(L)~=~-~{\rm Tr}~\rho_L~\log_2\rho_L~,
\ee
measures entanglement between the block and the rest of the chain.
Above $\rho_L$ is reduced density matrix of the subsystem of $L$ spins. In recent
years this entropy was studied extensively in ground states of quantum critical 
systems \cite{17b,17c,17d,17e,18a,18b,18c,JStatPhys}. At a quantum critical point, 
the entropy diverges like $\log L$ for large $L$ with a pre-factor determined by the 
central charge of an effective conformal field theory \cite{17a,17b,18a}. In the quantum 
Ising model at the critical $g_c=1$
\be
S^{\rm GS}(L) ~\simeq~ \frac16 \log_2 L~
\label{SGCcrit}
\ee 
for large $L$. Slightly away from the critical point, the entropy saturates at
a finite asymptotic value \cite{18a}
\be
S_{\infty}^{\rm GS} ~\simeq~ \frac16 \log_2 \xi 
\label{SGSoff}
\ee 
when the block size $L$ exceeds the finite correlation length $\xi$ in the
ground state of the system.

In a dynamical quantum phase transition the quantum state of the system
develops a finite correlation length $\hat\xi=\sqrt{\tau_Q}$. If this
dynamical correlation length were the only relevant scale of length, then 
one could expect the entropy of entanglement after a dynamical transition to be given by 
Eq.~(\ref{SGSoff}) with $\xi$ simply replaced by $\hat\xi$. However, as we could see
in Eq.~(\ref{att0smallk}), there are two scales of length, and -- strictly 
speaking -- there is no reason to expect that either of them alone is relevant 
in general. This is why we cannot rely on scaling arguments and the entropy has
to be calculated ``from scratch''.

We proceed in a similar way as in Ref. \cite{17b,17c,17e,18b,JStatPhys} and
define a correlator matrix for the block of $L$ spins 
\be
\Pi~=~
\left(
\begin{array}{ccc}
\alpha &,& \beta^{\dagger} \\  
\beta  &,& 1-\alpha  
\end{array}
\right)~,
\label{rhoblock}
\ee
where $\alpha$ and $\beta$ are $L\times L$ matrices of quadratic correlators
\bea
\alpha_{m,n} &\equiv& 
   \langle c_m c_n^\dagger \rangle ~=~ 
   \frac{1}{2\pi}\int_{-\pi}^\pi dk~|u_k|^2~e^{ik(m-n)}~
   \stackrel{\tau_Q\gg 1}{\approx}   \nonumber\\
&& \frac12\delta_{0,|m-n|}-\frac14\delta_{1,|m-n|}+
   \frac{e^{-\frac{(m-n)^2}{8\pi~\hat\xi^2}}}{2\sqrt{2}\pi~\hat\xi}~.
\label{alpha}
\eea
and 
\bea
\beta_{m,n} &\equiv& 
   \langle c_m c_n \rangle ~=~ 
   \frac{1}{2\pi i}\int_{-\pi}^\pi dk~u_kv_k^*~e^{ik(m-n)}~
   \stackrel{\ln\tau_Q\gg 1}{\approx}~
   \label{beta}\\
&& {\rm sign}(m-n)~
   \left[
   \frac14 \delta_{1,|m-n|}-
   \frac{e^{2i\tau_Q-\frac{i|m-n|^2}{4~\hat\xi l}-\frac{\pi|m-n|^2}{4~l^2}}}
        {2\sqrt{\pi~\hat\xi~l}}
   \sqrt{1-e^{-\frac{\pi|m-n|^2}{4~l^2}}}
   \right] ~,
\label{secondline}
\eea
where we used Eqs. (\ref{att0smallk}). Here $\hat\xi=\sqrt{\tau_Q}$ is the KZ correlation length and 
\be
l~=~\sqrt{\tau_Q}~\ln\tau_Q~
\label{l} 
\ee
is a dephasing length 
\footnote{
Indeed, for large $\tau_Q$, the non-ground-state part of $\beta_{m,n}$, i.e the second term in the square 
bracket in Eq. (\ref{secondline}), is negligible on distances $|m-n|\ll l$ as a result of dephasing the 
integral over $k$ in Eq. (\ref{beta}). Its integrand has the phase $\varphi_k$ in Eq. (\ref{att0smallk}) 
whose fast rotation with $k$ is driven mainly by the term $k^2\tau_Q\ln\tau_Q$. This term introduces 
a length scale $l_0=\sqrt{\tau_Q\ln\tau_Q}$ and, if it were the only length scale, the integral would be 
negligible or ``dephased'' for $|m-n|\ll l_0$. However, Eq. (\ref{att0smallk}) depends on $k$ also through 
the combination $\tau_Qk^2$ introducing $\hat\xi$ as the second scale. Due to conspiracy between the two 
scales, the integral is actually dephased on distances up to the net dephasing length $l=l_0^2/\hat\xi$ 
in Eq. (\ref{l})
}. 
When the linear quench is extended to $g\to-\infty$, as in Section \ref{IsingToInfty},
the dephasing length continues to grow like $l(t)=4t$. 

We note that $\alpha$ and $\beta$ are Toeplitz matrices with constant diagonals. The expectation values 
$\langle\dots\rangle$ are taken in the Bogoliubov vacuum state. Since this state is Gaussian, all higher 
order correlators can be expressed by the matrices $\alpha$ and $\beta$ - they provide complete 
characterization of the quantum state after the dynamical transition. Since the matrices depend on both 
scales $\hat\xi$ and $l$, both scales are necessary to characterize the Gaussian state.

\begin{figure}[t]
\begin{center}
\includegraphics[width=\columnwidth,clip=true]{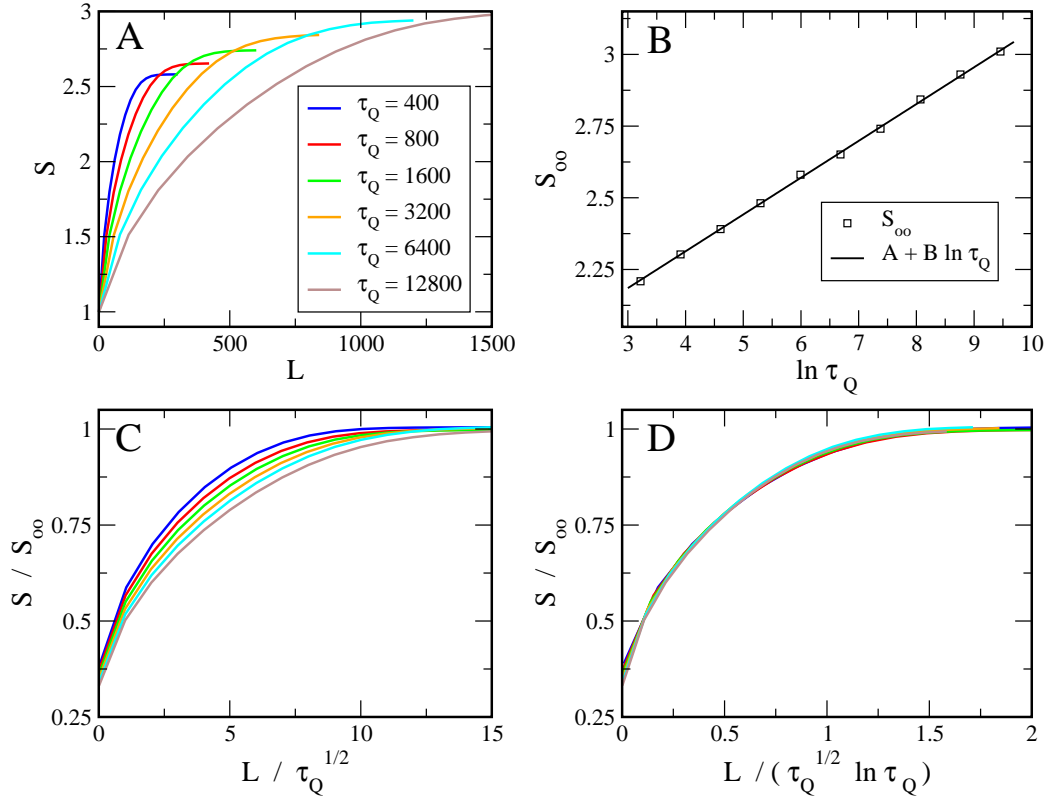}
\caption{
Panel A shows entropy of a block of $L$ spins after the dynamical phase
transition as a function of the block size $L$. The multiple plots
correspond to different values of the quench time $\tau_Q$. For all $\tau_Q$,
the entropy grows with the block size $L$ and saturates at a
finite value $S_\infty(\tau_Q)$ for large enough $L$. In Panel B, 
this asymptotic value of entropy is fitted with the function
$S_{\infty}(\tau_Q)=A+B\ln\tau_Q$. The best fit has $B=0.128\pm 0.004$
and $A=1.80\pm 0.05$. This $B$ is in reasonably good agreement with the
expected value of $B=\frac{\ln 2}{12}=0.120$. In Panels C and D, entropy 
$S(L,\tau_Q)$ is rescaled by the best fit to its asymptotic value
$S_\infty(\tau_Q)=A+B\ln\tau_Q$. With this rescaling one can focus
on how the entropy depends on the block size $L$. In Panel C, 
the block size is rescaled by $\hat\xi=\sqrt{\tau_Q}$ but the
rescaled plots do not overlap. However, as shown in Panel D, rescaling 
the block size $L$ by the second scale $l=\sqrt{\tau_Q}\ln\tau_Q$ 
makes the six plots collapse. (Figure from Ref. \cite{Cincio})
}
\label{figentropy}
\end{center}
\end{figure}

As noted in Ref.~\cite{17b,17c,17e,18b,JStatPhys}, the block entropy can be calculated as
\bea
S(L,\tau_Q)~=~-~{\rm Tr}~\rho~\log_2\rho~=~-~{\rm Tr}~\Pi~\log_2\Pi~.
\label{SLtauQ}
\eea
In this calculation we use Eq.~(\ref{solution}) and Eqs.~(\ref{alpha},\ref{beta}). 
The calculation involves numerical evaluation of the integrals in Eqs.~(\ref{alpha},\ref{beta}) 
and numerical diagonalisation of the matrix $\Pi$. Results are shown in Panel A 
of Fig.\ref{figentropy}. The entropy grows with the block size $L$ and saturates
at a finite value $S_\infty(\tau_Q)$ for large enough $L$. In Panel B, the 
asymptotic entropy is fitted with the (linear) function $S_{\infty}(\tau_Q)=A+B\ln\tau_Q$. 
On one hand, KZM suggests a simple replacement of the ground state correlation length $\xi$ 
in Eq.~(\ref{SGSoff}) by the dynamical KZ length $\hat\xi=\sqrt{\tau_Q}$ implying the asymptotic 
value
\be
S_{\infty}(\tau_Q)~\simeq~
\frac16\log_2\hat\xi~\simeq~
\frac{\ln 2}{12}~\ln\tau_Q~=~
0.120~\ln\tau_Q~.
\ee
On the other hand, the best fit gives $B=0.128\pm 0.004$ and $A=1.80\pm 0.05$. The best $B$ is
in reasonably good agreement with the expected value of $0.120$.

In Panels C and D of the same figure, the entropy $S(L,\tau_Q)$ is rescaled by 
its asymptotic value $S_\infty(\tau_Q)\approx A+B\ln\tau_Q$. After this transformation 
we can better focus on how the entropy depends on the block size $L$. A simple 
hypothesis would be that the entropy depends on $\hat\xi$ and saturates when $L\gg\hat\xi$. 
To check if this is true, in Panel C the block size $L$ is rescaled by 
$\hat\xi=\sqrt{\tau_Q}$ and this rescaling brings the plots close to overlap, although they 
do not overlap as well as one might have hoped. By contrast, as shown in Panel D, 
rescaling the block size $L$ by the dephasing length $l$ makes 
the multiple plots collapse. Thus we can conclude that the entropy saturates at
\be
S_\infty(\tau_Q)~\simeq~
\frac16~\log_2\hat\xi~~
{\rm when}~~ 
L~\gg~ l~
\label{conclS}
\ee
i.e. the entropy of a large block of spins is determined by the KZ dynamical correlation length 
$\hat\xi=\sqrt{\tau_Q}$, but the entropy saturates when the block size is greater than the dephasing 
length $l=\sqrt{\tau_Q}\ln\tau_Q$. A similar conclusion was also obtained in the numerical simulations 
in Ref. \cite{DynAfter}, where the non-integrable case with non-zero longitudinal magnetic field 
was considered as well.

\begin{figure}[t]
\begin{center}
\includegraphics[width=\columnwidth,clip=true]{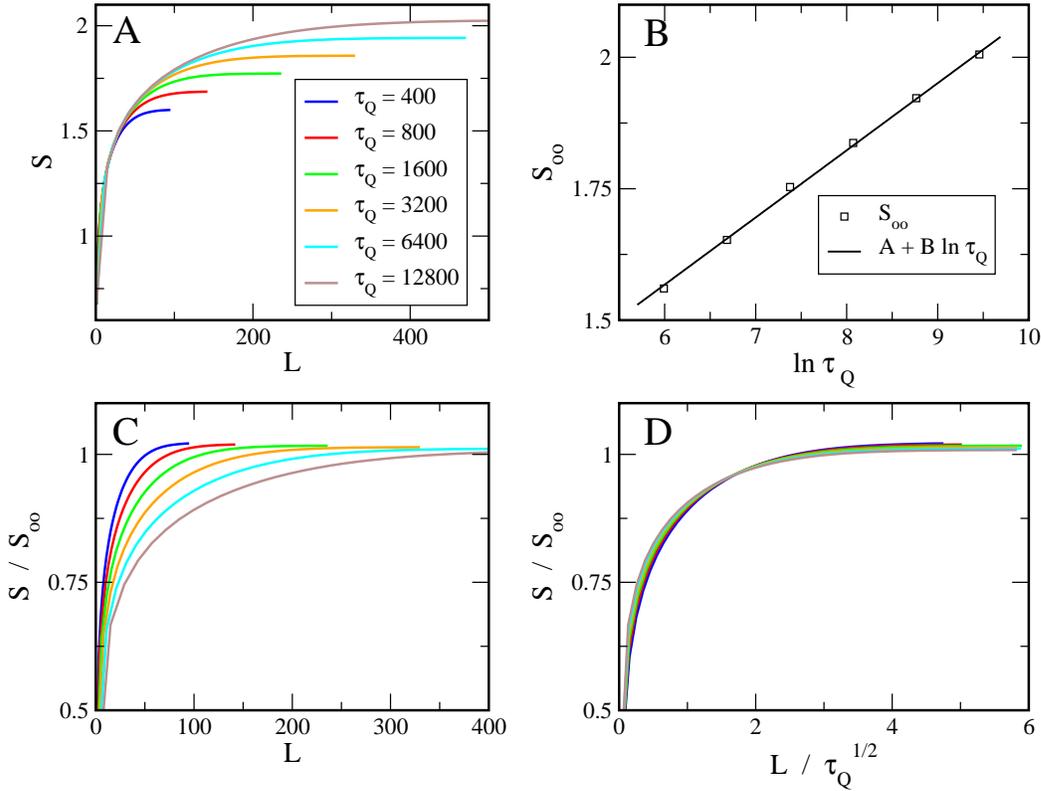}
\caption{
Panel A shows entropy of a block of $L$ spins during the dynamical phase
transition at the critical point $g=1$ as a function of the block size $L$. 
The multiple plots correspond to different values of the quench time 
$\tau_Q$. For all the quench times, the entropy grows with the block size 
$L$ and saturates at a finite value $S_\infty(\tau_Q)$ for large enough $L$. 
In Panel B this asymptotic value of entropy is fitted with the function
$S_{\infty}(\tau_Q)=A+B\ln\tau_Q$. The best fit has $B=0.126\pm 0.005$
and $A=0.80\pm 0.03$. This $B$ is in reasonably good agreement with the
expected value of $B=\frac{\ln 2}{12}=0.120$. In Panels C and D,
the entropy $S(L,\tau_Q)$ is rescaled by the best fit to its asymptotic 
value $S_\infty(\tau_Q)\approx A+B\ln\tau_Q$. With this rescaling one can focus
on how the entropy depends on the block size $L$. In Panel D, rescaling 
the block size $L$ by $\hat\xi=\sqrt{\tau_Q}$ makes the six
plots collapse.
(Figure from Ref. \cite{Cincio})
}
\label{figentropy-g1}
\end{center}
\end{figure}

According to KZM the correlation length $\hat\xi$ is determined in the impulse stage
when the system is crossing the critical point, while the second scale $l$
builds up by dephasing the excited state in the following adiabatic stage.
This scenario is confirmed by the block entropy at the moment when 
the system is crossing the critical point at $g_c=1$, see Figure \ref{figentropy-g1}.
The entropy saturates at
\be
S_\infty(\tau_Q)~\simeq~
\frac16~\log_2\hat\xi~~
{\rm when}~~ 
L~\gg~\hat\xi~.
\label{conclSgc}
\ee
Near the critical point, when the scale $l$ set up by dephasing just begins to build up, 
$\hat\xi$ is still the only relevant scale of length. The result (\ref{conclSgc}) confirms
the KZM replacement rule $\xi\to\hat\xi$ in Eq. (\ref{SGSoff}). As predicted by the
adiabatic-impulse-adiabatic approximation in Section \ref{KZargument}, the entropy is the same
as if the state were the ground state at $\hat\epsilon$.

\subsubsection{ Impurity of a block of spins after a quench }
\label{impurityIsing}

Since a fully analytic calculation of entropy does not seem possible, it 
is worthwhile to calculate another more easily tractable 
entanglement-related quantity. For example, an ``impurity'' of the 
correlator matrix $\Pi$ in Eq. (\ref{rhoblock})
\be
I(\Pi)~=~ {\rm Tr}~\Pi~(1-\Pi)
\ee
is zero only when the $L$ spins are in a pure state i.e. when all 
eigenvalues of $\Pi$ are either $0$ or $1$. It is maximal when all 
the eigenvalues are $\frac12$, or when the state is most entangled. 
Thanks to its simple quadratic form, the impurity can be calculated relatively 
easily. 

Simple calculation using the block structure of $\Pi$ in Eq.~(\ref{rhoblock}) 
and the Toeplitz property of the block matrices $\alpha$ and $\beta$ leads to
\bea
I\left(L\right) ~=~ 
2
\left( 
L\alpha_{0} - \sum_{j=1-L}^{j=L-1} (L-|j|) (\alpha_j^2 + |\beta_j|^2)      
\right),
\eea
where $\alpha_j = \alpha_{j,0}$ and $\beta_j = \beta_{j,0}$. Further calculation 
as in Ref. \cite{Cincio} shows that at the final $g=0$ the impurity saturates at
\be
I_\infty~\approx~
\frac{\ln\hat\xi}{\pi^2}~~
{\rm when}~~
L\gg l~,
\label{Iinfty}
\ee
in analogy to the entropy in Eq. (\ref{conclS}).

It is interesting to compare the dynamical impurity (\ref{Iinfty}) with the impurity in the 
ground state near the critical point. A simple calculation gives the asymptote of impurity 
at the critical point
\be
I^{\rm GS}(L)~\approx~
\frac{\ln L}{\pi^2}~.
\ee 
when $\ln L\gg1$. Near the critical point, the asymptote is valid when the block is much shorter
than the correlation length, $L\ll\xi$, and at larger $L$ the impurity saturates at 
\be
I^{\rm GS}_\infty~\approx~
\frac{\ln\xi}{\pi^2}~.
\label{IGS}
\ee
Again, in the same way as for the block entropy, the simple KZM replacement rule $\xi\to\hat\xi$ 
applied to Eq. (\ref{IGS}) gives the correct asymptotic value of the dynamical impurity in 
Eq.~(\ref{Iinfty}). As predicted by the adiabatic-impulse-adiabatic approximation in Section 
\ref{KZargument}, the dynamical impurity is the same as the impurity in the ground state at $\hat\epsilon$.

\subsubsection{ Correlation functions after a quench}
\label{correlationsIsing}

Correlation functions are of fundamental interest in the theory of phase transitions 
because they provide direct manifestation of their universal properties and are in 
general directly accessible experimentally. Here we present results for spin-spin correlation 
functions after a dynamical quantum phase transition, see Ref. \cite{Cincio,CherngLevitov}.

To begin with, we observe that for symmetry reasons the magnetization 
$\langle \sigma^x \rangle=0$, but the transverse magnetization
\bea
\langle \sigma^z_n \rangle~=~ 
\langle 1-2c^\dagger_nc_n \rangle~=~
2\alpha_0-1~\approx~
\frac{1}{2\pi\sqrt{2\tau_Q}}~
\eea
when $\tau_Q\gg 1$. This is what remains of the initial 
magnetization $\langle \sigma^z_n \rangle=1$ in the initial paramagnetic ground state 
at $g\to\infty$. As expected, when the linear quench is slow, then
the final magnetization decays towards $\langle \sigma^z_n \rangle=0$
characteristic of the ferromagnetic ground state at $g=0$. 

Final transverse spin-spin correlation function at $g=0$ is 
\bea
\label{CxxR}
C^{zz}_R &\equiv& 
\langle \sigma^z_n \sigma^z_{n+R} \rangle-  
\langle \sigma^z_n     \rangle
\langle \sigma^z_{n+R} \rangle~=~              
4\left(|\beta_R|^2-|\alpha_R|^2\right)~\approx~\nonumber\\
&&
\frac{e^{-\frac{\pi R^2}{2~l^2}}\left(1-e^{-\frac{\pi R^2}{4~l^2}}\right)}
     {\pi~\hat\xi~l}-\frac{e^{-\frac{R^2}{\pi~\hat\xi^2}}}{2\pi^2~\hat\xi^2}~, 
\eea
when $R>1$ and $\ln\tau_Q\gg 1$. This correlation function depends on both
$\hat\xi$ and the dephasing length $l$ in Eq. (\ref{l}), but its long range tail
\be
C^{zz}_R ~\sim~ e^{-\frac{\pi R^2}{2~l^2}} 
\ee
decays on the scale $l$.

In contrast, the ferromagnetic spin-spin correlation function 
\be
C^{xx}_R~=~
\langle \sigma^x_n \sigma^x_{n+R} \rangle~-~  
\langle \sigma^x_n \rangle
\langle \sigma^x_{n+R} \rangle~=~  
\langle \sigma^x_n \sigma^x_{n+R} \rangle            
\ee
cannot be evaluated so easily. As is well known, in the ground state $C^{xx}_R$ 
can be written as a determinant of an $R\times R$ Toeplitz matrix whose asymptote
for large $R$ can be obtained with the Szeg\"o limit theorem \cite{Schogo}. 
Unfortunately, in time-dependent problems the correlation function is not 
a determinant in general. However, below we circumvent this problem in an interesting range 
of parameters. 

\begin{figure}[t]
\begin{center}
\includegraphics[width=0.7\columnwidth,clip=true]{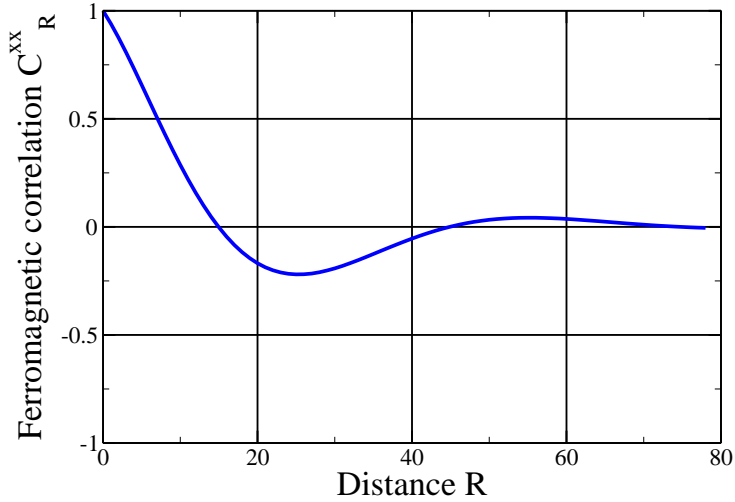}
\caption{
Ferromagnetic spin correlation in Eq. (\ref{Czz}). The oscillatory 
term means that consecutive kinks are anti-correlated -- they keep more or less 
the same distance $\simeq\hat\xi$ from each other forming something similar 
to a ...-kink-antikink-kink-antikink-... regular lattice with a lattice constant 
$\simeq\hat\xi$. However, fluctuations in the length of bonds in this lattice 
are comparable to the average distance itself giving the exponential decay of 
the correlator $C^{xx}_R$ on the same scale of $\simeq\hat\xi$. 
}   
\label{FigCherngLevitov}
\end{center}
\end{figure}

Using the Jordan-Wigner transformation, $C^{xx}_R$ can be expressed as 
\be
C^{xx}_R ~=~ \langle b_0 a_1 b_1 a_2 \ldots b_{R-1} a_R   \rangle ~.
\label{Czzstring}
\ee
Here $a_n$ and $b_n$ are Majorana fermions defined as $a_n = (c_n^{\dagger}+c_n)$ 
and $b_n = c_n^{\dagger}-c_n$. Here $a_n$ is hermitian and $b_n$ is anti-hermitian.
Using Eq. (\ref{alpha}) and Eq. (\ref{beta}) we obtain
\bea
\langle a_m b_n  \rangle &=& 2\alpha _ {n-m}+ 2 {\rm Re} \beta _{n-m} - \delta_{m,n}   \nonumber \\ 
\langle b_m a_n  \rangle &=& \delta_{m,n} - 2\alpha_ {n-m}+ 2 {\rm Re} \beta_{n-m}   \\ 
\langle a_m a_n  \rangle &=& \delta_{m,n} + 2 \imath {\rm Im} \beta_{m-n}  \nonumber \\
\langle b_m b_n  \rangle &=& \delta_{m,n} + 2 \imath {\rm Im} \beta_{m-n}  \nonumber 
\eea
Using Wick theorem, the average in Eq.~(\ref{Czzstring}) is a determinant of a matrix when
$\langle a_m a_n  \rangle=0$ and $\langle b_m b_n \rangle=0$ for $m\neq n$, or equivalently 
when ${\rm Im}\beta_{m-n}=0$ for $m\neq n$. Inspection of Eq.~(\ref{secondline}) shows that 
${\rm Im}\beta_{m-n}\approx 0$ when $|m-n|\ll l$. Consequently, when $R\ll l$ then 
we can neglect all ${\rm Im}\beta_{m-n}$ assuming that $ \langle a_m a_n \rangle  = 0 $ 
and $ \langle b_m b_n  \rangle  = 0 $ for $m\neq n$. The correlation 
function is a determinant of the Toeplitz matrix
$
\left[ \langle b_m a_{n+1} \rangle  \right]_{m,n=1,\ldots,R} 
$ 
within the quasiparticle horizon where $R\ll l$. Asymptotic behaviour of this Toeplitz determinant 
can be obtained \cite{Cincio} using standard methods in Refs. \cite{Schogo} and 
\cite{SenguptaPowellSachdev,CherngLevitov}:
\be
C^{xx}_R ~\propto~ 
\exp\left(-0.174~ \frac{R}{\hat\xi}   \right)~
\cos\left(  \sqrt{\frac{\ln 2}{2 \pi}}~ \frac{R}{\hat\xi} - \varphi_0 \right)
\label{Czz}
\ee
when $1 \ll R \ll l$ i.e. up to the dephasing length $l$, see Fig. \ref{FigCherngLevitov}. 

The final ferromagnetic correlation function (\ref{Czz}) exhibits decaying oscillatory 
behaviour on length scales much less than the scale $l$, but
both the wavelength of these oscillations and their exponentially decaying
envelope are determined by $\hat\xi$. As discussed in a similar situation 
by Cherng and Levitov \cite{CherngLevitov}, this oscillatory behaviour means that 
consecutive kinks keep more or less the same distance $\simeq\hat\xi$ from each 
other forming something similar to a ...-kink-antikink-kink-antikink-... lattice 
with a lattice constant $\simeq\hat\xi$. However, fluctuations in the length of bonds 
are comparable to the lattice constant itself and they give the (marginally underdamped) 
exponential decay of the correlator $C^{xx}_R$ on the same scale of $\simeq\hat\xi$.

\begin{figure}[t]
\begin{center}
\includegraphics[width=\columnwidth,clip=true]{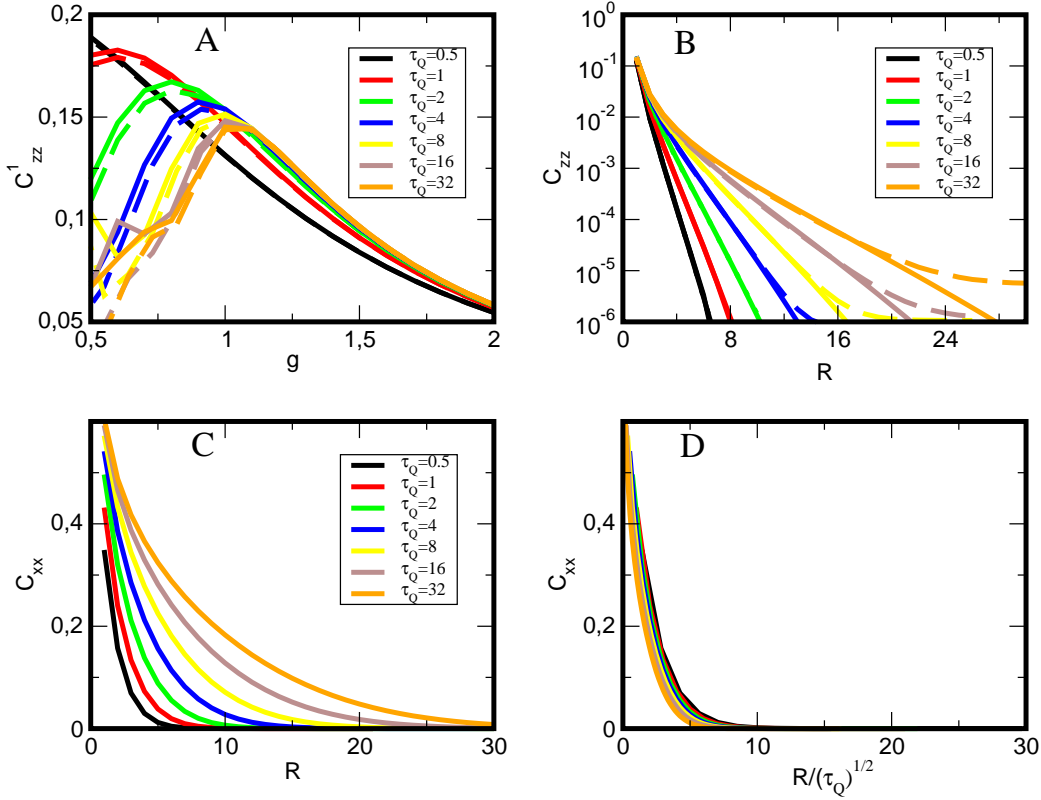}
\caption{
Panel A shows the dynamical transverse correlation $C^{zz}_1$ as a function 
of magnetic field $g$ in the linear quench. For each $\tau_Q$, we show both 
numerical (dashed) 
and analytical (solid) result. The plots overlap near the critical point at 
$g=1$ but diverge in the ferromagnetic phase when $g<1$ indicating a breakdown 
of our numerical simulations in this regime. Panel B shows analytic and numerical 
results for the dynamical transverse correlation function at the moment 
when the quench crosses the critical point at $g=1$. The transverse correlators 
overlap well confirming that our numerical simulations are still accurate 
at the critical point. Finally, in panel C, we show the dynamical ferromagnetic
correlation function $C^{xx}_R$ at $g=1$ and in panel D, we show the same
correlation function after rescaling $R/\hat\xi$. The rescaled plots
overlap quite well supporting the idea that near the critical point the KZ correlation
length $\hat\xi=\sqrt{\tau_Q}$ is the only relevant scale of length.
(Figure from Ref. \cite{Cincio})
}   
\label{figVidal}
\end{center}
\end{figure}

The gaps in analytic knowledge of the correlation functions were filled by numerical 
simulations with the real time TEDB (time evolving decimation block) algorithm \cite{Vidal} in Ref. \cite{Cincio}.
As illustrated in panel A of Figure \ref{figVidal}, the simulations were stable enough 
to cross the critical point and enter the ferromagnetic phase, but once in the 
ferromagnetic phase, the algorithm was breaking down due to the quasiparticle
horizon effect, see Section \ref{Horizon}. The numerics can be trusted at $g_c=1$, but it 
is not reliable when $g<1$. KZM can be tested at the critical point, but one cannot 
reliably follow the dephasing in the ferromagnetic phase. 

Panel B of Figure \ref{figVidal} shows the transverse correlation
$C^{zz}_R$ at $g_c=1$ for several values of $\tau_Q$. For each $\tau_Q$,  
both the numerical correlator and its analytic counterpart from Eq.~(\ref{CxxR}) 
are plotted and they approximately coincide. Equation (\ref{CxxR}) can be also
used to obtain analytically, but with some numerical integration, the exponential
tail of the transverse correlator when $\tau_Q\gg 1$:
\bea
C^{zz}_R 
&\approx&
\frac{0.44}{\tau_Q}~
\exp\left(-2.03 \frac{R}{\hat\xi} \right)
\label{CxxlargetauQ}
\eea 
accurate when $R\gg\hat\xi$. This tail decays on the KZ correlation length 
$\hat\xi$ which proves to be the relevant scale of length.

Panel C shows the ferromagnetic spin-spin correlation functions at $g_c=1$ for 
the same values of $\tau_Q$. They are roughly exponential and their correlation 
length seems to be set by $\hat\xi=\sqrt{\tau_Q}$. To verify this scaling hypothesis 
in panel D the same plots as in panel C are shown but with $R$ rescaled by $\hat\xi$. 
The rescaled plots overlap reasonably well confirming the expected 
$\hat\xi=\sqrt{\tau_Q}$ scaling. 

The correlation functions at the critical point $g_c=1$ are consistent with the prediction
of the adiabatic-impulse-adiabatic approximation in Section \ref{KZargument} that the
state in the impulse stage is approximately the ground state at $\hat\epsilon$
with correlations decaying exponentially on the correlation length $\hat\xi$.

\subsubsection{Fidelity between the state after a quench and the final ground state }

A fidelity $F=|\langle\psi|{\rm GS}\rangle|^2$ between the final state $|\psi\rangle$ 
and the desired final ground state $|{\rm GS}\rangle$ is a very sensitive measure of the 
quality of adiabatic quantum computation or adiabatic quantum state preparation. In 
the quantum Ising chain, the fidelity can be obtained analytically from the exact solution, 
see Ref. \cite{decoherence}. The fidelity $F$ is a probability that not a single pair of 
quasiparticles with opposite quasimomenta 
$(k,-k)$ is excited after a quench,
\be
F~=~\prod_{k>0}(1-p_k),
\label{fidelity}
\ee
where the product runs over the positive half-integer quasimomenta in Eq. (\ref{halfinteger}).
It is more convenient to calculate its logarithm
\be
\ln F ~=~\sum_{k>0}\ln(1-p_k) ~\approx~ \int_0^\pi dk~\ln(1-p_k)~, 
\ee
accurate for $N\gg\hat\xi$. If the transition terminates at $g=0$, then for
$\tau_Q\gg 1$ we have $p_k=\exp(-2\pi\tau_Q k^2)$ and
\bea
\ln F ~=~
-\frac{Nn_{\rm ex}}{2}
\frac{  \int_0^\infty ds~ \ln\left[1-e^{-s^2}\right]  }
     {  \int_0^\infty ds~ e^{-s^2}                    } ~=~
-1.3~n_{\rm ex}N~.
\label{Fsigma0}
\eea 
Here $n_{\rm ex}=\int_{-\pi}^\pi\frac{dk}{2\pi}p_k=1/2\pi\sqrt{2\tau_Q}$ is the
density of excitations in Eq. (\ref{nexIsing}). Thus, as might have been expected, the fidelity
decays exponentially with the chain size $N$ on the length scale $\hat\xi\simeq n_{\rm ex}^{-1}$, 
\be
F~=~e^{-1.3~n_{\rm ex}N}~,
\label{Fexp}
\ee
but with an unexpected coefficient $1.3$ greater than $1$.

It is instructive to compare the decay of fidelity in Eq. (\ref{Fexp}) to a simple 
Poissonian model where each of the $N$ bonds is either excited (with probability $n_{\rm ex}$) 
or not excited (with probability $1-n_{\rm ex}$) {\it independently} of other bonds. 
In this model the fidelity is a probability that none of the $N$ independent bonds 
is excited:
\be
F_{\rm Poisson}~=~(~1~-~n_{\rm ex}~)^N~\approx~e^{-n_{\rm ex}N}~
\label{Fmodel}
\ee 
when $n_{\rm ex}\ll 1$. Comparing Eqs. (\ref{Fexp}) and (\ref{Fmodel}) we find that the 
decay in Eq. (\ref{Fexp}) is faster than in the model of independent bonds. The faster
decay means anti-bunching correlations between the kinks randomly scattered along the spin 
chain in the final state. The same anti-bunching correlations are also responsible for the 
oscillations in the ferromagnetic correlation function in Eq. (\ref{Czz}) and 
Fig. \ref{FigCherngLevitov}.

The anti-bunching in Eq. (\ref{Fexp}) contrasts with the bunching apparent in fidelity 
between the ferromagnetic ground state at $0<|g|<1$, which has finite density of kinks 
$n_{\rm ex}$, and the fully polarized ferromagnetic ground state at $g=0$: 
\be
F_{\rm ferro}~=~e^{-\frac12n_{\rm ex}N}~,
\ee
see Ref. \cite{decoherence}. Here the coefficient is $\frac12$ meaning that the relevant 
density of uncorrelated excitations is one-half of the density of kinks. The kinks in the 
ferromagnetic ground state are bound into kink-antikink pairs randomly scattered along the 
chain. The bound pairs do not destroy the ferromagnetic long range order. This is most 
apparent when $g\ll1$ is perturbatively small and the ground state is a fully polarized 
ferromagnet but with an admixture of a small density of reversed spins. Each reversed spin 
is a tightly bound pair of kink and antikink which, however, does not affect the long range 
ferromagnetic order.

\subsubsection{ Geometric phase after a quench }
\label{Basu}

Connection between the geometric phase \cite{Berry} and quantum phase transitions
has been explored recently in Refs. \cite{BerryQPT}. The geometric phase can be used as
a tool to probe quantum phase transitions in many body systems. It is a witness
of a singular point in the energy spectrum arising in non-trivial geometric evolutions.
In Ref. \cite{GeometricPhase} the geometric phase of the final state at $g=0$ 
was calculated as 
\be
\Gamma ~=~ 2\pi N\left(n_{\rm ex}-1\right)-3\pi+
           \pi\left(\cos\pi n_{\rm ex}+\sin\pi N\cot\frac{\pi}{2N}\right)~.
\label{Berry}
\ee
It depends on the density of excited kinks $n_{\rm ex}$ in Eq. (\ref{nexIsing}).

\subsubsection{Generalized entanglement in the final state}

The generalized entanglement of a state $|\psi\rangle$ with respect to a Lie 
algebra $h$ with hermitian, orthonormal basis of generators $O_l$ enumerated by 
$l=1,...,M$ can be characterised by a purity of $|\psi\rangle$ relative to $h$
which is given by
\be
P_h\left( |\psi\rangle \right) ~=~ 
{\cal N} \sum_{l=1}^M \langle\psi|O_l|\psi\rangle^2~,
\ee
where ${\cal N}$ is a normalisation factor chosen so that $0\leq P_h\leq1$.

A fermionic system, like e.g. the quantum Ising model in the representation of
Jordan-Wigner fermions in Eq. (\ref{Hpm}), can be characterised by generalized 
entanglement with respect to the algebra $h$ of number preserving bilinear fermionic 
operators spanned by $c_i^\dag c_j$ with $1\leq i,j\leq N$. In Ref. \cite{Viola1}
it was shown that, near an isolated quantum critical point, the generalized entanglement 
of the ground state $|{\rm GS}(\epsilon)\rangle$ scales like
\be
P_h\left[ |{\rm GS}(\epsilon)\rangle \right] - 
P_h\left[ |{\rm GS}(0)\rangle \right]~\sim~
\xi^{-1}~,
\label{GenEntGS}
\ee
where $\xi$ is a correlation length in the ground state. The relative entanglement with 
respect to the critical point is inversely proportional to the diverging correlation 
length.

Moreover, as verified in Ref. \cite{Viola1}, in a linear quench across an isolated 
critical point we have
\be
P_h[|\psi(t)\rangle]-P_h[|{\rm GS}[\epsilon(t)]\rangle]~\sim~
\hat\xi^{-1}~
G(t/\hat t)~,
\label{GenEntDyn}
\ee
where $|\psi(t)\rangle$ is the state of the system during the transition,
$|{\rm GS}[\epsilon(t)]\rangle$ is the instantaneous ground state, and $G$ is a scaling function. 
The relative amount of entanglement with respect to the instantaneous ground state is inversely
proportional to the KZ length $\hat\xi$, which diverges in the adiabatic limit, and its time-dependence 
becomes universal when measured in units of $\hat t$ in Eq. (\ref{hatt}), as expected from KZM
in Section \ref{KZargument}. 

In particular, the simple replacement rule $\xi\to\hat\xi$ applied to the generalized entanglement 
in the ground state (\ref{GenEntGS}) gives
the correct dynamical result (\ref{GenEntDyn}) in accordance with the adiabatic-impulse-adiabatic
approximation in Section \ref{KZargument} where the state during the impulse stage is approximately
the ground state at $\hat\epsilon$ with correlation length $\hat\xi$.

\subsubsection{ Linear quench to $t\to+\infty$: quantum dephasing after KZM }
\label{IsingToInfty}

As we could see in Section \ref{correlationsIsing}, near the critical point $g_c=1$ 
the state of the system is close to the adiabatic ground state at $\hat\epsilon$ 
which can be characterised by a single KZ correlation length $\hat\xi=\sqrt{\tau_Q}$, 
but in the following adiabatic evolution to $g=0$ quantum dephasing develops the 
second longer scale $l=\sqrt{\tau_Q}\ln\tau_Q$. In this Section, as in Ref. \cite{CherngLevitov}, 
we extend the linear quench to $t\to+\infty$ to give the quantum dephasing enough time 
to complete its job, and see if there is any imprint of the KZ length $\hat\xi$ left 
in the final state at $t\to+\infty$ after the dephasing is completed.

The extended quench crosses both critical points $g_c=\pm 1$ and the
quasiparticle excitation probability in Eq. (\ref{pveck}) becomes
a sum of two Gaussians localised near the two Fermi points $k_F=0,\pi$:
\be
p_k~=~e^{-2\pi\tau_Q\sin^2k}~\stackrel{\tau_Q\gg1}{\approx}~
e^{-2\pi\tau_Q k^2} + e^{-2\pi\tau_Q (k-\pi)^2} ~.
\label{pk2Gaussians}
\ee
Consequently, the average density of excitations in Eq. (\ref{nexBZ}),
\be
n_{\rm ex}~=~\frac{1}{\pi\sqrt{2}}~\tau_Q^{-1/2}~,
\label{nexinfty}
\ee
is twice the density at $g=0$ in Eq. (\ref{nexIsing}). This $n_{\rm ex}$ is a density 
of $\uparrow$ spins excited in the final ground state 
$|\downarrow\downarrow\downarrow\dots\downarrow\rangle$ at $g\to-\infty$.

Another consequence of crossing two critical points and Eq. (\ref{pk2Gaussians}) is
a simplified quadratic correlator in Eq. (\ref{alpha}),
\bea
\alpha_{m,n} &=& \frac{1}{2\pi}\int_{-\pi}^\pi dk~p_k~e^{ik(m-n)}~=~
             \frac{1+(-1)^{m-n}}{2}~
             \frac{1}{\sqrt{2}\pi~\hat\xi}~
             e^{-\frac{(m-n)^2}{8\pi~\hat\xi^2}}~,
\label{alphainfty}
\eea
which is zero when $m-n$ is odd. The other correlator $\beta_{m,n}$ in Eq. (\ref{beta}) is
even simpler.

Indeed, the last adiabatic stage of the evolution, after crossing the second critical
point at $g_c=-1$, leaves plenty of time for quantum dephasing. When $g\ll -1$ the eigenstates 
of the Bogoliubov-de Gennes Eqs. (\ref{stBdG}) are $(u_k,v_k)\approx(1,0)$ and $(u_k,v_k)\approx(0,1)$ 
with eigenvalues $-\omega_k$ and $\omega_k$ respectively, where $\omega_k\approx2(g-\cos k)$.
Consequently, the adiabatic solution of Eqs. (\ref{dynBdG}) is
\be
u_k~\approx~\sqrt{p_k}~e^{i\frac{t^2}{\tau_Q}-2it\cos k-i\varphi_k}~,~~
v_k~\approx~\sqrt{1-p_k}~e^{-i\frac{t^2}{\tau_Q}+2it\cos k+i\varphi_k}~.
\ee      
As $t\to\infty$ the solutions $(u_k,v_k)$ accumulate phase factors $\exp{\mp i( 2t\cos k-t^2/\tau_Q ) }$ 
which oscillate strongly with $k$. These oscillations ``dephase'' the integral in Eq. (\ref{beta}) to zero:
\bea
\beta_{m,n}  &=& 
   \frac{1}{2\pi i}\int_{-\pi}^\pi dk~
   \sqrt{p_k(1-p_k)}~
   e^{2i\frac{t^2}{\tau_Q}-4it\cos k-2i\varphi_k}~
   e^{ik(m-n)}~\to~0~ 
\label{betainfty}
\eea
when $t\to\infty$.

However, when $t$ is large but finite, then the dephasing in Eq. (\ref{betainfty}) is effective 
only when $|m-n|\ll4t$, where the $4$ is twice the maximal group velocity of instantaneous
quasiparticles when $|g|\gg1$. This estimate means that the dephasing length $l$ in Eq. (\ref{l}) 
continues to grow like
\be
l(t)~=~4t~
\label{4t}
\ee
in the last adiabatic stage of the evolution after $g_c=-1$.

Notice that the correlator $\alpha_{m,n}$ depends on $\hat\xi$. Since in the absence of any non-zero 
$\beta_{m,n}$ the quadratic correlator $\alpha_{m,n}=\langle c_mc_n^\dag\rangle$ contains all the information 
about the final Gaussian state, then we can conclude that $\hat\xi$ remains a permanent imprint of KZM which 
has not been washed out by quantum dephasing. This imprint has several interesting manifestations:   

\begin{itemize}

\item 

Using Szeg\"o limit theorem, the ferromagnetic correlator can be found
as \cite{CherngLevitov}
\be
C^{xx}_R~\propto~
\frac{1+(-1)^R}{2}~
\exp\left(-0.174\frac{R}{\hat\xi}\right)
\cos\left(\sqrt{\frac{\ln 2}{2\pi}}\frac{R}{\hat\xi}-\varphi_\infty\right)
\label{Czzinfty}
\ee
when $\tau_Q\gg1$ and $1\ll R\ll l(t)$. As a consequence of crossing two critical 
points instead of one, this correlator is zero for odd $R$. When $t\to\infty$ then
$l(t)\to\infty$ and the correlator becomes accurate for all $R\gg1$.

As we can see, during the linear quench the ferromagnetic correlator evolves from 
the pure exponential decay at $g_c=1$, see Figs. \ref{figVidal}C and D, to the 
oscillating exponential decay at $g\to-\infty$, see Eq. (\ref{Czzinfty}). These 
oscillations are gradually developed by quantum dephasing during the adiabatic 
stages of the evolution. At $g=0$ the oscillations are limited to $R\ll\sqrt{\tau_Q}\ln\tau_Q$, 
but as $g\to-\infty$ they gradually extend to all (even) $R$. The dephasing relaxes the state 
to a crystal-like train of kinks and antikinks.

\item

The transverse magnetization is
\be
\langle\sigma^z_n\rangle ~=~2\alpha_{0,0}-1~=~-1+2n_{\rm ex}~,
\ee
i.e., the magnetization $-1$ in the final ground state at $g\to-\infty$ plus
twice the density of inverted spins in Eq. (\ref{nexinfty}).

\item

The transverse correlator is
\bea
\label{CxxRinfty}
C^{zz}_R ~=~ -4|\alpha_{0,R}|^2~\approx~
-\frac{ 1+(-1)^R }{2}~
\frac{ 2 ~ e^{-\frac{R^2}{4\pi~\hat\xi^2}} }
     { \pi^2 ~ \hat\xi^2}~,
\eea
compare Eqs. (\ref{CxxR},\ref{nexinfty},\ref{alphainfty}). This is roughly 
the $l\to\infty$ limit of the same correlator at $g=0$, compare Eq. (\ref{CxxR}).

\item

The block impurity introduced in Section \ref{impurityIsing} is
\be
I(\Pi)~\equiv~{\rm Tr}~\Pi(1-\Pi)~=~2{\rm Tr}\alpha(1-\alpha)~\approx~2n_{\rm ex}L
\label{Iext}
\ee
when $L\gg\hat\xi\gg1$ but $L\ll l(t)$. Here we used Eqs. 
(\ref{rhoblock},\ref{nexinfty},\ref{betainfty},\ref{alphainfty}). 
 
\item

The extensive impurity in Eq. (\ref{Iext}) implies that the block entropy is also 
extensive: $S(L,\tau_Q)=s L$ for $L\gg\hat\xi$. The entropy density $s$ 
is most conveniently evaluated as the entropy of the whole chain $S(N,\tau_Q)$ 
divided by $N$, see Ref. \cite{CherngLevitov}:
\bea
s &=&
\lim_{N\to\infty}
\frac{-2~{\rm Tr}~\alpha\log_2(1-\alpha)}{N} \nonumber\\
&=&
-\frac{2}{\ln 2}
\int_{-\pi}^\pi \frac{dk}{2\pi}~
\left[
p_k\log_2p_k+(1-p_k)\log_2(1-p_k)
\right]~\approx~
0.11~n_{\rm ex}~
\eea
accurate when $n_{\rm ex}\ll 1$.
Here we used Eqs. (\ref{rhoblock},\ref{SLtauQ},\ref{alphainfty},\ref{betainfty}) 
and evaluated the trace in the quasimomentum basis taking the continuous
$k$ limit for $N\to\infty$.

Notice that here the entropy of the whole chain $sN$ appears finite while the actual
quantum state of the isolated chain is pure. This is an artifact of sending 
$\beta_{m,n}\to0$ in Eq. (\ref{betainfty}) which is accurate only when 
$|m-n|\ll l(t)$ at a finite $t$. This condition limits validity of the extensive
entropy $S=0.11n_{\rm ex}L$ to $L\ll l(t)$.

\item

Two site entanglement at $g\to-\infty$ was studied in Ref. \cite{KSengupta2sitentanglement}. 
Both concurrence ${\cal C}^R$ and negativity ${\cal N}^R$ are zero when the 
distance $R$ between the two sites is odd. Only entanglement between even-neighbour 
sites can be non-zero. For large $\tau_Q$, these two measures of entanglement
scale as 
\be
{\cal C}^{2n}~\sim~\hat\xi^{-1}~,~~{\cal N}^{2n}~\sim~\hat\xi^{-2}~,
\ee
when $n$ is fixed, and they decay exponentially with $n/\hat\xi$ for large $n$. 
What is more, the concurrence and negativity are non-zero only when 
$\tau_Q>\tau_Q^{(2n)}$, where $\tau_Q^{(2n)}$ is a threshold that increases with $n$. 
When $\tau_Q\leq\tau_Q^{(2)}$ there is no two-site entanglement and the entanglement 
is entirely multipartite.

\end{itemize}

\subsubsection{ Transition in space } 
\label{IsingKZinspace}

\begin{figure}
\begin{center}
\includegraphics[width=0.99\columnwidth,clip=true]{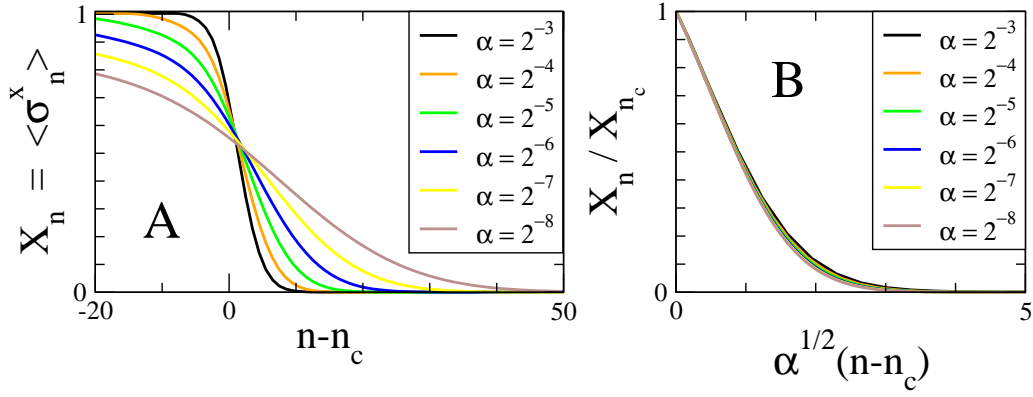}
\caption{ 
In A and B, spontaneous ferromagnetic magnetization as a function of 
$n-n_c$ and the rescaled $x=\sqrt{\alpha}(n-n_c)$ respectively. The collapse of the rescaled
plots in panel B demonstrates that the penetration depth of the ferromagnetic order
parameter into the paramagnetic phase is 
$\delta n\simeq\alpha^{-1/2}$ which is the same as the $\hat\xi$ predicted
in Eq. (\ref{tildexigeneral}).
(Figure from Ref. \cite{Ramsinhom})
} 
\label{Figpenetration}
\end{center}
\end{figure}

In this Section we support the general discussion in Section \ref{KZinspace} 
by a solution in the quantum Ising chain. We consider the ground state
of an open Ising chain in an inhomogeneous transverse field $g_n$ which 
can be linearised near the critical point at $n_c$ where $g_c=1$ as
\be
g_n~\approx~1~+~\alpha~(n-n_c)~,
\label{slope}
\ee
compare Eq. (\ref{nc}). The open chain is in the ferromagnetic phase where $n<n_c$ and 
in the paramagnetic phase where $n>n_c$. We want to know if the non-zero spontaneous ferromagnetic 
magnetization $X_n=\langle\sigma^x_n\rangle$ in the ferromagnetic phase penetrates across the 
critical point into the paramagnetic phase and what is the depth of this penetration? 

The quadratic Hamiltonian $H^+$ in Eq. (\ref{Hpm}) with open boundary conditions
is diagonalised to $H^+=\sum_m\omega_m\gamma_m^\dag\gamma_m$ by a Bogoliubov 
transformation $c_n=\sum_{m=0}^{N-1}(u_{nm}\gamma_m + v^*_{nm} \gamma_m^\dagger)$ 
with $m$ enumerating $N$ eigenmodes of stationary Bogoliubov-de Gennes equations
\bea
\omega_m u_{n,m}^\pm = 2 g_n u_{n,m}^\mp - 2 u_{n\mp 1,m}^\mp ~
\label{BdGinhom}
\eea
with $\omega_m\geq0$. Here $u_{nm}^\pm\equiv u_{nm} \pm v_{nm}$. We make a long 
wavelength approximation
$
u_{n\mp 1,m}^\mp \approx u_{n,m}^\mp \mp \frac{\partial}{\partial n}u_{n,m}^\mp
$
and obtain a long wavelength differential equation
\bea
\omega_m u_m^\pm = 2\alpha(n-n_c) u_m^\mp \pm 2 \partial_n u_m^\mp ~.
\label{lowBdG}
\eea
Its eigenmodes with $\omega_m>0$ are 
\bea
\omega_m = \sqrt{8m\alpha}~,
u_m(n) \propto \psi_{m-1}(x) + \psi_{m}(x) ~,
v_m(n) \propto \psi_{m-1}(x) - \psi_{m}(x) ~,
\label{spectrum}
\eea
where 
\be
x=\sqrt{\alpha}(n-n_c)
\ee
is a rescaled distance from the critical point, $\psi_{m\geq0}(x)$ are eigenmodes of 
a harmonic oscillator satisfying
$
\frac12
(-\partial_x^2+x^2)
\psi_m(x)=
(m+1/2)
\psi_m(x)~,
$
and $\psi_{-1}(x)\equiv 0$. 

The modes in Eq. (\ref{spectrum}) are localised near $n=n_c$ where $x=0$, in 
consistency with the linearisation in Eq. (\ref{slope}), and their typical width 
is $\delta x\simeq1$, or equivalently 
\be
\delta n~\simeq~\alpha^{-1/2}~.
\label{deltan}
\ee
When $\alpha\ll1$ then $\delta n\gg1$ and the long wavelength approximation 
in Eq. (\ref{lowBdG}) is justified. As $\delta n$ is the only scale of length
in the solution (\ref{spectrum}), it must determine the penetration depth of the 
ferromagnetic magnetization into the paramagnetic phase, see Fig. \ref{Figpenetration}.
Note that this $\delta n\simeq\hat\xi$ in the general Eq. (\ref{tildexigeneral}).  

What is more, the analytic solution (\ref{spectrum}) implies a finite energy gap 
in the spectrum of $H^+$,
\be
\hat\Delta~=~\omega_0~+~\omega_1~=~\sqrt{8\alpha}~\simeq\alpha^{1/2}~,
\ee
in agreement with the general scaling predicted in Eq. (\ref{tildeDeltageneral}).

\subsubsection{ Inhomogeneous transition } \label{IsingKZinhom}

\begin{figure}
\begin{center}
\includegraphics[width=0.8\columnwidth,clip=true]{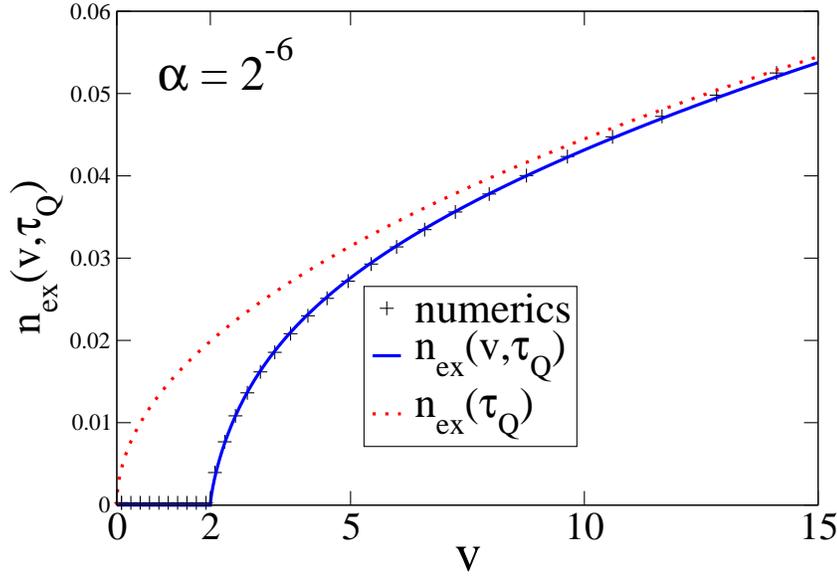}
\caption{ 
Comparison between the quasiparticle density predicted for an inhomogeneous
transition in Eq. (\ref{nexinhom}) (solid blue), the quasiparticle density
after a homogeneous transition in Eq. (\ref{nexIsing}) (dotted red), and 
numerical simulations on a lattice of $N=1000$ spins (crosses) at a fixed 
inhomogeneity $\alpha=2^{-6}$. 
(Figure from Ref. \cite{Ramsinhom})
}
\label{Figv}
\end{center}
\end{figure}

The general argument in Section \ref{KZinhom} can be also illustrated by a solution
in the quantum Ising chain. Consider a time-dependent transverse field $g_n(t)$ which 
can be linearised near the critical point $g_c=1$ as
\be
g_n(t)~\approx~1~+~\alpha~(n-vt)~.
\label{tslope}
\ee
A critical point at $n=vt$ moving with a constant velocity $v>0$ separates an expanding 
ferromagnetic phase for $n<vt$ from a shrinking paramagnetic phase for $n>vt$.
We want to know what is the density of quasiparticle excitations left behind the 
front in the ferromagnetic phase?

The time-dependent Bogoliubov-de Gennes equations are
\bea
i\frac{d}{dt} u_{n,m}^\pm = 2 g_n(t) u_{n,m}^\mp - 2 u_{n\mp 1,m}^\mp ~,
\label{exacttBdGinhom}
\eea
where $u_{nm}^\pm\equiv u_{nm} \pm v_{nm}$ are combinations of Bogoliubov modes in 
the transformation $c_n=\sum_{m=0}^{N-1}(u_{nm}\gamma_m + v^*_{nm} \gamma_m^\dagger)$. 
A long wavelength approximation
$
u_{n\mp 1,m}^\mp \approx u_{n,m}^\mp \mp \frac{\partial}{\partial n}u_{n,m}^\mp
$
leads to approximate long wavelength equations
\bea
i\partial_t 
\left(
\begin{array}{c}
u^+ \\
u^-
\end{array}
\right)
~=~
\left[
2\alpha(n-vt) \sigma^x +
2i \sigma^y \partial_n
\right]
\left(
\begin{array}{c}
u^+ \\
u^-
\end{array}
\right)~.
\label{tBdGinhom}
\eea
These equations were solved in Ref. \cite{Ramsinhom}. Their 
analytic solution predicts the density of quasiparticle excitations
\be
n_{\rm ex}(v,\tau_Q)~=~
\left\{
\begin{array}{cl}
\left(1-\frac{4}{v^2}\right)^{3/4} 
n_{\rm ex}(\tau_Q)  & 
{\rm ~~when~ } v>2~, \\
0                   & 
{\rm ~~otherwise~ } ~,
\end{array}
\right.
\label{nexinhom}
\ee
where $n_{\rm ex}(\tau_Q)=\frac{1}{2\pi\sqrt{2\tau_Q}}$ is the density 
after a uniform transition in Eq. (\ref{nexIsing}) with $\tau_Q=1/\alpha v$. 
The analytic solution is not consistent with the long wavelength approximation
in Eq. (\ref{tBdGinhom}) near $v=2$. This is why the exact equations 
(\ref{exacttBdGinhom}) were also solved numerically. Figure \ref{Figv} compares 
the analytic result (\ref{nexinhom}), the numerical results, and the result 
for a homogeneous transition in Eq. (\ref{nexinhom}).

As expected from the general scaling argument in \ref{KZinhom}, in the Ising 
chain there is a threshold velocity $\hat v\simeq1$ below which excitation 
of quasiparticles is suppressed with respect to the homogeneous KZM. Analytic
solution of the Ising chain gives the precise value $\hat v=2$. Not incidentally,
this is the fastest group velocity of quasiparticles at the critical
point whose dispersion relation is $\epsilon_k\approx2|k|$ for $|k|\ll\pi$. As 
we have seen in Section \ref{IsingKZinspace},
the spontaneous magnetization from the ferromagnetic phase behind the front
penetrates into the paramagnetic phase ahead of the front, see Fig. 
\ref{Figpenetration}. The paramagnetic spins near the critical front are biased 
to choose the same magnetization as in the ferromagnetic phase and in this way 
excitation of kinks (quasiparticles) behind the front is suppressed. The threshold 
$\hat v=2$ is the fastest velocity at which the penetrating order parameter can catch up
with the moving front. Indeed, the solution in Ref. \cite{Ramsinhom} shows
that when $v<2$ the ferromagnetic magnetization penetrates into the paramagnetic 
phase to a depth
\be
\delta n(v) ~\simeq~ 
\left(1-\frac{v^2}{4}\right)^{1/4}~
\alpha^{-1/2}~
\ee
which shrinks to zero when $v\to2^-$. Above $\hat v=2$ there is no bias
to suppress kink excitations.

\subsection{ Quench across a multicritical point of the XY chain }
\label{subsectionmulticritical}

\begin{figure}[t]
\begin{center}
\includegraphics[width=0.6\columnwidth,clip=true]{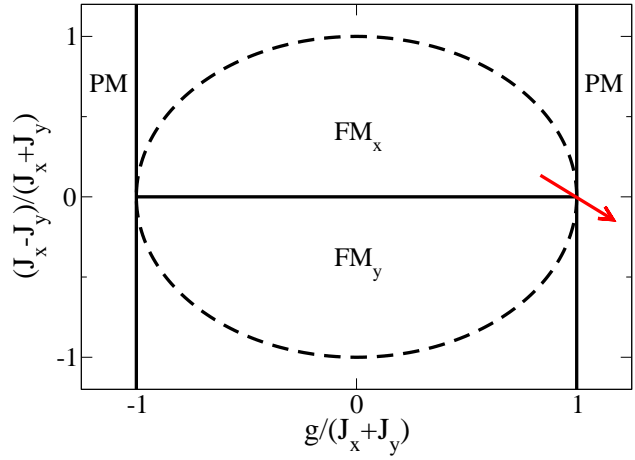}
\caption{ 
Phase diagram of the XY model (\ref{HXY}) in the plane of anisotropy
$(J_x-J_y)/(J_x+J_y)$ (vertical axis) and transverse field 
$g/(J_x+J_y)$ (horizontal axis). The vertical bold lines denote Ising transitions.
The system is also gapless along the horizontal critical line.
FM$_x$ (FM$_y$) marks a ferromagnetic phase with long range order in 
the $x$ ($y$) direction, and PM marks paramagnetic phases. The considered
multicritical point is located at $(1,0)$. The red arrow indicates direction
of the linear quench near the multicritical point.
}
\label{FigXY}
\end{center}
\end{figure}

The $XY$ model is a generalisation of the Ising model (\ref{Hsigma})
\be
H~=~-\sum_n
\left(
J_x\sigma^x_{n}\sigma^x_{n+1}+
J_y\sigma^y_{n}\sigma^y_{n+1}+
g\sigma^z_n
\right)~.
\label{HXY}
\ee
The Jordan-Wigner transformation (\ref{JordanWigner}) followed by a Fourier transform
$c_n=\sum_{k}c_ke^{ikn}/\sqrt{N}$ maps the $XY$ model to a one-dimensional quadratic 
fermionic Hamiltonian 
\be 
H~=~\sum_k
\left\{
2\left[g-(J_x+J_y)\cos k\right]c_k^\dag c_k+
i(J_x-J_y)\sin k~ c_{-k}c_k+{\rm h.c.}
\right\}~.
\ee
This Hamiltonian has the general form (\ref{HHk}) with
\be  
H_k~=~2
\left[
\begin{array}{cc}
g-(J_x+J_y)\cos k & i(J_x-J_y)\sin k \\ 
-i(J_x-J_y)\sin k    & -g+(J_x+J_y)\cos k
\end{array}
\right]~.
\label{Hkmulti}
\ee
The multicritical point is located at $J_x=J_y=\frac12g$, see Fig. \ref{FigXY}, where the Hamiltonian 
(\ref{Hkmulti}) is $H_k=2g(1-\cos k)\sigma^z$ with a quasiparticle spectrum $\omega_k=2g(1-\cos k)$. This critical 
spectrum is quadratic in $k$ for small $k$, hence $z=2$ is the dynamical exponent.

We make a linear quench across this critical point along a line in the parameter space 
where $J_x=\frac12g-\epsilon(t),J_y=\frac12g$ with the usual $\epsilon(t)=t/\tau_Q$,
see Fig. \ref{FigXY}. 
Along this quench line the Hamiltonian (\ref{Hkmulti}) is
\be 
H_k(t)~=~\epsilon(t)~\left[2(\sigma^z\cos k+\sigma^y\sin k)\right]~+~\left[2g(1-\cos k)\sigma^z\right]~.
\label{Hveckmulti}
\ee
At $k=0$ its spectrum is linear in $|\epsilon|$, $\omega_0=2|\epsilon|$, hence 
the critical exponents satisfy $\nu z=1$ and, consequently, $z=2,\nu=1/2$. 
With these exponents the simple KZM estimate (\ref{nex}) predicts 
$n_{\rm ex}\sim\tau_Q^{-1/4}$.

Equation (\ref{Hveckmulti}) corresponds to the general Eq. (\ref{Hveck}). The
orthonormalisation that follows Eq. (\ref{Hveck}) leads to Eq. (\ref{HabDelta})
with functions
\bea
\Delta(k) &=& 2g(1-\cos k)|\sin k|~,\\
b(k)      &=& 2g(1-\cos k)\cos k~.
\eea
Near the Fermi point $k_F=0$ we can expand $\Delta^2(k)\sim|k|^6$ and $b^2(k)\sim k^4$
and identify the exponents in Eqs. (\ref{zDelta},\ref{zb}) as
\be
z_\Delta~=~6~,~~z_b~=~4~.
\ee
Here the inequality $z_{\Delta}\leq z_b$ in Eq. (\ref{zDeltazb}) is not satisfied 
and the exact density
\be
n_{\rm ex}~\sim~\tau_Q^{-1/z_\Delta}~=~\tau_Q^{-1/6}
\label{16}
\ee
decays with a different exponent than the $1/4$ predicted by the simple KZM estimate
in Eq. (\ref{nex}). The anomalous scaling (\ref{16}) was obtained for the first 
time in Ref. \cite{multicritical}, see also \cite{HindusiReview}. More examples
of multicritical anomalous scaling can be found in Ref. \cite{Viola3}.

\subsection{ Kitaev model in 2D: quench across a gapless phase }
\label{quenchKitaev}

\begin{figure}[t]
\begin{center}
\includegraphics[width=0.8\columnwidth,clip=true]{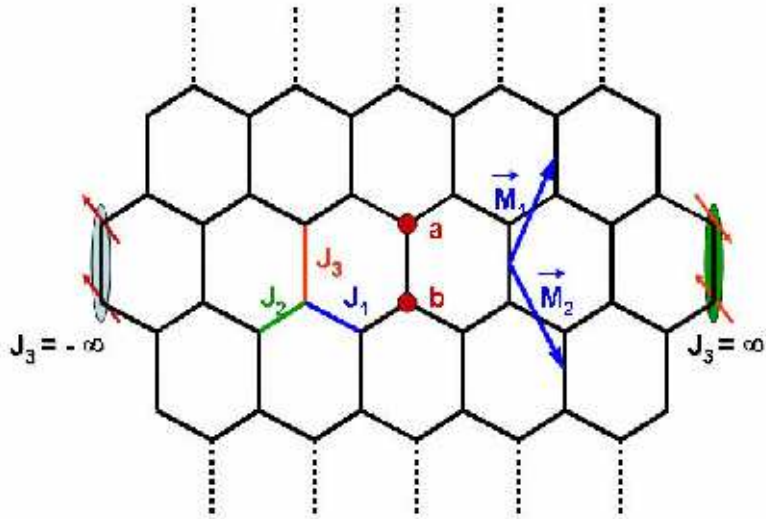}
\caption{
Schematic representation of the Kitaev model.
(Figure from Ref. \cite{Kitaev})
}   
\label{FigKitaev}
\end{center}
\end{figure}

The Kitaev model is a spin-$1/2$ model on a two-dimensional honeycomb lattice 
in Fig. \ref{FigKitaev} with a Hamiltonian \cite{KitaevModel}
\be
H~=~\sum_{j+l={\rm even}}
\left(
J_1~\sigma^x_{j,l}\sigma^x_{j+1,l}~+~
J_2~\sigma^y_{j-1,l}\sigma^y_{j,l}~+~
J_3~\sigma^z_{j,l}\sigma^z_{j,l+1} 
\right)~,
\ee
where $j$ and $l$ are row and column indices of the lattice, see 
Fig. \ref{FigKitaev}. It is an exactly solvable model with a gapless
phase for $|J_1-J_2|\leq J_3\leq J_1+J_2$. The model can be mapped
onto a non-interacting fermionic model by a suitable Jordan-Wigner 
transformation \cite{KitaevModel},
\be
H~=~i\sum_{\vec n}
\left(
J_1 b_{\vec n} a_{\vec n-\vec M_1} ~+~
J_2 b_{\vec n} a_{\vec n+\vec M_2} ~+~
J_3 D_{\vec n} b_{\vec n} a_{\vec n}
\right)~,
\label{HFKitaev}
\ee
where $a_{\vec n}$ and $b_{\vec n}$ are Majorana fermions sitting at the top
and bottom sites respectively of a bond labelled $\vec n$, 
vectors $\vec n=\sqrt{3}\hat i n_1+(\frac{\sqrt{3}}{2}\hat i+\frac32\hat j)n_2$
denote the midpoints of the vertical bonds shown in Fig. \ref{FigKitaev},
and $n_1,n_2$ are integers. The vectors $\vec n$ form a triangular reciprocal
lattice whose spanning vectors are $\vec M_1=\frac{\sqrt{3}}{2}\hat i+\frac32\hat j$
and $\vec M_2=\frac{\sqrt{3}}{2}\hat i-\frac32\hat j$. The operator $D_{\vec n}$ commutes 
with $H$ and can take values $\pm 1$ independently for each $\vec n$. The ground
state corresponds to $D_{\vec n}=1$ at every bond. Since $D_{\vec n}$
is a constant of motion, the dynamics of the model starting from the ground
state will never take the system out of the subspace with $D_{\vec n}=1$.

When $D_{\vec n}=1$ a Fourier transform brings the fermionic Hamiltonian 
(\ref{HFKitaev}) to $H=\sum_{\vec k}\psi_{\vec k}^\dag H_{\vec k} \psi_{\vec k}$,
where $\psi_{\vec k}^\dag=(a_{\vec k}^\dag,b_{\vec k}^\dag)$ are Fourier
transforms of $a_{\vec n}$ and $b_{\vec n}$, and the sum over $\vec k$ extends
over half the first Brillouin zone of the triangular lattice formed by vectors $\vec n$.
Here we quench
\be
J_3(t)~=~\epsilon(t)~=~\frac{t}{\tau_Q}
\ee 
from $t\to-\infty$ to $t\to+\infty$. The Hamiltonian $H_{\vec k}$ has the general
form of Eq. (\ref{HabDelta}) with
\bea
\Delta(\vec k) &=& 2[J_1\sin(\vec k\vec M_1)-J_2\sin(\vec k\vec M_2)]~,\\
b(\vec k)      &=& J_1\cos(\vec k\vec M_1)+J_2\cos(\vec k\vec M_2) ~,
\eea
$a(\vec k)=2$, $\sigma(\vec k)=\sigma^y$, and $\sigma_{\perp}(\vec k)=\sigma^x$.
The eigenvalues of $H_{\vec k}$ are $\pm\omega_{\vec k}$ and the quasiparticle spectrum 
$\omega_{\vec k}=\sqrt{4[\epsilon+b(\vec k)]^2+\Delta^2(\vec k)}$ is gapless in 
the gapless phase where $|J_1-J_2|\leq\epsilon\leq|J_1+J_2|$. For each $\epsilon$ 
in this range there is a Fermi point $\vec k_F$ such that $\Delta(\vec k_F)=0$
and $\omega_{\vec k_F}=0$. 

The set of all Fermi points for different $\epsilon$ is a one dimensional Fermi line 
where $\sin(\vec k\vec M_1)=\sin(\vec k\vec M_2)$. When $\tau_Q\gg1$ the excitation
probability $p_{\vec k}$ in Eq. (\ref{pveck}) is $1$ on the Fermi line and exponentially 
small everywhere except near this line, see Figure \ref{pk_Kitaev}. Expansion in 
$\vec k-\vec k_F$ near the Fermi line gives $z_{\Delta}=2,z_b=4$ when $J_1=J_2$ and 
$z_{\Delta}=2,z_b=2$ otherwise. In any case, the integration in Eq. (\ref{nexBZ}) gives
\be
n_{\rm ex}~\simeq~\tau_Q^{-\frac{d-m}{z_{\Delta}}}~=~\tau_Q^{-1/2}~.
\ee
This result was obtained for the first time in Ref. \cite{Kitaev}.
A sudden quench from a topologically ordered Hamiltonian to a Hamiltonian that does not
support topological order was considered in Ref. \cite{TopoQuench}.

\subsection{The random Ising chain:
            logarithmic dependence of excitation density on transition rate}
\label{RandomIsing}

An example of the random Ising chain is \cite{random,DFisher}
\be
H~=~-\sum_{n=1}^N \left( g~\sigma^z_n + J_n~\sigma^x_n\sigma^x_{n+1} \right)~.
\label{randomHsigma}
\ee 
with periodic boundary conditions $\vec\sigma_{N+1}~=~\vec\sigma_1$. Here $J_n$'s 
are random ferromagnetic couplings which, without loss of generality, can be assumed 
positive, $J_n>0$, and $g$ is a uniform transverse magnetic field. 
This model has two quantum critical points at 
$g_c=\pm\exp\left(\overline{\ln J_n}\right)$ separating a ferromagnetic phase, $|g|<g_c$, 
from two paramagnetic phases, $|g|>g_c$. 
The randomness is a relevant perturbation leading to a different universality class than 
the pure Ising model. 

When ensemble averaged quantities are considered, the critical exponent is $\nu=2$, 
instead of $\nu=1$ in the pure case, and the dynamical parameter $z$ diverges as
\be
z~\approx~\frac{1}{2|\epsilon|}
\label{zrandom}
\ee
near the critical point where $\epsilon\equiv(g-g_c)/g_c=0$. 

In a zero order approximation \cite{JDrandom,JDrandom2,decoherence}, one might take first the 
limit $\epsilon\to0$ which implies $z\to\infty$. In this limit, the general KZM 
estimate (\ref{hatxi}) implies
$
\hat\xi~\simeq~1~,
$ 
i.e., neither the correlation length $\hat\xi$ nor the excitation density 
$n_{\rm ex}\simeq\hat\xi^{-1}$ depend on the transition time $\tau_Q$. No matter how 
slow the transitions is, the density of excitations remains the same. This approximation 
demonstrates that there is no usual {\it power law} KZ scaling in the random Ising model.
However, it is too crude to exclude a weak logarithmic dependence.

A more accurate result is obtained with the adiabatic-impulse-adiabatic approximation 
which is central to KZM \cite{JDrandom,decoherence}, see Section \ref{KZargument}. The 
energy gap depends on $\epsilon$ as 
$
\Delta~\simeq~|\epsilon|^{z\nu}~\simeq~|\epsilon|^{1/|\epsilon|}~
$
while the transition rate is $|\dot\epsilon/\epsilon|=1/|t|=1/|\tau_Q\epsilon|$. The 
rate equals the gap at $\hat\epsilon$ when
\be
\frac{\alpha}{\tau_Q~\hat\epsilon}~=~\hat\epsilon^{1/\hat\epsilon}~,
\label{hatepsilonq}
\ee
where $\alpha\simeq1$ is a non-universal parameter. When $\ln\tau_Q\gg1$ an approximate 
solution is
\be
\hat\xi~\sim~
\hat\epsilon^{-2}~\sim~
\left( \ln \tau_Q \right)^2~.
\label{xilog}
\ee 
This logarithmic dependence on $\tau_Q$ is very weak as compared to any power law scaling. 
It means that no matter how slow the transition is the ensemble-averaged density of defects 
in the final ferromagnetic phase, 
\be
n_{\rm ex}~\simeq~\hat\xi^{-1}~\sim~(\ln\tau_Q)^{-2}~,
\label{nexrandom} 
\ee
remains roughly the same.

\begin{figure}[t]
\begin{center}
\includegraphics[width=0.6\columnwidth,clip=true]{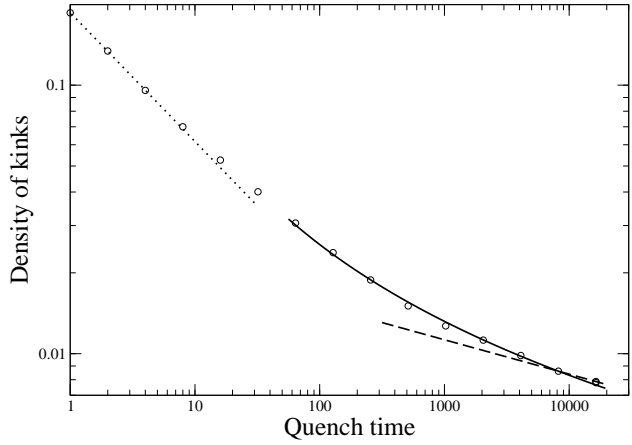}
\caption{ Density of kinks $n_{\rm ex}$ as a function of transition time $\tau_Q$ 
on a lattice of $N=512$ sites. Circles are final kink densities averaged over 4 
realizations of $J_n$. Error bars set the size of the circles. The solid line is the 
best fit $n_{\rm ex}(\tau_Q)=\frac{0.15}{\hat{\epsilon}^2(\tau_Q/3.4)}$ where 
$\hat\epsilon(\tau_Q/\alpha)$ is the solution of Eq.(\ref{hatepsilonq}). The dotted line 
is the best power law fit $d\sim\tau_Q^{-w}$ to the left-most 3 data points and the 
dashed line is the ``fit'' to the right-most 2 data points. The exponents 
are $w=0.48$ and $w=0.13$ respectively. The exponent $w=0.48$ for the fastest transitions 
is consistent with the exponent $\frac12$ in the pure Ising model. 
(Figure from Ref. \cite{JDrandom})
}
\label{fig_kinks}
\end{center}
\end{figure}

The prediction (\ref{xilog}) was tested by numerical simulation of the time-dependent
Bogoliubov-de Gennes equations
\bea
i\frac{d}{dt} u^\pm_{n,m} ~=~
2 g(t) u^\mp_{n,m} - 
J_{n-\frac12\mp\frac12,m}   
u^\mp_{n\mp1,m}~.
\label{tBdGrandom}
\eea
Here $u^\pm_{n,m}=u_{n,m}\pm v_{n,m}$ are combinations of Bogoliubov modes in the 
transformation $c_n=\sum_{m=0}^{N-1}(u_{nm}\gamma_m + v^*_{nm} \gamma_m^\dagger)$.
The random $J_n$'s were uniformly distributed in $(0,2)$ implying $g_c=2e^{-1}$. 
$g(t)=-g_c~t/\tau_Q$ was quenched from a large initial $g\gg g_c$ to the final $g=0$ 
where density of kinks was calculated. The initial conditions were stationary modes of 
the Bogoliubov-de Gennes equations (\ref{tBdGrandom}) at the initial large $g$ corresponding 
to the ground state of the system. 

Results of the simulations are collected in Fig. \ref{fig_kinks}. They confirm the logarithmic 
dependence in Eq. (\ref{xilog}) for large $\tau_Q$. However, for relatively fast transitions 
that freeze out at relatively large $\hat\epsilon$, the scaling $n_{\rm ex}\sim\tau_Q^{-1/2}$ 
characteristic for the pure Ising model is recovered, compare Eq. (\ref{nexIsing}). These 
transitions cannot feel the effect of disorder that is appreciable only when $\hat\epsilon$ is 
close enough to the critical point.

The density of kinks scales as $n_{\rm ex}\sim(\ln\tau_Q)^{-2}$. Since each kink contributes to 
the excitation energy, one might expect that the excitation energy density $\varepsilon$ at the 
final $g=0$ is proportional to $n_{\rm ex}$. Contrary to this simple expectation, the 
ensemble-averaged energy density was found to scale as
\be
\varepsilon~\sim~(\ln\tau_Q)^{-(3.4\pm0.2)}
\ee
with an exponent significantly greater than $2$, see Ref. \cite{JDrandom2}. The origin of the steeper 
exponent can be traced back to the uniform distribution of $J_n\in(0,2)$ which does not exclude arbitrarily 
weak bond strengths $J_n$. With increasing $\tau_Q$ the excited kinks tend to localize more and more on the 
weakest bonds, which are the easiest to excite, making 
the excitation energy decay faster than the kink density.

An alternative derivation of the scaling (\ref{nexrandom}) from the Landau-Zener theory, taking
into account distribution of minimal gaps at different values of $g$ during the quench, can be
found in Ref. \cite{JDrandom2}. This method originates from the theory of quantum annealing
\cite{QA}.

In summary, in the random Ising model, representing the infinite disorder universality class, 
density of excitations does not follow the usual KZM power law decay with the transition time, 
but the actual logarithmic dependence can still be obtained from the adiabatic-impulse-adiabatic 
approximation essential for KZM. 
This example and that in Ref. \cite{ZakrzDelande} suggest that disorder, when a relevant perturbation, 
makes excitation much easier than in the pure case, but we would need more evidence to support 
this claim as a general conclusion.

\subsection{ Topological insulators: anomalous excitation of edge states }
\label{Topological}

Topological insulators are an intriguing state of matter where edge transport exists
even in the presence of a bulk energy gap \cite{Kane}. The edge states responsible
for the transport are usually localised in the interface separating two topologically
different insulators, the simplest example being an integer quantum Hall effect
sample and the vacuum \cite{Hatsugai}. The topological edge states arise
in a variety of systems such as one dimensional spin models \cite{Affleck},
the integer and fractional quantum Hall effect \cite{IFQHE}, and have been experimentally
realised in topological insulators \cite{TIexp}. They also include
the anomalous half-integer quantum Hall effect in the honeycomb or square 
lattice \cite{AQHE} and the quantum Hall spin effect \cite{QHSE}. Depending on the symmetries
of the Hamiltonian and the dimension of the system topological insulators can be
classified in a periodic table \cite{TIPT}. One of the simplest systems is the one 
dimensional chain of Majorana fermions \cite{MajoranaF}, where the edge states are 
the Majorana fermions localised at the ends of the chain. Isolation of Majorana fermions 
would be an important step towards topological quantum computation \cite{topoQC}. 
Various proposals include the vortex core of two dimensional $p+ip$ superconductor 
\cite{pip}, tri-junctions of superconductor-topological insulator-superconductor 
\cite{tri}, and many others \cite{TIothers}.

Reference \cite{Delgado2} considers the one dimensional fermionic chain
\cite{MajoranaF}
\be
H~=~
\sum_j
\left(
-w a_j^\dag a_{j+1}+
\Delta a_j a_{j+1}-
\frac{\mu}{2}a_j^\dag a_j+
{\rm h.c.}
\right)
\label{HMajorana}
\ee
where $a_j$ are spinless fermionic lattice annihilation operators satisfying
$\{a_i,a_j^\dag\}=\delta_{ij}$. The fermions hopping between lattice sites with the
tunnelling frequency $w$ can be injected or removed from the wire in the form
of Cooper pairs. In the following we assume $\mu=0$ and imaginary $\Delta=i|\Delta|$ 
for simplicity.

In case of periodic boundary conditions, $a_{N+1}=a_1$, there are no edge
states and the system is translationally invariant. In the quasimomentum 
representation $a_j=\frac{1}{\sqrt{N}}\sum_k a_k e^{-ikj}$ with $k\in(-\pi,\pi]$,
the Hamiltonian becomes
\be
H~=~
\sum_k
\left(
a_k^\dag,a_{-k}
\right)
~H_k~
\left(
\begin{array}{c}
a_k \\
a_{-k}^\dag
\end{array}
\right)~,~~
H_k~=~
\left(
\begin{array}{cc}
~~-w \cos k~~      & ~~|\Delta|\sin k~~ \\
~~|\Delta|\sin k~~ & ~~w \cos k~~
\end{array}
\right)
\label{HMajoranak}
\ee
with energies of Bogoliubov quasiparticle excitations 
$\epsilon_k=2\sqrt{w^2\cos^2 k+|\Delta|^2\sin^2 k}$. These are the bulk excitations
of the topological insulator. For a fixed $|\Delta|>0$ they are gapless at a critical
point $w=0$ with the critical exponents $z=\nu=1$. 

Here we consider a linear quench 
\be
w(t)~=~\frac{t}{\tau_Q}
\ee
from $-|\Delta|$ to $|\Delta|$. Equation (\ref{HMajoranak}) becomes a set
of independent Landau-Zener problems for each $k$ like in Section \ref{LZargument}. 
When $\tau_Q\gg|\Delta|^{-2}$, then only quasiparticles with $k\approx0$ and $k\approx\pi$ 
get excited with probability $p_k=\exp(-\pi|\Delta|^2\tau_Q\sin^2 k/|\cos k|)$. In the thermodynamic
limit density of bulk excitations becomes an integral
\be
n_{\rm bulk}~=~
\int_{-\pi}^\pi\frac{dk}{2\pi}p_k~\approx~
\frac{1}{\pi|\Delta|}~\tau_Q^{-1/2}~.
\label{nbulk}
\ee
when $\tau_Q\gg|\Delta|^{-2}$. This is the usual KZ scaling expected for the critical exponents 
$z=\nu=1$.

\begin{figure}[t]
\begin{center}
\includegraphics[width=0.6\columnwidth,clip=true]{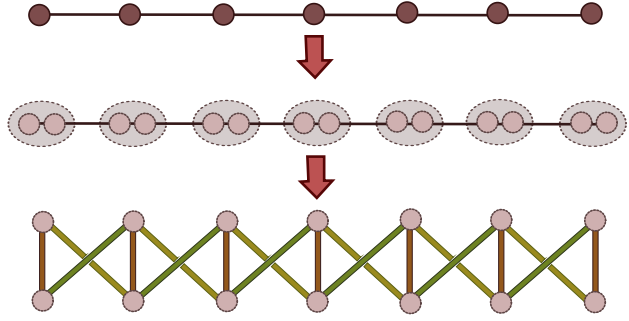}
\caption{ 
The fermionic chain (\ref{HMajorana}) is mapped to a Majorana ladder
by the transformation (\ref{majorana_definition}) which defines 2 Majorana
fermions on each lattice site. 
(Figure from Ref. \cite{Delgado2})
}
\label{majorana_ladder}
\end{center}
\end{figure}

So far everything was standard and, probably, not worth a special Subsection. 
However, as expected from a {\it topological} insulator, an open chain is 
different from the periodic chain because there are localised edge states. In
case of the open chain it is better to introduce hermitian Majorana fermions
\be
c_{2j-1}=
\frac{1}{\sqrt{2}}
\left(
e^{-i\pi/4}a_j^\dag+e^{i\pi/4}a_j
\right)~,~~
c_{2j}=
\frac{i}{\sqrt{2}}
\left(
e^{-i\pi/4}a_j^\dag-e^{i\pi/4}a_j
\right)~,
\label{majorana_definition}
\ee
satisfying $\{c_i,c_j\}=\delta_{i,j}$, which transform the Hamiltonian (\ref{HMajorana}) 
into
\be
H~=~i\sum_{j=1}^{N-1}
\left[
(w+|\Delta|)  c_{2j}c_{2j+1}+
(-w+|\Delta|) c_{2j-1}c_{2j+2}
\right]~.
\label{HMajoranac}
\ee
The Majorana fermions live on a virtual ladder whose $j$-th rung contains
two Majorana fermions $c_{2j-1},c_{2j}$ and corresponds to the $j$-th site
of the original chain, see Fig. \ref{majorana_ladder}. 

\begin{figure}
\begin{center}
\begin{minipage}{100mm}
\subfigure[]{
\resizebox*{4.5cm}{!}{\includegraphics{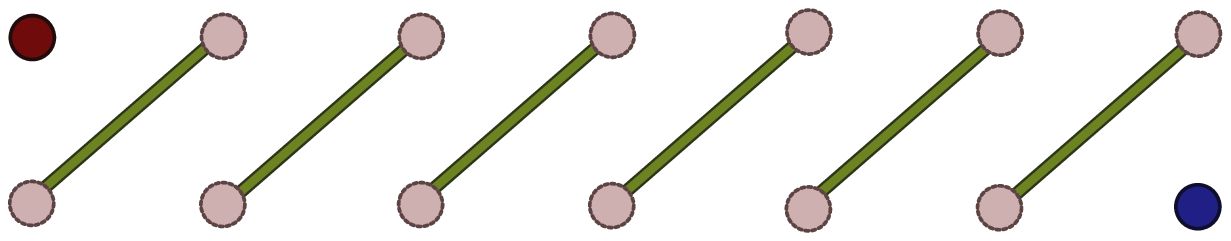}}}%
\hspace*{1cm}
\subfigure[]{
\resizebox*{4.5cm}{!}{\includegraphics{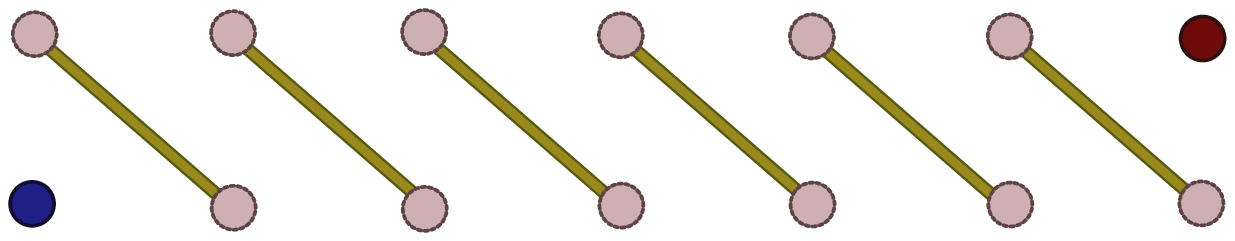}}}
\caption{
The gap-full paired bulk states and the unpaired edge zero modes for the initial 
$w=-|\Delta|$ (panel a) and the final $w=|\Delta|$ (panel b).
(Figure from Ref. \cite{Delgado2})
}
\label{majorana_states}
\end{minipage}
\end{center}
\end{figure}

At the initial $w=-|\Delta|$ the Hamiltonian 
(\ref{HMajoranac}) is a sum of $N-1$ products $2i|\Delta|c_{2j-1}c_{2j+2}$,
each of them having two eigenvalues $\pm|\Delta|$. In addition to them,
there are two Majorana operators $c_2,c_{2N-1}$ which, in case of periodic boundary
conditions, would couple through the term $2i|\Delta|c_{2N-1}c_{2}$
which is missing in the present open chain. Their Hilbert space has two states 
with zero energy. The paired and unpaired states are shown in 
Fig. \ref{majorana_states}a. 

In a similar way, at the final $w=|\Delta|$ there are $N-1$ products $2i|\Delta|c_{2j}c_{2j+1}$
with eigenenergies $\pm|\Delta|$ and two ``unpaired'' Majoranas
$c_1,c_{2N}$ with two zero energy states, see Fig. \ref{majorana_states}b. 
In both cases, the ``paired'' states are the gap-full bulk modes and the 
``unpaired'' zero energy states are the edge states created by opening the 
periodic chain.

The bulk and edge states can be written in a compact form with the help of the 
BCS state
$
|\Omega\rangle~=~
\prod_k
\left(
u_k+v_k a^\dag_{-k}a^\dag_k
\right)
|0\rangle~,
$
where $|0\rangle$ is a Fock vacuum for $a_j$ and the Bogoliubov coefficients are 
$u_k=\frac{1}{\sqrt2}\sqrt{1-\frac{w\cos k}{\epsilon_k}},
v_k=\frac{{\rm sign}(k)}{\sqrt2}\sqrt{1+\frac{w\cos k}{\epsilon_k}}$. 
This state is a zero energy state of the periodic chain with a Hamiltonian
$
H=\sum_{k>0}\epsilon_k
\left(
\gamma_{k,+}^\dag\gamma_{k,+}-
\gamma_{k,-}^\dag\gamma_{k,-}
\right)~,
$
where $\gamma_{k,+}=u_k^*a_k+v_ka^\dag_{-k}$ and $\gamma_{k,-}=-v_k^*a_k+u_ka^\dag_{-k}$. 
In case of the open chain and the initial $w=-|\Delta|$ the edge zero modes are 
$c_2|\Omega\rangle$ and $c_{2N-1}|\Omega\rangle$, and the bulk eigenstates of energies 
$\pm|\Delta|$ are $(\pm ic_{2j-1}+c_{2j+2})|\Omega\rangle$. At the final 
$w=|\Delta|$ the zero energy edge states are
$c_1|\Omega\rangle$ and $c_{2N}|\Omega\rangle$, and the bulk eigenstates
of energies $\pm|\Delta|$ are $(\pm ic_{2j}+c_{2j+1})|\Omega\rangle$.

We assume that the initial state at $w=-|\Delta|$ is one of the two unpaired edge states: the left zero mode $|L(-|\Delta|)\rangle=c_2|\Omega\rangle$ or the right one $|R(-|\Delta|)\rangle=c_{2N-1}|\Omega\rangle$. In adiabatic transition the left and right zero modes would evolve with increasing $w$ into \cite{MajoranaF,Delgado2}
\bea
|L(w)\rangle &\propto&
\left(c_2+r~c_6+\dots+r^{\frac{N+1}{2}}~c_{2N}\right)|\Omega\rangle~,\\
|R(w)\rangle &\propto&
\left(c_{2N-1}+r~c_{2N-5}+\dots+r^{\frac{N+1}{2}}~c_{1}\right)|\Omega\rangle~,
\eea 
where $r=(|\Delta|+w)/(|\Delta|-w)$ increases with $w$ from $r=0$ at the initial 
$w=-|\Delta|$, through $r=1$ at the critical $w=0$, to $r\to\infty$ at the final
$w=|\Delta|$. As $w$ increases, the edge states get 
increasingly de-localised until they become uniformly spread along the whole ladder at
the critical point. These de-localised states can be also written as 
\bea
|L(0)\rangle ~\propto~
\left(
\gamma^\dag_{-\pi,-} + \gamma_{0,+}^\dag
\right)|\Omega\rangle~,~~
|R(0)\rangle ~\propto~
\left(
\gamma^\dag_{-\pi,-} - \gamma_{0,+}^\dag
\right)|\Omega\rangle~.
\eea  
Each of these two states is an equal superposition of two gapless bulk
modes from the positive and negative energy branch. Thus no matter how slow,
the evolution at this point cannot be adiabatic and any further increase of 
$w$ above $0$ will not connect these states to the edge states at the
final $w=|\Delta|$. An initial edge state will end with probability $1/2$ 
in either the positive or negative energy band. Thus the probability to excite 
a defect is $1/2$ and does not depend on $\tau_Q$.

A similar analysis in Ref. \cite{Delgado1} in another topological insulator - 
the Creutz ladder \cite{Creutz} - concludes that, while the density
of bulk excitations decays like $\tau_Q^{-1/2}$, the edge modes are excited 
with a probability scaling as $\tau_Q^{-1.35}$. This is much steeper decay than 
for the bulk excitations in contrast to the Majorana chain, where the edge modes 
are excited with the probability $1/2$ independent of $\tau_Q$. Unlike
in the Majorana chain, the edge modes of the Creutz ladder remain localised at 
the critical point. These two examples demonstrate that the edge modes, whose
existence is ensured by topology, need special treatment in dynamical phase
transitions.

\subsection{ The Lipkin-Meshkov-Glick model: KZM and infinite coordination number }
\label{quenchLMG}

Most examples in this review are in one or at most two spatial dimensions, but in this 
Section we follow Ref. \cite{LMG} and consider the opposite extreme of infinite 
coordination number. In this limit it is hard to identify an analogue of the KZ correlation 
length $\hat\xi$ but, nevertheless, the density of excitations can still be obtained from
KZM. 

The specific model that we consider is the exactly solvable Lipkin-Meshkov-Glick model, 
\be
H~=~-\frac{2}{N}\sum_{i<j}
\left(
\sigma^x_i \sigma_j^x + \gamma \sigma^y_i \sigma^y_j
\right)
-g\sum_{i=1}^N \sigma^z_i~.
\ee 
It was introduced for the first time in the context of nuclear physics \cite{LMGmodel}
and then thoroughly studied in a number of papers \cite{LMGstudy}. Here $N\to\infty$ is 
a number of spins in the system, $\gamma\leq1$ is the anisotropy parameter, and $g$ 
is the transverse field. In a sense, the model is the infinite coordination number limit of 
the XY model or, when $\gamma=0$, the quantum Ising chain. 

The model has a second order quantum phase transition at $g_c=1$ with mean-field critical 
exponents. The magnetization in the $x$-direction (or the $xy$-plane in the isotropic case 
of $\gamma=1$) is
\be
m ~=~ (1-g^2)^{1/2}
\ee
in the ferromagnetic phase when $g\leq1$ and zero otherwise. The energy gap vanishes
at the transition as
\be
\Delta ~=~ \sqrt{(g-1)(g-\gamma)}
\ee
for $g>1$. The ferromagnetic ground state below $g=1$ is doubly degenerate for any 
$\gamma<1$. 

A sudden quench in the Ising version of this model, when $\gamma=0$, was considered
in Ref. \cite{LMGIsing}. Here we follow Ref. \cite{LMG} and consider a linear quench
\be
g(t)~=~-\frac{t}{\tau_Q}~
\ee
from the ground state at $g\to\infty$ to $g=0$. In the adiabatic limit we would expect 
final magnetization $m=1$ with all spins pointing in the same direction. After a 
transition with a finite rate a fraction $m_{\rm inc}$ of spins is reversed and the 
magnetization is {\it incomplete} 
\be
m~=~1-m_{\rm inc}~.
\ee 
This reduced magnetization costs a finite residual excitation energy per site 
$\frac{E_{res}}{N}$. When $m_{\rm inc}\ll1$ then $\frac{E_{res}}{N}\simeq m_{\rm inc}$. 

\begin{figure}
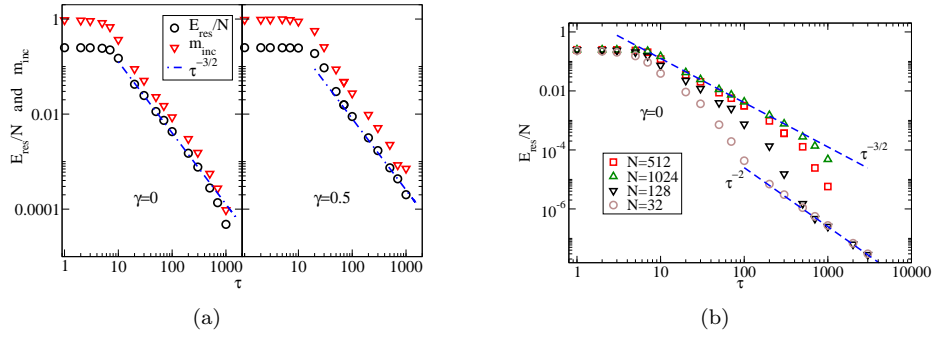

\begin{center}
\begin{minipage}{140mm}
\subfigure[]{
\resizebox*{5.5cm}{!}{\includegraphics{eres_mlac_gam_var.eps}}}%
\hspace*{1cm}
\subfigure[]{
\resizebox*{5.5cm}{!}{\includegraphics{E_res_vs_tau_lipkin_Nlarge.eps}}}
\caption{
In a, residual energy per spin and $m_{\rm inc}$ as a function of $\tau_Q$
in a system of $N=1024$ sites. For slow enough quenches the data are consistent
with the $\tau_Q^{-3/2}$ scaling.
In b, residual energy per spin as a function of $\tau_Q$ for different system
sizes ranging from $N=32$ to $N=1024$. For very large $\tau_Q$ the dependence
of energy on the transition rate crosses over to a steeper power law $\tau_Q^{-2}$,
as explained in Ref. \cite{LMG} by incomplete LZ anti-crossings.
(Figure from Ref. \cite{LMG})
}
\label{FigLMG}
\end{minipage}
\end{center}
\end{figure}

In Ref. \cite{LMG} the linear quench was evolved numerically with the results collected in 
Fig. \ref{FigLMG}a. When $\tau_Q\lesssim1$ we can see saturation at $m_{\rm inc}\approx1$.
These quenches are effectively instantaneous as there is simply not enough time to build 
up any ferromagnetic magnetization. However, when $\tau_Q\gg1$ then we observe a scaling 
consistent with
\be
\frac{E_{res}}{N} ~\sim~ m_{\rm inc} ~\sim~ \tau_Q^{-3/2} ~.
\label{LMG32}
\ee   
The exponent can be explained as follows.

Since the parity operator $\prod_{n=1}^N\sigma^z_n$ is a good quantum number and the initial 
ground state at large $g$ is the even-parity state fully polarized along the 
$z$-axis, then we can confine to the subspace of even parity. In this subspace,
when $g\gg1$ then the instantaneous gap to reverse two spins is $4g\approx \frac{4|t|}{\tau_Q}$.
On the other hand, the gap at the critical $g_c=1$ scales as 
\be
\Delta_c~\simeq~N^{-1/3}~
\ee 
when $N\to\infty$. Consequently, the transition between the ground state and the first excited state 
can be described by an effective LZ model with a time-dependent Hamiltonian
\be
H_{\rm LZ}~=~
\left(
\begin{array}{cc}
~-4\left(\frac{t}{\tau_Q}+g_c\right)~ &  ~\Delta_c~            \\
~\Delta_c~                            &  ~4\left(\frac{t}{\tau_Q}+g_c\right)~ 
\end{array}
\right)~
\label{LZLMG}
\ee
and the LZ excitation probability is 
\be
P~=~
\exp\left( -\frac{\pi}{4} \Delta_c^2 \tau_Q \right) ~\simeq~
\exp\left( -\frac{\pi}{4} N^{-2/3}   \tau_Q \right) 
\label{PLZLMG}
\ee
From this probability we can read that, for a given $\tau_Q$, the maximal size of a defect-free 
system is $N_{\rm free}\simeq\tau_Q^{3/2}$. In a similar way as in Section \ref{FiniteIsing} and 
Ref. \cite{KZIsing}, we can use its inverse $N_{\rm free}^{-1}$ as an estimate of the density
of reversed spins,
\be
m_{\rm inc} ~\simeq~ N_{\rm free}^{-1} ~\simeq~\tau_Q^{-3/2}~.
\ee 
This estimate explains the $\frac32$-scaling observed in Fig. \ref{FigLMG}a.

\subsection{ Bose-Hubbard model: 
             transition between gapped Mott insulator and gapless superfluid }
\label{KZinBH}

\begin{figure}
\begin{center}
\includegraphics[width=0.7\columnwidth,clip=true]{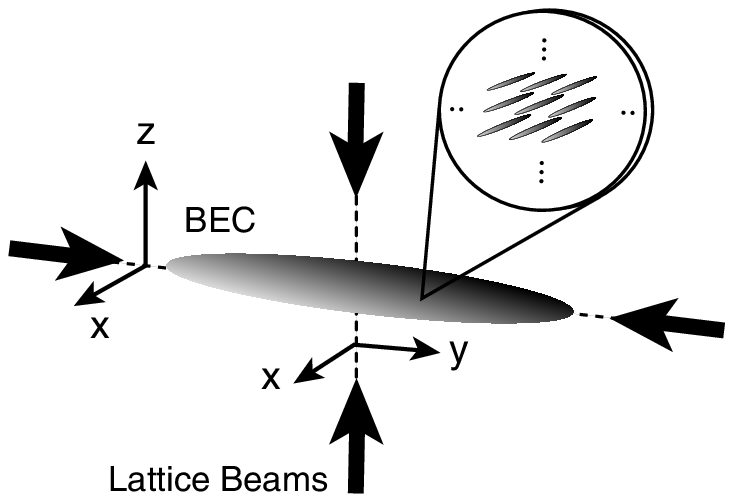}
\caption{ 
Schematic set up of the experiment in Refs. \cite{Greiner,GreinerPRL}.
A 2D lattice is formed by overlapping two optical standing waves along
the $y$-axis and the $z$-axis with a Bose-Einstein condensate in a harmonic trap.
The condensate is then confined to an array of several thousand narrow potential tubes.
(Figure from Ref. \cite{GreinerPRL})
} 
\label{FigGreiner}
\end{center}
\end{figure}

\begin{figure}[t]
\begin{center}
\includegraphics[width=0.95\columnwidth,clip=true]{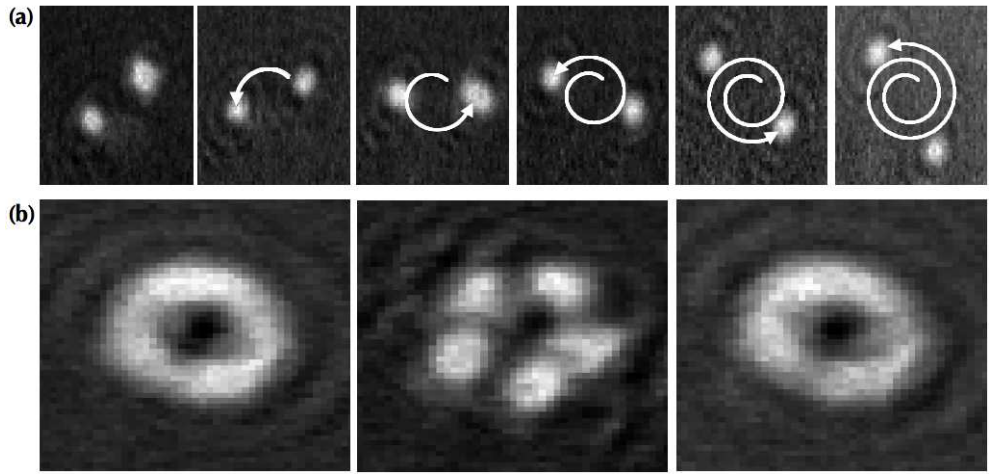}
\caption{ Painting an arbitrary and time-dependent potential by a rapidly moving laser beam.
          In a) a rotating two-site potential with Bose-Einstein condensates (BEC's), and in b)
          a toroidal BEC which is adiabatically converted into 5 disconnected spots and then
          back into toroidal form. (Figure from Ref. \cite{boshier})
}
\label{FigBoshier}
\end{center}
\end{figure}

The non-integrable Bose-Hubbard model is one of the two paradigmatic examples of systems with
a quantum phase transition \cite{Sachdev}, the other being the already described integrable 
quantum Ising chain. What may be even more important, the model was realised experimentally 
with ultracold atomic gases \cite{Greiner,GreinerPRL,Kasevich}, see Fig. \ref{FigGreiner}. Here we consider mainly 
its one-dimensional version 
\be
H~=~
-J \sum_{s=1}^N \left( a^\dag_{s+1}a_s +{\rm h.c.}\right)+
\frac{U}{2}\sum_{s=1}^N n_s(n_s-1)~,
\label{HBH}
\ee 
where $a_s$ are bosonic annihilation operators, $n_s=a_s^\dag a_s$ are number operators,
and $J$ is the tunnelling rate between nearest-neighbour lattice sites. For the sake
of convenience, we will use dimensionless units such that the on-site interaction strength 
is $U=1$, and assume periodic boundary conditions. In a realistic experiment with a few tens of lattice 
sites the dynamics should not be affected by the boundary conditions, if not to mention that 
the periodic boundary conditions should be directly accessible in a ring-shaped optical 
lattice \cite{amico} or by painting arbitrary and time-dependent potentials \cite{boshier}, 
see Fig. \ref{FigBoshier}.

The model is especially interesting in the thermodynamic limit when the number of lattice sites 
$N\to\infty$ and a number of particles is commensurate with $N$, i.e., average number of 
particles per site is an integer $n$. In this limit it has a quantum Berezinski-Kosterlitz-Thouless 
transition \cite{BKT} at $J_c\simeq n^{-1}$, see Ref. \cite{Kuhner2000}. When $J>J_c$ the 
system is in the gapless superfluid phase with (algebraically decaying) quasi-long-range order, 
and when $J<J_c$ it is in the gapfull Mott insulator phase. Here we consider dynamical transitions 
both from Mott insulator to superfluid and back. 

We begin with a linear ramp from Mott insulator to superfluid,
\be
J~=~\frac{t}{\tau_Q}
\label{JtauQ}
\ee
where $\tau_Q$ is the usual quench time. The evolution begins at $t=0$ from the
ground state of the Hamiltonian (\ref{HBH}) at $J=0$ which is the Mott state
\be
|n,n,n,\dots\rangle
\label{Mott}
\ee 
with a definite number of $n$ particles per site. Experimentally, a perfect linear 
ramp of the tunnelling rate can be obtained by logarithmically reducing amplitude of
optical lattice with an (optional) minor adjustment of interaction strength via 
the Feshbach resonance \cite{eddi}, see Figure \ref{FigFeshbach}, or by painting 
a time-dependent potential \cite{boshier}, see Figure \ref{FigBoshier}.

In a transition to the superfluid phase we are interested in correlation functions
\be
C_R(t)~=~\frac12\langle\psi(t)|a^\dag_{i+R}a_i+{\rm h.c.}|\psi(t)\rangle~.
\label{Cl}
\ee  
The correlations are related to the momentum distribution of atoms $n_k$ by a Fourier
transform $n_k=\frac{1}{N}\sum_R\exp(ikR)C_R$. They are also good observables
because $C_R$'s are conserved by the hopping term in the Hamiltonian (\ref{HBH}) and 
by the end of time evolution at large $J$, when the interaction is just a small 
perturbation to the hopping term, they take stable final values. In particular,
\be
K_1=1-C_1=\sum_k (1-\cos k) n_k
\label{K1general}
\ee
is a (normalized) kinetic hopping energy. Both the hopping energy in particular and 
the momentum distribution in general depend on the transition time $\tau_Q$. 

Since the model is not integrable, in the following Sections we review 
approximate solutions in different regimes of parameters.

\subsubsection{Slow transition from Mott insulator to superfluid }

We begin with the limit of slow transitions in infinite system. Numerical studies in this regime 
are extremely time consuming, therefore we concentrate mainly on KZM predictions. According to KZM, 
the state of the system after a slow transition has the characteristic length-scale $\hat\xi$ in 
Eq. (\ref{hatxi}), where $z$ and $\nu$ are critical exponents. For the Bose-Hubbard model the 
dynamical exponent $z=1$. The Mott insulator-superfluid transition (at an integer density 
of particles) in a $d$-dimensional Bose-Hubbard model belongs to the universality class of the
$(d+1)$-dimensional $XY$ spin model \cite{Fisher1989}. Therefore:

\begin{itemize}

\item

In one dimension this mapping implies that $\nu\to\infty$ as in the Berezinski-Kosterlitz-Thouless 
transition and 
\be
\hat\xi~\sim~\tau_Q~.
\ee
As a result, the hopping energy of excitations should scale as
\begin{equation}
K_1~\sim~\xi^{-2}~\sim~\tau_Q^{-2}~.
\label{kinetic}
\end{equation}
The exponent $-2$ means a steep dependence of the hopping energy on the quench time $\tau_Q$, which 
should make it easily discernible experimentally. 

\item

In two dimensions $\nu=0.67$ and
\be
\xi~\sim~\tau_Q^{0.40}~,~~K_1~\sim~\tau_Q^{-0.80}~.
\ee

\item

In three dimensions we have the mean-field exponent $\nu=1/2$ and
\be
\xi~\sim~\tau_Q^{1/3}~,~~K_1~\sim~\tau_Q^{-2/3}~.
\ee
This scaling was confirmed in Ref. \cite{Polkovnikov2005} by analytic calculations 
in the Bogoliubov theory of Ref. \cite{AltmanAuerbach}. 

\end{itemize}

The non-mean-field scalings predicted in 1D and 2D have not been verified numerically, 
but see Ref. \cite{XXZquench}.  

Slow transitions probe the universal critical exponents, but they require stability
of experimental set-up over long transition time $\tau_Q$. This is why it may be 
more practical to experiment first with faster transitions, where the point of freeze-out  
$\hat\epsilon$ may be not close enough to the critical point to capture the universal 
scalings predicted in this Section, but there are nevertheless (non-universal) scaling 
relations bearing witness to non-adiabaticity of the transition. In the following two 
Sections we consider two tractable regimes of parameters where such non-universal predictions 
can be made. In the next Section \ref{fastn1} we assume the density of one particle per site. 
This minimal commensurate density makes direct numerical simulations possible for fast quenches 
and limited lattice sizes. Numerical results can be further corroborated by approximate analytic 
solutions. In Section \ref{fastnlarge} we take the opposite extreme of large density when 
semiclassical methods become applicable and, finally, in Section \ref{SectionMeisner} we 
analyse a reverse transition from superfluid to Mott insulator.

\subsubsection{Fast transition from Mott insulator to superfluid at small density}
\label{fastn1}

In this Section we follow Ref. \cite{Cucchietti} and consider $n=1$ particle per site. This 
minimal commensurate density allows for some exact numerical simulations. For fast quenches 
it is enough to consider short times when the wave function remains close to the initial Mott 
state (\ref{Mott}) and can be approximated by a variational state
\begin{eqnarray}
|\psi(t)\rangle &=& a(t)|1,1,\dots\rangle + \nonumber\\
                & & b(t)
                    \left(
                            |0,2,1,1\dots\rangle+
                            |2,0,1,1\dots\rangle+\cdots
                    \right)/\sqrt{2N},
\label{wp}
\end{eqnarray}
where $|a|^2+|b|^2=1$ and dynamics of $a(t),b(t)$ is governed by 
\begin{equation}
i\frac{\partial}{\partial t}
\left(
\begin{array}{c}
a \\
b
\end{array}
\right)=
\left(
\begin{array}{cc}
0 & -t~\frac{2\sqrt{N}}{\tau_Q} \\
-t~\frac{2\sqrt{N}}{\tau_Q} & 1
\end{array}
\right)
\left(
\begin{array}{c}
a \\
b
\end{array}
\right).
\label{almostlz}
\end{equation}
A change of  basis 
$
(a',b')=e^{it/2}(a-b,-a-b)/\sqrt{2}
$
yields  
\begin{equation}
i\frac{\partial}{\partial t}
\left(
\begin{array}{c}
a' \\
b'
\end{array}
\right)=
\frac{1}{2}
\left(
\begin{array}{cc}
\frac{t}{\tau} & 1 \\
1 & -\frac{t}{\tau}
\end{array}
\right)
\left(
\begin{array}{c}
a' \\
b'
\end{array}
\right) \ \ \ , \ \ \ \tau= \frac{\tau_Q}{4\sqrt{N}}.
\label{LZtau}
\end{equation}
Here the time-dependent Hamiltonian is precisely the Landau-Zener Hamiltonian in 
Eq. (\ref{HLZ}) but, unlike in the standard LZ model where $t\in(-\infty,\infty)$, 
here the time evolution begins at $t=0$ from the instantaneous ground state right in the 
middle of the anti-crossing. This is the case (ii) considered in Section \ref{LZvKZ}.

\begin{figure}[t]
\begin{center}
\includegraphics[width=0.6\columnwidth,clip=true]{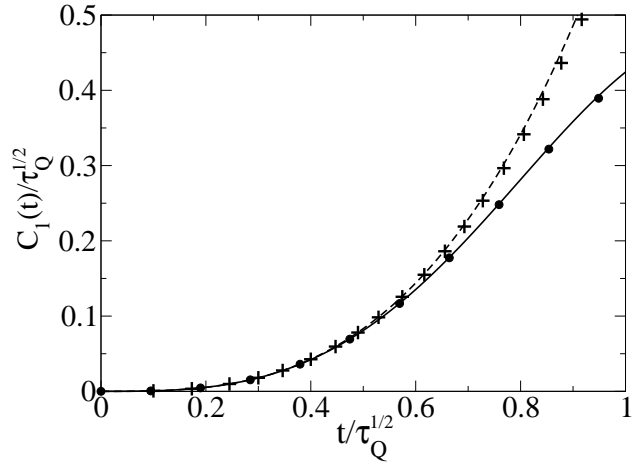}
\caption{Short time dynamics of $C_1(t)$. Numerics for $N=10$ lattice sites is given by 
	   the solid line ($\tau_Q=0.001$) and the large dots ($\tau_Q=0.1$).
	   The dashed line presents Eq. (\ref{k_t}). 
         (Figure from Ref. \cite{Cucchietti})
}
\label{k_t_plot}
\end{center}
\end{figure}

\begin{figure}[t]
\begin{center}
\includegraphics[width=0.6\columnwidth,clip=true]{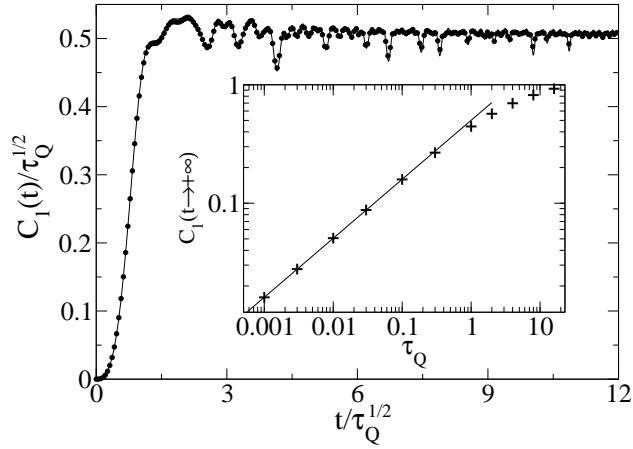}
\caption{Scaling properties of $C_1$ obtained numerically.
         Solid line: $\tau_Q=0.001$, dots: $\tau_Q=0.03$. 
         Inset: the solid line is a power law fit to data
	   for $0.001\le\tau_Q\le0.1$ giving $C_1(\infty)=0.501\tau_Q^{0.498}$.
	   All data is for $N=10$ and $J_{max}=600$.
         (Figure from Ref. \cite{Cucchietti})
}
\label{C1scaling}
\end{center}
\end{figure}

The quantity of interest is 
\be
C_1(t) ~=~ \frac{|b'(t)|^2-|a'(t)|^2}{\sqrt{N}}~.
\ee
An expansion of the exact solution in Eq. (\ref{PsmalltauQ}) for small $\tau_Q$ gives 
\begin{eqnarray}
\frac{C_1(t)}{\sqrt{\tau_Q}}= 
\frac{2}{3} 
\left[\frac{t}{\sqrt{\tau_Q}}\right]^3
\label{k_t}
\end{eqnarray}
to leading order in $t/\sqrt{\tau_Q}$. The expression (\ref{k_t}) suggests that simple
rescalings make the dependence of rescaled $C_1/\sqrt{\tau_Q}$ on rescaled time 
$t/\sqrt{\tau_Q}$ independent of the transition time $\tau_Q$. This prediction is 
confirmed in Fig. \ref{k_t_plot} where the numerical solutions $C_1(t)$ for
two widely different $\tau_Q=0.001,0.1$ collapse on each other. That a similar collapse 
in Fig. \ref{C1scaling} extends also to large $t$ comes as a bit of surprise.

The variational Ansatz (\ref{wp}) is accurate under two assumptions: 

\begin{itemize}

\item Fluctuations of occupation numbers around the average $n=1$ are small;

\item Particles can be displaced with respect to the initial Mott state (\ref{Mott}) 
      not more than to nearest-neighbour sites. 

\end{itemize}

The former assumption is self-consistent up to
\be
\hat t ~\simeq~ \tau_Q^{1/2}
\ee
when significant number fluctuations begin to develop. This time corresponds to
\be
\hat J ~\simeq ~ \tau_Q^{-1/2}
\ee
which is large, $\hat J\gg 1$, for fast transitions with $\tau_Q\ll1$. In a fast transition,
the number fluctuations do not have enough time to develop before $J$ enters the regime of 
strong tunnelling, $J\gg 1$. In this regime, the correlators $C_R$ are approximately constant,
see Fig. \ref{C1scaling}, 
so the asymptotic $C_R(\infty)$ can be accurately approximated by $C_R(\hat t)$ at $\hat J$ when 
the number fluctuations are still sufficiently small for the ansatz (\ref{wp}) to be accurate. 
Indeed, $C_1(\hat t)\simeq\frac23\sqrt{\tau_Q}$ in Eq. (\ref{k_t}) gives the correct scaling of 
the asymptotic $C_1(\infty)\sim\sqrt{\tau_Q}$, as confirmed by the numerical data in Fig. \ref{C1scaling}.

The latter assumption, which makes all correlators $C_R$ with $R>1$ vanish,
is relaxed in the Bogoliubov theory developed in Refs. \cite{AltmanAuerbach} and \cite{Cucchietti}. 
In consistency with the former assumption, the theory truncates the Hilbert space to states with
$0,1$ or $2$ particles per site only. The Mott state (\ref{Mott}) is a vacuum, a site with $2$ 
particles is occupied by a quasiparticle and an empty site by a quasihole. The quasiparticles and 
quasiholes are hard-core bosons, but the hard-core constraint is relaxed to make their Hamiltonian 
quadratic. After solving corresponding time-dependent Bogoliubov-de Gennes equations and making the 
approximation $C_R(\infty)=C_R(\hat t)$ we obtain
\begin{eqnarray}
C_1(\infty)&\approx& 3.9\times10^{-1}\sqrt{\tau_Q},\nonumber\\
C_3(\infty)&\approx&-3.5\times10^{-3}\sqrt{\tau_Q},\nonumber\\
C_5(\infty)&\approx& 1.4\times10^{-5}\sqrt{\tau_Q}, \nonumber
\end{eqnarray}
and $C_{2l}\sim \tau_Q$ for small $\tau_Q\ll1$. Bogoliubov theory predicts that dominant odd 
correlations $C_{2l+1}\sim\sqrt{\tau_Q}$ decay quickly with the distance $R=2l+1$.

\subsubsection{Fast transition from Mott insulator to superfluid at large density}
\label{fastnlarge}

\begin{figure}[b]
\begin{center}
\includegraphics[width=0.9999\columnwidth,clip=false]{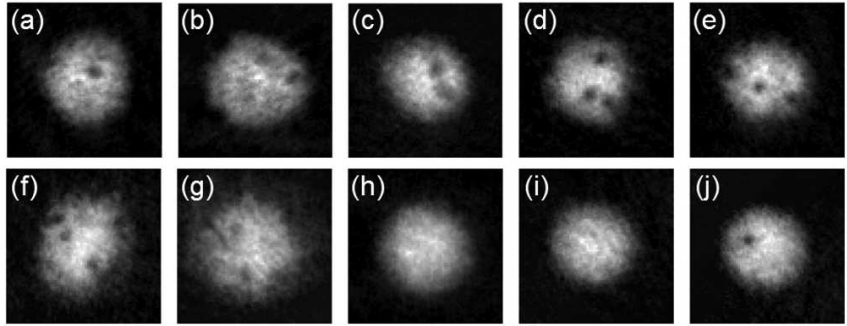}
\caption{ Images of a Bose-Einstein condensate taken after sudden merging of 3 initially disconnected condensates.
          The dark spots are vortices created in the process (Figure from Ref. \cite{Mercedes}).
}
\label{FigMercedes}
\end{center}
\end{figure}

In this Section we follow Ref. \cite{Meisner} and consider the opposite limit of large particle
density, $n\gg1$, in the initial Mott state (\ref{Mott}). The large density regime is tractable
by semiclassical methods, but it also makes the tight binding approximation leading to
the Hubbard model (\ref{HBH}) more problematic. For this reason most experiments simulating 
Bose-Hubbard model are carried out at densities close to unity \cite{Greiner}, but
experiments on number squeezing at large filling have been performed as well already at
the early stage of research into possible occurrence of the Mott transition \cite{Kasevich}.
Apart from this, for slow enough quenches high energy degrees of freedom neglected in 
the Hubbard model are likely to remain adiabatic during the transition. In this sense, the model 
is a first low energy approximation to a ring of initially disconnected Bose-Einstein condensates 
being merged into a toroidal trap as in the experiment in Fig. \ref{FigBoshier} b. In this Section
we will attempt to answer the question how does a winding number around the final toroidal condensate 
depend on the transition rate? The first experiment along these lines was done in Ref. \cite{Mercedes} 
where 3 initially disconnected condensates were suddenly merged into one by removing a ``Mercedes''
potential separating them. In many realisations of that experiment vortices were detected in the final 
condensate, see Fig. \ref{FigMercedes}.

For the sake of convenience, in this Section we rescale variables in the Bose-Hubbard Hamiltonian 
(\ref{HBH}) so that 
\begin{equation}
H = -J \sum_{s=1}^N \left( a_{s+1}^\dag a_s + {\rm h. c.} \right)
  + \frac{1}{2n} \sum_{i=1}^N n_s(n_s-1) ~
\label{HBHlargen}
\end{equation}
and the phase transition is at $J_c\simeq n^{-2}$. In the truncated Wigner method \cite{TW} employed 
in Ref. \cite{Meisner} the annihilation operators $a_s$ are represented by a complex field
$\phi_s$,
$
a_s\approx\sqrt{n}~\phi_s~,
$ 
which is normalised as
$
\sum_{s=1}^N |\phi_s|^2=N~,
$
and evolves with the time-dependent Gross-Pitaevskii equation 
\be
i\frac{d\phi_s}{dt}=
-J\left( \phi_{s+1} - 2\phi_s + \phi_{s-1} \right)+
|\phi_s|^2\phi_s~.
\label{GPE}
\ee
These approximations become accurate as $n\to\infty$ and the critical point
$J_c\simeq n^{-2}\to 0$. Quantum expectation values are estimated by averages over 
stochastic realisations of the field $\phi_s(t)$. For example, the correlation 
function (\ref{Cl}) becomes
\be
C_R ~=~
\frac{1}{2n} ~
\langle\psi(t)|a_{s+R}^\dag a_{s} + {\rm h.c.}|\psi(t)\rangle ~\approx~
\frac12
\overline{\left(\phi_s^*\phi_{s+R}+{\rm c.c.}\right)}~.
\label{CR}
\ee
Here the overline means average over stochastic realisations. All realisations of $\phi_s(t)$ 
evolve with the same deterministic Gross-Pitaevskii equation (\ref{GPE}), but they 
start from different random initial conditions distributed according to Wigner 
distribution of an initial quantum state. The initial Mott state (\ref{Mott}) translates 
into initial fields
\be
\phi_s(0)~=~e^{i\theta_s}~
\label{randomphases}
\ee
with independent random phases $\theta_s\in[-\pi,\pi)$. The Mott state has 
the same fixed number of $n$ particles at each site, translating into constant modulus
$|\phi_s(0)|^2=1$ and indeterminate quantum phases translating into random $\theta_s$.  

Our goal is to estimate average size of a random phase step 
$\Delta\theta_s=\theta_{s+1}-\theta_s$ 
between nearest-neighbour sites in a final state after transition. A random walk 
of phase around a chain of $N$ sites generates a net winding number whose average
magnitude scales as $\sqrt{N}$ times the average size of a phase step. We begin with 
a KZM-like estimate in a linearised Gaussian theory.

\begin{figure}[t]
\begin{center}
\includegraphics[width=0.8\columnwidth,clip=true]{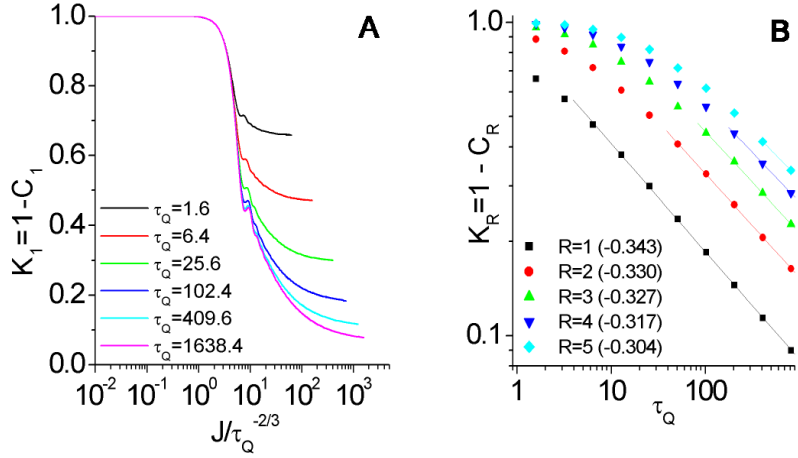}
\caption{ 
Kinetic hopping energy 
$K_1=1-C_1\approx\overline{1-\cos\Delta\theta_s}\approx\overline{\frac12\Delta\theta_s^2}$ 
as a function of rescaled $J/\hat J$ for different $\tau_Q$ is seen in A. When $J\ll 1$, all 
the plots overlap demonstrating that $\hat J=\tau_Q^{-2/3}$ is the relevant scale 
for $J\ll 1$. Individual plots depart from this small-$J$ ``common bundle'' at $J\simeq 1$, 
or equivalently $J/\tau_Q^{-2/3}\simeq \tau_Q^{2/3}$, when $K_1=1-C_1$ is expected to stabilise. 
In B, we show $K_R\equiv1-C_R$ at $J=10$ as a function of $\tau_Q$ for 
$R=1,...,5$. Data points for each $R$ were fitted with lines, their slopes giving 
exponents close to the $\frac13$-scaling predicted in Eq. (\ref{1/3-K1}) with error 
bars on their last digits. 
(Figure from Ref. \cite{Meisner})
}
\label{fighatJ}
\end{center}
\end{figure}

The Gross-Pitaevskii equation (\ref{GPE}) can be linearised in small fluctuations 
$\delta\phi_s$ around a uniform large background, $\phi_s=1+\delta\phi_s$, and 
$\delta\phi_s$ can be expanded in Bogoliubov modes as
$
\delta\phi_s=\sum_k \left( b_k u_k e^{iks}+ b_k^* v_k^* e^{-iks}\right)
$ 
with pseudomomentum $k$. For a constant $J$ we have 
$
b_k(t)=b_k(0)e^{-i\omega_k t}
$ 
with Bogoliubov frequencies
\be
\omega_k=2\sqrt{J(1-\cos k)\left[1+J(1-\cos k)\right]}
\label{BHomegak}
\ee
and stationary Bogoliubov modes
\be
u_k~=~-{\cal N}_k\left[1+2J(1-\cos k)+\omega_k\right],~
v_k~=~{\cal N}_k~.
\ee
Here ${\cal N}_k$ is such that $u_k^2-v_k^2=1$. In the Josephson regime, when $J\ll1$, we have 
$v_k\approx-u_k$ and a purely imaginary $\delta\phi_s$ in $\phi_s=1+\delta\phi_s$ is a phase 
fluctuation. However, for our random initial conditions (\ref{randomphases}), this linearisation 
is justified only for short wavelength modes of $\phi_s$, with $k\approx\pm\pi$, for whom the 
modes with longer wavelength appear to be the (locally) uniform large background. From now on 
we concentrate on the short wavelength modes because they determine the variance of the nearest-neighbour phase 
step $\Delta\theta_s$. 

When $k\approx\pm\pi$ and $J\ll1$ then $\omega_k\approx2\sqrt{2J}$. Early in the 
linear quench (\ref{JtauQ}) this $\omega_k$ is so small that the early evolution of the 
short wavelength modes is approximately impulse, i.e., their magnitude remains 
the same as in the initial Mott state and, consequently, 
$\overline{\Delta\theta_s^2}\simeq 1$ in this impulse stage. 
The impulse approximation breaks down at $\hat J$ when the transition 
rate $\dot\omega_k/\omega_k$ equals $\omega_k$, and the evolution becomes adiabatic. 
This happens at
\be
\hat J~\simeq~\tau_Q^{-2/3}~
\label{hatJ}
\ee 
which is self-consistent with the assumption of $J\ll1$ when $\tau_Q\gg 1$. 

The crossover from impulse to adiabatic evolution at $\hat J$ is the key ingredient 
of KZM. In the following adiabatic evolution after $\hat J$, but before $J\approx1$, 
short wavelength phase fluctuations scale as $\delta\phi_s\sim J^{-1/4}$ because 
the amplitudes $|b_k|$ do not change and the modes $u_k,v_k$ follow instantaneous
stationary Bogoliubov modes $u_k\approx-v_k\approx-1/2(2J)^{1/4}$. Consequently, 
$\Delta\theta_s$ has variance scaling as
$
\left.\overline{\Delta\theta_s^2}\right|_J \simeq
\overline{|\delta\phi_s|^2} \sim 
J^{-1/2}
$. 
Given the initial condition for the adiabatic stage at $\hat J$ that 
$\left.\overline{\Delta\theta_s^2}\right|_{\hat J}\simeq 1$, the phase fluctuations 
must shrink as
$
\left.\overline{\Delta\theta_s^2} \right|_J                          \simeq
\left.\overline{\Delta\theta_s^2} \right|_{\hat J}~(J/\hat J)^{-1/2} \simeq
\tau_Q^{-1/3} J^{-1/2}~
$
while $J\ll 1$. 

On the other hand, when $J\gg1$ then stationary modes $u_k\approx1$ and $v_k\approx0$ 
do not depend on $J$ and $\overline{\Delta\theta_s^2}$ does not depend on $J$ either.
This means that $\overline{\Delta\theta_s^2}$ must stabilise between the regimes of 
$J\ll1$ and $J\gg1$, i.e., around $J\simeq1$ where it saturates at its final value  
\be
\left.\overline{\Delta\theta_s^2}\right|_{J\gg 1}     ~\approx~
\left.\overline{\Delta\theta_s^2}\right|_{J\simeq 1}  ~\simeq~
\tau_Q^{-1/3}~
\label{the 1/3}
\ee
scaling with a power of $-1/3$. This variance determines e.g. the correlator $C_1$ in 
\be
K_1~=~1-C_1=1-\overline{\cos\Delta\theta_s}~\simeq~\tau_Q^{-1/3}~,
\label{1/3-K1}
\ee
for $\tau_Q\gg1$. The kinetic hopping energy per particle $K_1$
is expected to stabilise for $J\gg 1$, when the hopping term dominates 
over the non-linearity in Eq. (\ref{GPE}) and $K_1$ becomes an approximate 
constant of motion, see Fig. \ref{fighatJ}.

\begin{figure}[t]
\begin{center}
\includegraphics[width=0.7\columnwidth,clip=true]{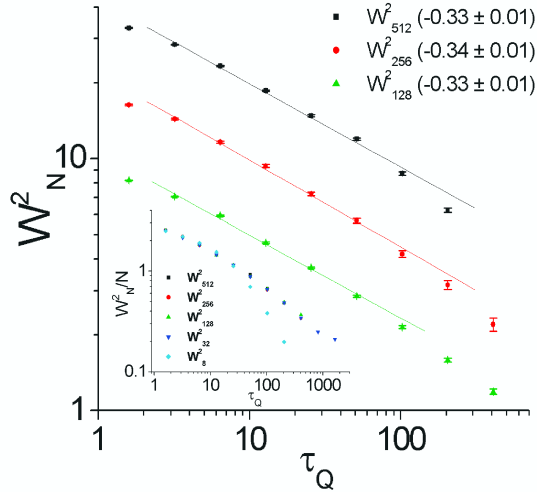}
\caption{ 
Variance of winding number $\overline{W_N^2}$ measured at $J=10$ as a function 
of $\tau_Q$ for lattice sizes $N=512,256,128$. Here point sizes equal error bars. 
The data points with $\tau_Q>2$ and $\overline{W_N^2}>2$ were fitted with the solid 
lines giving slopes close to the predicted $-\frac13$ in Eq. (\ref{W 1/3}). 
$\overline{W_N^2}$ is shown over a wider range of $\tau_Q$ to show the saturation 
for nearly instantaneous quenches, when $\tau_Q<2$, and the crossover to steeper 
slope when $\overline{W_N^2}<2$. The inset shows a rescaled 
$\overline{W^2_N}/N$ for $N=512,256,128,32,8$ to demonstrate that 
$\overline{W^2_N}\sim N$ in the KZ regime of $\tau_Q>2$ and $\overline{W_N^2}>2$. 
(Figure from Ref. \cite{Meisner})
}
\label{FigW1/3}
\end{center}
\end{figure}

The integer winding number is a phase accumulated after $N$ steps (divided by $2\pi$)
\be
W_N~=~
\frac{1}{2\pi}
\sum_{s=1}^N {\rm Arg}\left( \phi_{s+1}\phi^*_s \right)~,
\label{WN}
\ee
where ${\rm Arg}(...)\in(-\pi,\pi]$. A random walk of phase, with the variance of 
nearest neighbour phase differences scaling as in Eq. (\ref{the 1/3}), gives winding 
numbers with a variance
\be
\overline{W_N^2}~\simeq~N~\tau_Q^{-1/3}~.
\label{W 1/3}
\ee
There are two limits where this scaling must fail. For very fast quenches 
with $\tau_Q\ll 1$ phases are completely random between neighbouring sites, so 
$\overline{\Delta\theta_s^2}=\pi^2/3$, and $\overline{W_N^2}=N/12$. For quenches so 
slow that $\overline{W_N^2}<1$ the nature of the problem changes, leading to 
steeper falloff of $\overline{W_N^2}$ with $\tau_Q$. Between
these two limits the $\frac13$-scaling in Eq. (\ref{W 1/3}) for the winding number 
is confirmed by numerical results in Fig. \ref{FigW1/3}.

On one hand, the predicted $1/3$-scaling of the winding number is limited to relatively 
slow transitions with $\tau_Q\gg1$ so that $\hat J\ll1$ but, on the other hand, the transition 
must be fast enough for the truncated Wigner method to remain accurate. This means that the 
crucial $\hat J$ in Eq. (\ref{hatJ}) must be much greater than the critical $J_c\simeq n^{-2}$, 
or equivalently $\tau_Q\ll n^3$. For density of, say, $n\simeq 100$ atoms per site 
$\tau_Q$ can range over $6$ orders of magnitude.

\subsubsection{Fast transition from superfluid to Mott insulator at large density}
\label{SectionMeisner}

A reverse transition from the gapless superfluid to the gapped Mott insulator was 
analysed in Refs. \cite{Bishop2,SchutzholdSFMott}. Since at the final tunnelling rate
$J=0$ site occupation numbers $n_s$ are good quantum numbers, we will concentrate on time evolution
of the occupation numbers and their conjugate phase fluctuations. An initial superfluid ground state 
at large $J\gg n$ has large Poissonian number fluctuations $\sim\sqrt{n}$ at each site and small 
$\sim1/\sqrt{n}$ fluctuations of phase. By contrast, in the Mott ground state (\ref{Mott}) at the 
final $J=0$ the number fluctuations are zero and the phases are random. We can expect that a 
non-adiabatic transition from large $J\gg n$ to $J=0$ will end in an excited state with finite number 
fluctuations. The conserved magnitude of these final number fluctuations will depend on transition 
rate. 

In this Section we assume large density of $n\gg1$ particles per site when a polar decomposition 
$a_s=\sqrt{n_s}\exp(i\theta_s)$ with a phase operator $\theta_s$ makes sense. Deep in the 
superfluid regime we can define $n_s=n+\delta n_s$ and $\theta_s=\delta\theta_s$, where the 
$c$-number $n$ is the large particle density, while $\delta n_s\ll n$ and $\delta\theta_s$ are 
mutually conjugate small number and phase fluctuations respectively. Inserting the polar 
decomposition into an equation of motion $i\partial_ta_s=J(t)(a_{s+1}+a_{s-1})+n_sa_s$, expanding 
to leading order in the small quantum fluctuations, eliminating $\delta\theta_s$, and making 
Fourier transform of $\delta n_s$, we obtain a linearised Bogoliubov equation
\be
\left(
\frac{\partial}{\partial t} \frac{1}{J(t)} \frac{\partial}{\partial t} +
4(1-\cos k)[n+J(t)(1-\cos k)]
\right)
\delta n_k ~=~0~
\label{BHdeltan}
\ee  
with a time-dependent $J(t)$. 

Notice that in case of constant $J$ we obtain the Bogoliubov spectrum of quasiparticle excitations
$
\omega_k ~=~ \sqrt{4J(1-\cos k)\left[n+J(1-\cos k)\right]}~
$
which is identical with Eq. (\ref{BHomegak}) up to rescaled units. Here we are interested mainly
in the Josephson regime, $J\gg J_c\simeq 1/n$ and $J\ll n$, where the ground state is a number-squeezed 
state with a number variance
\be
\Delta n^2_{\rm GS}(J)~\simeq~\sqrt{nJ} 
\ee
and a phase variance $\Delta\theta^2_{\rm GS}\simeq 1/\Delta n^2_{\rm GS}$. In this regime
\be
\omega_k~\approx~\sqrt{4nJ(1-\cos k)}~\sim~(nJ)^{1/2}~.
\label{BHomegakJoseph}
\ee 
We assume that the transition is slow enough for most Bogoliubov modes to remain unexcited before they 
enter the Josephson regime, except for a narrow range of low frequency modes with small $|k|$ which 
becomes negligible with increasing transition time.  

In the Josephson regime, a mode $k$ becomes excited at a $\hat J$ when the transition rate $\dot J/J$ 
equals $\omega_k$. We consider two different functions $J(t)$:

\begin{itemize}

\item

Like in the previous Sections on the Mott to superfluid transition and in Ref. \cite{Bishop2}, 
we consider a linear
\be
J(t)~=~-\frac{t}{\tau_Q}~,
\ee
where $t$ runs from $-\infty$ to $0$. The transition rate $\dot J/J=1/|t|$ equals $\omega_k$ 
in Eq. (\ref{BHomegakJoseph}) at
\be
\hat J ~\sim~ n^{-1/3}\tau_Q^{-2/3}~.
\ee
After $\hat J$ the evolution becomes impulse and the final variance at $J=0$ remains the
same as in the ground state at $\hat J$:
\be
\Delta n^2~\sim~\Delta n^2_{\rm GS}(\hat J)~\sim~n^{1/3}\tau_Q^{-1/3}~.
\ee
Self-consistency requires $\hat J$ in the Josephson regime or, equivalently, 
$\frac{1}{n^2}\ll\tau_Q\ll n$. The transition has to be fast enough, $\tau_Q\ll n$,
for the crossover to the impulse stage to take place much above the critical point
$J_c$ where the Bogoliubov approximation would break down. 

\item

Since the tunnelling rate $J$ depends exponentially on the strength of an optical lattice, we also 
consider an exponential
\be
J(t)~=~J_0~e^{-t/\tau_Q}~,
\label{Jexp}
\ee
where $t\in(0,\infty)$, like in Ref. \cite{SchutzholdSFMott}. The transition rate $\dot J/J=1/\tau_Q$ 
equals $\omega_k$ in Eq. (\ref{BHomegakJoseph}) at
\be
\hat J ~\sim~ n^{-1}\tau_Q^{-2}~.
\ee
The final variance 
\be
\Delta n^2~\sim~\Delta n^2_{\rm GS}(\hat J)~\sim~\tau_Q^{-1}~
\label{tobeconfirmed}
\ee
does not depend on $n$ and its dependence on $\tau_Q$ is steeper than after a linear transition,
but $\hat J$ in the Josephson regime requires $n^{-1}\ll\tau_Q\ll1$, i.e., a more narrow range
of $\tau_Q$.

\end{itemize}

The adiabatic-impulse scaling argument above has two limitations: as usual, it does not predict 
any numerical pre-factors and, what is potentially more dangerous, it ignores any dependence of $\hat J$ 
on $k$. This is why we follow a more detailed calculation in Ref. \cite{SchutzholdSFMott} 
in the case of exponential transition. 

A new $k$-dependent time variable 
$\tau~=~-2(1-\cos k)J(t)\tau_Q$, running form $\tau\to-\infty$ to $\tau=0$, transforms Eq. (\ref{BHdeltan}) 
into a more universal
\be
\left[
\frac{\partial^2}{\partial\tau^2}+
\left(
1-\frac{2n\tau_Q}{\tau}
\right)
\right]
\delta n_k ~=~ 0~~.
\ee
This equation demonstrates that, after proper rescaling of time, different modes $\delta n_k$ 
evolve in the same way. Their evolution changes qualitatively at 
$\tau\simeq2n\tau_Q$ when it crosses over from pure oscillations with constant frequency, 
$\sim\exp(\pm i\tau)$, to damped oscillations with increasing frequency. This crossover 
corresponds roughly to entering the Josephson regime near $J\simeq n$. After the crossover
the damping makes number fluctuations $\Delta n$ shrink from the initial $\Delta n\simeq\sqrt{n}$ 
to its final value at $J=0$. 

More precisely, see Ref. \cite{SchutzholdSFMott}, 
\be
\delta n_k ~=~ \sqrt{n} ~ e^{-\pi n\tau_Q/2}W_{in\tau_Q,1/2}(2i\tau) ~ b_k ~+~
{\rm h.c.}~,
\ee
where $W$ is the Whittaker function and $b_k$ are bosonic Bogoliubov quasiparticles.
Here we work in the Heisenberg picture, where the state is a vacuum $|0\rangle$ annihilated by 
all $b_k$. The number fluctuations at the final $\tau=0$, corresponding to $J=0$, are
\be
\langle(\delta n_k)^2\rangle~\equiv~
\langle0|(\delta n_k)^2|0\rangle~=~
\frac{1-e^{-2\pi n\tau_Q}}{2\pi\tau_Q}~
+~{\cal O}\left[te^{-t/\tau_Q}(1-\cos k)\right]
~.
\label{deltank}
\ee
This limit is saturated and the number fluctuations freeze out when $t\gg\tau_Q$. 
Since the frozen $\langle(\delta n_k)^2\rangle$ is independent of $k$, we obtain non-zero frozen 
on-site variations
\be
\langle n_s^2\rangle-\langle n_s\rangle^2~=~
\langle(\delta n_s)^2\rangle~=~
\frac{1-e^{-2\pi n\tau_Q}}{2\pi\tau_Q}~,
\ee
but vanishing two-site correlations between $\delta n_s$ at different sites. After a very fast quench with 
$n\tau_Q\ll1$ the final variance is equal to $n$ like in the initial Poissonian ground state 
at $J\gg n$. These quenches are effectively impulse right from an initial $J\gg n$ and all the way down to 
$J=0$ - they are too fast for the initial state to adapt to the rapidly shrinking $J$. In the more 
interesting regime of $n\tau_Q\gg1$ the final frozen number fluctuations are
\be
\Delta n^2~=~\frac{1}{2\pi\tau_Q}~
\label{varnscaling}
\ee
in agreement with the scaling prediction in Eq. (\ref{tobeconfirmed}). The pre-factor
missing in Eq. (\ref{tobeconfirmed}) is $\frac{1}{2\pi}$.

When $J\approx0$ these uncorrelated number fluctuations evolve with evolution operator 
\be
e^{-in_s(n_s-1)t/2} ~\approx~
e^{-in^2t/2} e^{-in~\delta n_s t}~.
\label{J0evolution}
\ee 
and gradually translate into increasing phase fluctuations with a variance \cite{SchutzholdSFMott}
\be
\Delta\theta^2~=~\frac{t^2}{2\pi\tau_Q}~.
\ee
These phase fluctuations determine relative quantum depletion from the condensate, i.e.,
fraction of particles that are not condensed into a uniform condensate wave function
with a site-independent phase. For our Bogoliubov approximation to be self-consistent they 
need to remain small, $\Delta\theta^2\ll1$, at least down to $\hat J$ when the ground state 
phase fluctuations are $\Delta\theta^2\simeq\tau_Q$. Thus the range of validity of the scaling 
in Eq. (\ref{varnscaling}) is limited to
\be
n^{-1}~\ll~\tau_Q~\ll~1~.
\label{xbv}
\ee
Notice that in the simple scaling argument leading to Eq. (\ref{tobeconfirmed}) 
the condition (\ref{xbv}) is equivalent to $\hat J$ in the Josephson regime. 

We can conclude that the detailed calculation confirms the scaling $\Delta n^2\sim\tau_Q^{-1}$ 
predicted in Eq. (\ref{tobeconfirmed}), provides the missing pre-factor of $1/2\pi$, justifies 
{\it a posteriori} ignoring the non-adiabaticity in a narrow range of long wavelength modes, and 
limits validity of the scaling to the same range of $\tau_Q$ as the simple scaling argument based 
on the adiabatic-impulse approximation.

\begin{figure}[t]
\begin{center}
\includegraphics[width=0.95\columnwidth,clip=true]{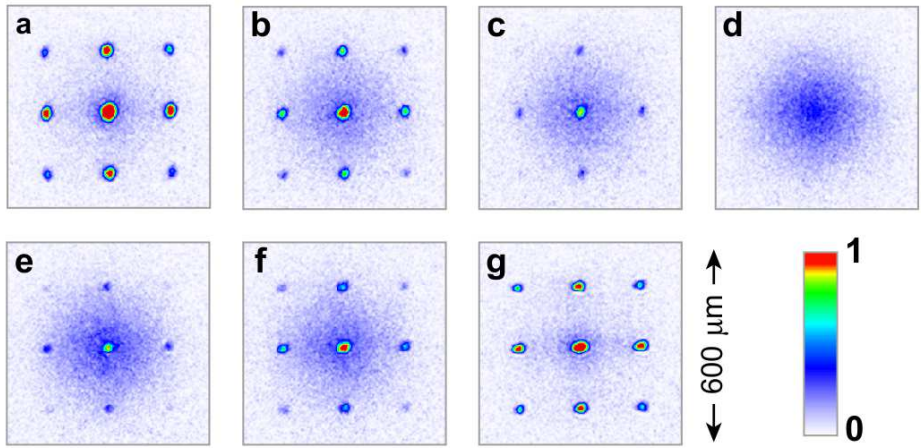}
\caption{ Dynamical evolution of the multiple matter wave interference pattern observed
          after sudden reduction of the tunnelling rate from the superfluid to the Mott
          phase. Panels a...g are absorption images taken after hold times:
          $0\mu s$, $100\mu s$, $150\mu s$, $250\mu s$, $350\mu s$, $400\mu s$ and
          $550\mu s$ respectively. A first distinct interference patten is visible (a),
          showing that initially the system can be described by a macroscopic
          condensate wave function. After $250\mu s$ the patten is lost (d) and after
          $550\mu s$ hold time it is almost perfectly restored.           
          (Figure from Ref. \cite{Greiner})
}
\label{FigRevivals}
\end{center}
\end{figure}

The vanishing final two-site correlator between different $\delta n_s$ has observable
implications for a two-site correlation function $C_R$. In the Heisenberg picture
at the final $J=0$, the annihilation operators evolve as $a_s(t)=\exp(-in_st)a_s(0)$ 
and the correlation function
\be
C_R(t)~=~\langle a_{s+R}^\dag(t) a_s(t) \rangle~=~C_R(0)~\langle\exp[it(n_{s+R}-n_s)]\rangle
\ee
decays to zero after a dephasing time $\simeq(\Delta n)^{-1}$. When the final $J$ is precisely
zero, then the decay is followed by periodic revivals every $2\pi$, but when the final $J$
is small but non-zero, then the revivals disappear after a dephasing time proportional to the width
$\simeq J^{-1}$ of the quasiparticle band at small $J$, see Ref. \cite{FischerRelaxation} for 
more details. The (disappearing) periodic revivals of the initial superfluid state 
were observed in the experiment \cite{Greiner} after a sudden quench to the Mott phase,
see Fig. \ref{FigRevivals}.

In the following Section we continue with bosonic atoms in a quasi-1D trap, but this time without 
the tight-binding approximation leading to the Bose-Hubbard model (\ref{HBH}).

\subsection{ Loading a 1D Bose gas into an optical lattice: 
             transition into the gapped phase of the sine-Gordon model}
\label{sinusGordon}

The parameter that defines properties of a Bose gas in one dimension 
\cite{LiebLiniger} is its interaction strength $\gamma=mg/\hbar^2\rho$, where $m$ 
is the mass of a particle, $g$ is the interaction strength in the contact interaction 
$g~\delta(x_1-x_2)$, and $\rho$ is the linear density of particles. In a spatially uniform 
system the spectrum of excitations is gapless for any $\gamma$ and the low energy physics 
is described by the Luttinger liquid \cite{LLHaldane}. The parameter $K$ of the Luttinger 
theory is $K\approx1+4/\gamma$ for strong and $K\approx\pi/\sqrt{\gamma}$ in for weak 
repulsion $\gamma$ \cite{LiebLiniger,KCazalilla}. In the extreme Tonks-Girardeau limit 
$\gamma\to\infty$ of strong repulsive interactions \cite{TonksGirardeau}, realised 
experimentally in Refs. \cite{Kinoshita,NewtonCradle,Paredes}, the particles behave like 
hard-core bosons which cannot occupy the same position in space.

\begin{figure}[t]
\begin{center}
\includegraphics[width=0.6\columnwidth,clip=true]{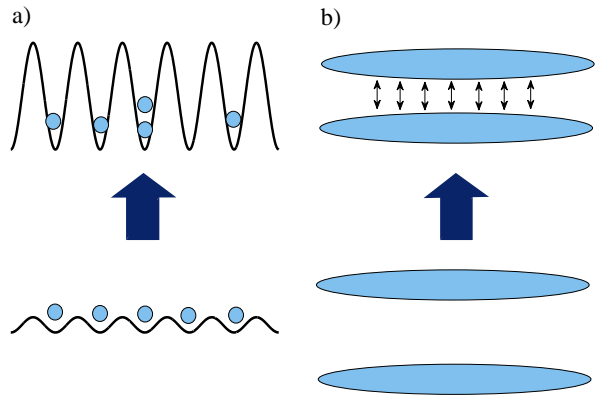}
\caption{ 
The two scenarios considered in Ref. \cite{DeGrandiPolkovnikov}.
Here we review case a) when a one dimensional Bose gas is loaded into an
optical lattice potential. A similar experiment was done in Ref. \cite{SGExp}.
(Figure from Ref. \cite{DeGrandiPolkovnikov})
}
\label{FigDeGrandiPolkovnikov}
\end{center}
\end{figure}

In Refs. \cite{DeGrandiPolkovnikov,infinitezimalPolkovnikov} the experimental set-up in 
Fig. \ref{FigDeGrandiPolkovnikov}a was considered, where by slowly turning on an optical lattice
the initial zero temperature Bose gas is loaded into a lattice potential. The lattice 
potential of increasing strength $\epsilon(t)$ results in a sine-Gordon Hamiltonian \cite{sine-Gordon}
\be  
H~=~\frac12\int dx
\left[
\Pi^2+(\partial_x\phi)^2-4\epsilon(t)\cos(2\sqrt{\pi K}~\phi)
\right]~,
\ee
where $\Pi(x)$ and $\phi(x)$ are mutually conjugate, and $K$ is the parameter
of the Luttinger liquid before turning on the perturbation. When $K<2$ the cosine term is
a relevant perturbation opening an energy gap in the spectrum of the gapless Luttinger liquid. 
In the repulsive regime when $1<K<2$, the massive excitations are solitons of mass 
\be
\Delta ~\simeq~ \epsilon^{\frac{1}{2-K}}~,
\label{soliton_mass}
\ee
for small $\epsilon$, see Refs. \cite{DeGrandiPolkovnikov} and \cite{Zamolodchikov}.
Since in a translationally invariant system the solitons can be excited only as 
topologically trivial pairs of solitons and antisolitons with opposite momenta $(k,-k)$, 
the spectrum of relevant excitations is $2\sqrt{\Delta^2+k^2}$. In the attractive regime 
when $K<1$, the spectrum includes also breathers that can be interpreted as 
soliton-antisoliton bound states. 

Since at the critical point $\Delta=0$ the excitation spectrum $2\sqrt{\Delta^2+k^2}$ is linear 
in $|k|$, the dynamical critical exponent is 
\be
z~=~1~.
\ee 
Consequently, the finite gap $\Delta$ translates into a finite correlation length 
$\xi\sim\Delta^{-1}$ in the ground state of the system. Given the scaling in Eq. (\ref{soliton_mass}), 
we identify the critical exponent 
\be
\nu~=~\frac{1}{2-K}~
\ee 
in $\xi\sim\epsilon^{-\nu}$. 

A linear ramp of the lattice amplitude, 
\be
\epsilon(t)~=~\frac{t}{\tau_Q}~,
\ee
with the time running from $t=0$ to $t\to\infty$, drives the system from the critical point at
$\epsilon=0$ into the gapped phase with $\epsilon>0$. This is an example of the general half-quench
analysed in Section \ref{halfKZ}. Given the exponents $z$ and $\nu$, we can use Eq. (\ref{halfnex}) 
to obtain the density of excited solitons and antisolitons
\be
n_{\rm ex}  ~\simeq~
\hat\xi^{-1}  ~\simeq~
\hat\epsilon^{\nu}  ~\simeq~
\tau_Q^{-1/(3-K)}~.
\label{nsolitons}
\ee 
The result (\ref{nsolitons}) is corroborated by the calculation in Ref. \cite{DeGrandiPolkovnikov} 
along the general lines of Section \ref{argumentPolkovnikov}. 

The non-adiabaticity of condensate loading into an optical lattice was also studied in 
Ref. \cite{ZakrzDelande} in presence of a harmonic trap potential and quasi-disorder.
The disorder makes the transition much less adiabatic than in the pure case, in coherence
with the example of the random Ising model in Refs. \cite{JDrandom,JDrandom2,decoherence} and 
Section \ref{RandomIsing}.

This Section completes our review of spinless bosons. In the next Section we consider
spin-$1$ Bose-Einstein condensates.

\subsection{ Spin-$1$ Bose-Einstein condensate: 
             transition from paramagnetic to ferromagnetic phase     }
\label{FerroKZ}

\begin{figure}[b]
\begin{center}
\includegraphics[width=0.70\columnwidth,clip=true]{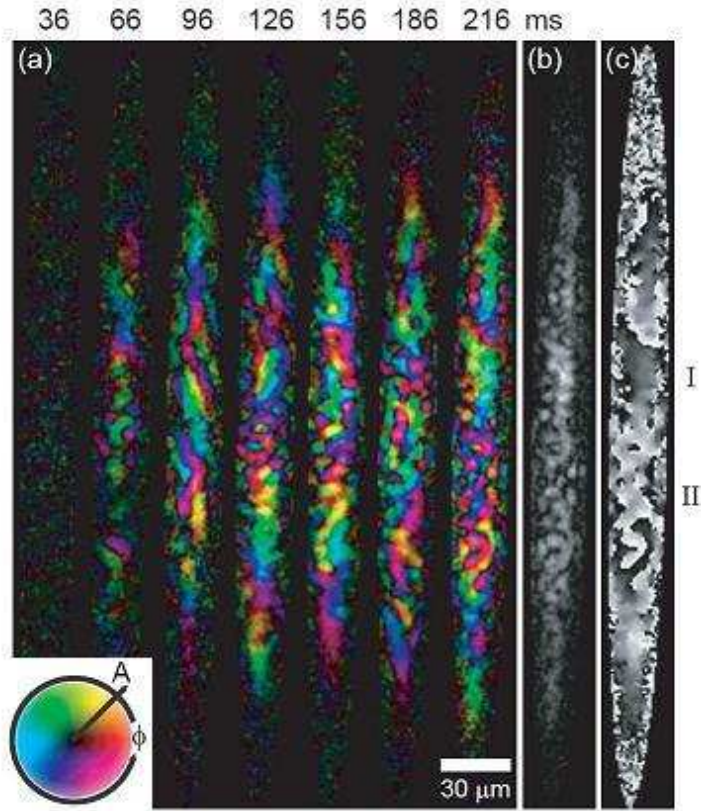}
\caption{ Images of a cigar-shaped spin-$1$ Bose-Einstein condensate after a sudden
          quench from the paramagnetic to the ferromagnetic phase. In a) transverse
          magnetization with magnetization density shown by brightness and its orientation 
          by colour. In b) magnetization density and in c) magnetization orientation respectively
          shown in grey scale. (Figure from Ref. \cite{ExpFerro}).
}
\label{FigFerro}
\end{center}
\end{figure}

A sudden quench in a spin-$1$ ferromagnetic Bose-Einstein condensate was realised in the
experiment of Ref. \cite{ExpFerro}, see Fig. \ref{FigFerro}, and analysed in Refs.
\cite{Lamacraft,Bodzioferro2,Bodzioferro}. Here, following Ref. \cite{Bodzioferro}, 
we consider for simplicity a one-dimensional homogeneous condensate in a finite box. Since the following 
analysis is on the mean-field level, most conclusions can be easily generalised to higher 
dimensions. In dimensionless variables, a system placed in a magnetic field $B$ along the 
$z$-axis has a mean-field energy functional
\be
E[\Psi]~=~\int_{\rm box} dz
\left[
\frac12 \frac{d\Psi^\dag}{dz} \frac{d\Psi^\dag}{dz} +
\frac{c_0}{2} (\Psi^\dag\Psi)^2 +
Q \Psi^\dag F_z^2 \Psi +
\frac{c_1}{2} \sum_\alpha (\Psi^\dag F_\alpha \Psi)^2 
\right]~.
\label{Espin1}
\ee
Here $\Psi^T=(\psi_1,\psi_0,\psi_{-1})$ describes the $m=0,\pm1$ condensate components,
the wave function is normalised $\int_{\rm box}dz \Psi^\dag\Psi=1$, and $F_{x,y,z}$ are spin-$1$
matrices. The first term in (\ref{Espin1}) is the kinetic energy, the second and the fourth
term describe spin-independent and spin-dependent atom interactions respectively, and
the third term is a quadratic Zeeman shift originating from atom interactions with the magnetic 
field $B$. 

A phase diagram of the model (\ref{Espin1}) is interesting when $c_1<0$, like for e.g. $^{87}$Rb atoms, 
and there is competition between the last two terms in Eq. (\ref{Espin1}). When we assume 
zero longitudinal magnetisation, $f_z\equiv\Psi^\dag F_z\Psi=0$, whose integral is a constant of motion, 
then the relevant parameter is
\be
\epsilon~=~\frac{Q}{n|c_1|}-2~,
\ee 
where $n=\Psi^\dag\Psi$ is the density. At $\epsilon=0$ there is a transition between the symmetric polar 
phase for $\epsilon>0$, where the ground state wave function is
\be
\Psi_P^T~\sim~(0,1,0)~,
\label{PsiP}
\ee
and the broken-symmetry ferromagnetic phase for $\epsilon<0$, where the degenerate ground states are
\be
\Psi_F^T~\sim~
\left(
e^{i\chi_1}~\sqrt{-2\epsilon}~,~
e^{i(\chi_1+\chi_{-1})/2}~2\sqrt{4+\epsilon}~,~
e^{i\chi_{-1}}~\sqrt{-2\epsilon}
\right)~.
\label{PsiF}
\ee
The relative phase $\chi_1-\chi_{-1}$ determines orientation of the transverse magnetisation 
$(f_x,f_y)=(\Psi^\dag F_x\Psi,\Psi^\dag F_y\Psi)$ in the $x-y$ plane. The ground state is degenerate with 
respect to this orientation allowing topological defects like textures in 1D, point vortices in 2D, and 
vortex lines in 3D. 

In the symmetric phase there are small thermal or at least quantum fluctuations around
the ground state (\ref{PsiP}) which can be expanded into Bogoliubov modes
\cite{spin1quantumfluctuations}. Near the critical point the gap in the quasiparticle 
spectrum scales as
\be
\Delta~\sim~\epsilon^{1/2}~.
\label{Deltanuz}
\ee
Since by definition $\Delta\sim\epsilon^{\nu z}$ for small $\epsilon$, we can identify 
$\nu z=1/2$. On the mean field level considered here the critical exponent $\nu=1/2$ 
and KZM predicts
\be
\hat\xi~\simeq~\tau_Q^{1/3}~,
\ee
compare Eq. (\ref{hatxi}), after a linear quench to the ferromagnetic phase. 

More precisely, see Refs. \cite{Lamacraft,Bodzioferro2,Bodzioferro}, what happens is that when the linear quench crosses the critical point $\epsilon=0$
the initial paramagnetic ground state (\ref{PsiP}) becomes a dynamically unstable false vacuum. The frequency of the $k=0$ Bogoliubov mode (contributing to the initial small fluctuations around this false vacuum) given by the gap in Eq. (\ref{Deltanuz}) becomes imaginary and the mode begins to grow quasi-exponentially on the timescale of $\hat t\simeq\tau_Q^{1/3}$ in a similar way as in the classical transitions considered in Section \ref{KZMclassical}. As $\epsilon$ enters deeper into the ferromagnetic phase, more long wavelength modes become unstable. This linear instability grows until time $\hat t$ when its growth is halted by the quartic term in the energy functional (\ref{Espin1}) and the linearisation in small fluctuations around the initial paramagnetic state breaks down. By this time, the long wavelength modes with $|k|$ up to $\hat\xi^{-1}$ have become unstable and exponentially amplified with respect to the initial small fluctuations back in the symmetric phase. This exponential amplification promotes $\hat\xi$ to the only relevant scale of length 
characterising the wave function $\Psi$ after the transition.

\begin{figure}
\begin{center}
\begin{minipage}{100mm}
\subfigure[]{
\resizebox*{5cm}{!}{\includegraphics{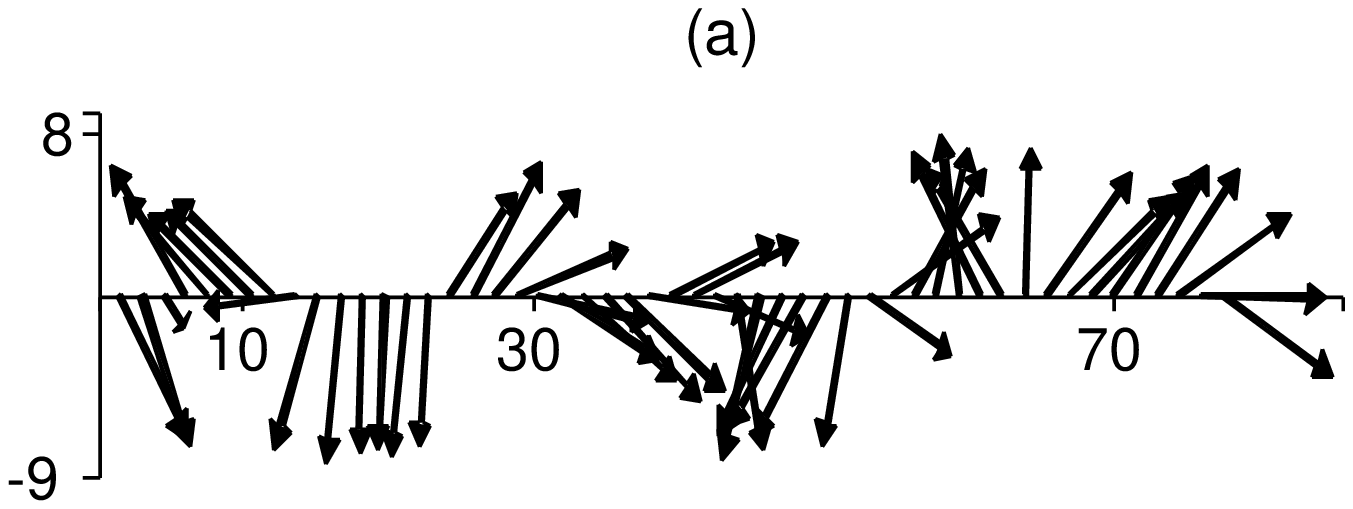}}}%
\subfigure[]{
\resizebox*{5cm}{!}{\includegraphics{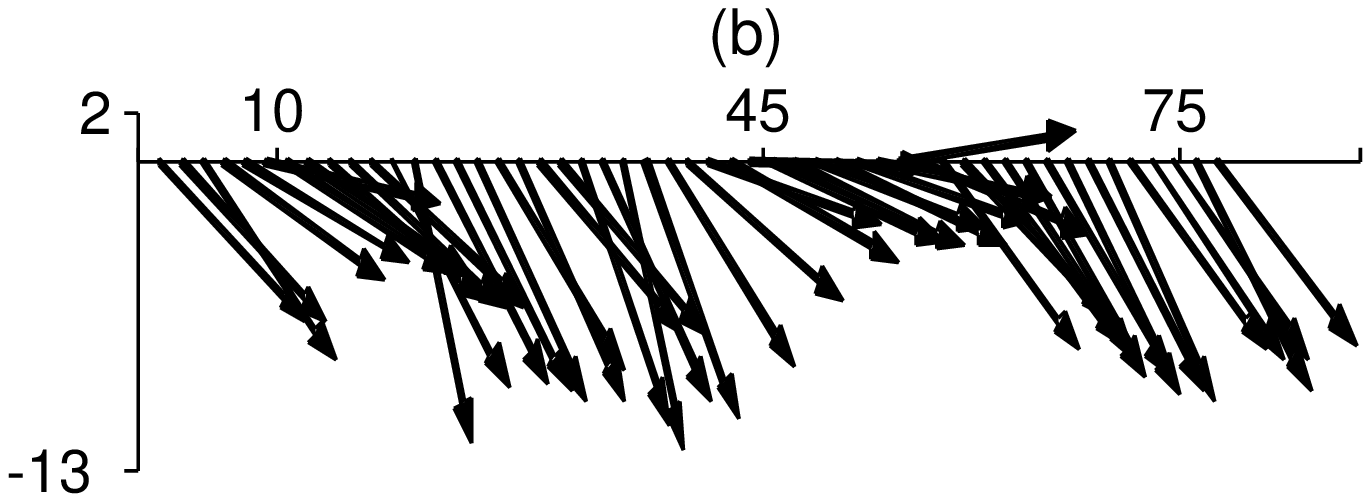}}}%
\caption{ Vector representation of the (magnified) transverse magnetisation 
$(f_x,f_y)\times 10^3$ at a) $\epsilon=-0.28$ and b) $\epsilon=-2$ during
a transition with $\tau_Q=10$.
(Figure from Ref. \cite{Bodzioferro})
}
\end{minipage}
\end{center}
\end{figure}

In the spirit of the truncated Wigner method, in the numerical simulations of Ref. \cite{Bodzioferro} 
the wave function $\Psi$ was initialised in the symmetric phase in the ground state (\ref{PsiP}) plus 
small Gaussian noise uncorrelated in space. The noise provides seed quantum/thermal fluctuations. Then 
the wave function was evolved by a linear quench $\epsilon(t)=-t/\tau_Q$ to the symmetry-broken phase. 
Figure \cite{spin1quantumfluctuations} shows two snapshots of the transverse magnetisation $(f_x,f_y)$ 
in the ferromagnetic phase. The magnetisation has orientation which is random, but correlated on the 
length scale $\hat\xi$. Topological textures in this random transverse magnetisation can be characterised 
by a winding number
\be
\frac{1}{2\pi}\int_{\rm box} dz~\frac{d}{dz}{\rm Arg}(f_x+if_y)~.
\ee  
Density of textures scales as $\hat\xi^{-1}$. Reference \cite{Bodzioferro2} considers formation
of topological textures in more than one dimnesion.

\begin{figure}[t]
\begin{center}
\includegraphics[width=0.95\columnwidth,clip=true]{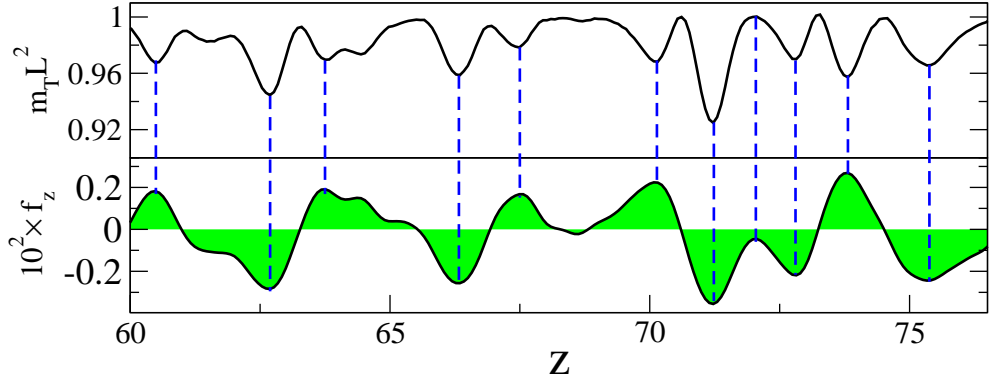}
\caption{ 
Snapshot of a magnetisation of the system at $t=2\tau_Q$ when $\epsilon=-2$ 
after a transition with $\tau_Q=10$. The top part shows transverse magnetisation 
$m_T=\sqrt{f_x^2+f_y^2}$ (times the box size squared $L^2$). The bottom
part shows the measurable longitudinal magnetisation $f_z$ (magnified by
a factor of $10^2$). The vertical dashed lines help to see coincidences
between minima of the transverse magnetisation and extrema of the longitudinal
magnetisation. Average size of the longitudinal domains was verified in 
Ref. \cite{Bodzioferro} to scale as $\hat\xi\simeq\tau_Q^{1/3}$.
(Figure from Ref. \cite{Bodzioferro})
}
\label{FigBodziodomains1}
\end{center}
\end{figure}

Another quantity measured in Ref. \cite{ExpFerro} was the longitudinal magnetisation. 
Both in the experiment and in the numerical simulations of 
Ref. \cite{Bodzioferro} the net longitudinal magnetisation was initially zero,
$\int_{\rm box}dz~f_z=0$, and fluctuations of local magnetisation $f_z$ were small.  
Conservation of the net longitudinal magnetisation allows for creation of a network of
magnetic domains with opposite $f_z$. The domains begin to form at the time
$\simeq\hat t$ of the quasi-exponential growth of the dynamical instability. The bottom
part of Figure \ref{FigBodziodomains1} shows one realisation of longitudinal domains. The top 
part of the same figure demonstrates anti-correlation between transverse magnetisation 
$m_T=\sqrt{f_x^2+f_y^2}$ and the longitudinal magnetisation. Both are correlated in space on 
the length scale $\hat\xi\simeq\tau_Q^{1/3}$.

\begin{figure}[t]
\begin{center}
\includegraphics[width=0.8\columnwidth,clip=true]{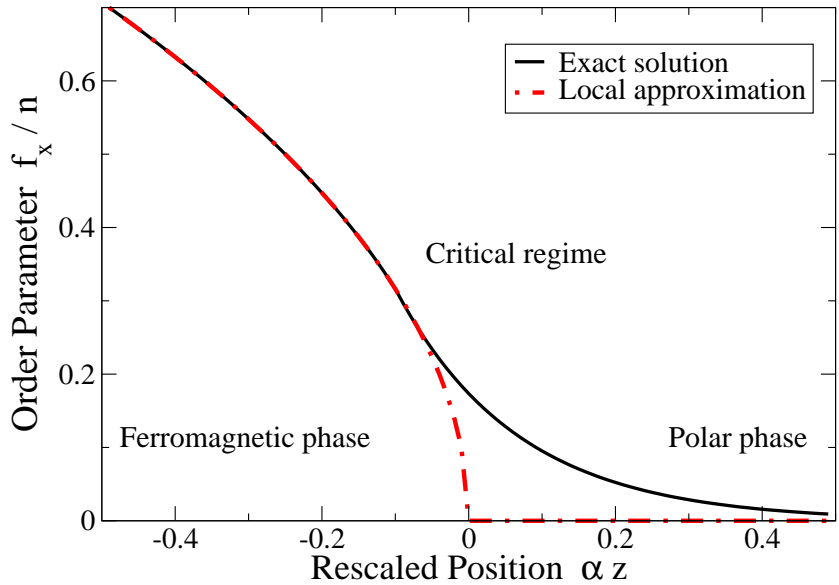}
\caption{ 
Typical condensate magnetisation $f_x$ (order parameter) in a spatial transition (\ref{alphaz}). 
The solid line is the exact solution and the dashed line in the local approximation. In the
critical regime near the critical point the exact solution deviates from the local approximation
and penetrates into the symmetric polar phase to a depth of $\hat\xi\simeq\alpha^{1/3}$.
(Figure from Ref. \cite{Bodzioferro})
}
\label{FigBodzioinspace}
\end{center}
\end{figure}

The model (\ref{Espin1}) provides also another illustration of KZM in space, see Sections 
\ref{KZinspace} and \ref{IsingKZinspace}. Reference \cite{Bodzioinspace} considers a transition 
in space with
\be
\epsilon(z)~=~\alpha~z~
\label{alphaz}
\ee  
from the symmetry-broken phase where $z<0$ to the symmetric polar phase where $z>0$.
Figure \ref{FigBodzioinspace} shows the transverse magnetisation $f_x(z)$ which is the order 
parameter. The dashed line is the magnetisation in the local approximation. It is non-zero in 
the broken-symmetry phase and vanishes as $f_x\sim\sqrt{-\epsilon}=\sqrt{\alpha z}$ when the 
critical point is approached, $z\to0^-$. The solid line is the exact solution which deviates 
from the local approximation at $\hat\epsilon\sim\alpha^{2/3}$ and penetrates into the
symmetric polar phase to a depth $\hat\xi\simeq\alpha^{-1/3}$. These scalings were verified
in Ref. \cite{Bodzioinspace} to be accurate for $\alpha\ll1$, as expected from the general
argument in Section \ref{KZinspace}.

\subsection{ Adiabatic sweep across a gapless regime }
\label{sweep}

The main result of KZM is that, in the thermodynamic limit, a sweep across a second order phase 
transition cannot be adiabatic, no matter how slow is the transition rate
\be
\delta~=~\tau_Q^{-1}~,
\ee
because the system is gapless at the critical point and the adiabatic theorem cannot apply. 
Consequently, there is a finite density of excitations $n_{\rm ex}$ (or density of excitation 
energy $\varepsilon$) which scales with a power of the transition 
rate $\delta$. The power generally increases with the dimensionality of the system making the 
density of excitations in higher dimensions decay faster in the adiabatic limit of $\delta\to0$,
compare e.g. Eq. (\ref{nex}). Lower dimensional systems are generally less adiabatic because 
they have more low energy states available for excitation. As pointed out in the seminal paper 
\cite{PolkovnikovNature}, the same observations must necessarily apply to all gapless systems, no 
matter what is their dimensionality, but their non-adiabaticity can be expected to be more dramatic
in lower dimensions. Gapless systems are quite generic as most systems with 
broken symmetries have gapless excitations like e.g. phonons, magnons, or spin waves.

In the zero temperature quantum limit, we can consider a linear sweep of a parameter 
$\kappa$ in a Hamiltonian,
\be
\kappa(t)~=~\kappa_i+\delta~ t~,
\label{kappa}
\ee 
starting from the ground state at an initial $\kappa_i$ and terminating at a final $\kappa_f$. 
The final density of the positive excitation energy depends on the transition rate,
$
\varepsilon(\delta)>0~.
$ 
According to Ref. \cite{PolkovnikovNature}, response of the system to the adiabatic sweep may 
fall into one of the following three regimes:
\begin{enumerate}
\item[A] When $\varepsilon(\delta)$ is an analytic function of $\delta$, then to 
         leading order in $\delta$ we have 
         \be
         \varepsilon(\delta)~\sim~\delta^2~.
         \ee   
         Here the linear term is zero because the positive excitation energy cannot 
         be made negative by a sweep in the opposite direction. This regime can be called
         {\it mean field} or {\it analytic}. Since $\varepsilon(\delta\to0)=0$,
         there is adiabatic limit in this regime. On a more microscopic level, the origin
         of the quadratic scaling lies in the Landau-Zener transition in Section \ref{LZvKZ},
         see cases (ii,iii,iv) and Eq. (\ref{P2n}) where the parameter driving the transition 
         has a discontinuous time derivative just like Eq. (\ref{kappa}). The discontinuity
         makes the excitation probability scale with a square of a transition rate $\tau_Q^{-2}\sim\delta^2$ 
         instead of the usual exponential decay, see a more detailed discussion in Ref. \cite{infinitezimalPolkovnikov}.        
\item[B] Even when $\varepsilon(\delta)$ is not analytic, it is still possible to 
         approach the adiabatic limit as
         \be
         \varepsilon(\delta)~\sim~|\delta|^a~
         \ee 
         where $a>0$ but $a\neq2$. This regime can be termed {\it non-analytic}. On the
         microscopic level it originates from a combination of the $\delta^2$ 
         scaling like in case A and large density of low energy excitations, see 
         Ref. \cite{infinitezimalPolkovnikov}.  
\item[C] Finally, in a {\it non-adiabatic} regime we have
         \be
         \varepsilon(\delta)~\sim~|\delta|^a~L^b~,
         \ee
         where $a,b>0$ and $L$ is the system size. Here the adiabatic limit does not exist for 
         infinite system size. Notice that it is not just the excitation energy, but the excitation 
         {\it energy density} that diverges in the thermodynamic limit.
\end{enumerate}
In a system with well defined quasiparticles the 
non-adiabaticity of the transition can be alternatively classified
by scaling of the density of excitations $n_{\rm ex}(\delta)$ when
$|\delta|\to0$. In general, the two classifications are different because
the dominant low energy quasiparticles have a non-trivial dispersion relation.

The existence of regimes B and C in the quantum limit was supported in Ref. \cite{PolkovnikovNature}
by solution of a generic low energy quadratic Hamiltonian,
\be
H~=~\frac12 \sum_q\left(q^2\phi_q^2+\kappa_q\Pi_q^2\right)~,
\ee 
where $\phi_q$ and $\Pi_q$ are conjugate coordinates and momenta respectively. 
In the context of superfluidity $\kappa_q$ is a compressibility. 
Here we choose
\be
\kappa_q~=~\kappa+\lambda q^2~
\ee
to cover all three regimes defined above. 
$\kappa$ is ramped as in Eq. (\ref{kappa}). For a large initial $\kappa_i$, the response of 
the system belongs to regime A with $\epsilon(\delta)\sim\delta^2$, but when $\kappa_i=0$ 
then 
\be
\varepsilon ~\sim~ \frac{|\delta|^{(d+1)/4}}{\lambda^{(d+1)/8}}~,
\ee
and the response belongs to regime B. Here $d$ is the number of dimensions.
The density of excitations $n_{\rm ex}(\delta)$ leads to a different classification.
When $\kappa_i$ is large, then the non-analytic regime B is realised for $d=1$ and 
the analytic regime A when $d\geq2$. However, when $\kappa_i=0$, then the system is 
non-analytic (B) when $d=2,3$, but {\it non-adiabatic} (C) in 1D. 

The crossover between different scalings can be further illustrated by e.g. 
Ref. \cite{linearFalicovKimball} where a near-adiabatic parameter change within a gapless 
metallic or gapped insulating phase of the Falicov-Kimball model is studied by the dynamical 
mean field theory \cite{DMFT}. The excitation energy density scales with a power of the quench 
rate $\delta$ whose exponent depends in general both on energy spectrum of the system and 
smoothness of the quench protocol. However, in the gapped insulator the exponent depends on 
the smoothness of the ramp only. By contrast, for a sufficiently smooth ramp protocol in the gapless 
metallic phase the exponent depends on the intrinsic spectrum of the system only, but this intrinsic 
behaviour is not observable when the ramp is not smooth enough. For instance, for a linear ramp 
there is a crossover to the $\delta^2$ scaling in case A. The quadratic scaling for energy density 
was also derived in Ref. \cite{prethprediction}.

\subsection{ Many-particle Landau-Zener problem: 
             adiabatic passage across a Feshbach resonance }
\label{manyLZ}

Non-adiabatic dynamics of a single-particle quantum system can often be described by 
the Landau-Zener problem where a probability of the transition from the initially 
occupied ground state to the excited state is exponentially small in the sweeping rate 
$\delta$. With more particles one generally encounters more LZ anti-crossings 
whose cumulative effect describes the behaviour of the system during the driving process. 
However, in many-body systems there is exponential density of energy levels
and it is not possible to divide the evolution into a series of independent
LZ anti-crossings: every anti-crossing takes finite time, but the frequency of anti-crossings
increases exponentially. In experiments with ultracold quantum gases, where 
macroscopically large numbers of particles are involved, the microscopic (LZ) treatment
is not practical, but it is often justified to use semiclassical treatments like the
truncated Wigner method \cite{TW}. In this context, non-adiabaticity in the semiclassical 
models has been discussed recently in 
Refs. \cite{sc5,sc6,sc7,sc8,sc9,sc10,ItinTorma,sc11,BEC2wells,
Bodzioinspace,Lamacraft,Bodzioferro,Meisner} and Sections \ref{SectionMeisner} and \ref{FerroKZ}.  

In this Section we consider a time-dependent Dicke model in dimensionless units with 
a time-dependent Hamiltonian
\be
H~=~
-\delta~t~b^\dag b~+~\delta~t~S^z~+~\frac{1}{\sqrt{N}}(b^\dag S^-+{\rm h.c.})~,
\label{HDicke}
\ee
where $S^\pm=S_x\pm iS_y$ are spin operators, the spin $S=N/2$, $b$ is a bosonic annihilation 
operator, and $\delta$ is the sweep rate. When $N=1$ we recover the standard LZ model in 
Eq. (\ref{HLZ}) and the excitation probability $P=\exp(-\pi/\delta)$ is exponentially
small in the adiabatic regime $\delta\ll1$. However, here we are interested in the opposite
extreme of a macroscopically large spin when $N\gg 1$. 

In the context of the adiabatic passage across a Feshbach resonance \cite{eddi} the 
Hamiltonian (\ref{HDicke}) is equivalent to
\be
H~=~
-\delta~t~b^\dag b~+~
\frac{\delta~ t}{2} \sum_{i=1}^N 
\left(c^\dag_{i,\uparrow}c_{i,\uparrow}+c^\dag_{i,\downarrow}c_{i,\downarrow}\right)~+~
\frac{1}{\sqrt{N}}
\sum_{i=1}^N
(b^\dag c_{i\downarrow}c_{i\uparrow}+{\rm h.c.})~,
\label{Hbcc}
\ee
where $c_{i\sigma}$ are fermionic annihilation operators, and $\sigma=\uparrow,\downarrow$ 
represents internal states of fermions, see the caption of Figure \ref{FigFeshbach}.
Initially at $t\to-\infty$ the system is prepared in the ground state with $2N$
fermionic atoms, and when $t\to+\infty$ its instantaneous ground state becomes the state
with $N$ bosonic molecules. In the adiabatic limit, the time-dependent Hamiltonian 
(\ref{Hbcc}) is meant to convert all atoms into molecules. However, as shown in 
Refs. \cite{sc8,sc10,ItinTorma} and outlined below, the sweep is never truly 
adiabatic, because it leaves behind a fraction of fermions which scales with
a power of $\delta$. 

In a semiclassical approximation to Eq. (\ref{HDicke}), we use the number-phase 
decomposition of the boson field, $b=\sqrt{Nn}~e^{i\varphi}$, and the polar 
representation of the spin variables: $S_z=\frac{N}{2}\cos\theta$, 
$S_x=\frac{N}{2}\sin\theta\cos\xi$, and $S_y=\frac{N}{2}\sin\theta\sin\xi$. Combining 
the angles as $\xi-\varphi+\pi\equiv\phi$, and using a conservation law 
$Nn=\frac{N}{2}(1-\cos\theta)$ to eliminate $\theta$, we obtain
$
H~=~-2N~\left(\delta~ t~n~+~n\sqrt{1-n}\cos\phi\right)~.
$
Finally, we rescale the Hamiltonian as
\be
H'~=~H/N~=~-\gamma~n~-~2n\sqrt{1-n}\cos\phi~,
\label{Hnphi}
\ee
where
\be
\gamma~=~2~\epsilon~ t'~,
\ee
$\epsilon=\delta/N$, and $t'=Nt$. Here $n$ and $\phi$ are mutually conjugate variables 
and $n\in[0,1]$ is a fraction of bosons. In the following we skip the primes.

\begin{figure}[t]
\begin{center}
\includegraphics[width=0.99\columnwidth,clip=true]{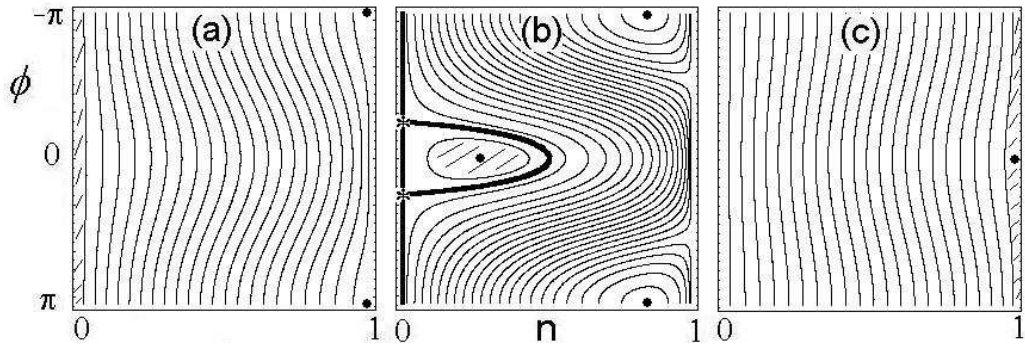}
\caption{ Phase portraits of the Hamiltonian (\ref{Hnphi}). Panels
a,b,c correspond to $\gamma=-6,-1.7,20$ respectively. The solid
dots are the fixed points, the asterisks mark the saddle points,
the solid line is the separatrix, and the shaded areas illustrate
the definition of the classical action, i.e., the area
enclosed by the integral in Eq. (\ref{Invariant}).
(Figure from Ref. \cite{ItinTorma})
}
\label{FigI}
\end{center}
\end{figure}

The equations of motion that follow from the Hamiltonian (\ref{Hnphi}) are 
$\dot n=-\partial_\phi H$ and $\dot\phi=\partial_nH$. Given an initial $n=n_-$ at 
$t\to-\infty$, we want to find $n=n_+$ after the sweep to $t\to\infty$. These two 
asymptotic values $n_\mp$ are related to an adiabatic invariant given by the action
\be
I~=~\int\frac{d\phi}{2\pi}~n~,
\label{Invariant}
\ee
where the integral is along a closed trajectory $n(t),\phi(t)$ in the phase space. 
More precisely, $I$ is the area (divided by $2\pi$) of the phase space enclosed by 
the trajectory $n(t),\phi(t)$ with the convention that, when moving 
along the trajectory, the enclosed area is on the left. For a fixed
$\gamma\to\mp\infty$, the trajectories become $\phi=\phi_0-\gamma t,n=n_{\mp}$ 
and the action is
\be
I_-~=~n_-~,~~I_+~=~1-n_+~,
\ee
compare Figs. \ref{FigI}a and \ref{FigI}c respectively, where the $I$'s are the 
shaded areas (divided by $2\pi$). 

In the adiabatic limit of $\epsilon\to0$ we expect the invariant $I$ to be conserved, 
$I_+=I_-$, and an initial state with $n_-=I_-$ bosons to be adiabatically 
converted into a final state with $n_+=1-I_+=1-I_-=1-n_-$ bosons. However, in 
a sweep with a finite rate $\epsilon$ we expect that $I$ changes by $\Delta I=I_+-I_->0$ 
and the final number of bosons, 
\be
n_+~=~1-n_--\Delta I~,
\label{nplus}
\ee 
is less than in the adiabatic limit by $\Delta I$.  
 
The change of the adiabatic invariant $\Delta I$ depends on $n_-$. In the truncated Wigner 
method \cite{TW}, the final $\Delta I(n_-)$ in Eq. (\ref{nplus}) has to be averaged over 
the initial Wigner function $W_-(n_-,\phi)$. Even in the initial ground state with no bosons 
the function $W_-(n_-)=2N\exp(-2Nn_-)$ has a finite spread $n_-\simeq1/N$. Thus we need
to consider not only $n_-=0$, but also small finite $n_-$. 
  
The classical phase space portrait of the Hamiltonian (\ref{Hnphi}) depends on 
a fixed $\gamma$. When $\gamma<-2$ or $\gamma>2$, there is only one fixed point
where $\partial_n H=0=\partial_\phi H$, see Figs. \ref{FigI}a and b.
At $\gamma=-2$ there is a bifurcation and in the range $-2<\gamma<2$ there are
two fixed points. In the same regime, there are two saddle points located where $n=0$ 
and $\cos\phi=-\gamma/2$, see Fig. \ref{FigI}b. The trajectory connecting these two 
saddles, called separatrix, separates rotating from oscillating motions. 
In the adiabatic limit, the major part of the total change of the classical 
action $\Delta I$ is generated near the separatrix and especially near the saddle 
points when they arise during the bifurcation at $\gamma=-2$.

The initial $n_-$ is subject to quantum fluctuations $\simeq N^{-1}$ so we need to consider 
finite but very small $n\simeq N^{-1}$. Approximating $\gamma=-2$ near the bifurcation,
expanding to leading order in small $n$, introducing new canonically conjugate
variables $P,Y$ and a new time variable $s$, we obtain an effective Hamiltonian
\be
H~=~\frac{P^2}{2}-s\frac{Y^2}{2}+\frac{Y^4}{2}~,
\ee
see Ref. \cite{ItinTorma} for more details.
The Hamiltonian implies the Panleve equation
$
\frac{d^2Y}{ds^2}~=~sY-2Y^3
$
whose asymptotes were found in Ref. \cite{Panleve}:
\bea
Y(s\to-\infty)&=&
\alpha(-s)^{-\frac14}
\sin\left(\frac23(-s)^{3/2}+\frac34\alpha^2\ln(-s)+{\rm const}\right)~,\\
Y(s\to+\infty)&=&
\pm\sqrt{\frac{s}{2}}
\pm\rho(2s)^{-\frac14}
\cos\left(
\frac{2\sqrt{2}}{3}s^{3/2}-\frac32\rho^2\ln(s)+{\rm const}
\right)~.
\eea
As $s\to\pm\infty$ the adiabatic invariant of the Panleve equation tends to 
$I_+=\frac{\alpha^2}{2}$ and $I_-=\frac{\rho^2}{2}$ respectively. Using a relation
between the constants $\alpha$ and $\rho$ in Ref. \cite{ItinTorma} and going back
to the variables of the Hamiltonian (\ref{Hnphi}) we obtain
\be
\Delta I(n_-)~=~
n_- -
\frac{\epsilon}{\pi}\ln\left(e^{\pi n_-/\epsilon}-1\right)-
\frac{2\epsilon}{\pi}\ln(2\sin(\pi\xi))
~,
\label{DeltaI}
\ee 
where $\xi\in[0,1)$ is a quasi-random variable. For a given initial $n_-$,
the final fraction of bosons is $n_+=1-n_--\Delta I(n_-)$.

There are two physically interesting ways of taking the adiabatic limit:
\begin{itemize}
\item When $\epsilon\to 0$ before $N\to\infty$, or more precisely $\epsilon\ll1/N$, 
      we obtain the asymptote
      \be
      n_+~\approx~1 - n_- + \frac{2\epsilon}{\pi}\ln[2\sin(\pi\xi)]~.
      \ee
      This final $n_+$ is a random variable because the random variable $n_-$ comes 
      from the initial Wigner function $W_-(n_-)=2N\exp(-2Nn_-)$ and $\xi$ from a uniform
      distribution between $0$ and $1$. The last $\xi$-term is zero on average, so
      it does not contribute to the average of $n_+$, but it increases its variance. 
      The amplification is relatively weak because $\epsilon\ll1/N$.
\item When $N\to\infty$ before $\epsilon\to 0$, or more precisely $1/N\ll\epsilon\ll1$, we 
      have
      \be
      n_+ ~\approx~
      1 -
      \frac{\epsilon}{\pi}\ln\left(\frac{\epsilon}{\pi n_-}\right) 
      ~.
      \ee
      Average of $n_+$ over the initial distribution $W_-(n_-)=2N\exp(-2Nn_-)$ is
      \be
      \langle n_+ \rangle ~\approx~
      1 ~-~
      \frac12~
      \epsilon\ln\epsilon~.
      \ee
      The small initial quantum fluctuations of $n_-\simeq1/N$ are amplified to
      much stronger fluctuations $\frac12\epsilon\ln\epsilon$.  
\end{itemize} 
In both cases we find that, unlike in the two-level LZ problem, the leading non-adiabatic effect 
scales with a power of $\epsilon$. There exists adiabatic {\it limit} when $\epsilon\to0$, but 
there is no adiabatic {\it regime} below any threshold value of $\epsilon$.

\subsection{ Summary }
\label{Summary1}

In Section \ref{DynamicsQPT} we reviewed accumulated evidence that adiabatic dynamics 
in a gapless isolated quantum system results in excitation which scales with a power
of the transition rate (but with an exception of a disordered system where
the excitation is only logarithmic in the rate). The leading motif was 
the adiabatic-impulse approximation essential for KZM. The main story left over 
some less universal features of the considered models, quite as well as some interesting 
ideas and problems that did not quite fit into the main narrative. Below we briefly
mention three recent examples.  

In Ref. \cite{GuAdiabatic} the adiabaticity of the linear quench across
a quantum phase transition was readdressed within the adiabatic perturbation theory. 
There turns out to be an upper estimate on the excitation probability of the system 
in terms of an adiabatic dimension $d_a$. The adiabatic dimension is the dimension of 
the fidelity susceptibility of the driving Hamiltonian \cite{daGu}. In the critical 
region, the quantum adiabatic dimension $d_a=2d+2z-2\Delta_V$, where $d$ is the dimension 
of the system, $z$ is the dynamical exponent, and $\Delta_V$ is the scaling dimension of 
the driving Hamiltonian \cite{daZanardi}. The upper estimate implies that the excitation is
negligible when the linear quench time $\tau_Q\gg L^{d_a}$, where $L$ is the linear size of 
the system. For instance, in the quantum Ising chain $d_a=2$ and the upper bound implies 
adiabaticity when $\tau_Q\gg L^2$, in agreement with the exact Eq. (\ref{nonadiabatic})
where $L=N$ is the number of spins. By contrast, in the Lipkin-Meshkov-Glick model 
the upper bound implies $\tau_Q\gg N^{4/3}$, while the exact adiabaticity condition in 
Eq. (\ref{PLZLMG}) is a weaker $\tau_Q\gg N^{2/3}$, so in this model the upper bound
is correct, but it overestimates the minimal $\tau_Q$ required to make the transition 
adiabatic. Reference \cite{infinitezimalPolkovnikov} introduces a family of generalized 
adiabatic susceptibilities $\chi_m$ enumerated by their order $m$. Their $\chi_2$
is the susceptibility considered in Ref. \cite{GuAdiabatic}, but they argue that
it is $\chi_4$ that describes the excitation probability and density of excitations
$n_{\rm ex}$ for slow linear quenches. In the Lipkin-Meshkov-Glick model they
obtain an accurate adiabatic condition $\tau_Q\gg N^{2/3}$ - the same as the exact 
Eq. (\ref{PLZLMG}). Both Refs. \cite{GuAdiabatic} and \cite{infinitezimalPolkovnikov} make 
a very interesting connection between the adiabaticity and the fidelity susceptibility.

This review is limited to isolated quantum systems, but it would not be quite physical 
to ignore the problem of adiabatic dynamics in open quantum systems. This problem has already 
been studied is some detail \cite{decFubini,decSchaller,decAmin,decPatane,decoherence},
but it is certainly far from being completely explored. The effect of classical and
quantum noise acting uniformly on the quantum Ising chain was considered in 
Refs. \cite{decFubini} and \cite{decSchaller} respectively. Numerical simulations
for a model of local noise acting on the disordered Ising chain were performed in 
Ref. \cite{decAmin}, and the effect of a static spin bath coupled locally to the
ordered Ising chain was considered in Ref. \cite{decoherence}. The static spin
bath was found to change the universality class of the quantum phase transition
to that of the disordered Ising chain in Section \ref{RandomIsing}. For weak coupling to the
spin bath and not too slow quenches the density of excitations scales like in the pure 
Ising chain in Eq. (\ref{nexIsing}), but for very slow transitions the scaling is replaced by 
a much slower logarithmic dependence like in Eq. (\ref{xilog}). Finally, in Ref. \cite{decPatane} 
the scaling theory was generalised to an open critical system and a quantum kinetic equation approach 
was formulated for adiabatic dynamics in the quantum critical region. It was found that for weak 
coupling and not too slow quenches the density of excitations is universal also in the presence of 
the external bath. Given the evidence at hand, we can conclude that a weak coupling to environment
does not alter KZM for not too slow quenches, but in the adiabatic limit we have $\hat\epsilon\to0$ 
and the KZM happens very close to the critical point, where the system is very susceptible to the 
influence of the environment. Moreover, thermalisation dynamics close to a quantum critical point
was considered in Ref. \cite{SolsBis}.  
  
An opposite situation is considered in Ref. \cite{decQuan}, where a central spin couples 
globally to the environment of the transverse Ising model subject to the linear quench like
in Section \ref{KZIsing}. The quenched environment monitors the state of the central spin
and its sensitivity is amplified by closeness to the phase transition. Decoherence  of
the central spin happens almost exclusively when the critical point of the environment is
traversed and is significantly enhanced by the non-equilibrium dynamics.

In the next part \ref{Relaxation} we consider relaxation of the excited state in the last
adiabatic stage of the evolution.

\section{ Apparent relaxation of an isolated quantum system after a sudden quench }
\label{Relaxation}

\subsection{ Introduction }
\label{Introduction2}

Once a system got excited, the question is what is going to be the fate of the excited state?
Does it relax to any stationary state or, maybe, even thermal state? This question 
has not been thoroughly investigated in the context of adiabatic linear quenches,
but with the exception of Section \ref{IsingToInfty} above where some of the consequences of 
quantum dephasing have been worked out in the integrable quantum Ising chain. The reason
partially is that in a linear quench it is hard to make a clear-cut distinction between the 
non-adiabatic process of exciting the system and the process of relaxation, because the
relaxation begins already during the non-adiabatic excitation. 

Nevertheless, the relaxation problem itself can be formulated in a clear-cut way when we assume that 
a system is initially prepared in the ground state of an initial Hamiltonian $H_0$ and then, at 
$t=0$, a {\it sudden} quench, faster than any time scale of the system, is made to a final 
Hamiltonian $H$:
\be
H_0 ~\stackrel{t=0}{\longrightarrow}~ H ~.
\ee
In this way, the ground state of $H_0$ all of a sudden becomes the excited initial state for adiabatic 
evolution with a time-independent $H$. At first sight, the discontinuous sudden quench may appear 
very different from the smooth linear quench: a slow linear quench makes an effort to
be adiabatic, while the sudden quench does not even pretend to be anything like adiabatic. 
In spite of this first impression, the difference is more quantitative than qualitative. As we 
know from Section \ref{KZargument}, even in the linear quench one can often use the 
adiabatic-impulse-adiabatic approximation. In this approximation the state of the system does not change 
during the impulse stage, when the parameter $\epsilon$ in the Hamiltonian evolves from
$\hat\epsilon$ to $-\hat\epsilon$, but the ground state at $\hat\epsilon$ survives to become 
the excited initial state for the adiabatic evolution after $-\hat\epsilon$. In this approximation
the linear quench is effectively a sudden quench between $\hat\epsilon$ and $-\hat\epsilon$. 
The quantitative difference is that in a slow linear quench the effective parameter jumps from 
$\hat\epsilon$ across a critical point $\epsilon=0$ to $-\hat\epsilon$ is small, while in a sudden quench 
it does not need to be small just as it does not need to cross any critical point. The relaxation after 
a sudden quench has been studied in many different integrable and non-integrable models. 
Section \ref{Relaxation} is an attempt to review some well established concepts as well as more controversial 
conjectures that were formulated in this relatively new area of research.

The first question that comes to mind is what is meant by relaxation in an isolated quantum system
at zero temperature? After all, the evolution of the initial pure state is unitary so the state
cannot relax to any mixed steady state. However, even when the state $|\psi\rangle$ 
of the whole system is pure, a state of its subsystem $\Omega$ is in general mixed 
because the subsystem is entangled with the rest of the system $\Omega^\perp$. 
The state of the subsystem must be described by a reduced density matrix 
$\rho_{\Omega}={\rm Tr}_{\Omega^\perp}|\psi\rangle\langle\psi|$ which is in general 
a mixed state. Thus a subsystem can be mixed even though the system as a whole 
remains pure. 

Having thus introduced mixedness, we can now attempt a definition of relaxation. A well defined 
question is: is there a mixed stationary state $\rho_\infty$ of the whole system such that 
\be
\lim_{t\to\infty}~\rho_{\Omega}(t)~=~{\rm Tr}_{\Omega^\perp}~\rho_\infty~
\ee 
for a finite subsystem $\Omega$? If yes, then expectation values of {\it local}
observables, with a finite support in the subsystem $\Omega$, relax to their 
expectation values in the steady state $\rho_\infty$. Even though the actual
pure state of the whole system $\rho(t)=|\psi(t)\rangle\langle\psi(t)|$ cannot relax, it 
appears to relax for local observables. There is no global relaxation, but there appears 
to be relaxation for local observers.

The local relaxation is not the most general way a relaxation of a pure state can be
defined. Suppose that we want the steady state $\rho_\infty$ to be a statistical 
description of a system like, e.g., a canonical or microcanonical ensemble. Even in 
classical physics a statistical description is not meant to capture all fine details
of a state but only its most ``coarse-grained'' properties mainly because a too detailed
description would not be tractable and thus not useful. From the point of view of 
observables, we expect good statistical description of a state to give accurate 
predictions of sufficiently coarse-grained observables, but we will not be surprised
when a measurement of a complicated fine-detailed observable reveals a difference 
between the actual state and, say, a canonical ensemble. The same is true for an isolated 
quantum system: we expect that coarse-grained observables $O$ can relax, 
\be
\lim_{t\to\infty}~\langle\psi(t)|O|\psi(t)\rangle~=~{\rm Tr}~\rho_\infty~O~,
\ee     
but a measurement of a sufficiently fine-detailed observable can reveal a difference between
the statistical $\rho_\infty$ and the actual pure state $|\psi(t)\rangle$. In a many-body 
system, for instance, simple few-body observables may be accurately described by $\rho_\infty$,    
while a complicated many-body observable can reveal that the statistical $\rho_\infty$
is not accurate. Thus the pure state $|\psi(t)\rangle$ does not relax, but its approximate 
coarse-grained description appears to relax. 

Having established what is meant by relaxation in an isolated quantum system, we can ask
what is the relaxed state $\rho_\infty$? A quick and in principle correct answer is: 
a diagonal ensemble. If a system does relax, then a diagonal infinite time average of 
its density matrix
\be
\overline{\rho}~\equiv~
\lim_{T\to\infty}~\frac{1}{T}\int_0^Tdt~|\psi(t)\rangle\langle\psi(t)|~=~
\sum_\alpha~p_\alpha~|\alpha\rangle\langle\alpha|~,
\label{overlinerho}
\ee
is the first candidate for $\rho_\infty$. Here $\alpha$ enumerates eigenstates of a non-degenerate $H$ and 
\be 
p_\alpha~=~|\langle\alpha|\psi(0)\rangle|^2~=~|\langle\alpha|\psi(t)\rangle|^2
\ee
is a conserved probability that the isolated system is in the eigenstate $\alpha$. Indeed, if a system relaxes 
to a steady state, then expectation values in the steady state must be the same as in the infinite time average 
$\overline{\rho}$. Unfortunately, the diagonal ensemble $\overline{\rho}$ may be disappointing as a statistical 
description, because it contains a lot of information about the initial state $|\psi(0)\rangle$ encoded in the 
microscopic initial probabilities $p_\alpha$. Thus $\overline{\rho}$ itself may be not the desired tractable 
statistical description $\rho_\infty$, but it is a good starting point to look for a more tractable $\rho_\infty$ 
as a coarse-grained description of $\overline{\rho}$. 

\begin{figure}
\begin{center}
\begin{minipage}{100mm}
\subfigure[]{
\resizebox*{5cm}{!}{\includegraphics{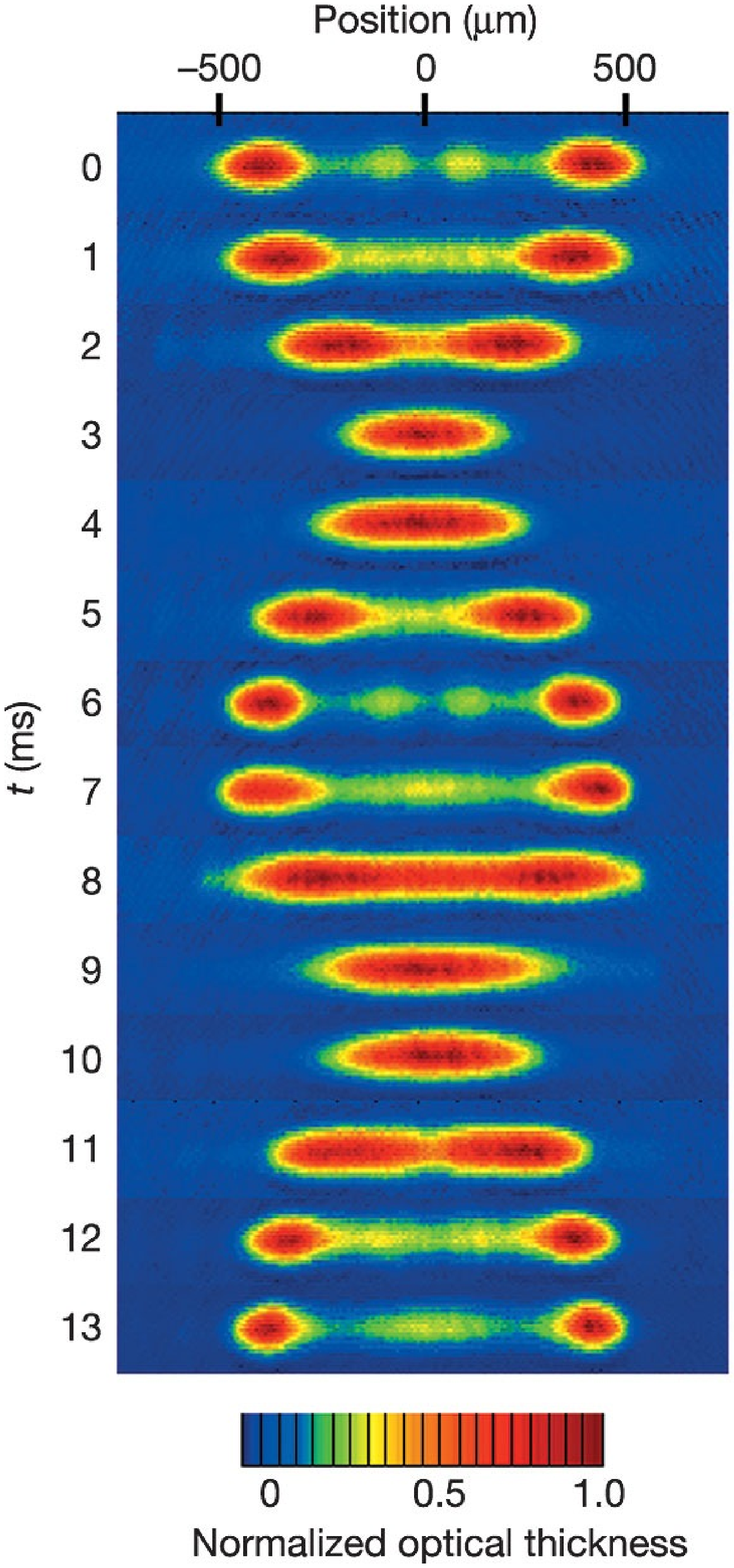}}}%
\subfigure[]{
\resizebox*{5cm}{!}{\includegraphics{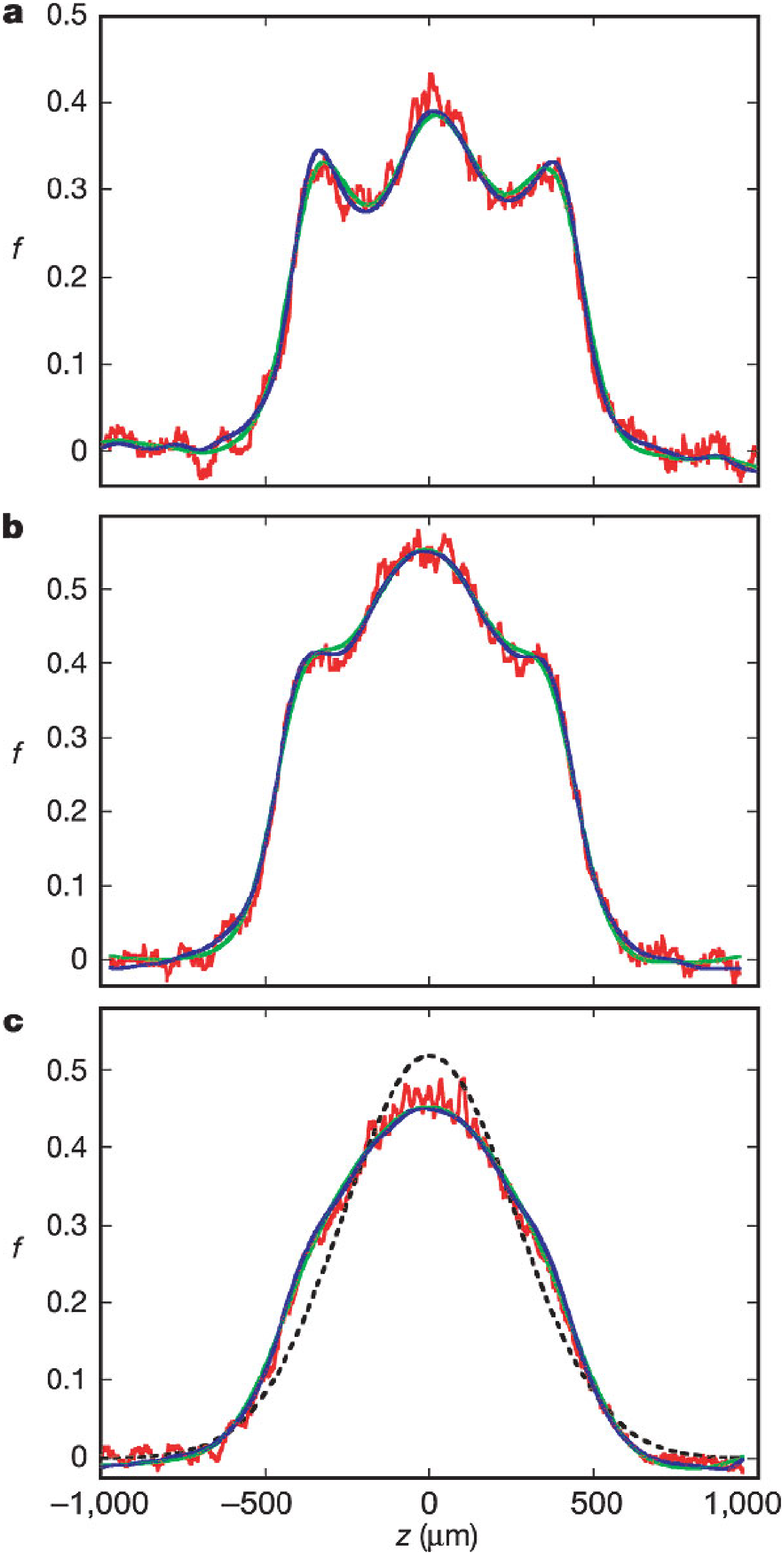}}}%
\caption{ 
Left panel: The Newton cradle experiment in Ref. \cite{NewtonCradle}. An atomic cloud of effectively
one-dimensional bosons confined in a harmonic trap is initially split into two 
clouds with opposite momenta. Then the two halves keep oscillating in the
trap and pass through each other without thermalisation of momentum distribution,
see the right panel. 
(Figure from Ref. \cite{NewtonCradle}) \\
Right panel:
The red curves in panels a,b,c are the actual momentum distributions measured for different
interaction strengths of one-dimensional hard-core bosons in Ref. \cite{NewtonCradle}.
The dashed line in panel c is the best Gaussian fit to the actual distribution. 
To the extent the actual distribution does not conform to a Gaussian, the atoms
have not thermalised. 
(Figure from Ref. \cite{NewtonCradle})
}
\end{minipage}
\end{center}
\label{FigNewtonCradle}
\end{figure}

By analogy to classical statistical physics, one can hypothesise that the steady state $\rho_\infty$ is 
the state of maximal entropy subject to the constraints imposed by integrals of motion $I_m$, where the 
integrals $I_m$ commute with $H$ and between themselves. The entropy is maximised by a generalised Gibbs 
ensemble (GGE)
\be
\rho^{\rm GGE}~=~{\cal N}~\exp\left(-\sum_m\lambda_m I_m\right)~,
\ee
where the numbers $\lambda_m$ are fixed by the conserved expectation values
${\rm Tr}\rho^{\rm GGE}I_m=\langle\psi(0)|I_m|\psi(0)\rangle$. The 
special case of only one integral $I_1=H$ is the canonical ensemble. However, unlike a classical 
system, any quantum system has as many integrals of motion as the dimension of its Hilbert 
space. Indeed, we can always choose the integrals to be projectors 
$I_\alpha=|\alpha\rangle\langle\alpha|$ on the eigenstates of $H$, or as integer powers 
of a finite Hamiltonian $I_m=H^m$ with $m=1,...,{\rm dim}(H)$. With the former choice the GGE becomes 
the microscopic diagonal ensemble $\overline{\rho}$, which may be too accurate to be tractable. The 
latter choice is also equivalent to the diagonal ensemble $\overline{\rho}$, but it may be a better 
starting point for further coarse-graining approximations, at least in the model considered in
Section \ref{ManmanaGGE} below, where the sum over $H^m$ can be accurately truncated to
a few lowest powers. Thus, unlike in a classical system, the problem with GGE is not how 
to find any integrals of motion, but how to find a subset of the abundant integrals 
of motion leading to a sufficiently accurate but still tractable GGE. Nevertheless, there
is a wide class of quadratic bosonic/fermionic Hamiltonians, where the number operators $n_\alpha$ 
of bosonic/fermionic quasiparticles provide such a small subset, see Sections \ref{GGE} 
and \ref{DephasingGGE} below, where these non-interacting systems are shown to relax to GGE
for local observables $O$. The absence of relaxation to a thermal state was experimentally observed 
in the seminal experiments on one-dimensional hard-core bosons \cite{Kinoshita,NewtonCradle}, see 
Fig. \ref{FigNewtonCradle}.

Thus the integrable quadratic Hamiltonians relax locally to a GGE, but what happens with an 
interacting system that cannot be mapped to any non-interacting Hamiltonian? It turns out that 
in certain circumstances a non-integrable system may relax to a microcanonical ensemble. 
This scenario requires two conditions to be met: the {\it eigenstate thermalisation hypothesis} 
(ETH) proposed in Ref. \cite{RefETH} and a narrow energy distribution $p_\alpha$ of the initial 
state in the eigenbasis of the final $H$. The narrow distribution often happens to be the case, 
see the examples in Sections \ref{ManmanaGGE} and \ref{ETH}. The ETH asserts that 
any few-body observable $O$ has the same expectation value in each eigenstate 
$|\alpha\rangle$ of a many-body Hamiltonian $H$ in a narrow energy window $\Delta E$. This 
not-quite-intuitive hypothesis is shown to be the case in the examples of non-integrable systems 
considered in Section \ref{ETH} while it is not true for the integrable models considered there. 
When we assume these two conditions, then for any few-body observable the diagonal ensemble (\ref{overlinerho}) 
is indistinguishable from a microcanonical ensemble
\be
\rho_\infty ~=~ 
{\cal N}~\sum_{\alpha,~|E_\alpha-\langle H\rangle|<\Delta E} 
|\alpha\rangle\langle\alpha|~
\label{micro}
\ee   
where $\Delta E$ is a narrow energy window around average energy $\langle H\rangle$. 
A coarse-grained few-body observable has the same expectation value in $\overline\rho$
as in $\rho_\infty$ because, thanks to ETH, it actually has the same expectation value
in any eigenstate in the energy window. In fact, assuming that ETH is exactly true, we 
can push it to its logical limit and claim that for any few-body observable the state
\be
\rho_\infty~=~|\alpha\rangle\langle\alpha|~,
\ee  
with $|\alpha\rangle$ being any individual eigenstate in the energy window $\Delta E$,
is as good an approximation to the actual diagonal ensemble $\overline\rho$ as the 
microcanonical ensemble. Indeed, since all expectation values $\langle\alpha|O|\alpha\rangle$
in the window are the same, there is no need to average over the window as in the
microcanonical ensemble (\ref{micro}). Thus each individual many-body eigenstate in the 
window provides a thermal (microcanonical) average for a few-body observable $O$. This 
thermal character of the individual eigenstates is hidden by their initial coherent 
superposition $|\psi(0)\rangle$, but it is revealed after the initial phase coherence 
is destroyed by relaxation or, more precisely, it appears to be destroyed for the simple 
(coarse-grained) few-body observables. 

The ETH is argued to hold in non-integrable quantum systems, while the GGE often is
an accurate description of the relaxed state in integrable quadratic systems, see Sections \ref{GGE}, 
\ref{DephasingGGE}, \ref{IsingGGE}, \ref{LuttingerGGE}, and \ref{HCBGGE} below. 
This is similar to classical physics, where integrable systems constrained by their
conservation laws do not thermalise, but non-integrable chaotic systems do thermalise \cite{FPU}. 
However, this mechanism of quantum thermalisation is qualitatively different than that of classical thermalisation. 
According to ETH, each many-body eigenstate is a thermal ensemble, while the ergodicity means 
that classical thermalisation requires probing essentially all states on a manifold of 
definite energy. A classical integrable system becomes chaotic when a non-integrable 
perturbation is stronger than certain threshold. A good example is the classic 
Fermi-Pasta-Ulam numerical experiment, where the perturbation is too weak to make the 
system chaotic \cite{FPU}. An interesting question, addressed in Section \ref{ETH}, is if
a similar threshold exists in a quantum system. Numerical simulations in small systems 
suggest that turning-off a non-integrable perturbation to an integrable model results
in a smooth departure from ETH, but it is not clear what happens in the thermodynamic limit.   

The microcanonical steady state required both ETH and a narrow initial energy distribution
$p_\alpha$. Since in an isolated system the distribution is constant, there is
no way it could relax to a canonical distribution $p_\alpha\sim\exp\left(-\beta E_\alpha\right)$ 
unless it happened to be canonical from the very beginning. Thus there is no global 
thermalisation, but one can still try to identify local or coarse-grained observables $O$ for 
which the global state appears to relax to the canonical ensemble. For instance, in a system of 
many identical particles one is often interested in a single particle momentum distribution $n_{\vec k}$.
This interest is well motivated by current ultracold atoms experiments, where it is routine to measure 
the momentum distribution in an atomic cloud expanding after opening a trap, see the example in the right 
panel of Fig. \ref{FigNewtonCradle}. Thermalisation of $n_{\vec k}$ is inhibited by integrability of the system 
like e.g. for quadratic Hamiltonians which relax locally to a non-thermal GGE. It can also be  
inhibited for particles interacting by two-body interactions in one dimension, where the conservation of 
momentum and kinetic energy in a collision of two particles does not allow for any changes in the momentum 
distribution. A spectacular recent quantum Newton cradle experiment \cite{NewtonCradle} demonstrates this 
inhibited relaxation in a quasi-one-dimensional regime, see Fig. \ref{FigNewtonCradle}. 
Integrability is also responsible for such phenomena as the ballistic character of transport \cite{Zotos} 
or the meta-stability of solitons \cite{Kamenev}.

Another interesting question, in addition to the nature of the final relaxed state, is 
the dynamics of relaxation to this state. Since the ground state of an initial 
$H_0$ may contain highly excited eigenstates of a final Hamiltonian $H$, the answer
to this question is not universal, because the usual universal low energy theories 
may not describe the relaxation accurately. This non-universality is illustrated
by examples in Section \ref{DynamicsRelaxation}, where it is shown that
the relaxation can be actually the fastest when $H$ is either at a critical point
or close to it. This is contrary to the usual notion of critical slowing down derived from 
a universal low energy theory. In spite of that, there are some remarkable universal 
features, like the quasiparticle horizon effect described in Section \ref{Horizon}.

Section \ref{Relaxation} is organised as follows. In Section \ref{Horizon} we review
the quasiparticle horizon effect. In Section \ref{GGE} the generalised Gibbs ensemble is 
introduced for quadratic Hamiltonians, and in Section \ref{DephasingGGE} it is shown
how a quadratic Hamiltonian prepared in an initial Gaussian pure state relaxes locally to GGE. 
The following Sections \ref{IsingGGE}, \ref{LuttingerGGE}, and \ref{HCBGGE} describe examples of quadratic systems 
relaxing to GGE, and discuss applicability of the GGE to different observables.    
In Section \ref{EisertGGE} we consider the non-integrable Bose-Hubbard model
in different limits when it is close to integrability and GGE is an accurate steady
state, but we also cite some evidence that it thermalises away from these integrable limits.
Section \ref{ManmanaGGE} provides an example of a system integrable by Bethe ansatz, where 
an accurate non-thermal GGE can be constructed out of integer powers of the Hamiltonian. In 
Section \ref{ETH} we review the microcanical ensemble in connection with the eigenstate 
thermalisation hypothesis (ETH). Finally, in Section \ref{DynamicsRelaxation} 
we mention some aspects of the dynamics of relaxation. We conclude in 
Section \ref{Summary2}.

\subsection{ Quasiparticle light cone effect in dephasing after a sudden quench }
\label{Horizon}

At $t=0$ a system is prepared in the ground state $|\psi_0\rangle$ of a Hamiltonian 
$H_0$ and then, after a sudden quench, it evolves unitarily with a different 
Hamiltonian $H$ at $t>0$. The initial $H_0$ is non-critical and it has a finite correlation
length $\xi_0$ and a finite gap $\Delta_0$. The sudden quench is faster than $\Delta_0^{-1}$.
The question is how do the correlations evolve in time? As shown in Ref. 
\cite{CalabreseCardyPRL}, the answer to this question is quite universal when $H$ is at 
a critical point. Particularly powerful analytic results can be obtained in one dimension when
the dynamical exponent is $z=1$ or, equivalently, there is a linear quasiparticle 
dispersion relation $\epsilon_k=v|k|$ with quasiparticle velocity $v$. This $1+1$
dimensional problem can be described asymptotically by a boundary conformal
field theory \cite{BCFT}, where the initial state is a boundary condition. 

The results from the conformal field theory suggest a simple physical picture.
The initial state $|\psi_0\rangle$ is a highly excited state as compared to the
ground state of the final Hamiltonian $H$. It is a source of quasiparticle
excitations. Quasiparticles originating from closely separated points, within
the correlation length $\xi_0$ in the ground state of $H_0$, are quantum
entangled. Once they are emitted, they behave semi-classically travelling
at speed $v$.  This simple picture has several important implications:

\begin{itemize}

\item Incoherent quasiparticles arriving at a given point from well separated
sources cause relaxation of most local observables at this point to their
expectation values in the ground state of $H$. There are exceptions like the local
energy density which is conserved. This relaxation is exponential 
$\sim\exp\left(-\pi xvt/2\tau_0\right)$, where $\tau_0\sim\Delta_0^{-1}$
is not universal, but $x$ is the bulk scaling dimension of a given observable.

\item Entangled quasiparticles arriving at the same time $t$ at points with
separation $R\gg\xi_0$ induce correlations between local observables at these
points. Since they travel at a definite speed $v$, there is a sharp light-cone
effect, i.e., the connected correlation functions do not change significantly
from their initial values until time $t\simeq R/2v$. The light-cone effect
is rounded off over the region $t-R/2v\simeq\tau_0$ because quasiparticles
remain entangled over this time scale. After $t\simeq R/2v$ the connected
correlations rapidly relax to time-independent values. At asymptotically
large separations $R$, but well within the light cone where $R\ll 2vt$,
they decay exponentially $\sim\exp\left(-\pi xR/2v\tau_0\right)$. This decay
is qualitatively different from the power law decay in the ground state of $H$ 
which is at the critical point. 

\item Entangled quasiparticles arriving at the same time $t\simeq R/2v$ at 
points separated by $R\gg\xi_0$ induce entanglement between these points. 
The entropy of a subsystem of size $L$, which is dominated by pairs of 
quasiparticles entangled across the boundary of the subsystem, initially
grows linearly with time as more and more quasiparticles cross the subsystem
boundary until it saturates at $t\simeq L/v$. In the thermodynamic limit,
in a system with periodic boundary conditions the saturation time is 
$t\simeq L/2v$, and in an open system the entropy saturates at $t\simeq L/v$. 
In a conformal field theory the saturation time is rounded off on the time 
scale $\Delta_0^{-1}$.

\item In the thermodynamic limit, the saturated entropy of the subsystem of 
size $L$ is linear in $L$ because the number of quasiparticles in the subsystem
is linear in $L$ and, after the saturation at $t={\cal O}(L/v)$, every quasiparticle 
which happens to be in the subsystem is more likely to be entangled with 
a quasiparticle outside than a quasiparticle inside the subsystem. 
Thus, there is no area law for the entropy, so the relaxation process cannot
be efficiently simulated by, say, the DMRG algorithm \cite{tDMRG} on a classical 
computer.

\end{itemize}

The light cone effect is even more rounded off when the critical $H$ is a lattice system. 
The saturation begins at $t_0\simeq R/2v_0$, where 
$v_0=\frac{\partial\epsilon_k}{\partial k}(k=0)$ is a long wavelength quasiparticle 
group velocity, but after $t_0$ the entropy (or correlation function) keeps varying
slowly with time because on a lattice there are quasiparticles slower than $v_0$.  

Beyond the conformal field theory, the light cone effect was confirmed in the 
quantum Ising model \cite{CalabreseCardyPRL}, the chain of coupled harmonic oscillators  
\cite{CalabreseCardyPRL}, the Heisenberg chain \cite{CalabreseFazio}, and
a number of other systems, see Ref. \cite{HorizonOther} and the following 
Sections. It was also confirmed away from criticality, where $v_0$ becomes the maximal group 
velocity of quasiparticles, see e.g.
Ref. \cite{MatheyPolkovnikov}. In general, a maximal velocity
of propagation of information in non-relativistic systems is guaranteed
by the Lieb-Robinson theorem \cite{LiebRobinson}.

By contrast, in a disordered Heisenberg chain the
entropy seems to grow logarithmically with time, at least as far as it can be inferred 
from numerical results \cite{CalabreseFazio}. This effect cannot be explained by diffusion of
quasiparticles replacing the ballistic motion in the pure case.

The light cone effect is the dominant feature of quantum relaxation after 
a sudden quench. The question what is the final relaxed state is the topic of the 
following Sections.

\subsection{ Generalised Gibbs ensemble (GGE) for a quadratic Hamiltonian }
\label{GGE}

Integrals of motion $I_\alpha$ mutually commute, $[I_\alpha,I_\beta]=0$, and commute with the 
Hamiltonian, $[I_\alpha,H]=0$. In a unitary evolution of an isolated quantum system,
$|\psi(t)\rangle=e^{-itH}|\psi(0)\rangle$, all moments of the integrals of motion are conserved,
\be
\langle\psi(t)| I_1^{r_1}I_2^{r_2} \dots |\psi(t)\rangle ~=~
\langle\psi(0)| I_1^{r_1}I_2^{r_2} \dots |\psi(0)\rangle ~,
\ee
because each product $I_1^{r_1}I_2^{r_2}\dots$ is also an integral of motion. 

When degeneracies in the spectrum of the Hamiltonian can be ignored, then 
the infinite time average of the density matrix of the system 
$\rho(t)=|\psi(t)\rangle\langle\psi(t)|$ is diagonal in the eigenbasis 
of the integrals of motion,
\be
\overline{\rho}~\equiv~
\lim_{T\to\infty} \frac{1}{T}\int^T_0 dt~\rho(t)~=~
\sum_{n_1,n_2,\dots}
p_{n_1,n_2,\dots}
|n_1,n_2,\dots\rangle\langle n_1,n_2,\dots|~,
\label{overlinerhodiag}
\ee
where $I_{\alpha}|n_1,n_2,\dots\rangle=n_\alpha|n_1,n_2,\dots\rangle$.
The diagonal $\overline{\rho}$ conserves all moments $I_1^{r_1}I_2^{r_2}\dots$,
\bea
{\rm Tr}~\overline{\rho}~I_1^{r_1}I_2^{r_2}\dots &=&
\lim_{T\to\infty} \frac{1}{T}\int^T_0 dt~
{\rm Tr}~\rho(t)~I_1^{r_1}I_2^{r_2}\dots                  \nonumber\\
&=&
\lim_{T\to\infty} \frac{1}{T}\int^T_0 dt~
\langle\psi(t)|I_1^{r_1}I_2^{r_2}\dots|\psi(t)\rangle \nonumber\\
&=&
\langle\psi(0)|I_1^{r_1}I_2^{r_2}\dots|\psi(0)\rangle~,
\eea
as expected from a candidate for a stationary state of an integrable system.

The diagonal average $\overline{\rho}$ can be rewritten in an equivalent form
\be
\overline{\rho}~=~
{\cal N}~
\exp
\left[
-\sum_{\alpha} \lambda_\alpha I_\alpha
-\sum_{\alpha\leq\beta} \lambda_{\alpha\beta} I_\alpha I_\beta
-\sum_{\alpha\leq\beta\leq\gamma} \lambda_{\alpha\beta\gamma} I_\alpha I_\beta I_\gamma -
\dots
\right]~,
\label{MGGE}
\ee
where the $\lambda$'s and the normalisation ${\cal N}$ are chosen so that 
${\rm Tr}~\overline{\rho}~I_1^{r_1}I_2^{r_2}\dots=
\langle\psi(0)|I_1^{r_1}I_2^{r_2}\dots|\psi(0)\rangle$ and 
${\rm Tr}\overline{\rho}=1$. The density matrix (\ref{MGGE}) has a form
of a most general Gibbs ensemble and as such it is manifestly the
density matrix that maximises the von Neumann entropy subject to the
constraints imposed by all the conserved moments 
$\langle\psi(0)|I_1^{r_1}I_2^{r_2}\dots|\psi(0)\rangle$. However, 
this exact representation of $\overline{\rho}$ may be not a useful coarse-grained 
{\it statistical} description because it contains as much information as the 
diagonal ensemble (\ref{overlinerhodiag}) itself.

Nevertheless, the exact form (\ref{MGGE}) is a good starting point for more tractable approximations. The 
crudest but still non-trivial approximation is known as a {\it generalised Gibbs ensemble} (GGE) \cite{HCB} 
\be
\overline{\rho}~\approx~
{\cal N}~
\exp
\left[
-\sum_{\alpha} \lambda_\alpha I_\alpha
\right]~\equiv~
\rho_{\rm GGE}~,
\label{rhoGGE}
\ee
where the $\lambda$'s are fixed by conservation of the first moments only:
${\rm Tr}~\overline{\rho}~I_\alpha~=~\langle\psi(0)|I_\alpha|\psi(0)\rangle$.
The mixed state $\rho_{\rm GGE}$ is the maximal entropy state subject to these 
constraints. The drastic approximation has a price of course: there are observables 
$O$ for which ${\rm Tr}~\rho_{\rm GGE}~O$ is a very bad approximation to the exact
${\rm Tr}~\overline{\rho}~O$. However, this is expected in statistical physics, where 
a sufficiently complex observable $O$ can reveal a difference between a coarse 
grained state like $\rho_{\rm GGE}$ and a microscopic state like $\overline{\rho}$,
but there is practically no difference for sufficiently coarse grained observables.
In the following we attempt to figure out what are the conditions for an observable
to be ``sufficiently coarse grained'' in case of the generalised Gibbs ensemble.

To be more specific, we assume that the integrals are numbers of bosonic or fermionic quasiparticles \cite{KollarEckstein,Barthel}, 
$I_\alpha=n_\alpha=\gamma^\dag_\alpha\gamma_\alpha$. Since $I_\alpha$'s mutually commute we can factorise (\ref{rhoGGE})
\be
\rho_{\rm GGE}~=~
\left(
\prod_\alpha
{\cal N}_\alpha
\exp\left[-\lambda_\alpha n_\alpha\right]
\right)~
| n_1,n_2,\dots \rangle\langle n_1,n_2,\dots |~.
\ee 
This diagonal matrix is a classical ensemble for the set of {\it independent}
random variables $n_\alpha$ with a product joint probability distribution
\be
p_{n_1,n_2,\dots}^{\rm GGE}~=~
\prod_\alpha
{\cal N}_\alpha
\exp\left[-\lambda_\alpha n_\alpha\right]~.
\ee 
In the GGE the occupation numbers $n_\alpha$ are independent and, in particular, uncorrelated. For an observable $O$ to have an 
accurate expectation value in GGE the observable cannot depend on correlations between different $n_\alpha$'s. 

In case of fermions, when $n_\alpha$ is either $0$ or $1$, this is also a sufficient 
condition because the expectation value $\langle n_\alpha \rangle$ plus normalisation
determine completely the probability distribution for $n_\alpha$. The distribution
can be written in e.g. the GGE form 
${\cal N}_\alpha\exp\left[-\lambda_\alpha n_\alpha\right]$ with the coefficients
${\cal N}_\alpha$ and $\lambda_\alpha$ determined by the two constraints.

In case of bosons, when $n_\alpha=0,1,2,\dots$, the expectation value 
$\langle n_\alpha \rangle$ and normalisation are not sufficient to determine the
probability distribution for $n_\alpha$. Thus the GGE ansatz   
$
{\cal N}_\alpha\exp\left[-\lambda_\alpha n_\alpha\right]~
$
may have wrong variance and higher moments even though it has the correct expectation 
value.

Thus we can conclude, that GGE is an accurate statistical description for observables
which do not depend on correlations between quasiparticle occupation numbers and, in 
case of bosons, higher moments of individual occupation numbers. This is the 
meaning of the mathematical conditions given in Ref. \cite{KollarEckstein}.

The GGE is an accurate statistical description of $\overline{\rho}$ for observables $O$
that do not depend on correlations between $n_\alpha$. Some correlations between
$n_\alpha$'s are allowed provided that their contribution is negligible in the thermodynamic 
limit. A good example are 1D hard core bosons which can be represented equivalently by 
non-interacting fermions. When perturbed by an alternating potential, as in Ref. \cite{HCB},  
the initial state contains correlations between pairs of occupation numbers $n_k$ and 
$n_{k+\pi}$ only. These sparse correlations make negligible contribution to {\it local} observables 
restricted to a finite region of real space. Another example is the Luttinger model equivalent
to non-interacting bosons, see Ref. \cite{Cazalilla} and Section \ref{LuttingerGGE}. A quench
of interaction strength in this translationally invariant Hamiltonian prepares an initial 
state with correlations between pairs of bosonic occupation numbers $n_k$ and $n_{-k}$ only.
Again, in the thermodynamic limit these sparse correlations have negligible contribution to 
local observables.

Moreover, with an additional assumption that the initial pure state is Gaussian and the 
Hamiltonian is quadratic in bosonic/fermionic annihilation operators, GGE can be shown
to be the unique steady state for local observables. This and the dynamics of local relaxation
to this stationary state is the subject of the next Section.

\subsection{ Local relaxation to GGE after a quench in a quadratic Hamiltonian }
\label{DephasingGGE}

In the last Section we discussed for what observables the time-averaged diagonal density matrix 
$\overline{\rho}$ in Eq. (\ref{overlinerhodiag}) can be accurately represented by the generalised 
Gibbs ensemble (GGE) in Eq. (\ref{rhoGGE}). Our interest in the time-averaged density 
matrix was motivated by the fact that {\it if} the long time limit of an expectation value 
${\rm Tr}~\rho(t)~O$ exists, then it is equal to the expectation value in the 
time-averaged density matrix,
\be
\lim_{t\to\infty} {\rm Tr}~\rho(t)~O ~=~ {\rm Tr}~\overline{\rho}~O~. 
\label{ttoinfty}
\ee
However, in an isolated quantum system a limit $\rho(t\to\infty)$ does not exist
\footnote{
Except for the trivial case of a time-independent $\rho$ diagonal in the eigenbasis of the Hamiltonian. 
}
, so the existence of the limit on the left hand side of (\ref{ttoinfty}) should not be taken for granted for an arbitrary 
observable $O$. This is why in this Section we investigate a more refined question if we can define a subset 
of observables $O$ for which 
\be
\lim_{t\to\infty} {\rm Tr}~\rho(t)~O ~=~ {\rm Tr}~\rho^{\rm GGE}~O~?
\ee
In other words, we are looking for a subset of observables for which a pure state $\rho(t)$ appears to 
relax to a mixed steady state $\rho^{\rm GGE}$.

Following Ref. \cite{Barthel}, we consider a general quadratic lattice Hamiltonian
\be
H~=~
\sum_{mn}
\left[
c_m^\dag V_{mn} c_n +
\frac12
\left(
c_m^\dag W_{mn} c_n^\dag+{\rm h.c.}
\right)
\right]
\label{Hquadratic}
\ee
where $c_m$ are bosonic or fermionic annihilation operators. The Hamiltonian
is diagonalised to $H=\sum_k\omega_k\gamma_k^\dag\gamma_k\equiv\sum_k\omega_kI_k$ 
by a Bogoliubov transformation
\be
c_m~=~\sum_k~ \left(u_{mk}\gamma_k~+~v_{m,k}^*\gamma_k^\dag\right) ~,
\label{cm}
\ee
where the index $k$ enumerates quasiparticle states. The quadratic Hamiltonian (\ref{Hquadratic}) 
has a Gaussian Bogoliubov vacuum ground state annihilated by all quasiparticle annihilation 
operators $\gamma_k$. Any Gaussian state $\rho$ is fully determined by its quadratic correlators
\bea
\alpha_{mn} ~=~ {\rm Tr}~\rho~c_mc_n^\dag~,~~
\beta_{mn}  ~=~ {\rm Tr}~\rho~c_mc_n~.
\label{alphabeta}
\eea
Here we assume that the initial pure state $\rho(0)$ is Gaussian. It can be prepared as a ground state
of a different quadratic Hamiltonian $H_0$ before the Hamiltonian is suddenly quenched to the
final $H$. The quadratic $H$ evolves $\rho(0)$ into a pure state $\rho(t)$ which is also Gaussian.

The lattice can be divided into a finite subsystem $\Omega$ and its environment $\Omega^\perp$. We are 
interested in {\it local} observables $O$ with a finite support in $\Omega$. To find expectation values 
of the local observables it is enough to know a reduced density matrix 
\be
\rho_{\Omega}(t)~=~{\rm Tr}_{\Omega^\perp}~\rho(t)~.
\ee
We want to show that, under certain conditions, the reduced density matrix tends to a steady state,
\be
\lim_{t\to\infty}\rho_{\Omega}(t)~=~
{\rm Tr}_{\Omega^\perp}~\rho^{\rm GGE}~, 
\ee
obtained by reduction of a {\it Gaussian} GGE
\be
\rho^{\rm GGE}~=~
{\cal N}
\exp\left[-\sum_k\lambda_k \gamma_k^\dag\gamma_k\right]~.
\label{GGEgammak}
\ee
Here $\lambda$'s are fixed by initial conditions
${\rm Tr}\rho^{\rm GGE}\gamma_k^\dag\gamma_k={\rm Tr}\rho(0)\gamma_k^\dag\gamma_k~\equiv~n_k$ for the 
integrals of motion $I_k=\gamma_k^\dag\gamma_k$. If the limit exists, then for all {\it local} observables 
$O$ in the finite $\Omega$ the state appears to relax to GGE. This is {\it apparent local relaxation} 
which should not be mistaken with any global relaxation of the pure state $\rho(t)$.

When $\rho$ is a Gaussian state fully characterized by the correlators (\ref{alphabeta}), 
then the reduced density matrix $\rho_\Omega$ is also a Gaussian state fully characterised 
by a subset of the same correlators (\ref{alphabeta}) with indices $m,n$ restricted to the 
subsystem $\Omega$. Thus all that we need to show for a Gaussian initial state is that 
\bea
\lim_{t\to\infty} \alpha_{mn}(t) ~=~
\alpha_{mn}^{\rm GGE}~,~~
\lim_{t\to\infty} \beta_{mn}(t)  ~=~
\beta_{mn}^{\rm GGE}~,~~
{\rm for}~~
m,n\in\Omega~.
\label{limitGGE}
\eea
Here
\bea
\alpha_{mn}(t) &=&
\sum_{k,k'}
{\rm Tr}~\rho(t)~
\left( u_{mk}\gamma_k~+~v_{m,-k}^*\gamma_{-k}^\dag         \right) 
\left( u^*_{nk'}\gamma_{k'}^\dag ~+~ v_{n,-k'}\gamma_{-k'} \right)~,
\label{alphat}\\ 
\beta_{mn}(t) &=&
\sum_{k,k'}
{\rm Tr}~\rho(t)~
\left( u_{mk}\gamma_k~+~v_{m,-k}^*\gamma_{-k}^\dag         \right) 
\left( u_{nk'}\gamma_{k'}~+~v_{n,-k'}^*\gamma_{-k'}^\dag   \right)~, 
\label{betat}
\eea
while the correlators in the GGE are ``diagonal'' in $k$
\bea
\alpha_{mn}^{\rm GGE} &=&
\sum_k\left[ u_{mk}u^*_{nk}(1\pm n_k) + v^*_{mk}v_{nk} n_k \right]~,\\
\beta_{mn}^{\rm GGE}  &=&
\sum_k\left[ u_{mk}v^*_{nk}(1\pm n_k) + v^*_{mk}u_{nk} n_k \right]~,
\label{alphabetaGGE}
\eea
with the upper/lower sign for bosons/fermions. They follow from diagonal expectation
values in GGE: 
\bea
&& {\rm Tr}\rho^{\rm GGE} \gamma_k\gamma_{k'}=0~, \\
&& {\rm Tr}\rho^{\rm GGE} \gamma_k\gamma_{k'}^\dag=\delta_{k,k'}(1\pm n_k)~.
\label{CorrDiag}
\eea
Equations (\ref{limitGGE}) are satisfied if a contribution of quasiparticle correlators
\bea
{\rm Tr}~\rho(t)~\gamma_k\gamma_{k'} 
&=& 
e^{-it(\omega_k+\omega_{k'})}~
{\rm Tr}~\rho(0)~\gamma_k\gamma_{k'}~, 
\label{gammagamma}\\
{\rm Tr}~\rho(t)~\gamma_k\gamma_{k'}^\dag 
&=&
e^{-it(\omega_k-\omega_{k'})}~
{\rm Tr}~\rho(0)~\gamma_k\gamma_{k'}^\dag ~~{\rm for}~k\neq k'~
\label{gammagammadag}
\eea
to correlators (\ref{alphat},\ref{betat}) vanishes when $t\to\infty$. 

This seems to be quite generic. Indeed, given that the indices $m,n$ are bounded to a finite $\Omega$, we 
need to consider a finite number of Bogoliubov coefficients $u_{mk},v_{mk}$ enumerated by $m\in\Omega$. 
In the thermodynamic limit of infinite lattice, when $t\to\infty$ then the phase factor 
$e^{-it(\omega_k+\omega_{k'})}$ in (\ref{gammagamma}) oscillates with $k$ and $k'$ fast enough 
to average out to zero the contribution of this term to the infinite sum over $k,k'$. The same 
is generically true for Eq. (\ref{gammagammadag}) when $k\neq k'$. More rigorous mathematical 
conditions, and some important exceptions, can be found in Ref. \cite{Barthel}.

After the dephasing in a finite subsystem $\Omega$ is completed, then for local observables in $\Omega$  
the quadratic quasiparticle correlators appear to be equal to quadratic correlators (\ref{CorrDiag}) 
in GGE (\ref{GGEgammak}). Since GGE is Gaussian and quadratic correlators determine a Gaussian state 
uniquely, then GGE is the unique (apparent) stationary state for local observables in $\Omega$.

Returning to the issue of relaxation, we can refine the above argument in a translationally invariant 
case, when $V_{mn}=V_{m-n}$ and $W_{mn}=W_{m-n}$ in the quadratic Hamiltonian (\ref{Hquadratic}) and 
the Bogoliubov modes in Eq. (\ref{cm}) become $u_{mk}=e^{ikm}u_k$ and $v_{mk}=e^{ikm}v_k$ with definite 
quasimomenta $k$ (we suppress vector notation in more than one dimension). A quench from a different translationally 
invariant quadratic Hamiltonian $H_0$ prepares an initial state $\rho(0)$ such that 
\be
{\rm Tr}~\rho(0)~\gamma_k\gamma_{k'} ~=~ \delta_{-k,k'}~\Delta_k ~,~~
{\rm Tr}~\rho(0)~\gamma_k\gamma_{k'}^\dag ~=~ \delta_{k,k'}~(1\pm n_k)~.
\ee
After simple algebra using the symmetry $\omega_k=\omega_{-k}$ we obtain
\bea
\alpha_{mn}(t) &=&
\alpha_{mn}^{\rm GGE} ~+~
\int_{-\pi}^\pi \frac{dk}{2\pi}
e^{ik(m-n)}
\left(
e^{-2it\omega_k}
u_kv_{-k}\Delta_k+
{\rm c.c.}
\right)~,\\
\beta_{mn}(t) &=&
\beta_{mn}^{\rm GGE} ~+~
\int_{-\pi}^\pi \frac{dk}{2\pi}
e^{ik(m-n)}
\left(
e^{-2it\omega_k}
u_ku_{-k}\Delta_k+
e^{2it\omega_k}
v_k^*v_{-k}^*
\Delta_k^*
\right)~.
\eea
In a small neighbourhood $(k_0-\delta k,k_0+\delta k)$ of a generic $k_0$ we can linearise in $(k-k_0)$ as
$\omega_k\approx\omega_{k_0}+v_{k_0}(k-k_0)$, where $v_{k_0}=\frac{d\omega_k}{dk}(k_0)$ is a group velocity
\footnote{
When $v_{k_0}=0$, then we can expand $\epsilon_k=\epsilon_{k_0}+A(k-k_0)^2$ 
and, when $u_k,v_k,n_k$ are not singular, the integrals decay like 
${\cal O}(\delta k/\sqrt{t})$ for $t\to\infty$. 
}
. Contributions to the above integrals from this neighbourhood are proportional to
\be
\int_{-\delta k}^{\delta k} \frac{d(k-k_0)}{2\pi}
e^{i(k-k_0)[m-n\pm 2v_{k_0}t]}~,
\ee
provided that $\delta k$ is much less than the shortest scale on which the functions $u_k,v_k,n_k$ 
can vary in $k$. If $u_k,v_k,n_k$ are not singular
\footnote{
The $u_k,v_k,n_k$ cannot be singular for fermions, when they are bounded by the constraints 
$|u_k|^2+|v_k|^2=1$ and $0\leq n_k\leq1$, but they can in principle be singular for bosons because 
the bosonic constraints $|u_k|^2-|v_k|^2=1$ and $0\leq n_k<\infty$ do not provide any upper bounds.  
}
, then for $t\to\infty$ this integral decays like ${\cal O}(\delta k/t)$ except when 
\be
|m-n|~\approx~2~v_{k_0}t~.
\label{quadraticdephasing}
\ee
The distance $2v_{k_0}t$ is a separation of a pair of quasiparticles with opposite momenta $(k_0,-k_0)$ 
created by the quench, see the quasiparticle horizon effect in Section \ref{Horizon}. When $t$ is long 
enough this separation becomes much longer than the subsystem $\Omega$ and the exception (\ref{quadraticdephasing}) 
does not apply to any $m,n\in\Omega$. This is another example of the quasiparticle horizon effect in 
Section \ref{Horizon}.

Thus the relaxation in Eqs. (\ref{limitGGE}) seems to be generic and the quadratic correlators tend to the
quadratic correlators in the GGE (\ref{GGEgammak}). Since both $\rho_\Omega(t\to\infty)$ and 
${\rm Tr}_{\Omega^\perp}\rho^{\rm GGE}$ are Gaussian, the equality of their quadratic correlators implies that 
both states are the same and, in particular, {\it all} local observables $O\in\Omega$ (and not only quadratic 
correlators) have the same expectation values in the two states. The GGE (\ref{GGEgammak}) is determined uniquely 
by the quadratic correlators plus the requirement that GGE is Gaussian. 

The Gaussian GGE (\ref{GGEgammak}) is determined by the initial quasiparticle occupation numbers $n_k$, so
it is not surprising that it correctly ``predicts'' the conserved $n_k$. The free lunch is that after the local 
relaxation GGE also correctly describes all local observables $O\in\Omega$. Thus we need to invest all $n_k$ to 
predict all local observables. This is lesser return than in the Gaussian canonical ensemble
\be
\rho^{\rm canonical}~=~
{\cal N}
\exp\left[-\beta H\right]~=~
{\cal N}
\exp\left[-\beta\sum_k\omega_k \gamma_k^\dag\gamma_k\right]~,
\label{GGEcanonical}
\ee
which is a special case of GGE (\ref{GGEgammak}) with $\lambda_k=\beta\omega_k$ and only one
adjustable parameter $\beta$ that we need to fix. However, this is the price for the integrability
of the quadratic Hamiltonian. In a sense, in the GGE (\ref{GGEgammak}) each non-interacting quasiparticle 
has its own inverse temperature $\beta_k=\lambda_k/\omega_k$.

We can conclude that in general a quadratic Hamiltonian initially prepared in a pure Gaussian state
relaxes locally to GGE. More precisely, expectation values of observables within a quasiparticle horizon
relax to their expectation values in GGE.

\subsection{ GGE and the transverse quantum Ising chain }
\label{IsingGGE}

The argument in Section \ref{DephasingGGE} applies to the transverse field 
quantum Ising chain (\ref{Hsigma}). A Jordan-Wigner transformation
maps the Hamiltonian (\ref{Hsigma}) to a translationally
invariant quadratic Hamiltonian (\ref{Hpm}). A sudden quench from an initial transverse field $g_i$ to a final $g_f>0$ 
prepares an initial Gaussian state which later relaxes {\it locally} to the GGE
(\ref{GGEgammak}), where $\gamma_k$ is an annihilation operator for a Bogoliubov quasiparticle 
at the final $g_f$, see Eqs. (\ref{Bog},\ref{stBdG},\ref{epsilonkIsing},\ref{uvplus}), and 
$\lambda_k$ are such that ${\rm Tr}~\rho^{\rm GGE}~\gamma^\dag_k\gamma_k={\rm Tr}~\rho(0)~\gamma^\dag_k\gamma_k$.

When expressed in terms of fermions, the quantum Ising model relaxes locally to GGE.
However, since the Jordan-Wigner transformation (\ref{JordanWigner}) between the original 
spin operators and the fermions is non-local, there are local spin operators which are not 
local in fermionic representation, the simplest example being $\sigma^x_n$. Thus 
in general only spin observables with a finite {\it fermionic} support relax to GGE.

The dephasing to GGE (\ref{GGEgammak}) was studied in Refs. \cite{IgloiRieger,SenguptaPowellSachdev}, 
although without any explicit reference to GGE for good chronological reasons. In particular, in 
Ref. \cite{SenguptaPowellSachdev} ferromagnetic correlation functions 
$C^{xx}_R={\rm Tr}~\rho^{\rm GGE}~\sigma^x_{n+R}\sigma^x_n$ were found in the final 
dephased state in the two limiting cases of $g_i=0$ and $g_i=\infty$. Here we list their 
tails for large $R$ only: 
\begin{itemize} 
\item When $g_i=0$
\be
C_R^{xx}~=~
\left\{
\begin{array}{ll}
\left(\frac{1+\sqrt{1-g_f^2}}{2}\right)^{R+1} &,~{\rm when}~g_f<1~,\\
\left(\frac12\right)^R                        &,~{\rm when}~g_f>1~,
\end{array}
\right.
\ee

\item When $g_i=\infty$
\be
C_R^{xx}~=~
\left\{
\begin{array}{ll}
\left(\frac{1}{2}\right)^R
\cos\left[R\arccos(g_f)\right] &,~{\rm when}~g_f<1~,\\
\left(\frac{1}{2g_f}\right)^R  &,~{\rm when}~g_f>1~,
\end{array}
\right.
\ee

\end{itemize}
The oscillatory correlation function in the ferromagnetic phase, $g_f<1$,  
obtained after dephasing from a fully disordered initial state, $g_i=\infty$, 
is a clear qualitative indication of a non-thermal steady state. Notice similarity
of this oscillatory correlation function to the correlators 
(\ref{Czz},\ref{Czzinfty}) obtained after a linear quench. 

Dephasing to the GGE instead of a thermal state does not mean that some quantities, 
which are not very sensitive to fine details of GGE, cannot behave like in a thermal
state. For instance, as shown in Ref. \cite{effTIsing}, time correlations of the order 
parameter decay exponentially at the same rate as in a canonical ensemble with an 
effective temperature determined by energy density pumped to the system by the 
sudden quench.  

We can conclude that after a sudden quench in the quantum Ising chain observables
which are local in fermionic representation relax to GGE.

\subsection{ GGE and Luttinger model (LM) }
\label{LuttingerGGE}

In this Section, following Ref. \cite{Cazalilla}, we consider local relaxation to GGE 
in the integrable Luttinger model (LM) \cite{LM}. This model describes low energy 
properties of a wide class of 1D Tomonaga-Luttinger liquids \cite{TLL}. The Hamiltonian 
of the LM is 
\bea
H_{\rm LM} &=& H_0~+~H_2~+~H_4~, \\
H_0        &=& v_Fp \sum_{p,\alpha} :\psi_\alpha^\dag(p)\psi_\alpha(p): ~,\\
H_2        &=& \frac{2\pi}{L} \sum_q g_2(q)~J_R(q)J_L(q)~,\\
H_4        &=& \frac{\pi}{L}  \sum_{q,\alpha} :J_\alpha(q)J_\alpha(-q):~,   
\eea
where the Fermi operators satisfy
$\{\psi_\alpha(p),\psi_\beta^\dag(p')\}=\delta_{p,p'}\delta_{\alpha,\beta}$
with $\alpha,\beta=L,R$ and anticommute otherwise. Anti-periodic boundary
conditions $\psi_\alpha(x+L)=-\psi_\alpha(x)$, where 
$\psi_\alpha(x)=\sum_p e^{is_\alpha p}\psi_\alpha(p)/\sqrt{L}$
with $s_R=-s_L=1$, result in a non-degenerate ground state. The quantised
momentum $p=2\pi(n-1/2)/L$ with integer $n$ is ``half-integer''. The current 
operators $J_\alpha(q)=\sum_p:\psi^\dag_\alpha(p+q)\psi_\alpha(p):$, where 
$q=2\pi m/L$ with integer $m$.

The currents obey the Kac-Moody algebra 
$[J_\alpha(q),J_\beta(q')]=\frac{qL}{2\pi}\delta_{q+q',0}\delta_{\alpha\beta}$
which allows one to introduce bosonic operators
\bea
b_0(q)      &=& -i(2\pi/|q|L)^{1/2} [\theta(q)J_R(-q)-\theta(-q)J_L(q)]~,\\
b_0^\dag(q) &=&  i(2\pi/|q|L)^{1/2} [\theta(q)J_R(q)-\theta(-q)J_L(-q)]~
\eea
for $q\neq0$. 
Moreover, there are two conserved operators $N=N_R+N_L$ and $J=N_R-N_L$,
where $N_\alpha=J_\alpha(0)$. The Hamiltonian $H_{\rm LM}$ is quadratic
in $b_0,b_0^\dag$ but not diagonal. It is diagonalised by the Bogoliubov
transformation
\bea
b(q) &=&
b_0(q) \cosh\varphi(q)+b_0^\dag(-q)\sinh(q) ~,\\
b^\dag(q) &=&
b_0(-q) \sinh\varphi(q)+b_0^\dag(q)\cosh(q) ~
\eea
with $\tanh\varphi(q)=g_2(2)/[v_F+g_4(q)]$ and becomes
\be
H_{\rm LM}~=~\sum_{q\neq0} v(q)|q|~b^\dag(q)b(q)~+~
             \pi v_N \frac{N^2}{L} ~+~
             \pi v_J \frac{J^2}{L} ~,
\label{HLMb}
\ee
where $v(q)=[(v_F+g_4(q))^2-g_2^2(q)]^{1/2}$, $v_N=v(0)e^{2\varphi(0)}$,
$v_J=v(0)e^{-2\varphi(0)}$. The integrals of motion are $N$, $J$, and the
bosonic occupation numbers $b^\dag(q)b(q)$.

We consider an interaction quench where both couplings $g_2(q)$ and $g_4(q)$
are suddenly switched on at $t=0$. Before the quench $H_{\rm LM}=H_0$ is a non-interacting
Fermi liquid and its ground state is a vacuum for the operators $b_0(q)$. 
After some algebra, see Ref. \cite{Cazalilla}, one obtains a one-body correlation function 
\be
C_{\psi_R}(x,t)~\equiv~
\langle 0|~e^{itH_{\rm LM}}\psi_R^\dag(x)\psi_R(0)e^{-itH_{\rm LM}}~|0\rangle~.
\ee
It is interesting to compare two asymptotes of the function in the thermodynamic limit: 
\be
C_{\psi_R}(x,t)~\approx~
\frac{i(R_0/2vt)^{\gamma^2}}{2\pi(x+ia)}~,~~{\rm when}~|x|\ll 2v(0)t~,
\ee
and 
\be
C_{\psi_R}(x,t)~\approx~
\frac{i|R_0/x|^{\gamma^2}}{2\pi(x+ia)}~,~~{\rm when}~|x|\gg 2v(0)t~,
\ee
where $\gamma=\sinh2\varphi(0)$. The correlation function relaxes from
the initial Fermi liquid form for $|x|\gg2v(0)t$ to a final non-Fermi liquid
form when $|x|\ll2v(0)t$, but the limiting form $C_{\psi_R}(x,t\to\infty)$
has a different exponent $\gamma^2$ than in the ground state of the LM. 

Thus in any bounded region $\Omega$ within the quasiparticle horizon, 
\be 
|x|~\ll~2v(0)t~,
\ee
the correlation functions are relaxed to a steady state. Given the quadratic 
form of the bosonic Hamiltonian (\ref{HLMb}) and the general discussion
in Section \ref{DephasingGGE}, it is not quite surprising to find that the 
correlators in the steady state can be obtained from a GGE
\be
\rho^{\rm GGE}~=~{\cal N}~\exp\left[-\sum_q \lambda(q)~b^\dag(q)b(q)\right]~,
\ee 
where $\lambda(q)$ are fixed by initials conditions.
We can conclude that any finite subsystem within the quasiparticle horizon appears 
relaxed to GGE.

\subsection{ GGE and hard-core bosons }
\label{HCBGGE}

The integrable hard-core bosons in one dimension \cite{HCBhistory} were realised experimentally 
in the seminal experiment \cite{Kinoshita}, where the momentum distribution in an expanding 
cloud of atoms was observed to have a stationary but non-thermal distribution.
Relaxation in this system was considered in Ref. \cite{HCB}, where it was
shown that the  correct momentum distribution $f_k$ for one-dimensional hard-core
bosons can be obtained from a GGE for an equivalent fermionic representation
of the hard-core bosons. However, in a subsequent paper \cite{Pustilnik} 
it was demonstrated that the stationary state preserves information 
not only on momentum distribution, but also on momentum correlations 
which are missing in the GGE description. Here we outline the argument
of Ref. \cite{Pustilnik}.

For $t<0$ a system of $N$ hard-core bosons was in the ground state in a harmonic trap potential,
\be
H_0~=~H~+~V_{\rm trap}~,
\ee
where
$
V_{\rm trap}~=~\int dx~V(x)\rho(x)~,
$
$\rho(x)$ is a density operator, and 
\be 
V(x)~=~\frac12m\omega^2_0 x^2
\label{omega0}
\ee 
is the trap potential. At $t=0$ the trap potential $V(x)$ is switched off,
\be 
\omega_0~\stackrel{t=0}{\longrightarrow}~0~,
\label{omegato0}
\ee
and the bosons expand with the translationally invariant Hamiltonian $H$.

It is convenient to make a Jordan-Wigner transformation
\be
\psi(x)~=~
\exp\left[i\pi\int_{-\infty}^x dy~\rho(y)\right]~\varphi(x)~,
\ee
where the operators $\psi(x)$ and $\varphi(x)$ correspond to spinless fermions and
bosons respectively:
$\{\psi(x),\psi^\dag(y)\}=[\varphi(x),\varphi^\dag(y)]=\delta(x-y)$. The free
Hamiltonian becomes
\be
H~=~
\int dx~
\psi^\dag(x)
\left[-\frac{1}{2m}\frac{\partial^2}{\partial x^2}\right]
\psi(x)~,
\ee
but the density operator retains its form
\be
\rho(x)~=~\varphi^\dag(x)\varphi(x)~=~\psi^\dag(x)\psi(x)~.
\ee
Since the fermionic occupation numbers in momentum space
\be
n_k~=~\psi_k^\dag\psi_k~,
\ee
where $\psi_k=(2\pi)^{-1/2}\int dx~e^{-ikx}\psi(x)$, mutually commute and
commute with $H=\int dk \frac{k^2}{2m}n_k$, they are the integrals of 
motion to be included in GGE, see the discussion in Section \ref{GGE}. Neither their expectation 
values $\langle n_k\rangle$ nor expectation values of their products depend on $t$. 
In particular, for $\delta n_k=n_k-\langle n_k\rangle$ the Wick theorem gives a correlator
\be
\langle \delta n_k \delta n_{k'} \rangle_t ~=~
-|\langle\psi^\dag_k\psi_{k'}\rangle_0|^2 ~\neq~0~.
\ee
Since the initial ground state in a harmonic trap is not translationally invariant, the 
right hand side is generally non-zero when $k'\neq k$.

For a harmonic trap $V(x)=\frac{x^2}{2ml^4}$ with $l=(m\omega_0)^{-1/2}$ the correlation 
function can be written as
\be
\langle\psi_k^\dag\psi_{k'}\rangle~=~
\sum_{n=0}^{N-1}\phi_n^*(k)\phi_n(k')~,
\ee
where $\phi_n(k)$ are eigenfunctions of the harmonic oscillator in momentum 
representation. When $N\gg1$ we can approximate, see Ref. \cite{Pustilnik},
\be
\langle n_k\rangle ~=~ \frac{R}{\pi}\sqrt{1-\frac{k^2}{k_F^2}}~,~~~
\langle \delta n_k \delta n_{k'} \rangle ~=~
-\frac{\sin^2[(k-k')R]}{\pi^2(k-k')^2}~,
\ee
where $k_Fl=R/l=\sqrt{2N}$ and $|k|,|k'|\ll k_F$. There are non-negligible
correlators between different fermionic occupation numbers. They originate from
the broken translational invariance of the initial state and are conserved after the 
opening of the trap (\ref{omegato0}).

In an experiment like Ref. \cite{Kinoshita} one can in principle measure {\it bosonic} 
correlation functions. Since the Jordan-Wigner transformation between bosons and fermions
is non-local, the bosonic correlations are in general different than their fermionic
counterparts. However, Ref. \cite{Pustilnik} demonstrates that bosonic occupation numbers 
$f_k$ tend to the conserved fermionic occupation numbers: $f_k(t\to\infty)=n_k$. 
Thus when $t\to\infty$ all the non-trivial initial correlations between the 
integrals of motion can be measured as bosonic momentum correlations.

When $t\to\infty$ GGE becomes $\rho^{\rm GGE}={\cal N}\exp(-\int dk~\lambda_kf_k)$.
By construction, this ensemble ``predicts'' correct expectation values 
${\rm Tr}\rho^{\rm GGE}f_k=n_k$, but it ignores any momentum correlations.
Thus, in accordance with the discussion in Section \ref{GGE}, GGE cannot be used to predict 
occupation numbers in momentum space. However, the general argument in Section \ref{DephasingGGE} 
proves that, after dephasing for $t\to\infty$, $\rho^{\rm GGE}={\cal N}\exp(-\int dk~\lambda_kf_k)$ 
accurately predicts observables within a finite {\it fermionic} support $\Omega$ in real space.

The local relaxation in Section \ref{DephasingGGE} cannot always be taken for granted as demonstrated by another example 
in Ref. \cite{MinguzziGangardt}. In that paper, instead of switching off the trapping potential, 
they consider a sudden quench of the trap frequency in Eq. (\ref{omega0}) from an initial $\omega_0$ 
to a lower frequency $\omega_1$:
\be 
\omega_0~\stackrel{t=0}{\longrightarrow}~\omega_1~.
\ee
This problem is solved exactly by a scaling transformation of the initial wave function. The wave function 
is coherently breathing with frequency $\omega_1/2$. There are periodic revivals of the initial state
every $\pi/\omega_1$ demonstrating the absence of dephasing at the revivals. This anomaly originates from the fact that 
the non-interacting Jordan-Wigner fermions in a harmonic potential have commensurate eigenenergies being integer 
multiples of $\omega_1$, so there is no way anything could irreversibly dephase when $t\to\infty$. Thus one of the basic 
assumptions in Section \ref{DephasingGGE} is not satisfied and there is no local relaxation to any GGE. 
However, one could argue that a small anharmonicity of the trapping potential would alter this conclusion
\footnote{
The limiting case of $\omega_1=0$, i.e. the opening of the trap (\ref{omegato0}) considered in this Section, 
results in a continuous dispersion $k^2/2m$ of the non-interacting Jordan-Wigner fermions.  
The integration over the continuous $k$ yields irreversible local relaxation when $t\to\infty$.

This limiting case suggests that even for $\omega_1>0$ there may be local relaxation in between the periodic 
revivals such that a local state $\rho_\Omega(t)$ oscillates between the initial $\rho_\Omega(0)$ at the 
revivals and ${\rm Tr}_\Omega\rho^{\rm GGE}$ in between. When $\omega_1\to0$ then the revival period 
$\pi/\omega_1$ stretches to infinity allowing enough time for irreversible local relaxation. 
}. 
%

\subsection{ GGE and the Bose-Hubbard model }
\label{EisertGGE}

The Bose-Hubbard model (\ref{HBH}) is not integrable, but it is close to integrability 
in some regimes or limits of parameters. It seems to relax to a non-thermal steady state when close 
to integrability, and thermalise otherwise. 

For instance, Ref. \cite{Eisert} considers a quench from $J=0$ to a large $J\gg n$ well in the superfluid 
phase. This quench jumps between two limits of the model where it is integrable. The state after the quench 
relaxes {\it locally} to a non-thermal steady state. It will be shown below that the steady state is a GGE. 
This is an example that even a non-integrable model can locally relax to a non-thermal steady state 
when it is close to integrability.

Before the quench, when the Hamiltonian is $H_0=\frac12\sum_jn_j(n_j-1)$, the initial state is
the non-Gaussian Mott ground state $|n,n,\dots\rangle$ with exactly $n$ bosons at each site. 
Suddenly, at $t=0$ a finite hopping rate $J$ is switched on. We assume $J\gg n$ so large that 
the interaction term can be neglected and
\be
H~\approx~-J\sum_{j=1}^N (a_{j+1}^\dag a_j+a_j^\dag a_{j+1})~=~
          -2J\sum_k \cos(k)a_k^\dag a_k~
\label{Hhopping}
\ee 
is just the hopping term, where $a_k=\sum_{j=1}^Na_je^{-ikj}/\sqrt{N}$ is an annihilation operator 
in pseudomomentum representation. $n_k=a_k^\dag a_k$ are the integrals of motion of this quadratic $H$. 
Since before the quench bosons were localised in the Mott state, their conserved occupation numbers are 
the same for all quasimomenta, ${\rm Tr}~\rho(t)~n_k~=n$.   

We are interested in local observables with support in a finite subsystem $\Omega$ of the lattice. 
The Hamiltonian (\ref{Hhopping}) is quadratic as assumed in Section \ref{DephasingGGE}, but the initial 
Mott state is not Gaussian. Nevertheless, Ref. \cite{Eisert} shows that in the thermodynamic limit 
$N\to\infty$ and $t\to\infty$ (here the order of limits is important) the reduced density matrix of 
a block $\Omega$ of $S$ sites converges to a product of Gaussian states
\be
\lim_{t\to\infty}
\lim_{N\to\infty}
\rho_\Omega(t)~=~
\prod_{j=1}^S \frac{e^{-\lambda~a^\dag_ja_j}}{1+n}
\ee
and $\lambda=\ln(1+1/n)$. This limiting steady state can be also obtained by reduction of the generalised 
Gibbs ensemble, 
\be
\lim_{t\to\infty}
\lim_{N\to\infty}
\rho_\Omega(t)~=~
{\rm Tr}_{\Omega^\perp}\rho^{\rm GGE}~, 
\ee
where the ensemble is 
\be
\rho^{\rm GGE}~=~
\frac{e^{-\sum_k\lambda~a^\dag_ka_k}}{(1+n)^N}~=~
\frac{e^{-\sum_j\lambda~a^\dag_ja_j}}{(1+n)^N}~=~
\prod_{j=1}^N \frac{e^{-\lambda~a^\dag_ja_j}}{1+n}~.       
\ee
This GGE is clearly different from a thermal state 
$\propto e^{-\beta\sum_k\omega_k~a^\dag_ka_k}$ because $\omega_k=2J(1-\cos k)$ is not a constant.
We can conclude that, in the integrable limit $J\to\infty$ of the non-integrable
Bose-Hubbard model, the non-Gaussian initial Mott state relaxes {\it locally} to 
a steady non-thermal GGE. 

\begin{figure}[t]
\begin{center}
\includegraphics[width=0.7\columnwidth,clip=true]{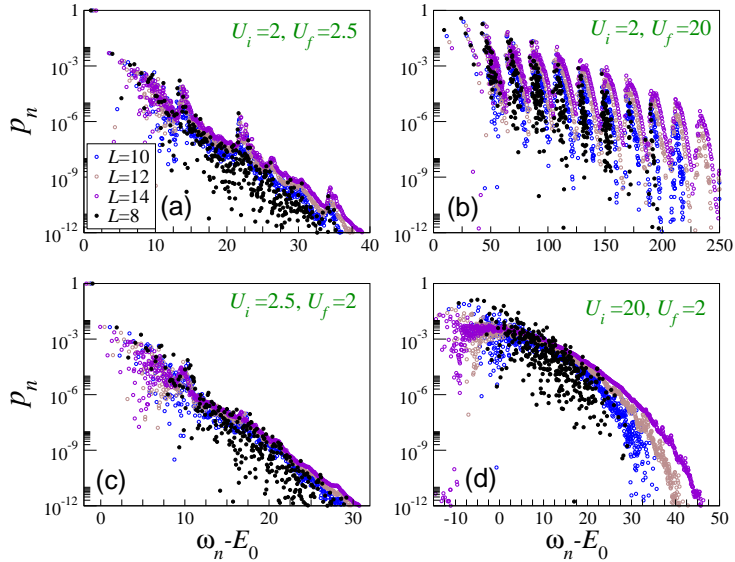}
\caption{ 
Here $U_i$ and $U_f$ are initial and final interaction strengths in the Bose-Hubbard Hamiltonian 
(\ref{HBH}) in a sudden quench $U_i\to U_f$. The critical point between the 
Mott and superfluid phases is $U_c=3.3$.
The plots show probabilities $p_n$ in a mixed time-averaged density
matrix after a quench, $\overline{\rho}=\sum_np_n|n\rangle\langle n|$,
versus the excitation energy $\omega_n-E_0$ of an eigenstate $|n\rangle$.
Panels a and c show roughly exponential dependence of $p_n$ on energy,
like in a canonical ensemble, while panels b and d indicate
strong memory of the initial conditions. The results come from numerical
calculations on a periodic lattice of $8,\dots,12$ sites with a density of 
$1$ particle per site.
(Figure from Ref. \cite{Roux})
}
\label{FigRoux}
\end{center}
\end{figure}

A similar problem was addressed in Ref. \cite{Roux} by numerical simulations of small lattices up to $12$ 
sites. Figure \ref{FigRoux} shows probability distributions $p_\alpha$ in the diagonal ensemble
$\overline{\rho}$ in Eq. (\ref{overlinerho}). A quench from the Mott-insulator to the superfluid phase 
prepares a clearly non-thermal diagonal ensemble, see Fig. \ref{FigRoux}d, but a small quench within the 
superfluid phase prepares a diagonal ensemble that looks more like a canonical ensemble, see Figs. \ref{FigRoux}a,c. 

A sudden quench in the opposite direction, i.e. from the superfluid to Mott-insulator phase, was 
studied by exact diagonalisation \cite{KollathLauchliAltman}, the adaptive time-dependent 
density matrix renormalisation group \cite{tDMRG}, and Bogoliubov theory \cite{AltmanAuerbach}. 
The model is trivially integrable at $J=0$, where unitary evolution factorises into a product over 
lattice sites $\prod_j e^{-in_j(n_j-1)/2}$. Since the quartic interaction $n_j(n_j-1)/2$ is an integer 
for any Fock state, the evolution is periodic in time with a period of $2\pi$. Any 
initial wave function revives periodically and there is no question of any relaxation
for $t\to\infty$. However, a non-zero hopping rate $J>0$ leads to relaxation of these periodic 
revivals or even overdamped relaxation, see Ref. \cite{FischerRelaxation}, Section \ref{fastnlarge} and 
the 2D experimental results in Fig. \ref{FigRevivals}. The numerical
data collected in Ref. \cite{KollathLauchliAltman} suggest that there are two distinct 
regimes: when $J$ is close to the Mott-superfluid transition, then for simple observables 
the final steady state is a thermal state with an effective temperature determined by energy
pumped into the system by the sudden quench, but when $J$ is perturbatively small,
then the steady state is not thermal and it retains memory of the initial state. The
last result is further supported by Fig. \ref{FigRoux}b from Ref. \cite{Roux}. 

The qualitatively different behaviour in the two regimes may be explained
by a simple model in Ref. \cite{AltmanAuerbach}. In this model, excitations of the Mott 
insulator state $|1,1,1,\dots\rangle$ are particles, created by $p^\dag_s$ and located 
at doubly occupied sites, and holes, created by $h^\dag_s$ and located at empty sites. 
When expanded to second order in the operators $p,h$ the Bose-Hubbard Hamiltonian (\ref{HBH}) 
becomes a quadratic Hamiltonian which can be diagonalised by a Bogoliubov transformation 
to a sum of non-interacting Bogoliubov quasiparticles,
$H_2=\sum_{k,\alpha}\omega_k \gamma^\dag_{k,\alpha}\gamma_{k,\alpha}$.  
This quadratic model, which neglects any interactions between quasiparticles 
$\gamma_{k,\alpha}$, is accurate for small $J$. Thus the general argument in Section \ref{DephasingGGE} 
predicts local relaxation to a non-thermal GGE with the integrals of motion 
$n_{k,\alpha}=\gamma^\dag_{k,\alpha}\gamma_{k,\alpha}$. 

In this model, thermalisation can occur due to quasiparticle interactions.
For instance, quartic terms of the form $\gamma^\dag_q\gamma_{k+q/2}\gamma_{-k+q/2}\gamma_0$ 
could make quasiparticle population equilibrate. Deep in the Mott regime, where the Mott gap 
$\simeq1$ in the quasiparticle spectrum $\omega_k$ is large as compared to the 
quasiparticle bandwidth $\simeq J$, this process is not compatible with energy conservation,
but it becomes increasingly effective as $J$ approaches the gapless critical point and it 
may explain the thermalisation observed close to criticality. However, a limited accuracy of 
the leading quadratic model was discussed already in Ref. \cite{AltmanAuerbach} and it is
also not known how these qualitative observations depend on a finite system size.

\subsection{ GGE and a system solvable by Bethe Ansatz }
\label{ManmanaGGE}

\begin{figure}[t]
\begin{center}
\includegraphics[width=0.9\columnwidth,clip=true]{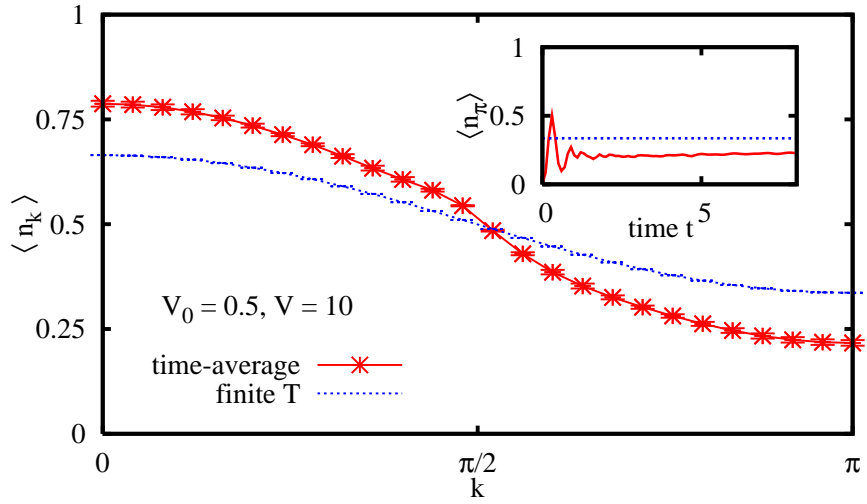}
\caption{ 
A time-averaged momentum distribution for the final $V=10$, i.e., momentum distribution 
in the time-averaged diagonal ensemble $\overline{\rho}$, and thermal momentum
distribution in a canonical ensemble. The inset shows a time dependence of
$\langle n_{\pi}\rangle$ together with a (dotted) thermal average. 
(Figure from Ref. \cite{Manmana})
}
\label{FigManmana3}
\end{center}
\end{figure}

Reference \cite{Manmana} considers a one-dimensional fermionic lattice Hamiltonian
\footnote{Transport was analysed in the system (\ref{HManmana}) in the first paper in Ref. \cite{ManmanaTransport}. }
\be
H~=~-\sum_{s}
\left(
c^\dag_{s+1}c_s+{\rm h.c.}
\right)+
V\sum_{s}n_sn_{s+1}
\label{HManmana}
\ee
with nearest neighbour interaction $V$ at half filling. This Hamiltonian is not quadratic and 
cannot be mapped to any quadratic Hamiltonian, but it is exactly solvable by Bethe 
ansatz. The model has a quantum critical point at $V_c=2$ separating a Luttinger liquid (when 
$V<2$) from a charge density wave insulator (when $V>2$). Ref. \cite{Manmana} considers open 
chains up to $100$ sites pushed out of equilibrium by sudden quenches from $V_0$ to $V$ and 
evolved in time using the Lanchos time evolution method \cite{tLanchos} and the adaptive 
DMRG \cite{tDMRG}. The quantity of interest is momentum distribution after time $t$:
\be
\langle n_k \rangle~=~
\frac{1}{L}
\sum_{m,n=1}^L e^{ik(m-n)}~ 
{\rm Tr}~\rho(t)~c_m^\dag c_n~.
\ee 
A general conclusion drawn from the numerical simulations in Ref. \cite{Manmana} is
that the momentum distribution relaxes to a steady state, but the steady state is
not a canonical ensemble, see Fig. \ref{FigManmana3}.

Since the Hamiltonian (\ref{HManmana}) cannot be mapped to any quadratic Hamiltonian, Section
\ref{DephasingGGE} cannot tell us what might be the small set of integrals of motion $I_\alpha$ in 
the GGE (\ref{rhoGGE}). In principle, any quantum system has as many integrals of motion as the 
dimension of its Hilbert space given by the projectors $I_\alpha=|\alpha\rangle\langle\alpha|$ on 
the eigenstates $|\alpha\rangle$ of its Hamiltonian. However, GGE build out of these projectors is 
equal to the time-averaged diagonal ensemble 
\be
\overline{\rho}~=~\sum_{\alpha}p_\alpha~ |\alpha\rangle\langle\alpha|~
\label{rhodiagonal}
\ee 
which may be not a tractable statistical description of the state. 

\begin{figure}[t]
\begin{center}
\includegraphics[width=0.9\columnwidth,clip=true]{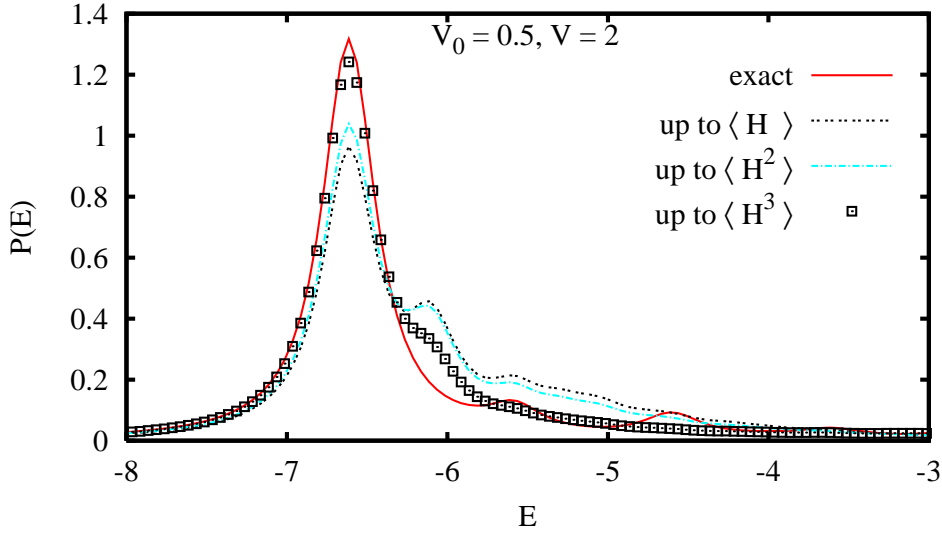}
\caption{ 
The exact energy distribution function $P(E)$ in the Hamiltonian after the quench
(red) together with its approximations by the distribution in Eq. (\ref{PE})
for increasing $M=1,2,3$. 
(Figure from Ref. \cite{Manmana})
}
\label{FigManmana5}
\end{center}
\end{figure}

Another universal set of integrals of motion are integer powers of a finite Hamiltonian: 
$I_m=H^m$ with $m=1,\dots,{\rm dim}(H)$. This set is as huge as the set of eigenstate 
projectors but, at least in the quench considered in Ref. (\ref{HManmana}), it can be accurately 
truncated to a small number of $M$ leading powers:
\be
\rho^{\rm GGE}~=~
{\cal N}
\exp\left(-\sum_{m=1}^M \lambda_m H^m\right)~
\label{GGEHn}
\ee
with $\lambda$'s fixed by the conserved moments ${\rm Tr}\rho^{\rm GGE}H^m={\rm Tr}\rho(0)H^m$. 
The ensemble (\ref{GGEHn}) can be rewritten as
$
\rho^{\rm GGE}=\sum_\alpha p(E_\alpha)~|\alpha\rangle\langle\alpha|
$, where
\be
p(E)~=~
{\cal N}
\exp\left(-\sum_{m=1}^M \lambda_m E^m\right)~.
\label{PE}
\ee
This form reveals that the essence of the generalised GGE (\ref{GGEHn}) simply is the assumption that 
$p_\alpha$ in Eq. (\ref{rhodiagonal}) is a smooth localized function of the energy $E_\alpha$, which 
can be accurately approximated by e.g. an exponent of a short polynomial in $E_\alpha$. Good quality of 
this approximation for reasonably small $M$, with $M=1$ being a canonical ensemble, is demonstrated in 
Fig. \ref{FigManmana5} which shows $p(E)$ multiplied by density of eigenstates $\rho(E)$, i.e.,
$P(E)=p(E)\rho(E)$.

\subsection{ Eigenstate thermalisation hypothesis (ETH) }
\label{ETH}

Since a state after a sudden quench often has a narrow energy distribution in a final 
Hamiltonian, see e.g. Fig. \ref{FigManmana5}, and the energy distribution $p(E)$ is conserved, 
it may be more accurate to describe a thermal state of an isolated quantum system by 
a microcanonical ensemble  
\be
p(E)~=~
{\cal N}
\left\{
\begin{array}{ll}
1,& {\rm when}~~|E-\langle H\rangle|<\Delta E~,\\ 
0,& {\rm otherwise}~,
\end{array}
\right.
\label{microcanonical}
\ee
Here $\langle H\rangle$ is average energy after a quench and $\Delta E$ is a finite energy window.
In this framework thermalisation of an observable $O$ means that the exact time-averaged 
diagonal ensemble $\overline{\rho}=\sum_\alpha p_\alpha |\alpha\rangle\langle\alpha|$
and the microcanonical ensemble (\ref{microcanonical}) give the same expectation value of $O$:
\be
\sum_\alpha p_\alpha \langle\alpha|O|\alpha\rangle~=~
{\cal N}
\sum_{
|E_\alpha-\langle H\rangle|<\Delta E
} 
\langle\alpha|O|\alpha\rangle~,
\label{equalityETH}
\ee
where $|\alpha\rangle$ are eigenstates of the Hamiltonian. 

One can imagine many different scenarios that might in principle imply the equality (\ref{equalityETH}),
but there is only one that does not depend on the initial state, except that $p_\alpha$'s have to be 
localised around the average energy $\langle H\rangle$. The {\it eigenstate thermalisation hypothesis} 
suggested in Ref. \cite{RefETH} proposes that the expectation value $\langle\alpha|O|\alpha\rangle$ of 
a {\it few-body} observable $O$ in an eigenstate $|\alpha\rangle$ with energy $E_\alpha$ of a large
{\it many-body} Hamiltonian is the same for all eigenstates in the energy window $E_\alpha\pm\Delta E$.      

ETH clearly implies the equality (\ref{equalityETH}) independently of the details of $p_\alpha$, provided 
that it is only localised around a given average energy. This is what is expected from thermalisation: 
the relaxed state should not depend on the initial state, except for its average energy.

The thermalisation mechanism implied by ETH is very simple: any eigenstate in the energy window 
gives the same expectation value of a few body observable $O$ or, even stronger, any single eigenstate 
in the window provides the same average as the microcanonical ensemble. This mechanism is very different 
from the ergodicity required of a classical system. In a non-integrable isolated quantum system, each 
eigenstate of the Hamiltonian implicitly contains a thermal state. The coherence between the eigenstates 
initially hides it, but then time evolution reveals it after dephasing.

At present, there are no general theoretical arguments supporting ETH, but there are some results for 
restricted classes of Hamiltonians: Deutsch in Ref. \cite{RefETH} showed that ETH holds for an integrable 
Hamiltonian perturbed by a matrix from a random Gaussian ensemble, nuclear shell calculations have shown 
that individual wave functions reproduce thermodynamic predictions \cite{shell}, some quantum systems whose 
classical versions are chaotic satisfy ETH in the semiclassical limit \cite{classchaotic}. More generally, 
ETH follows from the Berry conjecture \cite{RefETH,Berry} believed to hold in such systems \cite{semiclass}.  

\begin{figure}[t]
\begin{center}
\includegraphics[width=0.5\columnwidth,clip=true]{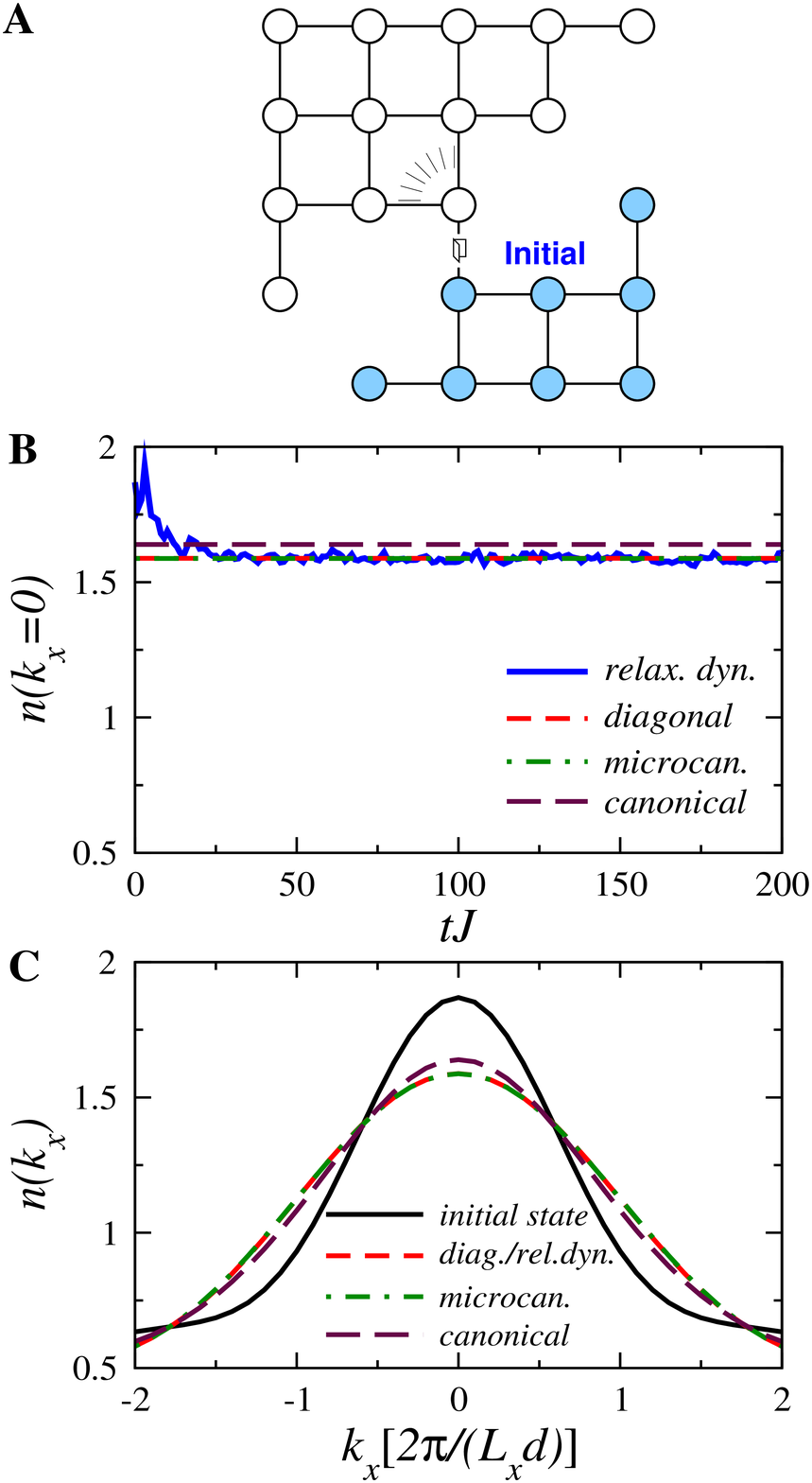}
\caption{ Panel A shows the lattice in the Hamiltonian (\ref{Hballoon}).
The bosons are initially prepared in the ground state of the sub-lattice
in the bottom-right corner and then, at $t=0$, released through the marked
link. Panel B shows how the momentum distribution $n(k_x=0)$ relaxes to
the value in the diagonal ensemble which is the same as predicted by the
microcanonical ensemble. Here 
$n(k_x,k_y)=L^{-2}\sum_{i,j}e^{-i\vec k(\vec r_i-\vec r_j)/L}$ with lattice
size $L=5$ and $n(k_x)=\sum_{k_y}n(k_x,k_y)$.
Panel C, compares the initial momentum distribution $n(k_x)$ with the steady
state diagonal ensemble which is indistinguishable from the microcanonical
distribution but significantly different from the canonical one.
(Figure from Ref. \cite{RigolNature})
}
\label{FigRigolNature1}
\end{center}
\end{figure}

In order to see if thermalisation occurs in a generic isolated quantum system and, 
in particular, if it occurs due to ETH, Ref. \cite{RigolNature} considers
a non-integrable system of $5$ hard core bosons on a finite lattice of $21$ sites 
in Fig. \ref{FigRigolNature1}a described by the Hamiltonian
\be
H~=~-J\sum_{\langle i,j \rangle}\left(b_i^\dag b_j+{\rm h.c.}\right)+
U\sum_{\langle i,j\rangle}n_in_j~
\label{Hballoon}
\ee 
with the nearest neighbour repulsion strength $U=0.1J$. The bosons are initially confined 
to the bottom-right part of the lattice in the ground state of a confined Hamiltonian. At 
$t=0$ they are allowed to tunnel to the initially empty top-left part. This set-up is 
a quantum analogue of an inflated balloon pierced inside a vacuum chamber to see that
the released air will soon uniformly fill the chamber and velocity distribution of air 
molecules will relax to the Maxwell velocity distribution. Reference \cite{RigolNature} 
considers momentum distribution. For instance, Fig. \ref{FigRigolNature1}b shows relaxation 
of the distribution at $k_x=0$ to the microcanonical ensemble. Fig. \ref{FigRigolNature1}c 
shows that the momentum distribution in the dephased diagonal ensemble is clearly different 
from the initial momentum distribution, indistinguishable from the microcanonical momentum 
distribution, but significantly different from the canonical one.

\begin{figure}[t]
\begin{center}
\includegraphics[width=0.95\columnwidth,clip=true]{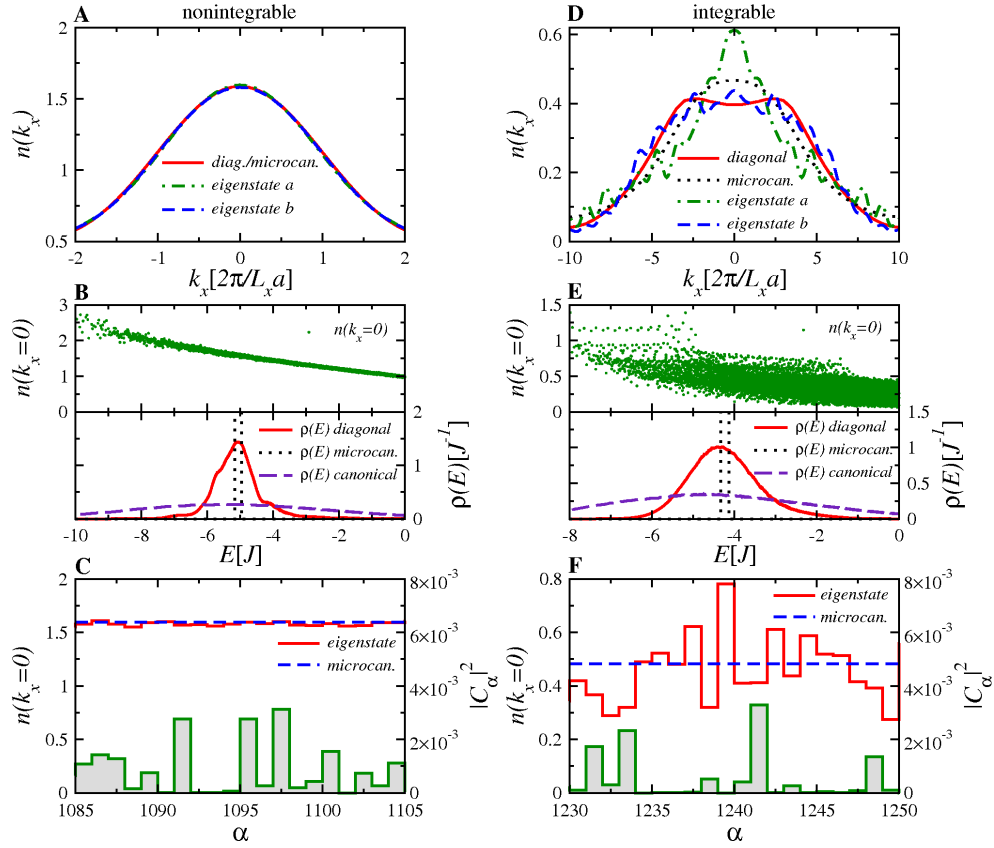}
\caption{ 
Left panels (A,B,C) correspond to the non-integrable system, and the right
panels (D,E,F) to the integrable one. Panels A and D compare momentum
distribution $n(k_x)$ with the microcanonical distribution, and with distributions
in two typical eigenstates close to the average energy. The four distributions
collapse in the non-integrable case (ETH holds), but they differ significantly 
in the integrable case (ETH does not hold). The top parts of panels B and E show 
expectation values of $n(k_x=0)$ in eigenstates at different energies. The 
non-integrable case (B) looks like a smooth curve, but the integrable case (E)
shows large fluctuations which do not resemble a smooth curve. The lower
parts of panels B and E show that energy distributions are very similar
in both non-integrable and integrable case. Finally, the upper parts of 
panels C and F show a focus on expectation values of $n(k_x=0)$ in the eigenstates 
$|\alpha\rangle$ close to the mean energy, and the lower parts of panels C and F
show probabilities $p_\alpha$ in these eigenstates. 
(Figure from Ref. \cite{RigolNature})
}
\label{FigRigolNature3}
\end{center}
\end{figure}

Figure \ref{FigRigolNature3} compares results for the {\it non-integrable} system
of 5 hard-core bosons on the lattice in Fig. \ref{FigRigolNature1}a, which was 
considered so far, and a similar {\it integrable} system of 5 hard-core bosons in 
a one-dimensional chain of $21$ sites. The bosons were initially prepared in the 
ground state of the 8-sites at one of the chain's ends, and then released 
by opening the link connecting the end to the rest of the chain. Results collected
in Fig. \ref{FigRigolNature3} show that while ETH holds in the non-integrable system,
it does not hold in the integrable one. 

Similar results were also obtained
in Ref. \cite{RigolSolo} for interaction quenches in a finite one-dimensional chain
with hard-core bosons whose integrability can be gradually broken by turning on the 
next-nearest-neighbour hopping and repulsion. When it is not integrable the system 
thermalises and ETH holds, but as the next-nearest-neighbour terms are turned off and 
the system tends to the integrable limit both the thermalisation and ETH break down.
The transition between the two regimes seems to be a smooth crossover, but it is not
clear how much this conclusion depends on the finite system size.

\subsection{ Dynamics of relaxation to a steady state }
\label{DynamicsRelaxation}

So far we have discussed mainly the nature of the asymptotic steady state. Finally, it is time 
for the dynamics of relaxation to this state. In this Section we briefly review two References 
\cite{prethermalization,relaxationXXZ} where the relaxation process was studied in some detail.

\begin{figure}[t]
\begin{center}
\includegraphics[width=0.95\columnwidth,clip=true]{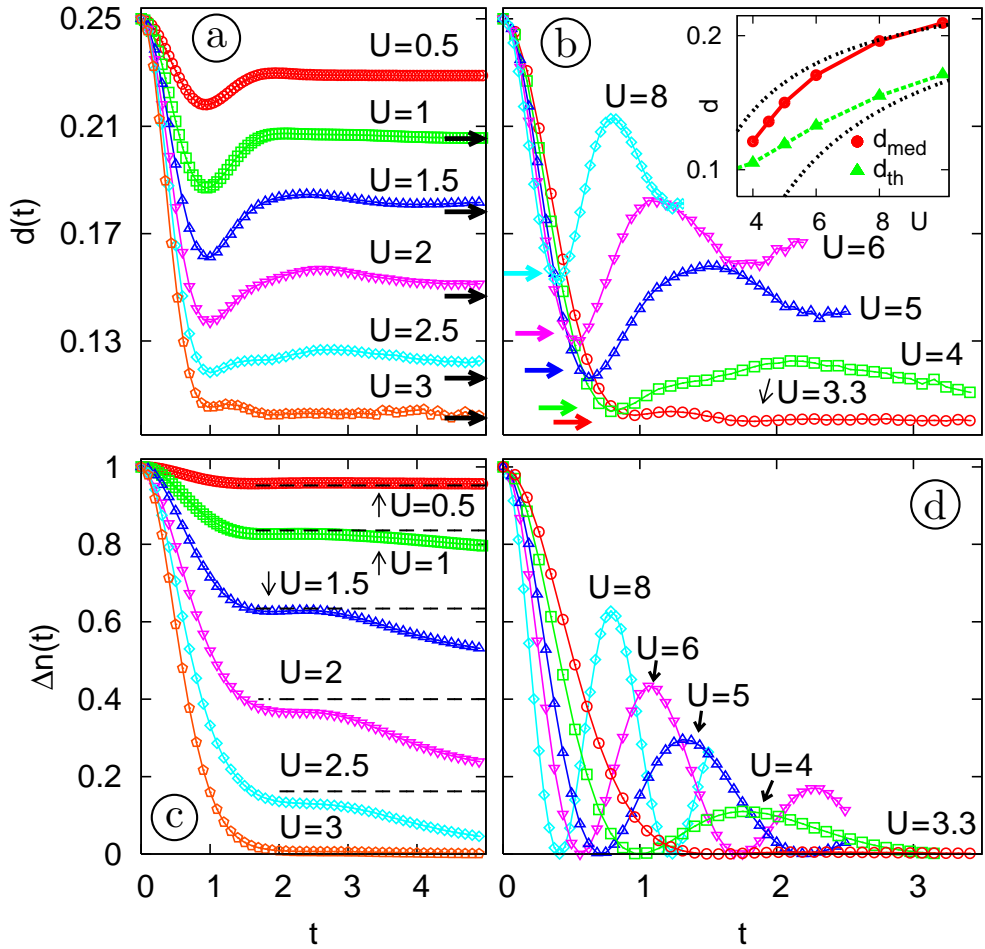}
\caption{ 
Double occupancy $d(t)$ and Fermi surface discontinuity $\Delta n(t)$
after quenches to $U\leq3$ (left panels) and $U\geq3.5$ (right panels).
The horizontal dashed lines in panel c indicate the quasi-stationary
value predicted in Ref. \cite{prethprediction}, and the horizontal arrows
in panel a mark corresponding thermal values. 
(Figure from Ref. \cite{prethermalization})
}
\label{FigPrethermalization}
\end{center}
\end{figure}

In the first of them \cite{prethermalization} they study a one-dimensional fermionic Hubbard 
model
\be
H~=~\sum_{ij\sigma}V_{ij}c_{i\sigma}^\dag c_{j\sigma}+
U\sum_i\left(n_{i\uparrow}-\frac12\right)
       \left(n_{i\downarrow}-\frac12\right)~.
\ee 
Here the hopping matrix $V_{ij}$ corresponds to a semi-elliptic density of states
$\rho(\epsilon)=\sqrt{4-\epsilon^2}/(2\pi)$. The system is initially in the
ground state of the non-interacting Hamiltonian, $U=0$ for $t<0$, and then
at $t=0$ the Coulomb repulsion is switched to a finite $U$. The relaxation following
this sudden quench was studied in Ref. \cite{prethermalization} within the time-dependent 
dynamical mean-field theory. Evolution of two observables, the double
occupancy $d(t)={\rm Tr}\rho(t)n_{i\uparrow}n_{i\downarrow}$ and the Fermi
surface discontinuity $\Delta n(t)$, is shown in Fig. \ref{FigPrethermalization}.
For both $U\gg1$ and $U\ll1$, the two quantities pass through a plateau where,
as predicted in Ref. \cite{prethprediction}, the system pre-thermalises in 
a quasi-steady state, before it finally relaxes to a thermal state. The size of the 
plateau depends on $U$. More detailed analysis shows that it is minimal at $U\approx 3.2$ 
marking a dynamical phase transition between the weak and strong coupling regimes where
the thermalisation is postponed until after the prethermalization on intermediate time-scales.
As remarked in Ref. \cite{prethermalization}, it is not clear whether and how 
this phenomenon is related to the existence of an equilibrium thermodynamic
phase transition at $U_c=4.76$, but the relaxation is the fastest at the crossover
between the strong and weak coupling regimes.

\begin{figure}[t]
\begin{center}
\includegraphics[width=0.95\columnwidth,clip=true]{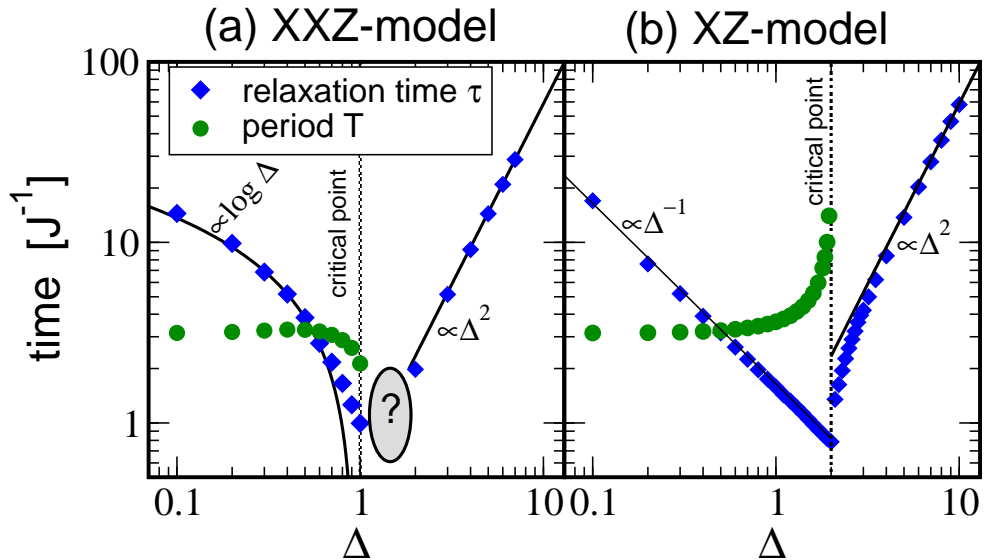}
\caption{ 
Relaxation time and oscillation period as functions of anisotropy $\Delta$
in the XXZ and XZ models. In both models the relaxation time is the shortest
at the critical point. 
(Figure from Ref. \cite{relaxationXXZ})
}
\label{FigRelaxationXXZ}
\end{center}
\end{figure}

A similar problem was considered in Ref. \cite{relaxationXXZ} in the antiferromagnetic
XXZ model (solvable by Bethe ansatz) and the antiferromagnetic XZ model (integrable 
by mapping to a quadratic Hamiltonian),
\bea
H_{XXZ}&=&\sum_s
\left(
\sigma^x_s\sigma^x_{s+1}+
\sigma^y_s\sigma^y_{s+1}+
\Delta\sigma^z_s\sigma^z_{s+1}
\right)~,\\
H_{XZ}&=&\sum_s
\left(
2\sigma^x_s\sigma^x_{s+1}+
\Delta\sigma^z_s\sigma^z_{s+1}
\right)~,
\eea
They studied unitary time evolution of the antiferromagnetic order starting from the highly 
non-equilibrium Neel state. The order vanishes exponentially with relaxation being oscillatory 
or non-oscillatory, depending on the anisotropy parameter $\Delta$. As demonstrated by the data 
collected in Fig. \ref{FigRelaxationXXZ}, the relaxation is the fastest near the quantum critical 
point contrary to the usual notion of critical slowing down. This effect can be explained by the 
gapped excitation spectrum away from the critical point. The relaxation is dominated by scattering 
between high-energy excitations introduced to the system through the highly excited initial state. 
The gap restricts the phase space available for scattering making relaxation slower. This leads to 
increasing relaxation time with increasing gap and minimal relaxation time near the critical point 
where the gap is zero.

In conclusion, in all models reviewed in this Section relaxation is the fastest either at a critical 
point or close to it. This is where either vanishing or small energy gap does not inhibit 
relaxation.

\subsection{ Summary } 
\label{Summary2}

In this part we focused on the nature of the apparent stationary states long time
after a sudden quench with a special emphasis on quadratic models. Dynamics of relaxation 
to the steady state was only touched in Section \ref{DynamicsRelaxation}. This is 
a non-trivial problem, especially for a large sudden quench, where not only low energy 
states are excited, but also high energy ones which are not described by any universal 
low energy effective theory. This problem should in general be less serious after a linear 
quench where, despite some similarity to a sudden quench in the adiabatic-impulse-adiabatic
approximation, only low energy modes are excited up to a cut-off set by the KZ
length $\hat\xi$, and this length itself can be accurately obtained from a universal low
energy theory. However, sudden quenches are interesting in their own right. After the 
pioneering theoretical work in the 1970's \cite{BarouchMcCoy}, they became recently the 
subject of dedicated experiments on oscillations and dephasing of the superfluid state 
after a sudden quench deep into the Mott insulator phase \cite{Greiner,Hofferberth}.

There are integrable models where the time evolution can be solved exactly, like 
the XY chain \cite{BarouchMcCoy,IgloiRieger,SenguptaPowellSachdev}, 1D hard-core bosons 
\cite{HCB,Pustilnik}, or the $\frac{1}{r}$ Hubbard chain \cite{KollarEckstein}, all of which 
are diagonalizable by the Jordan-Wigner transformation, but this type of integrability may 
result in very specific relaxation dynamics \cite{Brioli}. In models integrable by the Bethe 
ansatz it was not possible to extract dynamics in general, but with the notable exceptions of 
the Richardson and Lieb-Liniger models \cite{DynamicalBethe}. In view of these difficulties, it 
is necessary to develop approximate and/or numerical methods such as solutions of the 
dynamics of field theoretical models at the renormalisation group fixed points 
\cite{CalabreseCardyPRL,Cazalilla,DynamicsFixedPoint}, semiclassical theories \cite{SemiclassicalTheories},
exact diagonalisation \cite{RigolNature,LauchliKollath,Roux}, time-dependent dynamical 
renormalisation group (tDMRG) in one dimension
\cite{tDMRG,LauchliKollath,18c,Manmana,KollathLauchliAltman,tDMRGappl},
dynamical mean field theory in infinite dimensions \cite{prethermalization,FalicovKimball,DMFT}, 
and other approximate methods \cite{FischerRelaxation}.
Recently, new numerical schemes were proposed \cite{beyondDMRG} which go beyond
the standard tDMRG by focusing from the beginning on the dynamics of the interesting local observable 
rather than a quantum state as a whole. Some of the more universal results in the above papers
concerning the final relaxed steady states were reviewed in this part, but the less universal
dynamics of relaxation has been only touched. 

As explained in the introductory Section \ref{Introduction2}, it makes sense to define
a mixed steady state $\rho_\infty$ even for an isolated quantum system in a pure state, provided that 
the statistical description is used only for observables $O$ that are sufficiently local or 
coarse-grained. This observation encourages one to apply the usual concepts of thermodynamics to pure 
quantum states. The discussion in Section \ref{Introduction2} revealed the central role played by the 
diagonal part $\overline{\rho}=\sum_\alpha p_\alpha |\alpha\rangle\langle\alpha|$ of the density matrix  
in the eigenbasis $|\alpha\rangle$ of the time-independent Hamiltonian $H$ after the quench.  
Reference \cite{Sd} goes further and argues that also for a time-dependent Hamiltonian 
$H(t)$ it is only the diagonal part $\overline{\rho}(t)$ of the density matrix 
$\rho(t)=|\psi(t)\rangle\langle\psi(t)|$ in the instantaneous eigenbasis of $H(t)$ that
is relevant for thermodynamics, because any realistic measurement of a thermodynamic observable
$O$ actually measures its time average $\overline{O}={\rm Tr}\overline{\rho}O$. The time average 
dephases to zero any contribution from the off-diagonal elements of the density matrix $\rho(t)$,
and for sufficiently coarse-grained observables the dephasing may be fast enough to justify
using the time-dependent diagonal ensemble $\overline{\rho}(t)$ as a statistical description
of the actual underlying pure state $|\psi(t)\rangle$. This observation motivates definition
of the diagonal entropy
\be
S_d~=~-\sum_\alpha p_\alpha \log_2 p_\alpha~,
\label{Sd}
\ee  
which is in general non-zero, in contrast to the vanishing von Neumann entropy 
$S=-{\rm Tr}\rho(t)\log_2\rho(t)$ of the underlying pure state. The diagonal entropy does 
not change in adiabatic processes when transitions between instantaneous energy levels 
are negligible and the diagonal probabilities $p_\alpha$ do not change.
In Ref. \cite{Sd} it is shown that when the system is initially in thermal equilibrium
at temperature $T$, then $S_d$ can only increase or stay the same, in accordance with
the second law of thermodynamics. This and other properties \cite{Sd} make the diagonal
entropy a plausible candidate for the entropy of an isolated quantum system.  

Another Ref. \cite{heat} derives microscopic expression for the heat generated in an arbitrary 
process as the energy generated due to transitions between different instantaneous energy levels. 
When the initial density matrix is diagonal in the Hamiltonian eigenbasis, then the expression for 
heat contains only squares of the absolute values of the evolution operator which can be interpreted 
as transition probabilities between instantaneous energy levels. It can be shown \cite{heat} that 
when the initial density matrix is passive, i.e. its diagonal elements $p_\alpha$ decrease with increasing 
eigenenergy, then the heat is non-negative.

A closely related issue is considered in Ref. \cite{WorkDone}, where a probability distribution $P(W)$
is derived for a work $W$ done in an instantaneous quench starting at the critical point of a
quantum critical system. For instance, in the quantum Ising chain, a small sudden quench from the
critical transverse magnetic field $g=1$ to a nearby $g=1+\delta g$ does a work $W$ with the Gamma 
distribution 
\be
P(W)~\propto~W^{(\delta g/2\pi)^2-1}e^{-W} ~.
\ee
This distribution exhibits an interesting edge singularity with the exponent $(\delta g/2\pi)^2$ 
depending on the size $\delta g$ of the quench. Certainly, a lot of interesting research could be 
done along these lines and those of Ref. \cite{heat} especially in the context of the Jarzynski 
theorem \cite{Jarzynski}.

Another Reference \cite{Song} considers spin susceptibility in a sudden quench of the transverse
field $g$ in the quantum Ising chain. After the field is instantaneously switched from
an initial $g_i$ to a final $g_f$ the transverse magnetisation relaxes locally to a stationary
value $m(g_i,g_f)={\rm Tr}\overline{\rho}\sigma^z_j$. One can define susceptibility to the
initial field as $\chi_i=\partial m/\partial g_i$. Exact calculations in Ref. \cite{Song} show,
among other things, that close to the criticality, when $g_i\approx1$, the quench susceptibility
has a logarithmic divergence
\be
\chi(g_i,g_f)~\approx~-\frac{1}{\pi}\ln\left|g_i-1\right|~+~F(g_f)~,
\ee
where $F(g_f)$ is independent of $g_i$. Notice that the divergent susceptibility probes 
not only the ground state, which is known to undergo drastic changes at a quantum critical 
point, but also the excited states.

These are but a few examples which did not quite fit into the main story of 
Section \ref{Relaxation}, but which demonstrate that the subject is far from
closed and we can expect exciting new developments in the near future.

\section{ Open problems } 
\label{Summary}

The main open problems appear to be:

\begin{itemize}

\item Once the non-adiabaticity of quantum critical systems has been well established,
      we must at last face the problem how to prevent the unwanted excitation in the adiabatic 
      quantum state preparation. The non-linear quench or the inhomogeneous transition in 
      Sections \ref{non-linear} and \ref{KZinhom} respectively are the first proposals in 
      this direction, but we certainly still need more ideas and more work to overcome this 
      problem.

\item Due to the quasiparticle horizon effect, the entropy of entanglement in the excited
      state grows linearly with time and makes simulations with classical computers a formidable
      task. On the other hand, the excited high energy eigenstates render low energy 
      effective theories inaccurate. Thus an accurate solution of the relaxation dynamics in 
      a large non-integrable system remains a challenge.

\end{itemize}
These and other open problems may require another review in the near future.

\section*{ Acknowledgements } 

I would like to acknowledge stimulating interactions with Lukasz Cincio, Fernando Cucchietti,
Bogdan Damski, Krzysztof Sacha, Jakub Meisner, Marek Rams, Anatoli Polkovnikov, Haitao Quan, 
Marek Tylutki, Jakub Zakrzewski, Michael Zwolak, and Wojciech Zurek. This work was supported 
by Polish Government scientific funds (2009-2012) as a research project N202 124736 and 
by the Marie Curie ATK project COCOS (contract MTKD-CT-2004-517186).





\begin{thebibliography}{1000}
\markboth{Dynamics of a Quantum Phase Transition and Relaxation to a Steady State}{Advances in Physics}

\bibitem{BodzioReview} M. Lewenstein, A. Sanpera, V. Ahufinger, B. Damski, A. Sen(De), and U. Sen,
                       Adv. in Phys. {\bf 56} 243 (2007);
                       A.M. Rey, V. Gritsev, I. Bloch, E. Demler, and M.D. Lukin,
                       Rev . Lett 99, 140601 (2007).


\bibitem{Greiner} M. Greiner {\it et al.},
                  Nature {\bf 415}, 39 (2002);
                  Nature {\bf 419}, 51 (2002).

\bibitem{Kinoshita} T. Kinoshita, T. Wenger, and D.S. Weiss,
                    Science {\bf 305}, 1125 (2004).

\bibitem{NewtonCradle} T. Kinoshita, T. Wenger, and D. S. Weiss,
                       Nature {\bf 440}, 900 (2006).

\bibitem{Paredes} B. Paredes {\it et al.},
                  Nature 429, {\bf 277} (2004).

\bibitem{superexchange} S. Trotzky {\it et al.}, 
                        Science, 319, 295 (2008). 
                        
\bibitem{boshier} K. Henderson, C. Ryu, C. MacCormick, and M. G. Boshier,
                  New J. Phys. 11 (2009) 043030.                   

\bibitem{SGExp} E. Haller, R. Hart, M. J. Mark, J. G. Danzl, L. Reichs\"ollner, 
                M. Gustavsson, M. Dalmonte, G. Pupillo, H.-Ch. N\"agerl,
                arXiv:1004.3168.

\bibitem{IsingIon} A. Friedenauer, H. Schmitz, J. T. Glueckert, D. Porras, and T. Schaetz, 
                	Nature Physics {\bf 4}, 757 (2008); arXiv:0802.4072.

\bibitem{iontraps} I. Bloch, J. Dalibard, and W. Zwerger,
                   Rev. Mod. Phys. {\bf 80}, 885 (2008);
                   L.-M. Duan, E. Demler, and M.D. Lukin,
                   Phys. Rev. Lett. {\bf 91}, 090402 (2003);
                   A. Micheli, G.K. Brennen, and P. Zoller,
                   Nature Physics {\bf 2}, 341 (2006).

\bibitem{IsingNMR} J. Zhang, F. M. Cucchietti, C. M. Chandrashekar, M. Laforest, 
                   C. A. Ryan, M. Ditty, A. Hubbard, J. K. Gamble, R. Laflamme
                   Phys. Rev. A {\bf 79}, 012305 (2009).                  

\bibitem{QA} A. B. Finnila, M. A. Gomez, C. Sebenik, C. Stenson, and J. D. Doll,
             Chem. Phys. Lett. {\bf 219}, 343 (1994);
             T. Kadowaki and H. Nishimori,
             Phys. Rev. E {\bf 58}, 5355 (1998);
             J. Brooke, D. Bitko, T. F. Rosenbaum, and G. Aeppli,
             Science {\bf 284}, 779 (1999);
             G. E. Santoro, R. Martonak, E. Tosatti, and R. Car,
             Science {\bf 295}, 2427 (2002);
             A. Das and B. K. Chakrabarti,
             {\it Quantum Annealing and Related Optimization Methods},
             Lecture Notes in Physics,
             Springer Verlag, 2005;
             G. E. Santoro and E. Tosatti,
             J. Phys. A.: Math. Gen. {\bf 39}, R393 (2006);
             M. Sarjala, V. Pet\"aj\"a, and M. Alava,
             J. Stat. Mech. (2006) P01008.

\bibitem{Kasevich} C. Orzel, A. K. Tuchman, M. L. Fenselau, M. Yasuda, and M. A. Kasevich,
                   Science {\bf 291}, 2386 (2001); 
                   A. K. Tuchman, C. Orzel, A. Polkovnikov, and M. A. Kasevich,
                   Phys. Rev. A {\bf 74}, 051601 (2006);
                   W. Li, A. K. Tuchman, H.-C. Chien, M. A. Kasevich,
                   Phys. Rev. Lett. {\bf 98}, 040402 (2007).             
            
\bibitem{AdiabaticQC} E. Farhi {\it et al.},
                      Science {\bf 292}, 472 (2001);
                      R. Sch\"utzhold and G. Schaller,
                      Phys. Rev. A {\bf 74}, 060304 (2006).             

\bibitem{Sachdev} S. Sachdev, 
                  Quantum Phase Transitions (Cambridge University Press, Cambridge UK, 2001).

\bibitem{Feynman} R. Feynman, 
                  Found. Phys. {\bf 16}, 507 (1986).
                  
\bibitem{LZ} L.D. Landau and E.M. Lifshitz, 
             {\it Quantum Mechanics}, Pergamon, 1958; 
             C. Zener, Proc. R. Soc. A {\bf 137}, 696 (1932).

\bibitem{JW} P. Jordan and E. Wigner, 
             Z. Phys {\bf 47}, 631 (1928).
             
\bibitem{GreinerPRL} M. Greiner, I. Bloch, O. Mandel, T. W. Haensch, T. Esslinger,
                	 Phys. Rev. Lett. {\bf 87}, 160405 (2001). 
                	 
\bibitem{K} T. W. B. Kibble, 
            J. Phys. A {\bf 9}, 1387 (1976);
            Phys. Rep. {\bf 67}, 183 (1980); 
            Physics Today, 2007; 60 (9).

\bibitem{Z} W. H. Zurek, 
            Nature {\bf 317}, 505 (1985);
            Acta Physica Polonica B {\bf 24}, 1301 (1993);
            Phys. Rep. {\bf 276}, 177 (1996).
            
\bibitem{He4a} P.~C. Hendry {\it et al.},  
               Nature {\bf 368}, 315 (1994).

\bibitem{He4b} M.~E. Dodd {\it et al.}, 
               Phys. Rev. Lett. {\bf 81}, 3703 (1998).              

\bibitem{Rivers} N. Antunes and R. Rivers,
                 Phys. Rev. D {\bf 73}, 125003 (2006). 
                 
\bibitem{Bray} A.J. Bray, 
               Adv. Phys. {\bf 43}, 357 (1994).
               
\bibitem{He3} V.M.H. Ruutu {\it et al.}, 
              Nature {\bf 382}, 334 (1996);
              C. Ba\"urle {\it et al.}, 
              {\it ibid.} {\bf 382}, 332 (1996).

\bibitem{lowTc} R. Monaco {\it et al.}, 
                Phys. Rev. Lett. {\bf 89}, 080603 (2002);
                Phys. Rev. B {\bf 67}, 104506 (2003); 
                Phys. Rev. Lett. {\bf 96}, 180604 (2006);
                Phys. Rev. B {\bf 80}, 180501 (2009).
                
\bibitem{KZnum} P. Laguna and W.H. Zurek, 
                Phys. Rev. Lett. {\bf 78}, 2519 (1997);
                Phys. Rev. D {\bf 58}, 5021 (1998);
                A. Yates and W.H. Zurek, 
                Phys. Rev. Lett. {\bf 80}, 5477 (1998);
                G.J. Stephens {\it et al.}, 
                Phys. Rev. D {\bf 59}, 045009 (1999);
                N.D. Antunes {\it et al.}, 
                Phys. Rev. Lett. {\bf 82}, 2824 (1999);
                J. Dziarmaga, 
                Phys. Rev. Lett. {\bf 81}, 5485 (1998);
                E. Moro and G. Lythe,
                Phys. Rev. E {\bf 59}, R1303 (1999);
                J. Dziarmaga, P. Laguna and W. H. Zurek, 
                {\it ibid.} {\bf 82}, 4749 (1999);
                M.B. Hindmarsh and A. Rajantie, 
                {\it ibid.} {\bf 85}, 4660 (2000);
                G. J. Stephens, L. M. A. Bettencourt,
                and W. H. Zurek,
                {\it ibid.} {\bf 88}, 137004 (2002).
                
\bibitem{LC} I.L. Chuang {\it et al.}, 
             Science {\bf 251}, 1336 (1991);
             M.I. Bowick {\it et al.}, 
             {\it ibid.} {\bf 263}, 943 (1994).

\bibitem{highTc} R. Carmi and E. Polturak, 
                 Phys. Rev. Lett. {\bf 84}, 4966 (2000);
                 A. Maniv, E. Polturak, and G. Koren, 
                 {\it ibid.} {\bf 91}, 197001 (2003);
                 D. Golubchik, E. Polturak, and G. Koren,
                 Phys. Rev. Lett. {\bf 104}, 247002 (2010).
                 
\bibitem{ne}  S. Ducci, P.L. Ramazza, W. Gonzalez-Vi\~nas, F.T. Arecchi, 
              Phys. Rev. Lett. {\bf 83}, 5210 (1999); 
              S. Casado, W. Gonzalez-Vi\~nas, H. Mancini, S. Boccaletti, 
              Phys. Rev. E {\bf 63}, 057301 (2001);
              S. Casado at al., Eur. Phys. J. {\bf 146}, 87 (2007).

\bibitem{Anderson} C. N. Weiler, T. W. Neely, D. R. Scherer, A. S. Bradley, M. J. Davis, and B. P. Anderson,
                   Nature {\bf 455}, 948 (2008);
                   J.R. Anglin and W.H. Zurek, 
                   Phys. Rev. Lett. {\bf 83}, 1707 (1999).
                   
\bibitem{Bishop} J. Dziarmaga, A. Smerzi, W. H. Zurek, and A. R. Bishop,
                 Phys. Rev. Lett. {\bf 88}, 167001 (2002).

\bibitem{Meisner} J. Dziarmaga, J. Meisner, and W.H. Zurek,
                  Phys. Rev. Lett. {\bf 101}, 115701 (2008). 

\bibitem{Viola2} S. Deng, G. Ortiz, and L. Viola,
                 Eur. Phys. Lett. {\bf 84}, 67008 (2008).
                 
\bibitem{KZIsing} W.H. Zurek, U. Dorner and P. Zoller, 
                  Phys. Rev. Lett. {\bf 95}, 105701 (2005).
                 
\bibitem{Dziarmaga2005} J. Dziarmaga, 
                        Phys.Rev.Lett. {\bf 95}, 245701 (2005).
                        
\bibitem{Cincio} L. Cincio, J. Dziarmaga, M. M. Rams, and W. H. Zurek, 
                 Phys. Rev. A 75, 052321 (2007).

\bibitem{infinitezimalPolkovnikov} C. de Grandi, V. Gritsev, and A. Polkovnikov,
                                   Phys. Rev. B {\bf 81}, 012303 (2010); 
                                   Phys. Rev. B {\bf 81}, 224301 (2010).
                                   
\bibitem{nonlinSengupta} D. Sen, K. Sengupta, and S. Mondal,
                         Phys. Rev. Lett. {\bf 101}, 016806 (2008);
                         Phys. Rev. B {\bf 79}, 045128 (2009) .

\bibitem{nonlinPolkovnikov} R. Barankov and A. Polkovnikov, 
                            Phys. Rev. Lett. {\bf 101}, 076801 (2008).                                                            

\bibitem{DeGrandiPolkovnikov} C. De Grandi, R.A. Barankov, and A. Polkovnikov,
                              Phys. Rev. Lett. {\bf 101}, 230402 (2008).

\bibitem{Batrouni} G.G. Batrouni, V. Rousseau, R.T. Scalettar, M. Rigol, A. Muramatsu, P.J.H. Denteneer, and M. Troyer,
                   Phys. Rev. Lett. {\bf 89}, 117203 (2002);
                   G.G. Batrouni, H. R. Krishnamurthy, K. W. Mahmud, V.G. Rousseau, and R.T. Scalettar,
                   Phys. Rev. A {\bf 78}, 023627 (2008);
                   S.M. Pittman, G.G. Batrouni, R.T. Scalettar,
                   arXiv:0808.2809.

\bibitem{Dornerinhom} W.H. Zurek and U. Dorner, 
                      Phil. Trans. R. Soc. A {\bf 366}, 2953 (2008).
                      
\bibitem{Ramsinhom} J. Dziarmaga and M. M. Rams,
                    New J. Phys. {\bf 12}, 055007 (2010).                      

\bibitem{Bodzioinspace} B. Damski and W.H. Zurek, 
                        New J. Phys. {\bf 11}, 063014 (2009).

\bibitem{Karevski} T. Platini, D. Karevski, L. Turban, 
                   J. Phys. A: Math. Theor. {\bf 40}, 1467 (2007);
                   M. Collura, D. Karevski and L. Turban,
                   J. Stat. Mech. P08007 (2009). 

\bibitem{z2} J. Dziarmaga and M. M. Rams, 
             arXiv:1005.3763,
             A. Niederberger, M. M. Rams, J. Dziarmaga, F. M. Cucchietti, J. Wehr, and M. Lewenstein,
             arXiv:1004.1975.
             
\bibitem{Volovik} T. W. E. Kibble and G. E. Volovik,
                  JETP Letters {\bf 65}, 96 (1997);
                  J. Dziarmaga, P. Laguna, and W. H. Zurek,
                  Phys. Rev. Lett. {\bf 82}, 4749 (1999);
                  N. B. Kopnin and E. V. Thuneberg,
                  {\it ibid.} {\bf 83}, 116 (1999).             

\bibitem{VolovikRecent} W.H.Zurek, 
                        Phys. Rev. Lett. {\bf 102}, 105702(2009); 
                        A. del Campo, G. De Chiara, G. Morigi, M. B. Plenio, and A. Retzker, 
                        arXiv:1002.2524. 

\bibitem{Schaller} G. Schaller, 
                   Phys. Rev. A {\bf 78}, 032328 (2008). 

\bibitem{inhomoKarevski} T. Platini, D. Karevski, L. Turban, 
                         J. Phys. A: Math. Theor. {\bf 40}, 1467 (2007);
                         M. Collura and D. Karevski,
                         Phys. Rev. Lett. {\bf 104}, 200601 (2010).       

\bibitem{Bodzio1} B. Damski, Phys. Rev. Lett. {\bf 95}, 035701 (2005).

\bibitem{bodzioimpulse} B. Damski and W.H. Zurek,  
                        Phys. Rev. A {\bf 73}, 063405 (2006).
                        
\bibitem{eddi} E. Timmermans, P. Tommasini, M.  Hussein, and A. Kerman, 
               Phys. Rep. {\bf 315}, 199 (1999).                        
                        
\bibitem{LZmore} N.V. Vitanov and B.M. Garraway, 
                 Phys. Rev. A {\bf 53}, 4288 (1996);
                 N.V. Vitanov,
                 Phys. Rev. A {\bf 59}, 988 (1999).  
                 
\bibitem{Weber} E.T. Whittaker and G.N. Watson,
                {\it A Course of Modern Analysis},
                Cambridge University Press,
                Cambridge, UK, 1958.
                                                       
\bibitem{CherngLevitov} R. W. Cherng and L. S. Levitov, 
                        Phys. Rev. A {\bf 73}, 043614 (2006).
                 
\bibitem{XYHindusi} V. Mukherjee, U. Divakaran, A. Dutta, and D. Sen,
                    Phys. Rev. B 76, 174303 (2007).             

\bibitem{multicritical} U. Divakaran, V. Mukherjee, A. Dutta, and D. Sen,
                        J. Stat. Mech. P02007 (2009).

\bibitem{Viola3} S. Deng, G. Ortiz, and L. Viola,
                 Phys. Rev. B {\bf 80}, 241109 (2009). 
                 
\bibitem{gaplessline} U. Divakaran, A. Dutta, and D. Sen,
                      Phys. Rev. B {\bf 78}, 144301 (2008);
                      D. Chowdhury, U. Divakaran, and A. Dutta,
                      Phys. Rev. E 81, 012101 (2010).
                      
\bibitem{Kitaev} K. Sengupta, D. Sen, and S. Mondal,
                 Phys. Rev. Lett. {\bf 100}, 077204 (2008);
                 S. Mondal, D. Sen, and K. Sengupta,
                 Phys. Rev. B {\bf 78}, 045101 (2008);
                 U. Divakaran and A. Dutta,
                 Phys. Rev. B {\bf 79}, 224408 (2009).

\bibitem{KitaevModel} A. Kitaev,
                      Ann. Phys. (N.Y.) {\bf 321}, 2 (2006). 

\bibitem{HindusiReview} U. Divarkan, V. Mukherjee, A. Dutta, and D. Sen,
                        arXiv:0908.4004,                                                
                        published in "Quantum Quenching, Annealing and Computation", 
                        Eds. A. Das, A. Chandra and B. K. Chakrabarti, 
                        Lect. Notes in Phys., Springer, Heidelberg (2009). 

\bibitem{HindusiwariacjeLZ} V. Mukherjee, A. Dutta, and D. Sen,
                            Phys. Rev. B {\bf 77}, 214427 (2008);
                            V. Mukherjee and A. Dutta,
                            J. Stat. Mech. (2009) P05005;
                            U. Divakaran and A. Dutta,
                            Phys. Rev. B {\bf 79}, 224408 (2009);
                            U. Divakaran, A. Dutta, and D. Sen,
                            Phys. Rev. B {\bf 81}, 054306 (2010).

\bibitem{semiDirac} A. Dutta, R.R.P. Singh, and U. Divakaran,
                    Eur. Phys. Lett. {\bf 89}, 67001 (2010).

\bibitem{Balazs} B. Dora and R. Moessner,
                 Phys. Rev. B {\bf 81}, 165431 (2010).

\bibitem{c-term} D. Chowdhury, U. Divakaran, and A. Dutta,
                 Phys. Rev. E {\bf 81}, 012101 (2010). 

\bibitem{Polkovnikov2005} A. Polkovnikov, 
                          Phys. Rev. B {\bf 72}, R161201 (2005);
                          C. De Grandi and A. Polkovnikov,
                          arXiv:0910.2236,
                          contribution to "Quantum Quenching, Annealing and Computation",
                          Eds. A. Das, A. Chandra and B. K. Chakrabarti, 
                          Lect. Notes in Phys., Springer, Heidelberg (2009, to be published).

\bibitem{LSM} E. Lieb {\it et al.}, 
              Ann. Phys. (N.Y.) {\bf 16}, 406 (1961);
              S. Katsura, 
              Phys. Rev. {\bf 127}, 1508 (1962).

\bibitem{17b} G. Vidal, J.I. Latorre, E. Rico and A. Kitaev,
              Phys. Rev. Lett. {\bf 90}, 227902 (2003).

\bibitem{17c} N. Laflorencie,
              Phys. Rev. B {\bf 72}, R140408 (2005).

\bibitem{17d} G. Refael and J. E. Moore, 
              Phys. Rev. Lett. {\bf 93}, 260602 (2004);
              R. Santachiara, J. Stat. Mech. L06002 (2006).

\bibitem{17e} A. R. Its, B.-Q. Jin and V.E. Korepin, 
              J. Phys. A: Math. Gen. {\bf 38}, 2975 (2005).

\bibitem{18a} P. Calabrese and J. Cardy, 
              J. Stat. Mech. 0406, P002 (2004); 
              {\it ibid.} 0504, P010 (2005).

\bibitem{18b} P. Calabrese and J. Cardy, 
              Phys. Rev. Lett. {\bf 96}, 136801 (2006).

\bibitem{18c} G. De Chiara, S. Montangero, P. Calabrese and R. Fazio,
              J. Stat. Mech. P03001 (2006).  

\bibitem{JStatPhys} B.-Q. Jin and V. E. Korepin, 
                    J. Stat. Phys. 116, 79 (2004). 

\bibitem{17a} C. Holzhey, F. Larsen and F. Wilczek, 
              Nucl. Phys. B {\bf 424}, 443 (1994).

\bibitem{DynAfter} F. Pollmann, S. Mukerjee, A.G. Green, and J.E. Moore,
                   Phys. Rev. E {\bf 81}, 020101 (2010).

\bibitem{Schogo} P. J. Forrester and N. E. Frankel,
                 J. Math. Phys. {\bf 45}, 2003 (2004);
                 M. E. Fisher and R. E. Hartwig,
                 Adv. Chem. Phys. {\bf 15}, 333 (1968);
                 E. L. Basor and C. A. Tracy,
                 Phys. A {\bf 177}, 167 (1991);
                 F. Franchini and A. G. Abanov, 
                 J. Phys. A: Math. Gen. {\bf 38}, 5069 (2005); 
                 correction in J. Phys. A: Math. Gen. {\bf 39}, 14533 (2006).

\bibitem{SenguptaPowellSachdev} K.Sengupta, S. Powell, and S. Sachdev,
                                Phys. Rev. A {\bf 69}, 053616 (2004).

\bibitem{Vidal} G. Vidal, 
                Phys. Rev. Lett. {\bf 91}, 147902 (2003);  
                Phys. Rev. Lett. {\bf 93}, 040502 (2004).

\bibitem{decoherence} L. Cincio, J. Dziarmaga, J. Meisner, and M.M. Rams,
                      Phys. Rev. B {\bf 79}, 094421 (2009).

\bibitem{Berry} M.V. Berry,
                J. Phys. A {\bf 10}, 2083 (1977).

\bibitem{BerryQPT} A.C.M. Carollo and J.K. Pachos,
                   Phys. Rev. Lett. {\bf 95}, 157203 (2005);
                   S.L. Zhu,
                   Phys. Rev. Lett. {\bf 96}, 077206 (2006);
                   A. Hamma,
                   quant-ph/0602091;
                   S. Oh,
                   Phys. Lett. A {\bf 373}, 644 (2009).

\bibitem{GeometricPhase} B. Basu,
                         Phys. Lett. A {\bf 374}, 1205 (2010).
                         
\bibitem{Viola1} S. Deng, L. Viola, and G. Ortiz,
                 arXiv:0802.3941.                         

\bibitem{KSengupta2sitentanglement} K. Sengupta and D. Sen,
                                    Phys. Rev. A 80, 032304 (2009).

\bibitem{TopoQuench} D.I. Tsomokos, A. Hamma, W. Zhang, S. Haas, R. Fazio,
                     Phys. Rev. A 80, 060302(R) (2009).      

\bibitem{random} R. Shankar and G. Murthy, 
                 Phys. Rev. B {\bf 36}, 536 (1986);
                 B.M. McCoy and T.T. Wu, 
                 Phys. Rev. {\bf 176}, 631 (1968);
                 {\it ibid} {\bf 188}, 982 (1969);
                 R.B. Griffiths, 
                 Phys. Rev. Lett. {\bf 23}, 17 (1969);
                 B.M. McCoy, 
                 Phys. Rev. Lett. {\bf 23}, 383 (1969);
                 Phys. Rev. {\bf 188}, 1014 (1969).    

\bibitem{DFisher} D.S. Fisher, Phys. Rev. B {\bf 51}, 6411 (1995).        

\bibitem{JDrandom} J. Dziarmaga, 
                   Phys. Rev. B {\bf 74}, 064416 (2006);

\bibitem{JDrandom2} T. Caneva, R. Fazio, G. E. Santoro, 
                    Phys. Rev. B {\bf 76}, 144427 (2007). 

\bibitem{ZakrzDelande} J. Zakrzewski and D. Delande,
                       arXiv:0902.1117.
                                              
\bibitem{Kane} C.L. Kane, Science {\bf 314}, 1692 (2006).

\bibitem{Hatsugai} Y. Hatsugai, Phys. Rev. Lett. {\bf 71}, 3697 (1993).

\bibitem{Affleck} I. Affleck, T. Kennedy, E.H. Lieb, and H. Tasaki,
                  Phys. Rev. Lett. {\bf 59}, 799 (1987).

\bibitem{IFQHE} X. Wen, Advances in Physics {\bf 44}, 405 (1995).

\bibitem{TIexp} M. Koenig {\it et al.}, 
                Science {\bf 318}, 766 (2007);
                D. Hsieh {\it et al.},
                Nature {\bf 452}, 970 (2008),
                Science {\bf 323}, 919 (2009).
                
\bibitem{AQHE} V.P. Gusynin and S.G. Sharapov,
               Phys. Rev. Lett. {\bf 95}, 146801 (2005);
               N. Goldman {\it et al.},
               Phys. Rev. Lett. {\bf 103}, 035301 (2009).

\bibitem{QHSE} C.L. Kane and E.J. Mele, 
               Phys. Rev. Lett. {\bf 95}, 146802 (2005);
               Phys. Rev. Lett. {\bf 95}, 226801 (2005);
               B.A. Bernevig, T.L. Hughes, and S.-C. Zhang,
               Science {\bf 314}, 1757 (2006);
               L.Fu, C.L. Kane, and E.J. Mele,
               Phys. Rev. Lett. {\bf 98}, 106803 (2007);
               J.E. Moore and L. Balents,
               Phys. Rev. B {\bf 75}, 121306 (2007). 
 
\bibitem{TIPT} A.P. Schnyder, S.Ryu, A. Furusaki, and A.W.W. Ludwig,
               Phys. Rev. B {\bf 78}, 195125 (2008);
               Y. Kitaev,
               arXiv:0901.2686, 
                              Proceedings of the L.D.Landau Memorial Conference 
                              "Advances in Theoretical Physics", 
                              June 22-26, 2008, Chernogolovka, Moscow region, Russia.

\bibitem{MajoranaF} Y. Kitaev,
                    Phys. Usp. {\bf 44}, 131 (2001).

\bibitem{topoQC} A.Y. Kitaev,
                 Ann. Phys. {\bf 303}, 2 (2003);
                 C. Nayak, S.H. Simon, A. Stern, M. Freedman, and S. Das Sarma,
                 Rev. Mod. Phys. {\bf 80}, 1083 (2008).

\bibitem{pip} N. Read and D. Green,
              Phys. Rev. B {\bf 61}, 10267 (2000).

\bibitem{tri} L. Fu and C.L. Kane,
              Phys. Rev. Lett. {\bf 100}, 096407 (2008);
              Phys. Rev. Lett. {\bf 102}, 216403 (2009).

\bibitem{TIothers} S. Tewari, C. Zhang, S. Das Sarma, C. Nayak, and D.-H. Lee,
                   Phys. Rev. Lett. {\bf 100}, 027001 (2008);
                   P.A. Lee,
                   arXiv:0907.2681 (2009);
                   Y. Tanaka, T. Yokoyama, and N. Nagaosa,
                   arXiv:0907.2088;
                   K.T. Law, P.A. Lee, and T.K. Ng,
                   Phys. Rev. Lett. {\bf 103}, 237001 (2009).

\bibitem{Delgado2} A. Bermudez, L. Amico, and M.A. Delgado,
                   New J. Phys. {\bf 12}, 055014 (2010).

\bibitem{Delgado1} A. Bermudez, D. Patane, L. Amico, and M.A. Delgado,
                   Phys. Rev. Lett. {\bf 102}, 135702 (2009).

\bibitem{Creutz} M. Creutz, 
                 Phys. Rev. Lett. {\bf 83}, 2636 (1999).

\bibitem{LMG} T. Caneva, R. Fazio, and G.E. Santoro,
              Phys. Rev. B {\bf 78}, 104426 (2008) 

\bibitem{LMGmodel} H.J. Lipkin, N. Meshkov, and A.J. Glick,
                   Nucl. Phys. {\bf 62}, 188 (1965).

\bibitem{LMGstudy} R. Botet and R. Jullien,
                   Phys. Rev. B {\bf 28}, 3855 (1983);
                   F. Pan and J. Draayer,
                   Phys. Lett. {\bf 451}, 1 (1999);
                   J. Links, H. Zhou, R. McKenzie, and M. Gould,
                   J. Phys. A {\bf 36}, 63 (2003);
                   O. Castanos, R. Lopez-Pena, J. Hirsch, and E. Lopez-Moreno,
                   Phys. Rev. B {\bf 74}, 104118 (2006);
                   R. Unanyan and M. Fleischhauer,
                   Phys. Rev. Lett. {\bf 90}, 133601 (2003);
                   S. Dusuel and J. Vidal,
                   Phys. Rev. Lett. {\bf 93}, 237204 (2004);
                   F. Leyvraz and W. Hess,
                   Phys. Rev. Lett. {\bf 95}, 050402 (2005);
                   S. Dusuel and J. Vidal,
                   Phys. Rev. B {\bf 71}, 224420 (2005);
                   G. Ortiz, R. Somma, J. Dukelsky, and S. Rombouts,
                   Nucl. Phys. {\bf 707}, 421 (2005);
                   G. Chen and J. Liang,
                   New J. Phys. {\bf 8}, 297 (2006);
                   W. Heiss,
                   J. Phys. A {\bf 39}, 10081 (2006);
                   P. Ribeiro, J. Vidal, and R. Mosseri,
                   Phys. Rev. Lett. {\bf 99}, 050402 (2007);
                   Phys. Rev. E 78, 021106 (2008); 
                   G. Rosensteel, D. Rowe, and S. Ho,
                   J. Phys. A {\bf 41}, 025208 (2008). 

\bibitem{LMGIsing} A. Das, K. Sengupta, D. Sen, and B.K. Chakrabarti,
                   Phys. Rev. B {\bf 74} (2006) 144423.

\bibitem{amico} L. Amico, A. Osterloh, and F. Cataliotti, 
                Phys. Rev. Lett. {\bf 95}, 063201 (2005).

\bibitem{BKT} N. Goldenfeld,
              {\it Lectures on phase transitions and renormalisation group},
              Perseus Books, Reading, Massachusets, USA (1992);
              S.L. Sondhi, S.M. Girvin, J.P. Carini, and D. Shahar,
              Rev. Mod. Phys. {\bf 69}, 315 (1997).

\bibitem{Kuhner2000} T.D. K\"uhner, S.R. White, and H. Monien, 
                     Phys. Rev. B {\bf 61}, 12474 (2000); 
                     B. Damski and J. Zakrzewski,  
                     Phys. Rev. A {\bf 74}, 043609 (2006).
                     
\bibitem{Fisher1989} M.P.A. Fisher, P.B. Weichman, G. Grinstein, and D.S. Fisher, 
                     Phys. Rev. B {\bf 40}, 546 (1989).

\bibitem{AltmanAuerbach} E. Altman and A. Auerbach,
                         Phys. Rev. Lett. {\bf 89}, 250404 (2002).

\bibitem{XXZquench} F. Pellegrini, S. Montangero, G.E. Santoro, and R. Fazio,
                    Phys. Rev. B {\bf 77}, 140404 (2008);
                    E. Canovi, D. Rossini, R. Fazio, G.E. Santoro,
                    J. Stat. Mech. (2009) P03038.
                                          
\bibitem{Cucchietti} F. Cucchietti, B. Damski, J. Dziarmaga and W. H. Zurek,
                     Phys. Rev. A {\bf 75}, 023603 (2007).

\bibitem{Mercedes} D. R. Scherer, C. N. Weiler, T. W. Neely, and B. P. Anderson, 
                   Phys. Rev. Lett. {\bf 98}, 110402 (2007);
                   R. Carretero-Gonzalez, B.P. Anderson, P.G. Kevrekidis, D.J. Frantzeskakis, and C.N. Weiler,
                   Phys. Rev. A {\bf 77}, 033625 (2008).

\bibitem{TW}    D.F. Walls and G.J. Milburn,
                {\it Quantum Optics}, Springer-Verlag, Berlin 1994;
                M.J. Steel {\it et al.}, 
                Phys. Rev. A {\bf 58}, 4824 (1998);
                A. Sinatra {\it et al.},
                PRL {\bf 87}, 210404 (2001);
                J. Phys. B {\bf 35}, 3599 (2002);
                K. Goral {\it et al.},
                Opt. Express {\bf 8}, 92 (2001);
                M. J. Davis and S. A. Morgan,
                PRA {\bf 68}, 053615 (2003);
                M. J. Davis {\it et al.}
                J. Phys. B {\bf 37}, 2725 (2004);
                J. Phys. A {\bf 38}, 10259 (2005);
                M. Brewczyk {\it et al.},
                J. Phys. B {\bf 40}, R1 (2007);
                A. S. Bradley {\it et al.}, 
                PRA {\bf 77}, 033616 (2008);
                P.B. Blakie, A. S. Bradley, M.J. Davis, R.J. Ballagh, and C.W. Gardiner,
                Adv. Phys., 57, 363 (2008);
                A. Polkovnikov, 
                arXiv:0905.3384.

\bibitem{Bishop2} J. Dziarmaga, A. Smerzi, W. H. Zurek, and A. R. Bishop,
                  arXiv:cond-mat/0403607,
                  published in proceedings of NATO ASI {\it Patterns of Symmetry Breaking}, 
                  H. Arod\'z, J. Dziarmaga, and W.H. Zurek (Eds.),
                  Kluwer Academic Publishers (2002).

\bibitem{SchutzholdSFMott} R. Sch\"utzhold, M. Uhlmann, Y. Xu, and U.R. Fischer,
                           Phys. Rev. Lett. {\bf 97}, 200601 (2006);
                           U.R. Fischer, R. Sch\"utzhold, and M. Uhlmann,
                           Phys. Rev. A {\bf 77}, 043615 (2008);
                           R. Sch\"utzhold,
                           J. Low Temp. Phys. {\bf 153}, 228 (2008).

\bibitem{FischerRelaxation} U.R. Fischer and R. Sch\"utzhold,
                            Phys. Rev. A {\bf 78}, 061603 (2008).

\bibitem{LiebLiniger} E.H. Lieb and W. Liniger,
                      Phys. Rev. {\bf 130}, 1605 (1963).

\bibitem{LLHaldane} F.D.M. Haldane,
                    Phys. Rev. Lett. {\bf 47}, 1840 (1981). 

\bibitem{KCazalilla} M.A. Cazalilla,
                     J. Phys. B {\bf 37}, 1 (2004).

\bibitem{TonksGirardeau} M. Girardeau,
                         J. Math. Phys. {\bf 1}, 516 (1960);
                         C.N. Yang and Y.P. Yang,
                         J. Math. Phys. {\bf 10}, 1115 (1969);
                         L. Tonks,
                         Phys. Rev. {\bf 50}, 955 (1936).

\bibitem{sine-Gordon} S. Coleman,
                      Phys. Rev. D {\bf 11}, 2088 (1975).

\bibitem{Zamolodchikov} A. Zamolodchikov,
                        Int. J. Mod. Phys. A {\bf 10}, 1125 (1995).

\bibitem{ExpFerro} L.E. Sadler, J.M. Higbie, S.R. Leslie, M. Vengalattore, and D.M. Stamper-Kurn,
                   Nature {\bf 443}, 312 (2006).
                   
\bibitem{Lamacraft} A. Lamacraft, 
                    Phys. Rev. Lett. {\bf 98}, 160404 (2007).
                   
\bibitem{Bodzioferro2} M. Uhlmann, R. Sch\"utzhold, U. R. Fischer,
                       Phys. Rev. Lett. 99, 120407 (2007);
                       Phys. Rev. D {\bf 81}, 025017 (2010);
                       arXiv:1005.2649.                   

\bibitem{Bodzioferro} B. Damski and W. H. Zurek, 
                      Phys. Rev. Lett. 99, 130402 (2007);

\bibitem{spin1quantumfluctuations} K. Murata, H. Saito, and M. Ueda,
                                   Phys. Rev. A {\bf 75}, 013607 (2007).

\bibitem{PolkovnikovNature} A. Polkovnikov and V. Gritsev, 
                            Nature Physics {\bf 4}, 477,2008.

\bibitem{linearFalicovKimball} M. Eckstein and M. Kollar,
                               New J. Phys. {\bf 12}, 055012 (2010).

\bibitem{DMFT} J.K. Freedricks, V.M. Turkovski, and V. Zlati\'c,
               Phys. Rev. Lett. {\bf 97}, 266408 (2006);
               A. Hackl and S. Kehrein,
               J. Phys. C {\bf 21}, 015601 (2009).      

\bibitem{prethprediction} M. M\"ockel and S. Kehrein,
                          Phys. Rev. Lett. {\bf 100}, 175702 (2008);
                          Ann. Phys. {\bf 324}, 2146 (2009). 
                          
\bibitem{sc5} O. Zobay and B.M. Garraway, 
              Phys. Rev. A {\bf 61}, 033603 (2000);
              J. Liu {\it et al.},
              {\it ibid} {\bf 66}, 023404 (2002).

\bibitem{sc6} A. Ishkhanyan {\it et al.},
              Phys. Rev. A {\bf 73}, 043612 (2004).

\bibitem{sc7} I. Tikhonenkov {\it et al.},
              Phys. Rev. A {\bf 73}, 043605 (2006).

\bibitem{sc8} A. Altland, V. Gurarie, T. Kriechenbauer, and A. Polkovnikov,
              Phys. Rev. A {\bf 79}, 042703 (2009).

\bibitem{sc9} H. Pu {\it et al.},
              Phys. Rev. Lett. {\bf 98}, 050406 (2007);
              A.P. Itin and S. Watanabe,
              {\it ibid.} {\bf 99}, 223903 (2007);
              Phys. Rev. E {\bf 76}, 026218 (2007).

\bibitem{sc10} A.P. Itin {\it et. al},
               Physica D {\bf 232}, 108 (2007).

\bibitem{ItinTorma} A.P. Itin and P. T\"orma,
                    arXiv:0901.4778;
                    Phys. Rev. A {\bf 79}, 055602 (2009).

\bibitem{sc11} R.H. Dicke,
               Phys. Rev. {\bf 93}, 99 (1954). 

\bibitem{BEC2wells} C. Lee,
                    Phys. Rev. Lett. {\bf 102}, 070401 (2009);
                    T. Venumadhav, M. Haque, and R. Moessner,
                    Phys. Rev. B {\bf 81}, 054305 (2010).

\bibitem{Panleve} A.R. Its and A.A. Kapaev, 
                  Izv. Akad. Nauk SSSS, Ser. Mat. {\bf 51}, 878 (1987);
                  D.L. Vainshtein {\it et al.},
                  Plasma Phys. Rep. {\bf 25}, 299 (1999). 

\bibitem{GuAdiabatic} S.-J. Gu,
                      Phys. Rev. E {\bf 79}, 061125 (2009).

\bibitem{daGu} S.J. Gu and H.Q. Lin, 
               arXiv:0807.3491.

\bibitem{daZanardi} L.C. Venuti and P. Zanardi,
                    Phys. Rev. Lett. {\bf 99}, 095701 (2007). 

\bibitem{decFubini} A. Fubini, G. Falci, and A. Osterloh,
                    New J. Phys. {\bf 9}, 134 (2007).

\bibitem{decSchaller} S. Mostame, G. Schaller, and R. Sch\"utzhold,
                      Phys. Rev. A {\bf 76}, 030304 (2007).

\bibitem{decAmin} M.H.S. Amin, C.J.S. Truncik, and D.V. Averin,
                  Phys. Rev. A 80, 022303 (2009).

\bibitem{decPatane} D. Patane, A. Silva, L.Amico, R. Fazio, and G.E. Santoro,
                    Phys. Rev. Lett. {\bf 101}, 175701 (2008);
                    Phys. Rev. B {\bf 80}, 024302 (2009).

\bibitem{SolsBis} D. Patane, A. Silva, F. Sols, and L. Amico,
                  Phys. Rev. Lett. {\bf 102}, 245701 (2009).                  

\bibitem{decQuan} B. Damski, H. Quan, and W.H. Zurek,
                  arXiv:0911.5729.

\bibitem{RefETH} J.M. Deutsch,
                 Phys. Rev. A {\bf 43}, 2046 (1991);
                 M. Srednicki,
                 Phys. Rev. E {\bf 50}, 888 (1994).

\bibitem{FPU} E. Fermi, J. Pasta, and S. Ulam, 
              Los Alamos Report pp. LA-1940 (1955); 
              Focus Issue: The ``Fermi-Pasta-Ulam'' Problem - The First 50 Years,
              Chaos {\bf 15}, 015101 (2005);
              B.V. Chirikov,
              J. Nucl. Energy C {\bf 1}, 253 (1960);
              F.M. Izrailev, A.I. Khisamutdinov, and B.V. Chirikov,
              Los Alamos Report pp. LA-4440-TR (1970).

\bibitem{Zotos} X. Zotos,
                J. Low T. Phys., {\bf 126}, 1185 (2002).

\bibitem{Kamenev} D. M. Gangardt and A. Kamenev,
                  Phys. Rev. Lett. {\bf 104}, 190402 (2010).

\bibitem{CalabreseCardyPRL} P. Calabrese and J. Cardy,
                            J. Stat. Mech. P04010 (2005);
                            Phys.Rev.Lett. {\bf 96}, 136801 (2006);
                            J. Stat. Mech. P06008 (2007);
                            J. Stat. Mech. (2007) P10004.

\bibitem{BCFT} J.L. Cardy, 
               Nucl. Phys. B {\bf 240}, 514 (1984);    
              
\bibitem{tDMRG} S.R. White and A.E. Feiguin,
                Phys. Rev. Lett. {\bf 93}, 076401 (2004);
                A.J. Daley, C. Kollath, U. Sch\"olwock, and G. Vidal,
                J. Stat. Mech. P04005 (2004);
                D. Gobert {\it et al.},
                Phys. Rev. E {\bf 71}, 036102 (2005).               

\bibitem{CalabreseFazio} G. De Chiara, S. Montangero, P. Calabrese, and R. Fazio,
                         J.Stat.Mech. 0603 (2006) P001.

\bibitem{HorizonOther} L. Amico, A. Osterloh, F. Plastina, R. Fazio, and G.M. Palma,
                       Phys. Rev. A {\bf 69}, 022304 (2004);
                       C. Kollath, U. Schollwoeck, J. von Delft, and W. Zwerger,
                       Phys. Rev. A {\bf 71}, 053606 (2005);
                       C. Kollath, U. Schollwoeck, and W. Zwerger,
                       Phys. Rev. Lett. {\bf 95}, 176401 (2005);
                       T.S. Cubitt and J.I. Cirac,
                       quant-ph/0701053;
                       M. Polini and G. Vignale,
                       Phys. Rev. Lett. {\bf 98}, 266403 (2007);
                       T. Antal, Z. Racz, A. Rakos, and G.M. Schutz,
                       Phys. Rev. E {\bf 59}, 4912 (1999);
                       D. Karevski,
                       Eur. Phys. J B {\bf 27}, 147 (2001);
                       Y. Ogata,
                       Phys. Rev. E {\bf 66}, 066123 (2002);
                       W.H. Aschbacher and C.-A. Pillet,
                       J. Stat. Phys. {\bf 112}, 1153 (2003);
                       T. Platini and D. Karevski,
                       Eur. Phys. J B {\bf 48}, 225 (2005);
                       W.H. Aschbacher and J.-M. Barbaroux,
                       Lett. Math. Phys. {\bf 77}, 11 (2006);
                       T. Platini and D. Karevski,
                       J. Phys. A {\bf 40}, 1711 (2007); 
                       W.H. Aschbacher,
                       Lett. Math. Phys. {\bf 79}, 1 (2007).               

\bibitem{MatheyPolkovnikov} L. Mathey and A. Polkovnikov,
                            Phys. Rev. A {\bf 81}, 033605 (2010).

\bibitem{LiebRobinson} E. H. Lieb and D. W. Robinson,
                       Comm. Math. Phys. {\bf 28}, 251 (1972).
                       
\bibitem{HCB} M. Rigol, A. Muramatsu, and M. Olshanii,
              Phys. Rev. A {\bf 74}, 053616 (2006);
              M. Rigol, V. Dunjko, V. Yurovsky, and M. Olshanii,
              Phys. Rev. Lett. {\bf 98}, 050405 (2007).                       

\bibitem{KollarEckstein} M. Kollar and M. Eckstein,
                         Phys. Rev. A {\bf 78}, 013626 (2008).

\bibitem{Barthel} T. Barthel and U. Schollw\"ock,
                  Phys. Rev. Lett. {\bf 100}, 100601 (2008).

\bibitem{Cazalilla} M.A. Cazalilla,
                    Phys. Rev. Lett. {\bf 97}, 156403 (2006);
                    Phys. Rev. A, {\bf 80}, 063619 (2009). 

\bibitem{IgloiRieger} F. Igloi and H. Rieger,
                      Phys. Rev. Lett. {\bf 85}, 3233 (2000).

\bibitem{effTIsing} D. Rossini, A. Silva, G. Mussardo, and G.E. Santoro,
                    Phys. Rev. Lett. {\bf 102}, 127204 (2009).

\bibitem{LM} J.M. Luttinger, 
             J. Math. Phys. {\bf 4}, 1154 (1963);
             E.H. Lieb and D.C. Mattis,
             {\it ibid.} {\bf 6}, 304 (1965);
             A. Luter and I. Peschel,
             Phys. Rev. B {\bf 9}, 2911 (1974)

\bibitem{TLL} F.D.M. Haldane,
              Phys. Rev. Lett. {\bf 45}, 1358 (1980);
              J. Phys. C {\bf 14}, 2585 (1981);
              Phys. Rev. Lett. {\bf 47}, 1840 (1981);
              T. Giamarchi,
              {\it Quantum Physics in One Dimension},
              Oxford University Press, Oxford 2004;
              A.O. Gogolin, A.A. Nersesyan, and A.M. Tsvelik,
              {\it Bosonization and Strongly Interacting Systems},
              Cambridge University Press, Cambridge 1998.

\bibitem{HCBhistory} M. Girardeau, 
                     J. Math. Phys. {\bf 1}, 516 (1960);
                     A. Lenard,
                     {\it ibid.} {\bf 5}, 930 (1964);
                     H.G. Vaidya and C.A. Tracy,
                     Phys. Rev. Lett. {\bf 42}, 3 (1979).
                     
\bibitem{Pustilnik} D.M. Gangardt and M. Pustilnik,
                    Phys. Rev. A. {\bf 77}, 041604 (2008).                     

\bibitem{MinguzziGangardt} A. Minguzzi and D. M. Gangardt,
                           Phys. Rev. Lett 94, 240404 (2005).
                           
\bibitem{Eisert} M. Cramer, C.M. Dawson, J. Eisert, and T.J. Osborne,
                 Phys. Rev. Lett. {\bf 100}, 030602 (2008);
                 M. Cramer, A. Flesch, I.P. McCulloch, U. Schollwock, and J. Eisert,
                 Phys. Rev. Lett. 101, 063001 (2008).

\bibitem{Roux} G. Roux,
               Phys. Rev. A {\bf 79}, 021608 (2009).

\bibitem{KollathLauchliAltman} C. Kollath, A.M. L\"auchli, and E. Altman,
                               Phys. Rev. Lett. {\bf 98}, 180601 (2007).

\bibitem{Manmana} S.R. Manmana, S. Wessel, R.M. Noack, and A. Muramatsu,
                  Phys. Rev. Lett. {\bf 98}, 210405 (2007);
                  Phys. Rev. B {\bf 79}, 155104 (2009).

\bibitem{ManmanaTransport} S. Mukerjee, V. Oganesyan, and D. A. Huse,
                           Phys. Rev. B {\bf 73}, 035113 (2006);
                           V. Oganesyan and D. A. Huse,
                           Phys. Rev. B {\bf 75}, 155111 (2007);
                           V. Oganesyan, A. Pal, and D. A. Huse,
                           Phys. Rev. B {\bf 80}, 115104 (2009);

\bibitem{tLanchos} P. Schmitteckert,
                   Phys. Rev. B {\bf 70}, 121302 (2004);
                   S.R. Manmana, A. Muramatsu, and R.M. Noack,
                   AIP Conf. Proc. {\bf 789}, 269 (2005);
                   M. Hochbruck and C. Lubich,
                   BIT {\bf 39}, 620 (1999);
                   R.M. Noack and S.R. Manmana,
                   AIP Conf. Proc. {\bf 789}, 93 (2005);
                   N. Laflorencie and D. Poilblanc,
                   Lect. Notes Phys. {\bf 645}, 227 (2004).

\bibitem{shell} M. Horoi, V. Zelevinsky, and B.A. Brown,
                Phys. Rev. Lett. {\bf 74}, 5194 (1995).

\bibitem{classchaotic} A.I. Shnirelman,
                       Usp. Mat. Nauk. {\bf 29}, 181 (1974);
                       A. Voros,
                       {\it Stochastic Behavior in Classical and Quantum Hamiltonian
                        Systems}, Springer Verlag, Berlin, 1979;
                       Y.C. de Verdiere,
                       Comm. Math. Phys. {\bf 102}, 497 (1985);
                       S. Zelditch,
                       Duke Math J. {\bf 55}, 919 (1987).

\bibitem{semiclass} E.J. Heller and B.R. Landry,
                    J. Phys. A {\bf 40}, 9259 (2007). 

\bibitem{RigolNature} M. Rigol, V. Dunjko, and M. Olshanii,
                      Nature {\bf 452}, 854 (2008).

\bibitem{RigolSolo} M. Rigol,
                    Phys. Rev. Lett. {\bf 103}, 100403 (2009).

\bibitem{prethermalization} M. Eckstein, M. Kollar, and P. Werner,
                            Phys. Rev. Lett. {\bf 103}, 056403 (2009);
                            Phys. Rev. B 81, 115131 (2010).

\bibitem{relaxationXXZ} P. Barmettler, M. Punk, V. Gritsev, E. Demler, and E. Altman,
                        Phys. Rev. Lett. {\bf 102}, 130603 (2009);
                        arXiv:0911.1972.
                        
\bibitem{BarouchMcCoy} P. Mazur,
                       Physica {\bf 43}, 533 (1969);
                       P. Pfeuty,
                       Ann. of Phys. {\bf 57}, 79 (1970);
                       E. Barouch, B. McCoy, and M. Dresden,
                       Phys. Rev. A {\bf 2}, 1075 (1970);
                       E. Barouch, B. McCoy,
                       Phys. Rev. A {\bf 3}, 786 (1971);
                       Phys. Rev. A {\bf 3}, 2137 (1971).

\bibitem{Hofferberth} S. Hofferberth {\it et al.},
                      Nature {\bf 449}, 324 (2007).

\bibitem{Brioli} G. Brioli, C. Kollath, and A.M. L\"auchli,
                 arXiv:0907.3731.

\bibitem{DynamicalBethe} N. Kitanine, J.M. Maillet, N.A. Slavnov, and V. Terras,
                         Nucl.Phys. B {\bf 729}, 558 (2005); 
                         J.-S. Caux and P. Calabrese,
                         Phys.Rev. A {\bf 74}, 031605 (2006); 
                         J.-S. Caux, P. Calabrese, and N.A. Slavnov,
                         J. Stat. Mech. (2007) P01008; 
                         A. Faribault, P. Calabrese, and J.-S. Caux,
                         J. Stat. Mech. (2007) P01008. 
                         V. Gritsev, T. Rostunov, and E. Demler,
                         J. Stat. Mech. (2010) P05012.

\bibitem{DynamicsFixedPoint} J. Berges, S. Borsnyi, and C. Wetterich,
                             Phys. Rev. Lett. {\bf 93}, 14202 (2004);
                             E. Bettelheim, A.G. Abanov, and P. Wiegman,
                             Phys. Rev. Lett. {\bf 97}, 246402 (2006);
                             A.A. Burkov, M.D. Lukin, and E. Demler,
                             Phys. Rev. Lett. {\bf 98}, 200404 (2007); 
                             V. Gritsev, E. Demler, M. Lukin, and A. Polkovnikov,
                             Phys. Rev. Lett. {\bf 99}, 200404 (2007);
                             A. Iucci and M. Cazalilla, 
                             Phys. Rev. A {\bf 80}, 063619 (2009);
                             J. Sabio and S. Kehrein,
                             New J. Phys. {\bf 12}, 055008 (2010). 

\bibitem{SemiclassicalTheories} A. Polkovnikov, S. Sachdev, and S.M. Girvin,
                                Phys. Rev. A {\bf 66}, 053607 (2002);
                                A. Polkovnikov, 
                                Phys. Rev. A {\bf 68}, 033609 (2003);
                                A. Polkovnikov, 
                                arXiv:0905.3384.

\bibitem{LauchliKollath} A.M. L\"auchli and C. Kollath,
                         J. Stat. Mech. P05018 (2008).

\bibitem{tDMRGappl} D. Gobert, C. Kollath, U. Schollw\"ock, and G. Sch\"utz,
                    Phys. Rev. E {\bf 71}, 036102 (2005);
                    P. Barmettler {\it et al.},
                    Phys. Rev. A {\bf 78}, 012330 (2008);
                    Phys. Rev. Lett. {\bf 102}, 130603 (2009);
                    T. Barthel, C. Kasztelan, I.P. McCulloch, and U. Schollw\"ock,
                    Phys. Rev. A {\bf 79}, 053627 (2009);
                    A. Flesch {\it et al.},
                    Phys. Rev. A {\bf 78}, 033608 (2008).

\bibitem{FalicovKimball} M. Eckstein and M. Kollar,
                         Phys. Rev. Lett. {\bf 100}, 120404 (2008).         

\bibitem{beyondDMRG} S.R. White and I. Affleck,
                     Phys. Rev. B {\bf 77}, 134437 (2008);
                     M.B. Hastings and L.S. Levitov,
                     arXiv:0806.4283;
                     T. Barthel, U. Schollw\"ock, and S.R. White,
                     Phys. Rev. B {\bf 79}, 245101 (2009);
                     M.B. Hastings, 
                     J. Math. Phys. {\bf 50}, 095207 (2009);
                     M.C. Banuls, M.B. Hastings, F. Verstraete, and J.I. Cirac,
                     Phys. Rev. Lett. {\bf 102}, 240603 (2009).

\bibitem{Sd} R. Barankov and A. Polkovnikov,
             arXiv:0806.2862.

\bibitem{heat} A. Polkovnikov,
               Phys. Rev. Lett. {\bf 101}, 220402 (2008).

\bibitem{WorkDone} A. Silva,
                   Phys. Rev. Lett. {\bf 101}, 120603 (2008).

\bibitem{Jarzynski} C. Jarzynski,
                    Phys Rev Lett {\bf 78}, 2690 (1997).

\bibitem{Song} Ying Li, M.X. Huo, and Z. Song,
               Phys. Rev. B {\bf 80}, 054404 (2009).
                                
                                      
\end{thebibliography}
\end{document}